\documentclass[a4paper,onecolumn,11pt]{quantumarticle}
\pdfoutput=1

\usepackage{tikz}
\usetikzlibrary{decorations,decorations.pathmorphing,decorations.markings,calc,snakes,positioning,fixedpointarithmetic,shapes,shapes.geometric,patterns}

\usepackage{etoolbox}
\AtBeginEnvironment{tikzpicture}{\catcode`$=3 } 

\tikzset{alignmid/.style={baseline={([yshift=-.5ex]current bounding box.center)}}} 
\tikzset{every picture/.append style=alignmid}

\tikzset{
bottomzigzag/.style={postaction={draw,decorate, decoration={zigzag,amplitude=1pt,segment length=3pt,raise=1pt}}},
zigzag/.style={draw,decorate, decoration={zigzag,amplitude=1pt,segment length=3pt}},
rc/.style=rounded corners,
}

\tikzset{
    -|/.style={to path={-| (\tikztotarget)}},
    |-/.style={to path={|- (\tikztotarget)}},
}

\usepackage{ifthen}
\usepackage{xparse,pgffor}

\tikzset{
mark/.code={
\tikzset{postaction={/network/mark/.cd,#1,/tikz/.cd,decorate,decoration={name=markings,mark=at position \netmarkpos with{
\begin{scope}[netmarktrafo]
\netmarkcode
\end{scope}
}}}}
\def\netmarkpos{0.5}
},
}

\def\netmarkpos{0.5}
\def\netmarkposoff{0cm}
\def\netmarkcode{}

\tikzset{
netmarktrafo/.style={},
netmarkstyle/.style={solid,semithick,sharp corners},
}

\pgfkeys{
/network/mark/.cd,
sty/.code={
\tikzset{netmarkstyle/.style={#1}}
},
asty/.code={
\tikzset{netmarkstyle/.append style={#1}}
},
f/.style={asty=fill},
p/.code={ 
\def\netmarkpos{#1}
},
poff/.code={ 
\def\netmarkposoff{#1cm}
},
e/.code={ 
\def\netmarkpos{\pgfdecoratedpathlength-0.005cm-\netmarkposoff}

\tikzset{netmarktrafo/.append style={shift={(-\netmarkwidth,0)}}}
},
s/.code={ 
\def\netmarkpos{0.005cm+\netmarkposoff}

\tikzset{netmarktrafo/.append style={shift={(\netmarkwidth,0)},xscale=-1,yscale=-1}}
},
a/.code={ 
\def\netmarkpos{\pgfdecoratedpathlength-0.005cm}

\tikzset{netmarktrafo/.append style={xscale=-1,shift={(-\netmarkwidth,0)}}}
},
b/.code={ 
\def\netmarkpos{0.005cm}

\tikzset{netmarktrafo/.append style={xscale=-1,shift={(\netmarkwidth,0),yscale=-1}}}
},
-/.code={ 
\tikzset{netmarktrafo/.append style={xscale=-1}}
},
r/.code={ 
\tikzset{netmarktrafo/.append style={yscale=-1}}
},
sideoff/.code={ 
\tikzset{netmarktrafo/.append style={shift={(0,#1)}}}
},
slab/.code={ 
\def\netmarkwidth{0}
\def\netmarkcode{
\node[inner sep=0.04cm,netmarkstyle,draw=none] (mylabelwidthtest) at (0,0){\phantom{#1}};
\path let \p1=(mylabelwidthtest.north east), \p2=(mylabelwidthtest.south east), \n1 = {max(abs(\y1),abs(\y2))} in node[inner sep=0.04cm,netmarkstyle] at (0,\n1) {#1};
}
},
lab/.code={ 
\def\netmarkwidth{0}
\def\netmarkcode{
\node[inner sep=0.04cm,anchor=\netmarkanchor] (mylabelwidthtest) at (0,0) {\phantom{#1}};
\draw[white] (mylabelwidthtest.\pgfdecoratedangle)--(mylabelwidthtest.\pgfdecoratedangle+180);
\node[inner sep=0.04cm,anchor=\netmarkanchor,netmarkstyle] at (0,0) {#1};
}
},
rotlab/.code={ 
\def\netmarkwidth{0}
\def\netmarkcode{
\node[inner sep=0.04cm,fill=white,transform shape,rotate=90,anchor=\netmarkrotanchor,netmarkstyle] (mydecorationnodename) at (0,0) {#1};
}
},
mrotlab/.style={ 
rotlab={$\scriptstyle{#1}$}
},
arr/.code={ 
\def\netmarkwidth{0.04}
\def\netmarkcode{\draw[netmarkstyle] (-0.04,0.08)--(0.04,0)--(-0.04,-0.08);}
},
arrbar/.code={ 
\def\netmarkwidth{0.08}
\def\netmarkcode{\draw[netmarkstyle] (-0.08,0.08)--(0,0)--(-0.08,-0.08) (0.04,0.08)--(0.04,-0.08);}
},
rarr/.code={ 
\def\netmarkwidth{0.04}
\def\netmarkcode{\draw[netmarkstyle] (-0.04,-0.08)arc(90-180:90:0.08);}
},
dot/.code={ 
\def\netmarkwidth{0.08}
\def\netmarkcode{\draw[netmarkstyle] (0,0)circle(0.08);}
},
flag/.code={ 
\def\netmarkwidth{0.06}
\def\netmarkcode{\draw[netmarkstyle] (-0.06,0)--(0,0.09)--(0.06,0)--cycle;}
},
darr/.code={ 
\def\netmarkwidth{0.08}
\def\netmarkcode{\draw[netmarkstyle] (-0.04,0)--(0.04,0)--(-0.04,0.08)--cycle;}
},
circ/.code={
\def\netmarkwidth{0.1}
\def\netmarkcode{\draw[netmarkstyle] (0,0) circle (0.1);}
},
hcirc/.code={ 
\def\netmarkwidth{0.1}
\def\netmarkcode{\draw[netmarkstyle] (-0.1,0) arc (180:0:0.1);}
},
diamond/.code={
\def\netmarkwidth{0.1}
\def\netmarkcode{\draw[netmarkstyle] (-0.1,0)--(0,-0.1)--(0.1,0)--(0,0.1)--cycle;}
},
bar/.code={ 
\def\netmarkwidth{0.05}
\def\netmarkcode{
\draw[netmarkstyle] (0,-0.08cm-0.5*\pgflinewidth)--(0,0.08cm+0.5*\pgflinewidth);
}
},
dbar/.code={ 
\def\netmarkwidth{0.13}
\def\netmarkcode{
\draw[netmarkstyle] (-0.04cm,-0.08cm-0.5*\pgflinewidth)--(-0.04cm,0.08cm+0.5*\pgflinewidth) (0.04cm,-0.08cm-0.5*\pgflinewidth)--(0.04cm,0.08cm+0.5*\pgflinewidth);
}
},
hbar/.code={ 
\def\netmarkwidth{0.05}
\def\netmarkcode{
\draw[netmarkstyle] (0, 0.5*\pgflinewidth)--++(0,0.12);
}
},
mill/.code={
\def\netmarkwidth{0.16}
\def\netmarkcode{
\draw[netmarkstyle] (0,-0.5*\pgflinewidth)--++(-0.08,-0.08)--++(0,0.08);
\draw[netmarkstyle] (0,0.5*\pgflinewidth)--++(0.08,0.08)--++(0,-0.08);
}
},
three dots/.code={
\def\netmarkwidth{0.2}
\def\netmarkcode{
\fill (-0.12,0) circle (0.5*0.05) (0,0) circle (0.5*0.05) (0.12,0) circle (0.5*0.05);
}
},
}

\tikzset{wid/.style={minimum width=#1cm}}
\tikzset{hei/.style={minimum height=#1cm}}

\tikzset{sx/.style={xshift=#1cm}}
\tikzset{sy/.style={yshift=#1cm}}

\tikzset{box/.style={draw,rectangle}}
\tikzset{fbox/.style={draw,rectangle, line width=1.1}}
\tikzset{roundbox/.style={draw,rectangle,rounded corners}}
\tikzset{froundbox/.style={draw,rectangle, rounded corners, line width=1.1}}
\tikzset{rounddiamond/.style={draw,diamond,rounded corners}}
\tikzset{dot/.style={draw, shape=circle, fill=black, scale=0.5}}

\tikzset{
netbox/.code={
\node[draw,netbdstyle] (\atomname) at (0,0) {#1};
\coordinate (\atomname-r) at (\atomname.east);
\coordinate (\atomname-l) at (\atomname.west);
\coordinate (\atomname-t) at (\atomname.north);
\coordinate (\atomname-b) at (\atomname.south);
\coordinate (\atomname-tr) at (\atomname.north east);
\coordinate (\atomname-br) at (\atomname.south east);
\coordinate (\atomname-tl) at (\atomname.north west);
\coordinate (\atomname-bl) at (\atomname.south west);
},
}

\tikzset{bdlw/.code={\tikzset{mybdstyle/.style={draw, line width=#1}}}}
\tikzset{bdcol/.code={\tikzset{mybdstyle/.append style={#1}}}}

\newcommand\setelements[1]{
\pgfkeys{/network/atom/.cd,#1}
}

\newcommand\atoms[2]{
\foreach \name/\keys in {#2}{
\expandafter\atom\expandafter{\keys,#1}{\name}
}
}

\newcommand\atom[2]{
\def\atomname{#2}
\tikzset{
nettrafo/.style={},
netatompos/.style={},
netdeco/.style={},
netpostdeco/.style={},
}

\pgfkeys{/network/atom/.cd,#1}

\begin{scope}[netatompos] 
\begin{scope}[nettrafo] 
\netshapecoords 
\fill[netbackstyle] \netshapepath;
\clip \netshapepath;
\tikzset{netdeco}
\draw[netbdstyle] \netshapepath;
\end{scope}
\tikzset{netpostdeco} 
\end{scope}

}

\pgfkeys{
/network/atom/.cd,
square/.code={

\def\netshapepath{(-\tempsize,-\tempsize)rectangle (\tempsize,\tempsize)}
\def\netshapecoords{
\node[rectangle,wid=2*\tempsize,hei=2*\tempsize,inner sep=0,transform shape](\atomname)at(0,0){};
\coordinate(\atomname-c) at (0,0);
\coordinate(\atomname-r) at (\tempsize,0);
\coordinate(\atomname-l) at (-\tempsize,0);
\coordinate(\atomname-t) at (0,\tempsize);
\coordinate(\atomname-b) at (0,-\tempsize);
\coordinate(\atomname-br) at (\tempsize,-\tempsize);
\coordinate(\atomname-tr) at (\tempsize,\tempsize);
\coordinate(\atomname-bl) at (-\tempsize,-\tempsize);
\coordinate(\atomname-tl) at (-\tempsize,\tempsize);
}},
circ/.code={

\def\netshapepath{(0,0)circle(\tempsize)}
\def\netshapecoords{
\node[circle,wid=2*\tempsize,hei=2*\tempsize,inner sep=0,transform shape](\atomname)at(0,0){};
\coordinate(\atomname-c) at (0,0);
\coordinate(\atomname-r) at (\tempsize,0);
\coordinate(\atomname-l) at (-\tempsize,0);
\coordinate(\atomname-t) at (0,\tempsize);
\coordinate(\atomname-b) at (0,-\tempsize);
}},
triang/.code={

\def\netshapepath{(-30:\tempsize)--(90:\tempsize)--(-150:\tempsize)--cycle}
\def\netshapecoords{
\node[regular polygon,regular polygon sides=3,wid=2*\tempsize,inner sep=0,transform shape](\atomname)at(0,0){};
\coordinate(\atomname-c) at (0,0);
\coordinate(\atomname-cr) at (-30:\tempsize);
\coordinate(\atomname-cl) at (-150:\tempsize);
\coordinate(\atomname-ct) at (90:\tempsize);
\coordinate(\atomname-mb) at (-90:0.5*\tempsize);
\coordinate(\atomname-mr) at (30:0.5*\tempsize);
\coordinate(\atomname-ml) at (150:0.5*\tempsize);
}},
diamond/.code={

\def\netshapepath{(0,-\tempsize)--(\tempsize,0)--(0,\tempsize)--(-\tempsize,0)--cycle}
\def\netshapecoords{
\node[rotate=45,rectangle,wid=sqrt(2)*\tempsize,hei=sqrt(2)*\tempsize,inner sep=0,transform shape](\atomname)at(0,0){};
\coordinate(\atomname-c) at (0,0);
\coordinate(\atomname-r) at (\tempsize,0);
\coordinate(\atomname-l) at (-\tempsize,0);
\coordinate(\atomname-t) at (0,\tempsize);
\coordinate(\atomname-b) at (0,-\tempsize);
}},
pentagon/.code={

\def\netshapepath{(-126:\tempsize)--(-54:\tempsize)--(18:\tempsize)--(90:\tempsize)--(162:\tempsize)--cycle}
\def\netshapecoords{
\node[regular polygon,regular polygon sides=5,wid=2*\tempsize,inner sep=0,transform shape](\atomname)at(0,0){};
\coordinate(\atomname-c) at (0,0);
\coordinate (\atomname-mb)at(-90:{\tempsize*cos(36)});
\coordinate (\atomname-mbr)at(-18:{\tempsize*cos(36)});
\coordinate (\atomname-mtr)at(54:{\tempsize*cos(36)});
\coordinate (\atomname-mtl)at(126:{\tempsize*cos(36)});
\coordinate (\atomname-mbl)at(-162:{\tempsize*cos(36)});
\coordinate (\atomname-cbr)at(-54:\tempsize);
\coordinate (\atomname-cr)at(18:\tempsize);
\coordinate (\atomname-ct)at(90:\tempsize);
\coordinate (\atomname-cl)at(162:\tempsize);
\coordinate (\atomname-cbl)at(-126:\tempsize);
}},
halfcirc/.code={

\def\netshapepath{(\tempsize,0)arc(0:180:\tempsize)--++(0,-0.04)-|cycle}
\def\netshapecoords{
\node[circle,wid=2*\tempsize,hei=2*\tempsize,inner sep=0,transform shape](\atomname)at(0,0){};
\coordinate(\atomname-c) at (0,0);
\coordinate(\atomname-r) at (\tempsize,0);
\coordinate(\atomname-l) at (-\tempsize,0);
\coordinate(\atomname-t) at (0,\tempsize);
\coordinate(\atomname-b) at (0,0);
}},
void/.code={
\def\netshapepath{}
\def\netshapecoords{
\coordinate(\atomname) at (0,0);
\coordinate(\atomname-c) at (0,0);
}},
labbox/.code={
\def\netshapepath{(0,0)}
\def\netshapecoords{}
\tikzset{netpostdeco/.append style={netbox=#1}}
},
}

\pgfkeys{
/network/atom/.cd,
p/.code={\tikzset{netatompos/.append style={shift={(#1)}}}},
xscale/.code={\tikzset{nettrafo/.append style={xscale=#1}}},
yscale/.code={\tikzset{nettrafo/.append style={yscale=#1}}},
scale/.code={\tikzset{nettrafo/.append style={scale=#1}}},
rot/.code={\tikzset{nettrafo/.append style={rotate=#1}}},
hflip/.style={xscale=-1},
vflip/.style={yscale=-1},
big/.style={scale=3/2},
small/.style={scale=2/3},
huge/.style={scale=9/4},
tiny/.style={scale=4/9},
flat/.style={yscale=2/3},
fflat/.style={yscale=4/9},
}

\tikzset{
netbdstyle/.style={line width=0.15em}, 
netdecstyle/.style={},
netpostdecstyle/.style={},
netbackstyle/.style={white},
}
\pgfkeys{
/network/atom/.cd,
bdstyle/.code={\tikzset{netbdstyle/.style={#1}}},
bdastyle/.code={\tikzset{netbdstyle/.append style={#1}}},
decstyle/.code={\tikzset{netdecstyle/.style={#1}}},
decastyle/.code={\tikzset{netdecstyle/.append style={#1}}},
postdecstyle/.code={\tikzset{netpostdecstyle/.style={#1}}},
postdecastyle/.code={\tikzset{netpostdecstyle/.append style={#1}}},
backstyle/.code={\tikzset{netbackstyle/.style={#1}}},
backastyle/.code={\tikzset{netbackstyle/.append style={#1}}},
style/.style={bdstyle=#1, decstyle=#1, postdecstyle=#1},
astyle/.style={bdastyle=#1, decastyle=#1, postdecastyle=#1},
whiteblack/.style={decstyle=white,backstyle=black},
nobd/.style={bdstyle={line width=0}},
}

\tikzset{
netbscope/.code={\begin{scope}[#1]},
netescope/.code={\end{scope}},
}

\pgfkeys{
/network/atom/.cd,
bscope/.code={\tikzset{netdeco/.append style={netbscope={#1}}}},
escope/.code={\tikzset{netdeco/.append style={netescope}}},
dec/.style 2 args={bscope={#1},#2,escope}, 
vbar/.style={dec={rotate=90}{hbar}},
dcross/.style={dec={rotate=45}{hbar},dec={rotate=-45}{hbar}},
cross/.style={hbar,vbar},
sect6/.style={sect=60},
sect8/.style={sect=45},
sect10/.style={sect=36},
}

\def\regdec#1{\pgfkeys{/network/atom/.cd,#1/.code={\tikzset{netdeco/.append style={net#1}}}}}
\regdec{all}\regdec{rhalf}\regdec{rquart}\regdec{brquart}\regdec{dot}\regdec{spiral}\regdec{swirl}\regdec{hstripe}\regdec{hbar}\regdec{rrey}
\pgfkeys{/network/atom/.cd,sect/.code={\tikzset{netdeco/.append style={netsect=#1}}}}

\tikzset{
netall/.code={\fill[netdecstyle] (-0.3,-0.3)rectangle (0.3,0.3);}, 
netrhalf/.code={\fill[netdecstyle] (0,-0.3)rectangle (0.3,0.3);}, 
netrquart/.code={\fill[netdecstyle] (0.075,-0.3)rectangle (0.3,0.3);}, 
netbrquart/.code={\fill[netdecstyle] (0,0)rectangle (0.3,-0.3);}, 
netsect/.code={\fill[netdecstyle] (0,0)--(0,-0.3)arc(-90:-90+#1:0.3)--cycle;}, 
netdot/.code={\fill[netdecstyle] (0,0)circle(0.07);}, 
netspiral/.code={\draw[netdecstyle] plot [variable=\t,domain=0:4] ({0.075*\t*cos(pi*(\t-0.5) r)},{0.075*\t*sin(pi*(\t-0.5) r)});}, 
netswirl/.code={\fill[netdecstyle] plot [variable=\t,domain=0:2] ({0.15*\t*cos(pi*(\t-0.5) r)},{0.15*\t*sin(pi*(\t-0.5) r)}) arc(-90:-450:0.3)--cycle;}, 
nethstripe/.code={\fill[netdecstyle] (-0.3,-0.05)rectangle(0.3,0.05);}, 
nethbar/.code={\draw[netdecstyle] (-0.3,0)--(0.3,0);}, 
netrrey/.code={\draw[netdecstyle] (0,0)--(0.3,0);} 
}

\pgfkeys{
/network/atom/.cd,
postbscope/.code={\tikzset{netpostdeco/.append style={netbscope={#1}}}},
postescope/.code={\tikzset{netpostdeco/.append style={netescope}}},
postdec/.style 2 args={postbscope={#1},#2,postescope}, 
arc/.code={\tikzset{netpostdeco/.append style={netarc=#1}}},
lab/.code={\tikzset{netpostdeco/.append style={netlab={#1}}}},
shadecirc/.code={\tikzset{netpostdeco/.append style=netshadecirc}},
shaderect/.code={\tikzset{netpostdeco/.append style={netshaderect={#1}}}},
debug/.code={\tikzset{netpostdeco/.append style=netdebug}},
markline/.code 2 args={\tikzset{netpostdeco/.append style={netmarkline={#1}{#2}}}},
postcirc/.code={\tikzset{netpostdeco/.append style=netpostcirc}},
}

\tikzset{
netlab/.code={
\pgfkeys{/network/atom/lab/.cd,#1}
\node[netpostdecstyle] at (\ifdefined\netlabpos\netlabpos\else\netlabang:\netlabdist\fi) {\netlabwrap{\netlabtext}};
},
netarc/.code args={#1:#2:#3}{
\draw[netpostdecstyle] (#1:#3) arc (#1:#2:#3);
},
netshadecirc/.code= {
\fill[opacity=0.4,netpostdecstyle] (0,0)circle(0.4);
},
netpostcirc/.code= {
\draw[netpostdecstyle] (0,0)circle(0.15);
},
netshaderect/.code= {
\fill[rc,opacity=0.4,netpostdecstyle] ($-1*(#1)$) rectangle (#1);
},
netdebug/.code= {
\node[red] at (0,0){\atomname};
},
netmarkline/.code 2 args= {
\draw (\atomname)edge[mark={#2}]++(#1);
},
}

\def\netlabwrap#1{#1}
\pgfkeys{
/network/atom/lab/.cd,
t/.code={\def\netlabtext{#1}}, 
p/.code={\def\netlabpos{#1}}, 
ang/.code={\def\netlabang{#1}},
dist/.code={\def\netlabdist{#1}},
wrap/.code={\def\netlabwrap##1{#1}}, 
}

\setelements{
vertex/.style={tiny,circ,all},
}
\def\qubita{purple}
\def\qubitb{green}

\usepackage[numbers,sort&compress]{natbib}

\usepackage[utf8]{inputenc}
\usepackage[english]{babel}
\usepackage[T1]{fontenc}
\usepackage{amsmath}
\usepackage{hyperref}
\usepackage{amsmath,amsthm,amsfonts,amssymb}
\usepackage{graphicx}
\usepackage{tikz}
\usetikzlibrary{shapes.geometric} 
\usetikzlibrary{decorations.markings}
\IfFileExists{tikzlibrarybraids.code.tex}{\usetikzlibrary{braids}}{}
\usepackage{physics}

\usepackage{comment}
\usepackage{float}
\usepackage{subfig}
\usepackage{xcolor}
\tikzstyle{midarrow} = [decoration={markings, mark= at position 0.6 with {\arrow{Stealth}}}, postaction={decorate}]
\usepackage{graphicx}
\newcommand{\rrangle}{\rangle\hspace{-0.8mm}\rangle}
\newcommand{\llangle}{\langle\hspace{-0.8mm}\langle}
\usepackage{physics}
\usepackage{amssymb}

\usepackage{mathtools}

\usepackage{color}

\newtheorem{theorem}{Theorem}

\newcommand{\sTr}{\text{s}\Tr}
\newcommand{\CC}{\mathbb{C}}
\newcommand{\g}{\mathbf{g}}
\newcommand{\h}{\mathbf{h}}
\renewcommand{\k}{\mathbf{k}}
\newcommand{\f}{\mathbf{f}}
\newcommand{\B}{\mathcal{B}}
\newcommand{\C}{\mathcal{C}}
\newcommand{\Z}{\mathcal{Z}}
\newcommand{\wg}{\chi}
\newcommand{\hg}{\chi}
\newcommand{\Span}{\mathrm{Span}}
\newcommand{\G}{\mathcal{G}}
\renewcommand{\S}{\mathcal{S}}
\renewcommand{\Vec}{\mathcal{V}ec}
\newcommand{\Rep}{\text{Rep}}
\newcommand{\mycomment}[1]{}
\newcommand{\wh}{\widehat}
\newcommand{\wt}{\widetilde}
\newcommand{\sVec}{\text{s}\mathcal{V}\text{ec}}
\newcommand{\nnb}{\nonumber}
\renewcommand{\Tr}{\mathrm{Tr}}
\newcommand{\Bos}{\text{Bos}}
\newcommand{\zz}{\mathbb{Z}}
\newcommand{\ovl}{\overline}
\newcommand{\SWAP}{\text{SWAP}}
\definecolor{FGreen}{RGB}{58,158,92}
\definecolor{lightred}{RGB}{223, 123, 84}
\definecolor{lightbrown}{RGB}{242,216,199}
\definecolor{invisiblegray}{RGB}{238,238,237}
\definecolor{lightgreen}{RGB}{222, 235, 218}
\definecolor{lightcyan}{RGB}{153, 191, 216}
\definecolor{lightcyan2}{RGB}{208, 224, 235}
\definecolor{cyan2}{RGB}{144,201,226}
\definecolor{lightyellow}{RGB}{252, 238, 183}
\definecolor{OrangeRed}{RGB}{223,123,84}
\definecolor{CyanGreen}{RGB}{81,177,184}
\definecolor{rose}{RGB}{226,45,139}
\definecolor{Ycolor}{RGB}{248,236,104}
\definecolor{Hcolor}{RGB}{255,255,0}
\definecolor{Xcolor}{RGB}{255,136,136}
\definecolor{Zcolor}{RGB}{204,255,204}
\definecolor{Lred}{RGB}{251,211,210}
\definecolor{Lgreen}{RGB}{204,255,204}
\definecolor{Lblue}{RGB}{165,206,253}
\definecolor{tickcolor}{RGB}{181,0,250}
\definecolor{Teal}{RGB}{62,138,141}
\newcommand{\ZXrEmpty}{0.13}
\newcommand{\dashedpattern}{[dash pattern=on 2pt off 1.5pt]}
\newcommand{\dottedpattern}{[dash pattern=on 0.45pt off 0.8pt]}
\newcommand{\midarrow}[1]{[decoration={markings, mark=at position #1 with {\arrow{>}}}, postaction={decorate}]}
\newcommand{\dottedmidarrow}[1]{[decoration={markings, mark=at position #1 with {\arrow{>}}}, postaction={decorate},dash pattern=on 0.45pt off 0.6pt]}

\newcommand{\ToDo}[1]{{\color{rose}[To-Do:\ #1]}}
\newcommand{\YR}[1]{{\color{Teal}[YR:\ #1]}}
\newcommand{\andi}[1]{{\color{blue}[Andi:\ #1]}}
\newcommand{\resub}[1]{#1}

\usepackage{tikz}
\usepackage{lipsum}

\begin{document}

\title{Graphical Calculus for Fermionic Tensors}
\author{Yuanjie Ren}
\affiliation{Department of Physics, Massachusetts Institute of Technology, Cambridge, Massachusetts 02139, USA}
\author{Kaifeng Bu}
\affiliation{Department of Mathematics, The Ohio State University, Columbus, Ohio 43210, USA}
\affiliation{Department of Physics, Harvard University,  Cambridge, Massachusetts 02138, USA}
\author{Andreas Bauer}
\affiliation{Department of Mechanical Engineering, Massachusetts Institute of Technology, Cambridge, MA 02139, USA}
\maketitle

\begin{abstract}
We introduce a graphical calculus, consisting of a set of fermionic tensors with tensor-network equations, which can be used to perform various computations in fermionic many-body physics purely diagrammatically.
The indices of our tensors primarily correspond to fermionic modes, but also include qubits and fixed odd-parity states.
Our graphical calculus extends the ZX calculus for systems involving qubits.
We apply the calculus in order to represent various objects, operations, and computations in physics, including fermionic Gaussian states, the partial trace of Majorana modes, purification protocols, the fermionization isometry, and the construction of fermionic codes.
\end{abstract}
\tableofcontents
\section{Introduction}
Tensor-network methods have been used extensively in theoretical many-body physics~\cite{Verstraete_2008,Bridgeman_2017,gorantla2024tensornetworksnoninvertiblesymmetries,Or_s_2014,Vidal_2008,Swingle_2012,PhysRevLett.69.2863}.
A tensor network is a graph whose vertices are dressed with tensors, and whose edges corresponds to contractions between pairs of indices of these tensors.
While tensor networks may be most well-known for their numerical application (using matrix product states), they also have been successfully applied to analytically solve problems in quantum many-body physics.
Particularly fruitful in the context of quantum information have been diagrammatic calculi for qubit systems, such as the ZX, ZH, and ZW-calculus \cite{hadzihasanovic2015,vilmart2018,Backens_2019,vandewetering2020,Po_r_2023,shaikh2024}. 
These calculi have been useful in quantum information science, such as for $T$-count reduction \cite{Kissinger_2020}, quantum circuit simplification and compilation \cite{Duncan_2020,PhysRevX.10.041030}, lattice surgery \cite{de_Beaudrap_2020}, measurement-based quantum computation \cite{duncan2012}, and spacetime error correction~\cite{Bombin_2024,path_integral_qec,bauer2024xyfloquetcodesimple,Teague2023}

Even though the primary focus of quantum information processing has always been on qubits, fermions and Majorana operators have also received some attention recently \cite{Bravyi_2010,Mudassar_2024, PhysRevLett.133.230603,schuckert2024,aghaee2024interferometricsingleshotparitymeasurement,lyu2024fermionicgaussiantestingnongaussian,lyu2024displacedfermionicgaussianstates}.
For example, Refs.~\cite{Bravyi_2010,Mudassar_2024} delve into the general theory of Majorana codes. 
Refs.~\cite{PhysRevLett.133.230603,schuckert2024,aghaee2024interferometricsingleshotparitymeasurement} outline how the current and future Majorana-based quantum computation can be implemented. 
Furthermore, Refs.~\cite{lyu2024fermionicgaussiantestingnongaussian,lyu2024displacedfermionicgaussianstates} introduce the classical simulation of fermion Gaussian states and a measure of non-Gaussianity for fermions. 
Despite these advances, there is no commonly used and well-developed graphical calculus for native fermions to date.

In this work, we propose such a diagrammatic calculus for fermionic systems, which is inspired by the ZX calculus for qubits.
Our diagrams look the same as tensor-network diagrams for qubits or quantum spin systems, but are interpreted as fermionic tensor-network diagrams \cite{Kraus2009, MERAF3, FermionicPhases, tensor_types, gao2024, mortier2024, Shukla_2020}. Mathematically, these are similar to string diagrams in the symmetric monoidal category of super vector spaces.
We give a set of elementary fermionic tensors and simple diagrammatic rules to manipulate their tensor-network diagrams.
Instead of only describing fermionic modes, our diagrammatic calculus also contains the ZX calculus for qubits, and allows us to describe qubit-fermion interactions.
The elementary degrees of freedom (corresponding to the ``fermionic bond dimensions'' used in our diagrams) consist of a fermionic mode, a qubit, and an odd-parity state or ``fermion'' that is needed to emulate odd-parity tensors.

Instead of completeness of our set of tensors or rules, we focus on the practical applications of our formalism.
{%
Our goal is to provide a practical tool with sound rewrites; we do not give a completeness proof.}

We explicitly show how our diagrammatic calculus naturally captures a variety of important fermionic constructions,
including Gaussian states and operators, odd-Majorana partial-trace channels, purification protocols, or the embedding of a qubit into a pair of fermion modes.
We also show how bosonization in different dimensions can be represented as tensor networks using our elementary tensors.
Finally, we demonstrate how our calculus can be applied to quantum error correcting codes by constructing a fermionic version of the Floquet code \cite{Hastings_2021,Davydova_2023,Kesselring_2024,bauer2025xyfloquetcodesimple,ren2023topologicalquantumcomputationassisted}.

To a certain extent, the ZW calculus can be used to describe fermionic systems, as shown in Refs.~\cite{deFelice2019,Po_r_2023}.
However, the diagrams in the ZW calculus are ordinary tensor-network diagrams, mathematically corresponding to string diagrams in the category of vector spaces, not super-vector spaces.
In order to emulate fermionic exchange statistics, Refs.~\cite{deFelice2019,Po_r_2023} introduce the fermionic braiding as an extra elementary tensor.
In contrast, our formalism is directly based on fermionic tensor networks, where the fermionic exchange statistics are included implicitly.
Our diagrams are fully specified by the list of tensors and contracted index pairs, independent of how we draw these contractions on a 2D piece of paper.
As a consequence, many of the rules in Refs.~\cite{deFelice2019,Po_r_2023} become trivial.
Another different diagrammatic approach to fermion systems is the Quon language \cite{Liu_2017,kang20252dquonlanguageunifying,feng2025quonclassicalsimulationunifying}, which has brought new insights to quantum resource theory.
The main differences from our calculus are that the lines in the Quon language represent Majorana operators instead of fermionic modes, and that the diagrams are embedded inside some 2-dimensional background manifold.
Both the Quon language and the ZW calculus~\cite{deFelice2019,Po_r_2023} assign a global flow of time, which we choose not to in this work. The absence of time flow allows for a more flexible layout of the tensor networks, which is useful, for example, for describing 2D and 3D bosonization.

The paper is structured as follows:
In Section~\ref{sec_z2_tensor_networks}, we introduce general fermionic tensor networks, including $\mathbb{Z}_2$ grading, reordering signs, tensor product, and contraction.
In Section~\ref{sec_fermionic_calculus}, we introduce the elementary degrees of freedom, tensors, and diagrammatic rules of our calculus.
In Section~\ref{sec_examples_and_applications}, we illustrate our calculus at hand of a variety of examples.

\section{Fermionic tensor networks}\label{sec_z2_tensor_networks}
In this section, we introduce fermionic tensor networks in general.

\subsection{Fermionic tensors and their basic operations}
A fermionic tensor \cite{mortier2024, tensor_types, liquid_intro} is specified by the following:
\begin{itemize}
\item A set of indices $I$.
\item For each index $i\in I$, a direction $t_i\in \{+,-\}$, which is either \emph{ingoing} ($t_i=-$) or \emph{outgoing} ($t_i=+$).
\item An ordering $O$ of the indices.
\item For each index $i\in I$, a \emph{even bond dimension} $d_i^{\text{even}}\in \mathbb N_0$, as well as an \emph{odd bond dimension} $d_i^{\text{odd}}\in \mathbb N_0$.
We will sometimes refer to the pair $(d_i^{\text{even}},d_i^{\text{odd}})$ simply as the \emph{bond dimension}, and denote it as $d_i^{\text{even}}|d_i^{\text{odd}}$.
\item A map $T$ that associates to each \emph{index configuration} $\{x_i\}_{i\in I}$ with $0\leq x_i<d_i^{\text{even}}+d_i^{\text{odd}}$ a \emph{tensor entry} $T(\{x_i\}_{i\in I}) = T_{\{x_i\}_{i\in I}}$,
\begin{equation}
\begin{aligned}
T: \bigtimes_{i\in I} \{0,\ldots,d_i^{\text{even}}+d_i^{\text{odd}}-1\}&\rightarrow \mathbb C\;,\\
\vec x &\rightarrow T_{\vec x}\;.
\end{aligned}
\end{equation}
\item For each index configuration $x_i$,
we define its \emph{parity} $|x_i|\in \zz_2\simeq\{0,1\}$ as
\begin{equation}
|x_i|
=\begin{cases}
0&\text{if } 0\leq x_i<d_i^{\text{even}}\\
1&\text{if } d_i^{\text{even}}\leq x_i<d_i^{\text{even}}+d_i^{\text{odd}}
\end{cases}
\end{equation}
The tensor must to have \emph{total even parity}.
That is, the entries of $T$ are non-zero only for index configurations $\{x_i\}_{i\in I}$ such that the sum over all individual parities $|x_i|$ is even:
\begin{equation}
\sum_{i\in I} |x_i|\neq 0 \mod 2 \quad\Rightarrow\quad T_{\{x_i\}}=0\;.
\end{equation}
\end{itemize}
Adopting notation from Ref.~\cite{mortier2024}, we can specify the direction of each index and their ordering by writing a sequence of rounded ket and bra symbols after the tensor entries $T$, such as
\begin{equation}
T_{x_0 x_1 x_2} |x_0) (x_2| \;|x_1)\;.
\end{equation}
The ordering of indices is then reflected in the ordering of the ket and bra symbols.
Ket symbols correspond to outgoing indices, and bra symbols correspond to ingoing indices.
Note that fermionic tensors contain ordinary tensors if we take all indices to have purely even bond dimensions of the form $d|0$.
We will reserve ordinary ket and bra symbols with angle brackets for such purely even indices, including qubits of dimension $2|0$.
Note that each index corresponds to a super vector space, which is commonly denoted by $\mathbb{C}^{d_i^{\text{even}}|d_i^{\text{odd}}}$.
The tensor is simply an element in the even sector of the tensor product of all the super vector spaces, where the tensor product is $\zz_2$-graded.
Instead of ket and bra symbols, it is also common in the literature to use \emph{Grassmann variables} $\theta$,
\begin{equation}
T_{x_0 x_1 x_2} \theta_i^{x_0} \overline\theta_k^{x_2} \;\theta_j^{x_1}\;.
\end{equation}

After defining fermionic tensors, let us introduce their basic operations.
The first operation is the \emph{index transposition}, given by (1) flipping the order of two consecutive indices $i,j$, and (2) multiplying the tensor entries $T$ by $(-1)^{|x_i||x_j|}$, for example,
\begin{equation}
T_{x_0x_1x_2} |x_0) (x_2| \; |x_1) = (-1)^{|x_0||x_2|} \cdot T_{x_0x_1x_2}  (x_2|\; |x_0) |x_1)\;.
\end{equation}
Index transposition is its own inverse, and we commonly regard two tensors related by index transposition as ``the same tensor with respect to a different ordering'', hence the equality sign in the equation above.
We could write this as a symbolic rule,
\begin{equation}
\label{eq:permutation}
|x) |y) = (-1)^{|x||y|}|y) |x)\;,
\end{equation}
and similar for bra instead of ket indices.
This is also the same as the exchange rules of Grassmann variables,
\begin{equation}
\theta_i \theta_j=-\theta_j\theta_i
\quad\Rightarrow\quad
\theta_i^{x_i} \theta_j^{x_j}=(-1)^{|x_i||x_j|} \theta_j^{x_j}\theta_i^{x_i}
\end{equation}
Since we can arbitrarily change the index ordering while accordingly updating the array $T$, we could fix the ordering to be the same as the ordering in which the indices occur in the subscript of the array $T$.
However, the contraction of a fermionic tensor network will require us to change the ordering as a subroutine, which is why we consider different orderings in the first place.

The second operation on fermionic tensors $T_1$ and $T_2$ is their \emph{tensor product}:
It is given by (1) taking the union of the two sets of indices, (2) concatenating the two index orderings, and (3) taking the ordinary tensor product of the tensor entries $T$.
Note that when concatenating the orderings, we may put either the indices of $T_1$ or those of $T_2$ first.
Both are equal, because the total parity of each tensor is equal, and so exchanging all indices of $T_1$ with those of $T_2$ does not give rise to a fermionic reordering sign.

The third operation is the \emph{contraction} of an outgoing index $i\in I$ and an ingoing index $j\in I$ of a tensor $T$.
Such a contraction is well defined only if the fermionic bond dimensions of $i$ and $j$ are equal, $d_i=d_j$.
It can be calculated by the following procedure:
(1) Use the permutation rule in Eq.~\eqref{eq:permutation} to change the ordering such that the two indices are consecutive, in the order $(j||i)$.
(2) Contract the two indices $i$ and $j$ of the array $T$ as usual and remove them from the ordering.
In the proof of $X$-spider fusion in Appendix~\ref{sect_proof_of_soundness}, we will see an explicit implementation of this convention inside a nontrivial graphical rewrite.

The fourth operation is the \emph{blocking} of indices:
Namely, we can take two indices $i$ and $j$ of a fermionic tensor and consider them together as a single index $k$.
The bond dimension of $k$ is then given by the $\zz_2$-graded product
\begin{equation}
d_k^{\text{even}} = d_i^{\text{even}}d_j^{\text{even}} + d_i^{\text{odd}}d_j^{\text{odd}}\;,\quad
d_k^{\text{odd}} = d_i^{\text{even}}d_j^{\text{odd}} + d_i^{\text{odd}}d_j^{\text{even}}\;.
\end{equation}
This corresponds to taking the tensor product of two super vector spaces.
We can only block two indices if they are both ingoing or both outgoing.
In order to block two indices, we (1) permute the orderings such that they are consecutive, and (2) block them as usual.
Note that we choose the ordering of the two indices to be different when blocking ingoing and outgoing indices:
\begin{equation}
|ij)\coloneqq| i)| j),\quad (ij|\coloneqq( j|( i| = (-1)^{|x_i||x_j|} ( i | ( j|.
\end{equation}
This ensures that contracting indices commutes with blocking.

The fifth and last operation on fermionic tensors is the \emph{Hermitian conjugation}.
This is given by (1) swapping each index from ingoing to outgoing and vice versa, (2) inverting the ordering, and (3) taking the complex conjugate of the tensor entries, for example
\begin{equation}
T_{xyz} |x) (z| \;|y)\;
\rightarrow
\ovl{T_{xyz}} (y|\; |z) (x|\;.
\end{equation}

Finally, we note that for every fermionic bond dimension, we can define two special tensors: The identity tensor $\operatorname{id}$, and the fermion parity $P$:
\begin{align}
\operatorname{id}_{ab} = \delta_{a=b} |a)(b|\;,\qquad
P_{ab} = \delta_{a=b} (-1)^a |a)(b| = \delta_{a=b} (b| |a)\;.
\end{align}

\subsection{Fermionic tensor networks and their diagrams}

After introducing fermionic tensors, let us discuss fermionic tensor networks.
A (fermionic) tensor network is a computation that takes a set of (fermionic) tensors as input and computes one resulting tensor by (1) making copies of the input tensors, (2) taking the tensor product of all the copies, and (3) contracting some index pairs of the resulting tensor.
A (fermionic) tensor network can be represented by a \emph{tensor-network diagram} as follows.
Each input tensor is represented by some shape, and each of its indices is represented by a line sticking out from the shape, such as
\begin{align}
        \begin{tikzpicture}[baseline={(current bounding box.center)}]
    \def\r{0.3}
    \foreach \angle in {-20, 60, 140, 220} {
        \draw (0,0) -- ++(\angle:2.6*\r);
    }
    \node at ($(-20:3.6*\r)$) {$1$};
    \node at ($(60:3.6*\r)$) {$2$};
    \node at ($(140:3.6*\r)$) {$3$};
    \node at ($(220:3.5*\r)$) {$4$};
      \filldraw[fill=white, draw=black] (0,0) circle (0.85*\r);
      \node at (0.5*\r, -2*\r) {$\cdots$};
      \node at (0,0) {$T$};
    \end{tikzpicture}
    \end{align}
In order to distinguish the indices, we associate each index with a specific location on the boundary of the shape where the corresponding line starts.
In some cases, however, it can be hard to distinguish the indices like this in the drawing, for example if the shape is a circle as above.
In these cases, we put a purple ``tick'' between two of the indices, which gives us a way to distinguish the indices by going clockwise round the shape, starting at the tick,
\footnote{
Note that the clockwise ordering has nothing to do with the index ordering of the tensor as such:
This index ordering belongs to the internal data of the tensor and is not visible in the tensor network.
Another possibility for distinguishing the indices is to add a different marker (like an arrow, tick, or dot, or number) next to where each index enters the shape.
In this case, the position where the line terminates on the shape does not matter.
}
\begin{align}
\begin{tikzpicture}[baseline={([yshift=-2.5pt]current bounding box.center)}]   
\def\r{0.5}
\def\lw{1}
\def\FontSize{0.9}
    \def\del{0.07}
    \def\xdel{1.5}
    \def\radius{1.5}
    \foreach \angle in {135, 180, 225} {
        \draw
            (0,0) -- ++ (\angle:\radius*\r);
    }
        \foreach \angle in {-45, 0,45 } {
        \draw
            (0,0) -- ++ (\angle:\radius*\r);
    }
     \draw (0,0) -- ++ (-90:1.4*\r);
     \draw[white] (0,0) -- ++ (90:1.4*\r);
    \draw[tickcolor,line width=\lw] (0,0) -- (0,0.8*\r);
    \filldraw[fill=white,line width=\lw] (0,0) circle (0.3*\r);
\end{tikzpicture}
=
\begin{tikzpicture}[baseline={([yshift=-2.5pt]current bounding box.center)}]   
\def\r{0.5}
\def\lw{1}
\def\FontSize{0.9}
    \def\del{0.07}
    \def\xdel{1.5}
    \def\radius{1.5}
    \foreach \angle in {135, 180, 225} {
        \draw
            (0,0) -- ++ (\angle:\radius*\r);
    }
        \foreach \angle in {-45, 0,45 } {
        \draw
            (0,0) -- ++ (\angle:\radius*\r);
    }
     \draw (0,0) -- ++ (-90:1.4*\r);
    \node at ($(-40:1.8*\r)$) {$3$};
    \node at ($(0:1.8*\r)$) {$2$};
    \node at ($(40:1.8*\r)$) {$1$};
    \node at ($(140:1.8*\r)$) {$7$};
    \node at ($(180:1.8*\r)$) {$6$};
    \node at ($(220:1.8*\r)$) {$5$};
     \node at ($(-0*\r,-1.75*\r)$) {$4$};
      \node at ($(-0*\r,1.75*\r)$) {{\color{white}$4$}};
    \filldraw[fill=white,line width=\lw] (0,0) circle (0.3*\r);
\end{tikzpicture}
\end{align}
Note that the ticks as such have nothing to do with the fermionic nature of the tensors, and could be used in the same way for ordinary (e.g. qubit) tensor-network diagrams. It just happens that the elementary $X$ and $Z$-spiders of the qubit $ZX$ calculus are invariant under index permutation, which is why these tensors do not need ticks.

In order to draw a tensor-network diagram, we draw one copy of the shape for each copy of every input tensor.
For every contraction between two indices of two tensor copies, we connect the two lines sticking out of the corresponding shapes.
Furthermore, we use two special shapes for the identity and the fermion parity:
The identity is drawn by simply connecting its input and output via a line (with no shape), and the fermion parity is drawn as a small black dot,
\footnote{Note that neither the identity nor the fermion parity tensor require a tick since its two indices can be distinguished by the fact that one is ingoing and the other is outgoing.}
\begin{align}
\operatorname{id}_{ab}=
\begin{tikzpicture}[baseline={([yshift=-2.5pt]current bounding box.center)}]   
\def\r{0.5}
\def\lw{1}
\def\x{1.6}
  \draw(\x*\r,0)[decoration={markings, mark=at position 0.55 with {\arrow{>}}}, postaction={decorate}]  --(0,0) ;
\end{tikzpicture}
\;,
\quad \quad \quad \quad 
P_{ab}=
\begin{tikzpicture}[baseline={([yshift=-2.5pt]current bounding box.center)}]   
\def\r{0.5}
\def\lw{1}
\def\x{1.1}
  \draw(\x*\r,0)[decoration={markings, mark=at position 0.6 with {\arrow{>}}}, postaction={decorate}]  --(0,0) ;
    \draw(0,0)[decoration={markings, mark=at position 0.6 with {\arrow{>}}}, postaction={decorate}]  --(-\x*\r,0) ;
            \filldraw[fill=black,draw=black](0,0)circle (0.125*\r);
\end{tikzpicture}
\;=
\begin{tikzpicture}[baseline={([yshift=-2.5pt]current bounding box.center)}]   
\def\r{0.5}
\def\lw{1}
\def\x{1.3}
  \draw(\x*\r,0)[decoration={markings, mark=at position 0.6 with {\arrow{>}}}, postaction={decorate}]  --(0.3*\r,0) ;
    \draw(-0.3*\r,0)[decoration={markings, mark=at position 0.6 with {\arrow{>}}}, postaction={decorate}]  --(-\x*\r,0) ;
          \draw[white,line width=\lw] (0,0) -- (0,-0.6*\r);
          \draw[tickcolor,line width=\lw] (0,0) -- (0,0.6*\r);
            \filldraw[fill=Zcolor,draw=black](0,0)circle (0.3*\r);
            \node at (0,0) { $\pi$};
\end{tikzpicture}
\;.\label{eq_fermion_parity_identity}
\end{align}

Note that it does not matter how we draw the diagram, that is, where we draw the tensor shapes, and which path the line connecting two indices takes.
All that matters is which index of which tensor is contracted with which index of which other tensor.
That is, the diagram is fully specified by a list of tensors and a list of index pairs to contract.
For example, it does not matter if indices cross over or under another, if an index crosses itself, or whether an index passes above or below a tensor:
\footnote{This may be unexpected, since if we view the line as the worldline of a fermionic particle, the diagram would correspond to a topological twist yielding a factor of $-1$.}
\begin{align}
\begin{tikzpicture}[baseline={([yshift=-2.5pt]current bounding box.center)}]   
\def\r{0.5}
\def\x{1.7}
\def\ZXr{0.25}
\draw(0,0)--(\x*\r,\x*\r);
\filldraw[fill=white,draw=white](0.5*\x*\r,0.5*\x*\r)circle (0.2*\r);
\draw(0,\x*\r)--(\x*\r,0);
\end{tikzpicture}
\ =\ 
\begin{tikzpicture}[baseline={([yshift=-2.5pt]current bounding box.center)}]   
\def\r{0.5}
\def\x{1.7}
\def\ZXr{0.25}
\draw(0,0)--(\x*\r,\x*\r);
\draw(0,\x*\r)--(\x*\r,0);
\end{tikzpicture}
\ =\ 
\begin{tikzpicture}[baseline={([yshift=-2.5pt]current bounding box.center)}]   
\def\r{0.5}
\def\x{1.7}
\def\ZXr{0.25}
\draw(0,\x*\r)--(\x*\r,0);
\filldraw[fill=white,draw=white](0.5*\x*\r,0.5*\x*\r)circle (0.2*\r);
\draw(0,0)--(\x*\r,\x*\r);
\end{tikzpicture}\;,\text{ or }\quad
\begin{tikzpicture}[baseline={([yshift=-2.5pt]current bounding box.center)}]
\def\r{0.5}
\def\x{1}
\def\h{1.1}
   \newcommand{\halfpart}{ 
\draw (0,0) .. controls (2*\r,0.*\r) and  (\x*\r+0.5*\r,\h*\r) .. (\x*\r,\h*\r);
}
\halfpart
\begin{scope}[shift={(2*\x*\r,0)}]
\begin{scope}[xscale=-1]
\halfpart
\end{scope}
\end{scope}
\end{tikzpicture}
=\ 
\begin{tikzpicture}[baseline={([yshift=-2.5pt]current bounding box.center)}]
\def\r{0.5}
\def\x{1}
\draw(0,0)--(2*\x*\r,0);
\end{tikzpicture}
\;,\text{ or }\quad
\begin{tikzpicture}[baseline={([yshift=-2.5pt]current bounding box.center)}]   
\def\r{0.5}
\def\lw{1}
\def\x{1.5}
\def\X{1.5}
\def\FontSize{0.9}
    \def\h{0.6}
    \def\H{1}
    \def\y{0.6}
     \draw (0,0) .. controls (0,0.2*\r) and (-0.5*\r,\h*\r) .. (-\X*\r,\h*\r);
       \draw(0,0) .. controls (0,-0.2*\r) and (-0.5*\r,-\h*\r) .. (-\X*\r,-\h*\r);
                 \node[rotate=90] at (-0.9*\X*\r,0.05*\r) {$\cdots$};
          \draw (0,0) .. controls (0,0.2*\r) and (0.5*\r,\h*\r) .. (\X*\r,\h*\r);
       \draw (0,0) .. controls (0,-0.2*\r) and (0.5*\r,-\h*\r) .. (\X*\r,-\h*\r);
        \node[rotate=90] at (0.9*\X*\r,0.05*\r) {$\cdots$};
       \draw[white,line width=\lw] (0,0) -- (0,-0.5*1.414*\r);
        \filldraw[fill=gray!50,line width=\lw] (0,0) circle (0.3*\r);
   \newcommand{\halfpart}{
   \draw (0,\y*\r) .. controls (-0.9*\r,\y*\r) and  (-\x*\r+1*\r,-\H*\r) .. (-\x*\r,-\H*\r);
   }
   \halfpart
\begin{scope}[xscale=-1]
\halfpart
\end{scope} 
\end{tikzpicture}
=
\begin{tikzpicture}[baseline={([yshift=-2.5pt]current bounding box.center)}]   
\def\r{0.5}
\def\lw{1}
\def\x{1.3}
\def\X{1.3}
\def\FontSize{0.9}
    \def\h{0.6}
    \def\H{1}
    \def\y{0.6}
     \draw (0,0) .. controls (0,0.2*\r) and (-0.5*\r,\h*\r) .. (-\X*\r,\h*\r);
       \draw (0,0) .. controls (0,-0.2*\r) and (-0.5*\r,-\h*\r) .. (-\X*\r,-\h*\r);
                 \node[rotate=90] at (-0.9*\X*\r,0.05*\r) {$\cdots$};
          \draw (0,0) .. controls (0,0.2*\r) and (0.5*\r,\h*\r) .. (\X*\r,\h*\r);
        \draw (0,0) .. controls (0,-0.2*\r) and (0.5*\r,-\h*\r) .. (\X*\r,-\h*\r);
        \node[rotate=90] at (0.9*\X*\r,0.05*\r) {$\cdots$};
       \draw[white,line width=\lw] (0,0) -- (0,-0.5*1.414*\r);
        \filldraw[fill=gray!50,line width=\lw] (0,0) circle (0.3*\r);
   \draw (\x*\r,-\H*\r) -- (-\x*\r,-\H*\r);
\end{tikzpicture}
\end{align}
All in all, the fermionic tensor network diagrams look just the same as ordinary tensor networks.
The only difference is that for ordinary tensor networks it is not necessary to distinguish between input and output indices.
Recall that the reason why this is needed in the fermionic case is in order to specify the internal ordering of the two contracted indices.

The fact that fermionic tensor networks can be drawn in such a way is not a triviality, but can be attributed to some algebraic relations that the contraction and tensor product fulfil.
For example, the tensor product is commutative and associative, and contractions commute with tensor products and with another.
For more details, we refer the reader to Ref.~\cite{tensor_types}.
Note again that we can only contract an index pair if their fermionic bond dimensions agree, and if one of them is ingoing and the other one is outgoing.
We will often use different line styles (such as solid, dotted, dashed) to distinguish the different fermionic bond dimensions of indices.

\section{Fermionic ZX calculus}\label{sec_fermionic_calculus}
Let us now introduce a concrete set of fermionic bond dimensions, tensors, an tensor-network equations that form our fermionic ZX calculus. 

\subsection{Elementary degrees of freedom}
Before we describe the elementary tensors and their rules, let us discuss the elementary fermionic bond dimensions that the lines in the tensor networks correspond to, and the associated elementary degrees of freedom.
In the standard ZX calculus, all lines correspond to qubits with bond dimension 2.
In our fermionic tensor-network calculus, we will use three different types of degrees of freedom, corresponding to three different fermionic bond dimensions.
\begin{itemize}
\item Most importantly, we use \emph{fermionic modes}, which correspond to the fermionic bond dimension $1|1$, or the super vector space $\mathbb C^{1|1}$.
That is, these degrees of freedom have two states, the state $0$ where the mode is not occupied by fermion, and the state $1$ where the mode is occupied with a fermion.
In a tensor-network diagram, we will represent these bond dimensions as solid lines 
$
\begin{tikzpicture}[baseline={([yshift=-2.5pt]current bounding box.center)}]   
\def\r{0.5}
\def\lw{1}
\def\x{1.6}
  \draw\midarrow{0.55}(\x*\r,0)--(0,0) ;
\end{tikzpicture}$ or $
\begin{tikzpicture}[baseline={([yshift=-2.5pt]current bounding box.center)}]   
\def\r{0.5}
\def\lw{1}
\def\x{1.6}
  \draw(\x*\r,0)--(0,0) ;
\end{tikzpicture}$, where the one without arrow assignment means either direction works and such a situation will occur for lines with an open end in a graphical equation. 
\item We also use qubits, corresponding to the fermionic bond dimension $2|0$ or the super vector space $\mathbb C^2\simeq \mathbb C^{2|0}$
That is, there are two states, both of which are unoccupied with a fermion.
We use these to represent joint fermion-qubit operations, and also as auxiliary bonds to build purely fermionic tensors.
The corresponding indices will be represented as dashed lines $\begin{tikzpicture}[baseline={([yshift=-2.5pt]current bounding box.center)}]   
\def\r{0.5}
\def\lw{1}
\def\x{1.6}
  \draw\dashedpattern(\x*\r,0)--(0,0) ;
\end{tikzpicture}$ in the diagrams.
\item We use \emph{odd-parity states}, which correspond to the fermionic bond dimension $0|1$ or super vector space $\mathbb C^{0|1}$.
These can be thought as fermionic modes that are always occupied by a fermion.
In this sense, an odd-parity state ``is a fermion''.
We draw odd-parity states as dotted lines $\begin{tikzpicture}[baseline={([yshift=-2.5pt]current bounding box.center)}]   
\def\r{0.5}
\def\lw{1}
\def\x{1.6}
  \draw\dottedpattern(\x*\r,0)--(0,0) ;
\end{tikzpicture}$ in the diagrams.
Note that adding an odd-parity state index to a tensor will not change the number of coefficients needed to specify the tensor.

However, it does change the total parity of the remaining tensor indices from even to odd.
So while an index of fermionic bond dimension $1|0$ is truely trivial and can be omitted from the diagram, doing this for the $0|1$ odd-parity lines leads to some subtleties as we discuss later.
\end{itemize}

\subsection{Elementary tensors}
Let us now list the elementary tensors that we use in our calculus.
\begin{itemize}
\item {\bfseries Even $X$-spider}. 
   For every number $m$ of fermion-mode indices and every choice of ingoing or outgoing for each index, we can define the even $X$-spider as the tensor where all index configurations with overall even parity have entry $1$, and all other entries are $0$ (which is automatic by the even-parity constraint of any fermionic tensor):
    \begin{align}
\begin{tikzpicture}[baseline={([yshift=-2.5pt]current bounding box.center)}]   
  \def\r{0.5}
  \def\ticklength{0.5}
  \def\ZXr{0.25}
\def\centralradius{0.06}
\def\lw{1}
\def\FontSize{0.9}
    \def\del{0.07}
    \def\xdel{1.5}
    \def\radius{1.15}
      \draw (0,0) -- ++ (225:\radius*\r);  
     \draw (0,0) -- ++ (0:\radius*\r);  
     \draw\midarrow{0.5} (0:0.67*\radius*\r) -- ++ (0:0.01*\radius*\r);  
       \draw\midarrow{0.5} (225:0.53*\radius*\r) -- ++ (45:0.01*\radius*\r);  
    \foreach \angle in { 180, -45} {
        \draw[decoration={markings, mark=at position 0.75 with {\arrow{>}}}, postaction={decorate}] 
            (0,0) -- ++ (\angle:\radius*\r);
    }
        \foreach \angle in {45,-90 } {
        \draw[decoration={markings, mark=at position 0.75 with {\arrow{<}}}, postaction={decorate}] 
            (0,0) -- ++ (\angle:\radius*\r);
    }
       \draw[tickcolor,line width=\lw] (0*\r,0) -- ++(0,1.3*\ticklength*\r);  
    \node at ($(-40:1.6*\r)$) {$x_2$};
    \node at ($(0:1.6*\r)$) {$x_1$};
    \node at ($(40:1.6*\r)$) {$x_0$};
\node at ($(60:1.45*\r)$) {$\cdots$};
    \node at ($(180:1.6*\r)$) {$x_5$};
    \node at ($(220:1.6*\r)$) {$x_4$};
     \node at ($(-0*\r,-1.55*\r)$) {$x_3$};
     \filldraw[fill=Xcolor,draw=black](0,0)circle (\ZXr*\r); 
\end{tikzpicture}
&=\delta_{\sum_{i=0} x_i=0/2\mathbb{Z}} \bigg[\cdots |x_5)(x_4|(x_3|\ |x_2)|x_1)(x_0|\bigg]\;.
\end{align}

It is important to note that while the $X$-spider for qubits is invariant under index permutations, the fermionic $X$-spider is not:
If we exchange two indices, we get additional sign factors when we match up the orderings of the permuted and the original tensor.
As a consequence, we need to put a purple tick in order to distinguish the indices as discussed in the previous section.
The purple tick serves as a starting point for a canonical counterclockwise index ordering.
{These exchange signs are absent for qubit indices. In particular, the sign in Eq.~\eqref{eq_odd_parity_rewiring} comes from exchanging two odd-parity indices.}
\item {\bfseries Odd $X$-spider}. 
Similarly, we define the odd $X$ spider to be:
\begin{align}
\begin{tikzpicture}[baseline={([yshift=-2.5pt]current bounding box.center)}]   
  \def\r{0.5}
  \def\ticklength{0.5}
  \def\ZXr{0.25}
  \def\centralradius{0.06}
\def\lw{1}
\def\FontSize{0.9}
    \def\del{0.07}
    \def\xdel{1.5}
    \def\radius{1.15}
     \draw\dottedmidarrow{0.3} (45:\radius*\r) -- (0,0);
     \draw (0,0) -- ++ (225:\radius*\r);
       \draw\midarrow{0.5} (225:0.53*\radius*\r) -- ++ (45:0.01*\radius*\r);  
    \foreach \angle in { 180, 0, -45} {
        \draw[decoration={markings, mark=at position 0.75 with {\arrow{>}}}, postaction={decorate}] 
            (0,0) -- ++ (\angle:\radius*\r);
    }
    \foreach \angle in {-90 } {
        \draw[decoration={markings, mark=at position 0.75 with {\arrow{<}}}, postaction={decorate}] 
            (0,0) -- ++ (\angle:\radius*\r);
    }
    \node at ($(-40:1.6*\r)$) {$x_2$};
    \node at ($(0:1.6*\r)$) {$x_1$};
    \node at ($(40:1.6*\r)$) {$x_0$};
\node[rotate=45] at ($(135:1.4*\r)$) {$\cdots$};
    \node at ($(180:1.6*\r)$) {$x_5$};
    \node at ($(220:1.6*\r)$) {$x_4$};
     \node at ($(-0*\r,-1.55*\r)$) {$x_3$};
     \filldraw[fill=Xcolor,draw=black](0,0)circle (\ZXr*\r); 
     \filldraw[fill=black,draw=black](0,0)circle (\centralradius*\r);
\end{tikzpicture}
&=\delta_{\sum_{i=1} x_i=1/2\mathbb{Z}} \bigg[\cdots |x_5)(x_4|(x_3|\ |x_2)|x_1)\bigg]\;.
    \end{align}
{%
The central dot is redundant when the odd-parity line is present. We keep it to make the odd parity more visible and to use the same dotted spider when odd-parity lines are dropped.}
The odd $X$ spider does not need a purple tick as the odd-parity line can be used to define the starting point of the counter-clockwise ordering.
This has the neat side effect that the position of the odd-parity line does not matter:
Swapping a fermion-mode in configuration $a$ with the odd-parity line, and swapping the starting point of the counter-clockwise ordering past $a$ both yield a factor of $(-1)^a$, and these factors cancel.
As we will see below, this implies that we can drop the odd-parity lines of the odd $X$ spiders if we neglect global factors of $-1$.
\item $Z$-{\bfseries spider}. 
Suppose there are even number $m\in 2\mathbb{N}$ of fermionic edges with ingoing/outgoing directions, and an arbitrary number $n\in \mathbb N$ of qubit edges.  We define the $Z$ spider for fixed $x\in \mathbb{F}_2^{m}$ and $y\in \mathbb{F}_2^n$ as follows: 
 \begin{align}
\begin{tikzpicture}[baseline={([yshift=-2.5pt]current bounding box.center)}]   
\def\r{0.5}
\def\h{.6}
\def\w{0.45}
\def\x{1.5}
\def\width{1.2}
\def\lw{1}
\def\radius{10}
\def\ticklength{0.6}
  \def\ZXr{0.28}
    \draw[tickcolor,line width=\lw] (0,0) -- ++(0*\r,0.9*\ticklength*\r);
         \draw\midarrow{0.6} (0,0) .. controls (0,0.2*\r) and (-0.5*\r,\h*\r) .. (-\x*\r,\h*\r);
       \draw\midarrow{0.6} (0,0) .. controls (0,-0.2*\r) and (-0.5*\r,-\h*\r) .. (-\x*\r,-\h*\r);
                 \node[rotate=90] at (-0.925*\x*\r,0.05*\r) {$\cdots$};
          \draw\dashedpattern (0,0) .. controls (0,0.2*\r) and (0.5*\r,\h*\r) .. (\x*\r,\h*\r);
       \draw\dashedpattern (0,0) .. controls (0,-0.2*\r) and (0.5*\r,-\h*\r) .. (\x*\r,-\h*\r);
        \node[rotate=90] at (0.9*\x*\r,0.05*\r) {$\cdots$};
           \node at (-2.05*\r,0) {$x$};
            \node at (1.95*\r,0) {$y$};
            \draw[fill=Zcolor,draw=black] (-\ZXr*\r,-\ZXr*\r) rectangle (\ZXr*\r,\ZXr*\r);
                       \node at (0,0) {$z$};
\end{tikzpicture}
:=(\delta_{x,\vec{0}}\delta_{y,\vec{0}}+z\delta_{x,\vec{1}}\delta_{y,\vec{1}})\cdots|x_2)|x_1)\cdots |y_2\rangle |y_1\rangle ,\quad \forall z\in \mathbb{C}.
 \end{align}
 If $z=e^{i\alpha}$ is a phase, we use a circle instead of a rectangle for the tensor, and use $\alpha$ instead of $z$ as a label:
\begin{align}
\begin{tikzpicture}[baseline={([yshift=-2.5pt]current bounding box.center)}]   
\def\r{0.5}
\def\h{.6}
\def\w{0.45}
\def\x{1.5}
\def\width{1.2}
\def\lw{1}
\def\radius{10}
\def\ticklength{0.6}
  \def\ZXr{0.25}
    \draw[tickcolor,line width=\lw] (0,0) -- ++(0*\r,0.9*\ticklength*\r);
         \draw\midarrow{0.6} (0,0) .. controls (0,0.2*\r) and (-0.5*\r,\h*\r) .. (-\x*\r,\h*\r);
       \draw\midarrow{0.6} (0,0) .. controls (0,-0.2*\r) and (-0.5*\r,-\h*\r) .. (-\x*\r,-\h*\r);
                 \node[rotate=90] at (-0.925*\x*\r,0.05*\r) {$\cdots$};
          \draw\dashedpattern (0,0) .. controls (0,0.2*\r) and (0.5*\r,\h*\r) .. (\x*\r,\h*\r);
       \draw\dashedpattern (0,0) .. controls (0,-0.2*\r) and (0.5*\r,-\h*\r) .. (\x*\r,-\h*\r);
        \node[rotate=90] at (0.9*\x*\r,0.05*\r) {$\cdots$};
   \filldraw[fill=Zcolor,draw=black](0,0)circle (1.2*\ZXr*\r); 
           \node at (0,0) {$_{\alpha}$};
           \node at (-2.05*\r,0) {$x$};
            \node at (1.95*\r,0) {$y$};
\end{tikzpicture}
:=(\delta_{x,\vec{0}}\delta_{y,\vec{0}}+e^{i\alpha}\delta_{x,\vec{1}}\delta_{y,\vec{1}})\cdots |x_2)|x_1)\cdots |y_2\rangle |y_1\rangle .
\end{align}
There is no $Z$ spider for odd numbers $m\in 2\mathbb{N}+1$, since such a $Z$ spider would not have even (nor odd) parity.

\item {\bfseries $W$-tensor}. 
For every $n\in \mathbb{N}$, $x\in \mathbb{F}_2^n$, and $y\in \mathbb{F}_2$, define the \emph{$W$-tensor} as a tensor with $n$ output indices, and one input index  \cite{deFelice2019,Po_r_2023}
\begin{align}
\begin{tikzpicture}[baseline={([yshift=-2.5pt]current bounding box.center)}]
\def\r{0.35}
\def\s{1.5}
\draw[decoration={markings, mark=at position 0.63 with {\arrow{>}}}, postaction={decorate}] (2*\r,0) -- (0.6*\r,0);
\node at (-2.6*\r,0) {$x$};
\node at (2.6*\r,0) {$y$};
 \draw [decoration={markings, mark=at position 0.65 with {\arrow{>}}}, postaction={decorate}](0, 0) .. controls (-1*\r, 1*\r) and (-1.5*\r, 1*\r) .. (-2*\r, 1*\r);
  \draw[decoration={markings, mark=at position 0.65 with {\arrow{>}}}, postaction={decorate}] (0, 0) .. controls (-1*\r,- 1*\r) and (-1.5*\r, -1*\r) .. (-2*\r, -1*\r);
  \node[rotate=90] at (-1.7*\r, 0.05*\r) {$\cdots$};
   \newcommand{\rightangle}{
   \filldraw[fill=white] (0,0) -- (\s*\r,0) -- (0,\s*\r) -- cycle;   
   }
\begin{scope}[shift={(0.6*\r,0*\r)}]
 \begin{scope}[rotate=135]
\rightangle
 \end{scope}
 \end{scope}
\end{tikzpicture}
:=
\delta_{\sum_i\overline{x}_i,y} |x_nx_{n-1}\cdots x_1 )(y|,
\end{align}
where $\overline{x}\in \{0,1\}\subset \mathbb Z$ denotes the reinterpretation of the binary number $x\in \mathbb{F}_2 \simeq \{0,1\}$ as an integer. That is, the tensor entry is $1$ if either all indices are in the $|0)$ configuration, or if the input index is $|1)$ and exactly one output index is $|1)$. For all other configurations, the tensor entry is $0$.
We also define a dual $W$ tensor where input and output indices are swapped, and the definition is the same otherwise.
\item In addition to the tensors above, we also consider the building blocks of the ordinary $ZX$ calculus for qubits. These include the Hadamard gate
$
\begin{tikzpicture}[baseline={([yshift=-2.5pt]current bounding box.center)}]
\def\d{0.33}
\def\r{0.085}
\def\rr{1.5}
\def\lw{1}
\def\ticklength{0.6}
\def\h{0.35}
\def\s{0.0707}
     \draw\dashedpattern (-\h*\rr,0)--(0,0);
     \draw\dashedpattern(\h*\rr,0)--(0,0);
       \filldraw[fill=yellow] (-\s*\rr,-\s*\rr) rectangle (\s*\rr,\s*\rr);
\end{tikzpicture}
$, as well as the X spider
$
\begin{tikzpicture}[baseline={([yshift=-2.5pt]current bounding box.center)}]
\def\d{0.33}
\def\r{0.5}
\def\x{1}
 \def\ZXr{0.25}
 \draw\dashedpattern(-\x*\r,0)--(0,0);
 \draw\dashedpattern(\x*\r,0)--(0,0);
\filldraw[fill=Xcolor,draw=black](0,0)circle (\ZXr*\r);
\end{tikzpicture}$.
Note that the $Z$-spider of the ordinary $ZX$ calculus is a special case of the fermionic $Z$ spider above with $m=0$, i.e., without any fermion-mode but only qubit indices.
However, our fermionic $X$ spider is not a generalization of the qubit $X$ spider.
\end{itemize}
\subsection{Equivalence rules}
In this section we  list elementary rules for the calculus.
{%
A nonzero global prefactor on a pure-state vector, or a positive global prefactor on an unnormalized density matrix, disappears upon normalization. For a POVM or quantum instrument, this applies only to a common positive prefactor for all outcomes; relative outcome weights and prefactors in exact tensor equalities must be kept.}
Note that we sometimes omit the arrow directions (ingoing/outgoing) for the open indices of an equation.
In this case, the equation holds for any choice of arrow directions, as long as the directions match on both sides.
\begin{itemize}
\item {\bfseries Identity}.
\begin{align}
\begin{tikzpicture}[baseline={([yshift=-2.5pt]current bounding box.center)}]   
\def\r{0.5}
\def\ZXr{0.25}
\def\x{2}
\def\lw{1}
\def\ticklength{0.5}
\draw\midarrow{0.55}(\x*\r,0) --(0,0);
\end{tikzpicture}
=
\begin{tikzpicture}[baseline={([yshift=-2.5pt]current bounding box.center)}]   
\def\r{0.5}
\def\ZXr{0.25}
\def\x{1.5}
\def\lw{1}
\def\ticklength{0.53}
\def\y{3}
\draw\midarrow{0.6}(\x*\r,0)--(0,0);
\draw\midarrow{0.55}(\y*\r,0)--(\x*\r,0);
      \draw[white,line width=\lw] (\x*\r,0) -- ++(0,-\ticklength*\r);
      \draw[tickcolor,line width=\lw] (\x*\r,0) -- ++(0,\ticklength*\r);
\filldraw[fill=Zcolor,draw=black](\x*\r,0)circle (\ZXr*\r);
\end{tikzpicture}
=
\begin{tikzpicture}[baseline={([yshift=-2.5pt]current bounding box.center)}]   
\def\r{0.5}
\def\ZXr{0.25}
\def\x{1.5}
\def\lw{1}
\def\ticklength{0.5}
\def\y{3}
\draw\midarrow{0.6}(\y*\r,0)--(\x*\r,0);
\draw\midarrow{0.55}(\y*\r+\x*\r,0)--(\y*\r,0);
      \draw[white,line width=\lw] (\x*\r,0) -- ++(0,\ticklength*\r);
      \draw[white,line width=\lw] (\y*\r,0) -- ++(0,-\ticklength*\r);
      \draw[tickcolor,line width=\lw] (\y*\r,0) -- ++(0,\ticklength*\r);
\filldraw[fill=Xcolor,draw=black](\y*\r,0)circle (\ZXr*\r);
\end{tikzpicture}
=
\begin{tikzpicture}[baseline={([yshift=-2.5pt]current bounding box.center)}]   
\def\r{0.5}
\def\x{1.}
\draw\midarrow{0.6}(\x*\r,0)--(-\x*\r,0);
\end{tikzpicture}
\label{eq_fusion_and_odd-parity_line_pair}
\end{align}

\item {\bfseries Odd-parity rewiring}.
\begin{align}
\begin{tikzpicture}[baseline={([yshift=-2.5pt]current bounding box.center)}]   
\def\r{0.5}
\def\x{1.5}
\draw\dottedmidarrow{0.65}(0,0)--(0,\x*\r);
\draw\dottedmidarrow{0.65}(\x*\r,0)--(\x*\r,\x*\r);
\end{tikzpicture}
\ =\ (-1)\ \ 
\begin{tikzpicture}[baseline={([yshift=-2.5pt]current bounding box.center)}]   
\def\r{0.5}
\def\x{1.7}
\draw\dottedmidarrow{0.7}(0,0)--(\x*\r,\x*\r);
\draw\dottedmidarrow{0.7}(\x*\r,0)--(0,\x*\r);
\end{tikzpicture}
=\ 
\begin{tikzpicture}[baseline={([yshift=-2.5pt]current bounding box.center)}]   
\def\r{0.5}
\def\x{1.5}
\draw\dottedmidarrow{0.65}(0,0)--(0,\x*\r);
\draw\dottedmidarrow{0.65}(\x*\r,0)--(\x*\r,\x*\r);
\end{tikzpicture}
\ \simeq\ 
\begin{tikzpicture}[baseline={([yshift=-2.5pt]current bounding box.center)}]   
\def\r{0.5}
\def\x{1.7}
\draw\dottedmidarrow{0.7}(0,0)--(\x*\r,\x*\r);
\draw\dottedmidarrow{0.7}(\x*\r,0)--(0,\x*\r);
\end{tikzpicture}
\label{eq_odd_parity_rewiring}
\end{align}

This rule implies that if we do not care about a global prefactor of $-1$ (which is often justified for diagrams describing quantum mechanics), it doesn't matter how the odd-parity lines are contracted.
In other words, we can simplify our diagrammatic calculus by dropping the odd-parity lines, and we do so at various places.
We do note, however, that there are pedagogical reasons for keeping the odd-parity lines.
(1) Our diagrams are closely related to string diagrams for the symmetric monoidal category of super vectorspaces.
The odd-parity lines do correspond to the object $\mathbb C^{0|1}$ which is not the monoidal unit, so it is natural to represent it by a non-trivial line.
(2) Physically, an odd-parity line connecting two distant places $a$ and $b$ in a diagram representing some process signifies that a fermion does need to be transported between $a$ and $b$.
It is not possible to implement such a process by independent operations at $a$ and $b$, which the dropping of the odd-parity lines would seem to suggest.
{%
For contracted odd-parity indices, the pairing and the orientation within each pair determine the sign; we nevertheless treat these indices on the same footing as the other indices, using the conventions of Section~\ref{sec_z2_tensor_networks}.}
\item {\bfseries Tick rotation}.
\begin{align}

\end{align}
where a pair of solid-dashed line with dots in the middle represents an arbitrary even number of solid lines  plus an arbitrary number of dashed lines around each $Z$-spider.
\item {\bfseries $X$-spider fusion.}
\begin{align}
%
\end{align}
In the global-phase-agnostic case, the two fusion rules involving odd $X$-spiders can be rewritten as
\begin{align}
%
\;.
\label{eq_parity_surrounding_X_spider}
\end{align}
\item {\bfseries FQ triangle annihilation}. 
\begin{align}
\sqrt{2}\ 
%
\;.
\label{eq_bialgebra}
\end{align}
\item {\bfseries ZX-calculus}.  
The rules of the ordinary qubit $ZX$ calculus~\cite{vandewetering2020, gorantla2024tensornetworksnoninvertiblesymmetries} are also part of our fermionic calculus. In addition to the rules below shown, there are a few more moves which are obtained from exchanging $Z$ and $X$-spiders:
\begin{align}

\end{align}
\item {\bfseries $W$-tensor rules.}
Finally, there are a few rules involving the $W$ tensor. Below we adopt the convention that the ticks are on the right-hand side of the two-leg $Z$-spider (a phase gate) along its arrow direction. Because of this fixed convention, we do not put ticks on Z-spider living on unoriented lines because those lines represents two possible directions.
\begin{align}
%
\qquad \text{(WL)}
\label{eq_W_node_part3}
\end{align}
where $\beta:=1/(1+\alpha)$.
In the notation of \cite{Po_r_2023}, the first five displayed rules are the fermionic analogues of $(\mathrm{Pcy})$, $(\mathrm{AD})$, $(\mathrm{Sym})$, $(\mathrm{Aso})$, and $(\mathrm{WW})$. They generate the higher-valent binary $W$-nodes by iteration. The remaining three rules are specific to the graded setting: $(\mathrm{FBZW})$ replaces the qudit $(\mathrm{BZW})$ rule, because a literal $3$-leg fermionic $Z$-spider is not a valid generator here; $(\mathrm{WT})$ rewrites the transpose of $W$ through a yellow box; and $(\mathrm{WL})$ is the basic $W$-loop elimination.
\end{itemize}

\begin{theorem}[Soundness of the rewrite rules]\label{thm_soundness_of_rewrite_rules}
All graphical equations listed in this subsection are sound, namely both sides of each equation represent the same tensor.
\end{theorem}
Proofs for selected rules are collected in Appendix~\ref{sect_proof_of_soundness}.

\section{Examples and applications}\label{sec_examples_and_applications}
\subsection{Simple operators and states}
In this section, we list a few simple states and operations of fermionic systems, and show how to represent them in terms of our graphical calculus.
{%
These states and operations are building blocks for fermionic quantum circuits. Their graphical representations let us use the same rewrite rules for operators and states.}

\begin{itemize}
\item {\bfseries Majorana operators}. 
Majorana operators (see Appendix~\ref{sect_clifford_and_majorana}) can be represented using our graphical calculus in a very simple way --- they are just an $X$ tensor with two fermion-mode indices and one odd-parity index:
\begin{align}
\begin{tikzpicture}[baseline={([yshift=-2.5pt]current bounding box.center)},every node/.style={scale=0.9}]
\def\r{0.5}
\def\lw{1}
\def\ZXr{0.25}
\def\x{1.5}
\def\ticklength{0.5}
\draw\dottedpattern(0,0) -- (0,\r);
\draw[white](0,0) -- (0,-\r);
\draw[decoration={markings, mark=at position 0.45 with {\arrow{>}}}, postaction={decorate},line width=0.5pt] (\x*\r,0)--(0,0);
\draw[decoration={markings, mark=at position 0.63 with {\arrow{>}}}, postaction={decorate},line width=0.5pt](0,0) -- (-\x*\r,0);
\filldraw[fill=white, draw=black] (0,0) circle (1.5*\ZXr*\r);
\node[font=\small] at (0,0) {$\gamma_j$};
\end{tikzpicture}
=
\begin{tikzpicture}[baseline={([yshift=-2.5pt]current bounding box.center)}]
\def\r{0.5}
\def\lw{1}
\def\ZXr{0.25}
\def\x{1.2}
\def\xx{2.2}
\def\rela{1.1}
\def\centralradius{0.1}
\draw\dottedpattern(0,0) -- (0,0.8*\r);
\draw[white](0,0) -- (0,-0.8*\r);
\begin{scope}[xscale=-1]
\draw[decoration={markings, mark=at position 0.45 with {\arrow{<}}}, postaction={decorate},line width=0.5pt] (\x*\r,0)--(0,0);
\draw\midarrow{0.45} (-\x*\r,0)--(0,0) ;
\filldraw[fill=Xcolor, draw=black] (0,0) circle (\ZXr*\r);
\filldraw[fill=black,draw=black](0,0) circle (\centralradius*\r);
\end{scope}
\end{tikzpicture}
,\quad \quad \quad 
\begin{tikzpicture}[baseline={([yshift=-2.5pt]current bounding box.center)},every node/.style={scale=0.7}]
\def\r{0.5}
\def\lw{1}
\def\ZXr{0.25}
\def\x{1.5}
\def\ticklength{0.5}
\draw\dottedpattern(0,0) -- (0,\r);
\draw[white](0,0) -- (0,-\r);
\draw[decoration={markings, mark=at position 0.45 with {\arrow{>}}}, postaction={decorate},line width=0.5pt] (\x*\r,0)--(0,0);
\draw[decoration={markings, mark=at position 0.63 with {\arrow{>}}}, postaction={decorate},line width=0.5pt](0,0) -- (-\x*\r,0);
\filldraw[fill=white, draw=black] (0,0) circle (1.5*\ZXr*\r);
\node[font=\small] at (0,0) {$\gamma_j'$};
\end{tikzpicture}
=\ (-1)\ 
\begin{tikzpicture}[baseline={([yshift=-2.5pt]current bounding box.center)}]
\def\r{0.5}
\def\lw{1}
\def\ZXr{0.25}
\def\ticklength{0.5}
\def\x{1.2}
\def\centralradius{0.1}
\draw\dottedpattern(0,0) -- (0,0.8*\r);
\draw[white](0,0) -- (0,-0.8*\r);
\draw[decoration={markings, mark=at position 0.3 with {\arrow{>}}}, postaction={decorate},line width=0.5pt] (1.4*\x*\r,0)--(0,0);
\draw[decoration={markings, mark=at position 0.63 with {\arrow{>}}}, postaction={decorate},line width=0.5pt](0,0) -- (-\x*\r,0);
\filldraw[fill=Xcolor, draw=black] (0,0) circle (\ZXr*\r);
\filldraw[fill=black,draw=black](0,0) circle (\centralradius*\r);
\filldraw[fill=black,draw=black](0.75*\r,0) circle (0.11*\r);
\end{tikzpicture}
\label{eq_majorana_operaotr_def}
\end{align}
In the global-phase-agnostic setting, we can equivalently write
\begin{align*}
\begin{tikzpicture}[baseline={([yshift=-2.5pt]current bounding box.center)},every node/.style={scale=0.9}]
\def\r{0.5}
\def\lw{1}
\def\ZXr{0.25}
\def\x{1.5}
\draw[decoration={markings, mark=at position 0.45 with {\arrow{>}}}, postaction={decorate},line width=0.5pt] (\x*\r,0)--(0,0);
\draw[decoration={markings, mark=at position 0.63 with {\arrow{>}}}, postaction={decorate},line width=0.5pt](0,0) -- (-\x*\r,0);
\filldraw[fill=white, draw=black] (0,0) circle (1.5*\ZXr*\r);
\node[font=\small] at (0,0) {$\gamma_j$};
\end{tikzpicture}
&\simeq
\begin{tikzpicture}[baseline={([yshift=-2.5pt]current bounding box.center)}]
\def\r{0.5}
\def\lw{1}
\def\ZXr{0.25}
\def\x{1.2}
\def\xx{2.2}
\def\rela{1.1}
\def\centralradius{0.1}
\begin{scope}[xscale=-1]
\draw[decoration={markings, mark=at position 0.45 with {\arrow{<}}}, postaction={decorate},line width=0.5pt] (\x*\r,0)--(0,0);
\draw\midarrow{0.45} (-\x*\r,0)--(0,0) ;
\filldraw[fill=Xcolor, draw=black] (0,0) circle (\ZXr*\r);
\filldraw[fill=black,draw=black](0,0) circle (\centralradius*\r);
\end{scope}
\end{tikzpicture}
\quad \quad \quad\quad \quad 
\begin{tikzpicture}[baseline={([yshift=-2.5pt]current bounding box.center)},every node/.style={scale=0.7}]
\def\r{0.5}
\def\lw{1}
\def\ZXr{0.25}
\def\x{1.5}
\draw[decoration={markings, mark=at position 0.45 with {\arrow{>}}}, postaction={decorate},line width=0.5pt] (\x*\r,0)--(0,0);
\draw[decoration={markings, mark=at position 0.63 with {\arrow{>}}}, postaction={decorate},line width=0.5pt](0,0) -- (-\x*\r,0);
\filldraw[fill=white, draw=black] (0,0) circle (1.5*\ZXr*\r);
\node[font=\small] at (0,0) {$\gamma_j'$};
\end{tikzpicture}
\simeq
\begin{tikzpicture}[baseline={([yshift=-2.5pt]current bounding box.center)}]
\def\r{0.5}
\def\lw{1}
\def\ZXr{0.25}
\def\x{1.2}
\def\centralradius{0.1}
\draw[decoration={markings, mark=at position 0.3 with {\arrow{>}}}, postaction={decorate},line width=0.5pt] (1.4*\x*\r,0)--(0,0);
\draw[decoration={markings, mark=at position 0.63 with {\arrow{>}}}, postaction={decorate},line width=0.5pt](0,0) -- (-\x*\r,0);
\filldraw[fill=Xcolor, draw=black] (0,0) circle (\ZXr*\r);
\filldraw[fill=black,draw=black](0,0) circle (\centralradius*\r);
\filldraw[fill=black,draw=black](0.75*\r,0) circle (0.11*\r);
\end{tikzpicture}\;.
\end{align*}
where $\gamma'_j\equiv \gamma_{j+n}$.  One can easily see that this aligns with the usual algebraic relation of the fermion parity operator $P_j\simeq\gamma_j\gamma_j'$.
    \item {\bfseries Fermionic $\ket 1$ state.} The $\ket 1$ state of a fermionic mode is simply an $X$ tensor with a single fermion-mode index, and one odd-parity index:
    $\begin{tikzpicture}[baseline={([yshift=-2.5pt]current bounding box.center)}]  
\def\r{0.5}
  \def\ZXr{0.25}
  \def\centralradius{0.08}
  \draw\midarrow{0.65}(0,0)--(-1.25*\r,0);
  \draw\dottedpattern(0,0)--(1.25*\r,0);
   \filldraw[fill=Xcolor,draw=black](0,0)circle (1.2*\ZXr*\r); 
   \filldraw[fill=black,draw=black](0,0)circle (\centralradius*\r);
  \end{tikzpicture}=|1)\in \mathbb{C}^{0|1}$. 
  Again, we see that the odd-parity index can be used to effectively represent tensors with odd parity.
    \item {\bfseries Creation and annihilation operators.}
    The creation and annihilation operators can be expressed easily in terms or graphical calculus.
    They are given by combining the $W$ tensor with the fermionic $\ket 1$ state:
    \begin{align}
\begin{tikzpicture}[baseline={([yshift=-2.5pt]current bounding box.center)},every node/.style={scale=0.9}]   
\def\r{0.63}
  \def\ZXr{0.25}
  \def\x{0.8}
  \draw(0,-\x*\r)--(0,\x*\r);
  \draw\dottedpattern(0,0)--(-\x*\r,0);
   \draw[fill=white] (-\ZXr*\r,-\ZXr*\r) rectangle (\ZXr*\r,\ZXr*\r);
   \node at (0,0) {$a$};
\end{tikzpicture}
=
\begin{tikzpicture}[baseline={([yshift=-2.5pt]current bounding box.center)},every node/.style={scale=0.72}]   
\def\r{0.63}
  \def\s{0.85}
  \def\ZXr{0.2}
  \def\centralradius{0.06}
  \def\x{1}
  \def\delta{0.28}
      \newcommand{\rightangle}{
   \filldraw[fill=white] (0,0) -- (\s*\r,0) -- (0,\s*\r) -- cycle;   
   }
  \draw(0,-1.2*\x*\r)--(0,0);
  \draw(\delta*\r,0)--(\delta*\r,\x*\r);
  \draw\midarrow{0.625}(-\delta*\r,0).. controls (-\delta*\r, 3*\delta*\r) and (-1.*\r+0.1*\r,\delta*\r+0.3*\r) .. (-1.05*\r,\delta*\r);
  \draw\dottedpattern(-1.1*\r,\delta*\r) .. controls (-1.1*\r-0.3*\r,\delta*\r-0.2*\r) and (-1.4\x*\r ,0).. (-1.82*\x*\r,0);
\filldraw[fill=Xcolor,draw=black](-1.1*\r,\delta*\r)circle (\ZXr*\r);
\filldraw[fill=black,draw=black](-1.1*\r,\delta*\r)circle (\centralradius*\r);
  \begin{scope}[shift={(0*\r,-0.5*\r)}]
  \begin{scope}[rotate=45] 
   \rightangle
  \end{scope}
  \end{scope}
\end{tikzpicture}
    \end{align}
    More generally, the $W$-node defines an associative and super-commutative super-algebra over $\mathbb{C}^{1|1}$, generated by $\mathcal{A}=\langle \mathbf{1},a\rangle $ and $a^2=0$, or more explicitly:
\begin{align}
W:\quad &\ket{11}\mapsto 0,\quad \ket{01}\mapsto \ket{1}\quad \ket{10}\mapsto \ket{1}\quad \ket{00}\mapsto \ket{0}\\
&a^2=0\quad\quad\   \mathbf{1}a=a\quad \quad\  a\mathbf{1}=a\quad \quad \ \mathbf{11}=\mathbf{1}.
\end{align}
 The creation and annihilation operators correspond to the regular representations of this algebra.
 \item \textbf{Embedding qubits into fermions}. 
 It is possible to embed a qubit into two fermionic modes, using the relation of fermionic bond dimensions $2|0\subset 2|2 = 1|1\otimes 1|1$, see for example Ref.~\cite{Kitaev_2006}.
 Namely, we can map the two qubit basis states $\ket 0$ and $\ket 1$ onto the two even-parity states $|0)|0)$ and $|1)|1)$ of the two modes.
 In terms of our diagrammatic calculus, this map $\mathcal F_K$ is simply given by the $Z$ spider with two fermion-mode indices and one qubit index:
\begin{align}
    \mathcal{F}=
\begin{tikzpicture}[baseline={([yshift=-2.5pt]current bounding box.center)}]   
\def\r{0.6}
\def\h{0.5}
\def\H{0.3}
\def\x{1}
\def\X{0.6}
\def\lw{1}
\def\ticklength{0.5}
  \def\ZXr{0.17}
     \draw[tickcolor,line width=\lw] (0*\r,0*\r) -- ++(0,1*\ticklength*\r);   
    \draw(0,0) .. controls (0,-0.2*\r) and (0.5*\r,-\h*\r) .. (\x*\r,-\h*\r);
            \draw(0,0) .. controls (0,0.2*\r) and (0.5*\r,\h*\r) .. (\x*\r,\h*\r);
             \draw\dashedpattern(0,0)--(-1.15*\r,0);
 \filldraw[fill=Zcolor,draw=black](0*\r,0*\r)circle (\ZXr); 
\end{tikzpicture}
\label{eq_kitaev_map}
\end{align}
 This map transforms Pauli operators on the qubit into Majorana operators on the pair of fermion modes, for example:
 \begin{align}
\begin{tikzpicture}[baseline={([yshift=-2.5pt]current bounding box.center)}]   
\def\r{0.66}
\def\h{0.5}
\def\H{0.3}
\def\x{1}
\def\X{0.6}
\def\lw{1}
\def\ticklength{0.43}
  \def\ZXr{0.22}
     \draw[tickcolor,line width=\lw] (0*\r,0*\r) -- ++(0,1*\ticklength*\r);   
    \draw\midarrow{0.68}(0,0) .. controls (0,-0.2*\r) and (0.5*\r,-\h*\r) .. (\x*\r,-\h*\r);
            \draw\midarrow{0.68}(0,0) .. controls (0,0.2*\r) and (0.5*\r,\h*\r) .. (\x*\r,\h*\r);
             \draw\dashedpattern(-0.5*\ZXr,0)--(-2.13*\r,0);
    \filldraw[fill=Xcolor,draw=black](-1.1*\r,0*\r)circle (1*\ZXr*\r); 
     \node at (-1.1*\r,0) {$_\pi$};
 \filldraw[fill=Zcolor,draw=black](0*\r,0*\r)circle (\ZXr*\r); 
\end{tikzpicture}
=
\begin{tikzpicture}[baseline={([yshift=-2.5pt]current bounding box.center)}]   
\def\r{0.66}
\def\h{0.5}
\def\H{0.3}
\def\x{1}
\def\X{0.6}
\def\lw{1}
\def\rr{0.2}
\def\ticklength{0.43}
\def\centralradius{0.06}
  \def\ZXr{0.22}
     \draw\dashedpattern(-\ZXr,0)--(-1.16*\r,0);
     \draw[tickcolor,line width=\lw] (0*\r,0*\r) -- ++(0,1*\ticklength*\r);   
    \draw\midarrow{0.6}(0,0) .. controls (0,-0.2*\r) and (0.5*\r,-\h*\r) .. (\x*\r,-\h*\r);
\draw\midarrow{0.6}(0,0) .. controls (0,0.2*\r) and (0.5*\r,\h*\r) .. (\x*\r,\h*\r);
  \draw\midarrow{0.7}(\x*\r,\h*\r)--++(0.75*\r,0);
  \draw\midarrow{0.7}(\x*\r,-\h*\r)--++(0.75*\r,0);
 \filldraw[fill=Zcolor,draw=black](0*\r,0*\r)circle (\ZXr*\r); 
 \draw\dottedpattern (\x*\r,\h*\r) -- (\x*\r,-\h*\r);
  \filldraw[fill=Xcolor,draw=black](\x*\r,\h*\r)circle (\ZXr*\r); 
  \filldraw[fill=Xcolor,draw=black](\x*\r,-\h*\r)circle (\ZXr*\r); 
  \filldraw[fill=black,draw=black](\x*\r,\h*\r)circle (\centralradius*\r);
  \filldraw[fill=black,draw=black](\x*\r,-\h*\r)circle (\centralradius*\r);
\end{tikzpicture}
\quad \quad \quad 
\begin{tikzpicture}[baseline={([yshift=-2.5pt]current bounding box.center)}]   
\def\r{0.66}
\def\h{0.5}
\def\H{0.3}
\def\x{1}
\def\X{0.6}
\def\lw{1}
\def\ticklength{0.43}
  \def\ZXr{0.22}
     \draw[tickcolor,line width=\lw] (0*\r,0*\r) -- ++(0,1*\ticklength*\r);   
    \draw\midarrow{0.68}(0,0) .. controls (0,-0.2*\r) and (0.5*\r,-\h*\r) .. (\x*\r,-\h*\r);
            \draw\midarrow{0.68}(0,0) .. controls (0,0.2*\r) and (0.5*\r,\h*\r) .. (\x*\r,\h*\r);
             \draw\dashedpattern(-0.5*\ZXr,0)--(-2.13*\r,0);
    \filldraw[fill=Zcolor,draw=black](-1.1*\r,0*\r)circle (1*\ZXr*\r); 
     \node at (-1.1*\r,0) {$_\pi$};
 \filldraw[fill=Zcolor,draw=black](0*\r,0*\r)circle (\ZXr*\r); 
\end{tikzpicture}
=
\begin{tikzpicture}[baseline={([yshift=-2.5pt]current bounding box.center)}]   
\def\r{0.66}
\def\h{0.5}
\def\H{0.3}
\def\x{1}
\def\X{0.6}
\def\lw{1}
\def\rr{0.2}
\def\ticklength{0.43}
  \def\ZXr{0.22}
     \draw\dashedpattern(-\ZXr,0)--(-1.13*\r,0);
     \draw[tickcolor,line width=\lw] (0*\r,0*\r) -- ++(0,1*\ticklength*\r);   
  \draw(0,0) .. controls (0,-0.2*\r) and (0.5*\r,-\h*\r) .. (\x*\r,-\h*\r);
\draw(0,0) .. controls (0,0.2*\r) and (0.5*\r,\h*\r) .. (\x*\r,\h*\r);
  \filldraw[fill=black,draw=black] (0.9*\x*\r,0.97*\h*\r) circle (0.09*\r);
  \draw\midarrow{0.55}(\x*\r,\h*\r)--++(0.75*\r,0);
  \draw\midarrow{0.55}(\x*\r,-\h*\r)--++(0.75*\r,0);
 \filldraw[fill=Zcolor,draw=black](0*\r,0*\r)circle (\ZXr*\r); 
\end{tikzpicture}
=
\begin{tikzpicture}[baseline={([yshift=-2.5pt]current bounding box.center)}]   
\def\r{0.66}
\def\h{0.5}
\def\H{0.3}
\def\x{1}
\def\X{0.6}
\def\lw{1}
\def\rr{0.2}
\def\ticklength{0.43}
  \def\ZXr{0.22}
     \draw\dashedpattern(-\ZXr,0)--(-1.13*\r,0);
     \draw[tickcolor,line width=\lw] (0*\r,0*\r) -- ++(0,1*\ticklength*\r);   
  \draw(0,0) .. controls (0,-0.2*\r) and (0.5*\r,-\h*\r) .. (\x*\r,-\h*\r);
\draw(0,0) .. controls (0,0.2*\r) and (0.5*\r,\h*\r) .. (\x*\r,\h*\r);
  \filldraw[fill=black,draw=black] (0.9*\x*\r,-0.97*\h*\r) circle (0.09*\r);
  \draw\midarrow{0.55}(\x*\r,\h*\r)--++(0.75*\r,0);
  \draw\midarrow{0.55}(\x*\r,-\h*\r)--++(0.75*\r,0);
 \filldraw[fill=Zcolor,draw=black](0*\r,0*\r)circle (\ZXr*\r); 
\end{tikzpicture}
 \end{align}
 where Eq.~\eqref{eq_beta_mpi} is used in the first equation and the $Z$-spider fusion is used in the second one.
    \item \textbf{Kitaev chain ground state}.
    {%
On a ring, a spin structure specifies periodic or antiperiodic boundary conditions for the fermions. The bounding spin structure is the antiperiodic choice~{\cite{Aasen_2019}}.}
    Consider the Kitaev chain \cite{A_Yu_Kitaev_2001} at its topologically non-trival fixed point with bounding spin structure\footnote{{On this ring, a spin structure is combinatorially represented as a $\mathbb Z_2$-valued 0-cycle on the 1D cellulation, and there are two homology classes. In our convention, each point marks a sign flip of the fermion: the bounding spin structure corresponds to an odd number of points and can be represented by a single point, while the non-bounding spin structure corresponds to an even number of points and can be represented by the empty set. The position of the single point in the bounding spin structure is the position of the tick.}}, which is defined {on a ring of $n$ fermionic modes, with Hamiltonian}
    \begin{align}
        H={i}\sum_{0\leq j<n} \gamma_{j+1}\gamma_j'\;.
    \end{align}
    {Here, the convention $\gamma_n=-\gamma_0$ implements the bounding spin structure.}
{%
With this sign, the interior terms favor the $+1$ eigenspace of $X_{j+1}X_j$ under the Jordan--Wigner transformation.}
    Its ground state is simply an $X$ spider with $n$ indices.
    This can be easily seen by looking at the Jordan-Wigner transformation given by Eq.~\eqref{eq_DJW} {$D_{JW}:(i\gamma_{j+1}\gamma_j')\mapsto -X_{j+1}X_j$ for an interior bond}.
    The position of the tick pointing downward is related to the combinatorial representation of the spin structure. The local ground state projector, {$\frac12(1+i\gamma_j'\gamma_{j+1})$}, can be represented diagrammatically using two fermionic $X$ spiders:
\begin{align}
{\frac12}\,
\begin{tikzpicture}[baseline={([yshift=-2.5pt]current bounding box.center)}]   
\def\r{0.55}
\def\h{.5}
\def\w{1}
\def\x{1.4}
\def\y{1.1}
\def\Y{1.1}
\def\width{1.2}
\def\lw{1}
\def\ticklength{0.5}
\def\ZXr{0.25}
\draw[tickcolor,line width=\lw] (0,0) -- ++(-\ticklength*\r,0);
\draw[tickcolor,line width=\lw] (\x*\r,0) -- ++(\ticklength*\r,0);
\draw\midarrow{0.7}(0,0)--(0,\y*\r);
\draw\midarrow{0.35}(0,-\Y*\r)--(0,0);
\draw\midarrow{0.7}(\x*\r,0)--(\x*\r,\y*\r);
\draw\midarrow{0.35}(\x*\r,-\Y*\r)--(\x*\r,0);
\draw\midarrow{0.55}(0,0)--(\x*\r,0);
\filldraw[fill=Xcolor,draw=black](0*\r,0*\r)circle (\ZXr*\r);
\filldraw[fill=Xcolor,draw=black](\x*\r,0*\r)circle (\ZXr*\r);
 \filldraw[fill=black,draw=black](0,-0.47*\Y*\r)circle (0.07*\r);
\node[left,inner sep=2pt] at (0,0.8*\y*\r) {$u$};
\node[left,inner sep=2pt] at (0,-0.8*\Y*\r) {$x$};
\node[right,inner sep=2pt] at (\x*\r,0.8*\y*\r) {$v$};
\node[right,inner sep=2pt] at (\x*\r,-0.8*\Y*\r) {$y$};
\node[above,inner sep=2pt] at (0.5*\x*\r,0) {$c$};
\end{tikzpicture}\;.
\end{align}
{%
Let $x,u$ be the input and output indices at site $j$ on the left, $y,v$ those at site $j+1$ on the right, and $c\in\mathbb F_2$ the contracted index. In the order $|v)|u)(x|(y|$, the parity dot cancels the sign from reordering the fermionic indices. The normalized diagram therefore has coefficients
\begin{align*}
\frac12\sum_{c\in\mathbb F_2}\delta_{u,x+c}\delta_{v,y+c}
&=\frac12\bigl(\delta_{u,x}\delta_{v,y}+\delta_{u,x+1}\delta_{v,y+1}\bigr).
\end{align*}
All additions in the indices are modulo two. These are the coefficients of $(1+X_{j+1}X_j)/2$, or $(1+i\gamma'_j\gamma_{j+1})/2$ by Eq.~\eqref{eq_DJW}.}
    One can show graphically that the ground state ($X$ spider) is indeed invariant under the local ground-state projectors.
    \begin{align}
{\frac12}\,
\begin{tikzpicture}[baseline={([yshift=-2.5pt]current bounding box.center)}]   
\def\r{0.75}
\def\h{.5}
\def\w{1}
\def\x{1.}
\def\y{0.7}
\def\Y{0.6}
\def\width{1.2}
\def\lw{1}
\def\ticklength{0.4}
\def\ZXr{0.2}
\begin{scope}[shift={(-0.5*\x*\r,0)}]
\draw[tickcolor,line width=\lw] (0,0) -- ++(-\ticklength*\r,0);
\draw[tickcolor,line width=\lw] (\x*\r,0) -- ++(\ticklength*\r,0);
\draw\midarrow{0.8}(0,-\Y*\r)--(0,\y*\r);
\draw\midarrow{0.8}(\x*\r,-\Y*\r)--(\x*\r,\y*\r);
\draw\midarrow{0.55}(0,0)--(\x*\r,0);
\filldraw[fill=Xcolor,draw=black](0*\r,0*\r)circle (\ZXr*\r);
\filldraw[fill=Xcolor,draw=black](\x*\r,0*\r)circle (\ZXr*\r);
 \filldraw[fill=black,draw=black](0,-0.7*\Y*\r)circle (0.06*\r);
    \end{scope}
    \begin{scope}[shift={(0,-\Y*\r-\h*\r)}]
     \draw[tickcolor,line width=\lw] (0,0) -- ++(0*\r,-\ticklength*\r);
           \draw\midarrow{0.9} (0,0) .. controls (-0.2*\r,0) and (-0.5*\x*\r, 0.2*\h*\r) .. (-0.5*\x*\r, \h*\r);
         \draw\midarrow{0.9} (0,0) .. controls (0.2*\r,0) and (0.5*\x*\r,0.2*\h*\r) .. (0.5*\x*\r, \h*\r);
               \node[rotate=90] at (-0.8*\x*\r,-0.1*\r) {$\cdots$};
               \node[rotate=90] at (0.87*\x*\r,-0.1*\r) {$\cdots$};
           \filldraw[fill=Xcolor,draw=black](0*\r,0*\r)circle (\ZXr*\r);
         \end{scope}
\end{tikzpicture}
=
\begin{tikzpicture}[baseline={([yshift=-2.5pt]current bounding box.center)}]   
\def\r{0.75}
\def\ZXr{0.2}
\def\h{.5}
\def\p{1}
\def\y{1.2}
\def\dy{0.58}
\def\delta{0.17}
\def\width{1.2}
\def\lw{1}
\def\ticklength{0.4}
\def\x{1.3}
\def\y{1.2}
       \draw[tickcolor,line width=\lw] (0,0) -- ++(0,-\ticklength*\r);
           \draw\midarrow{0.65} (0,0) .. controls (0,0.2*\r) and (-0.5*\r,\h*\r) .. (-\x*\r,\h*\r);
       \draw\midarrow{0.65} (0,0) .. controls (0,-0.2*\r) and (-0.5*\r,-\h*\r) .. (-\x*\r,-\h*\r);
                 \node[rotate=90] at (-0.925*\x*\r,0.05*\r) {$\cdots$};
          \draw\midarrow{0.65} (0,0) .. controls (0,0.2*\r) and (0.5*\r,\h*\r) .. (\x*\r,\h*\r);
       \draw\midarrow{0.65} (0,0) .. controls (0,-0.2*\r) and (0.5*\r,-\h*\r) .. (\x*\r,-\h*\r);
        \node[rotate=90] at (0.9*\x*\r,0.05*\r) {$\cdots$};
       \draw\midarrow{0.7}(0,0) .. controls (-0.5*\delta*\r,0) and (-1.5*\delta*\r,0.5*\x*\r) .. (-1.5*\delta*\r,\x*\r);
             \draw\midarrow{0.7} (0,0) .. controls (0.5*\delta*\r,0) and (1.5*\delta *\r,0.5*\x*\r) .. (1.5*\delta*\r,\x*\r);
        \filldraw[fill=Xcolor,draw=black](0,0)circle (\ZXr*\r);
\end{tikzpicture}\;,
    \end{align}
    where $\cdots$ represents  some arbitrary number of legs, and hence the action works on any position $j$. All that is needed in the derivation above is the {$X$-spider fusion, tick rotation, and self-contraction rules}. {The self-contraction contributes a factor $2$, cancelling the prefactor $1/2$ and leaving the original $X$ spider.}
\item \textbf{Scattering process}
Consider the scattering process for two Majorana operators used in the Quon language, see Table~I of Ref.~\cite{kang20252dquonlanguageunifying}.
The physical operator for such a scattering process depends on whether the two Majorana operators correspond to the same mode or to two different modes.
If they correspond to the same mode, then the operator is just a 2-index $Z$ spider with a specific phase.
If they belong to two different modes (such as $k$ and $k+1$, then the operator can be represented using two 3-index $X$ spiders which are contracted via a 2-index $Z$ spider with a specific phase:
\begin{align}
    \frac{1+e^{i\theta}}{2}+\frac{1-e^{i\theta}}{2} i\gamma_{k+1}'\gamma_{k}
    =(-1) \frac{1+e^{i\theta}}{2}\ 
\begin{tikzpicture}[baseline={([yshift=-2.5pt]current bounding box.center)}]   
\def\r{0.5}
\def\h{.7}
\def\w{1}
\def\x{2.5}
\def\y{1}
\def\Y{1}
\def\width{1.2}
\def\lw{1}
\def\ticklength{0.6}
\def\ZXr{0.25}
\draw[tickcolor,line width=\lw] (0,0) -- ++(-\ticklength*\r,0);
\draw[tickcolor,line width=\lw] (\x*\r,0) -- ++(-0.707*\ticklength*\r,+0.707*\ticklength*\r);
\draw\midarrow{0.65}(0,0)--(0,\y*\r);
\draw\midarrow{0.5}(0,-\Y*\r)--(0,0);
\draw\midarrow{0.65}(\x*\r,0)--++(0,\y*\r);
\draw\midarrow{0.5}(\x*\r,-\Y*\r)--++(0,\Y*\r);
\draw\midarrow{0.6}(0.5*\x*\r,0)--(0,0);
\draw\midarrow{0.6}(\x*\r,0)--(0.5*\x*\r,0);
 \draw[tickcolor,line width=\lw] (0.5*\x*\r,0) -- ++(0,\ticklength*\r);
 \draw[fill=Zcolor] (-\ZXr*\r+0.5*\x*\r,-\ZXr*\r) rectangle (\ZXr*\r+0.5*\x*\r,\ZXr*\r);
   \node at (0.5*\x*\r,0) {$_{z}$};
\filldraw[fill=Xcolor,draw=black](0*\r,0*\r)circle (\ZXr*\r);
\filldraw[fill=Xcolor,draw=black](\x*\r,0*\r)circle (\ZXr*\r);
\end{tikzpicture},
\quad \quad 
z=i\frac{1-e^{i\theta}}{1+e^{i\theta}}\;.
\label{scattering_process}
\end{align}
\end{itemize}
\subsection{Gaussian operators}
{%
Gaussian states and operators describe free-fermion systems. Expressing them in terms of our elementary tensors allows their contractions and operator products to be computed within the same graphical calculus.}
In this section, we show how Gaussian states or operators (see Appendix~\ref{sec_gaussian_operators}), or general Gaussian tensors~\cite{tensor_types} can be described in terms of our graphical calculus.
We will also show how to diagrammatically perform index contractions over Gaussian tensors, and multiplication of particle-number-conserving Gaussian operators.
The relation between $W$-based graphical calculi and Pfaffian-type tensor contractions is not new. In particular, Ref.~\cite{arxiv230208767} showed that ZW-style graph states can be used to represent Pfaffians and planar perfect matchings, using explicit fermionic swaps, and also proved completeness of that ruleset for the corresponding Pfaffian computations. A related bosonic picture, based on Hafnians, was later formalized in the Fock-$W$ fragment of the Focked-up ZX calculus in Ref.~\cite{shaikh2024}. The point of the present section is therefore not to claim the first graphical description of Pfaffian/Hafnian-type structures, but to show how the Gaussian tensors relevant here can be represented and manipulated inside our natively fermionic calculus, without introducing explicit fermionic swap generators. We also emphasize that, unlike Ref.~\cite{arxiv230208767}, we do not prove a completeness theorem for Gaussian or Pfaffian computations in the present manuscript.
{%
The contraction example below illustrates the idea of reducing a Gaussian network to a standard form: when the required Schur complement is defined, graphical rewrites turn the contracted network into a scalar multiple of $T(\beta)$, with the updated matrix $\beta$ given by a Schur complement. The result is again expressed using the same $W$-tensor construction. This gives a concrete example of how the rules can be used to reduce and compare Gaussian tensor networks.}


For any anti-symmetric complex $n\times n$ matrix $\alpha$, we can define a \emph{Gaussian tensor} $T(\alpha)$ as the following fermionic tensor with $n$ fermion-mode indices: 
\begin{align}
T(\alpha)=\sum_{x\in \mathbb{F}_2^n}\operatorname{Pf}(\alpha|_x)\ (x|,\label{eq_gamma_alpha}
\end{align}
where we have, without loss of generality,  made all arrows of the open edges incoming, corresponding to the fermionic bra vector $(x|$ (see Eq.~\eqref{four_leg_gaussian_tensor}). Making them all outgoing or mixed is an equally good convention. 
In the above expression, $x$ is the length-$n$ bitstring corresponding to the index configuration, and $\alpha|_x$ denotes the $|x|\times |x|$-dimensional submatrix of $\alpha$ which contains only the rows and columns $i$ for which $x_i=1$.
$\operatorname{Pf}(\ )$ denotes the Pfaffian of a matrix. {For an anti-symmetric matrix, the Pfaffian is the signed sum over pairings of its indices of the corresponding products of matrix entries.}
Gaussian pure states, density matrices and unitaries on $n$ modes are examples of Gaussian tensors with $n$, $2n$, and $2n$ indices, respectively.
See Appendix~\ref{sec_gaussian_operators} for more details. 

Gaussian tensors can either represent the state vector of a Gaussian pure state, the matrix of a Gaussian density matrix, or the matrix representing a Gaussian operator.

We can represent a Gaussian tensor as a tensor-network diagram in our graphical calculus as follows:
Each of the $n$ indices is identified with the input index of one of $n$ $W$-tensors.
Each of the $n$ $W$-tensors has $n$ output indices, and each pair of $W$-tensors shares a contracted index pair.
The contraction is not direct, but via a 2-index $Z$ tensor.
The $Z$ tensor connecting the $W$-tensors at indices $i$ and $j$ tensor has an amplitude given by $\alpha_{ij}$ (in case the tick of the $Z$ tensor is located such that it points from $i$ to $j$, otherwise the amplitude is $\alpha_{ji}=-\alpha_{ij}$).
The tensor is formed by the $n$ uncontracted input indices of the $n$ $W$-tensors.
The diagram below shows the representation for $n=4$ with open edges labeled by $\{0,1,2,3\}$.
\begin{align}
\begin{tikzpicture}[baseline={([yshift=-2.5pt]current bounding box.center)},every node/.style={scale=0.72}]
\def\r{0.38}
  \def\ZXr{0.25}
\def\x{4.3}
\def\X{5}
\def\y{5.5}
\def\Y{5.8}
\def\sq{0.6}
\def\s{1.7}
\def\lw{1}
\def\ticklength{1}
  \def\ZXr{0.3}
  \coordinate (P4) at (-0.28*\x*\r,0.28*\x*\r);
\coordinate (P5) at (-0.28*\x*\r,-0.28*\x*\r);
       \draw[tickcolor,line width=\lw] (-0.707*\x*\r,0) -- ++(-\ticklength*\r,0);   
       \draw[tickcolor,line width=\lw] (0.707*\x*\r,0) -- ++(\ticklength*\r,0);  
         \draw[tickcolor,line width=\lw] (0,0.707*\x*\r) -- ++(0,-\ticklength*\r);  
           \draw[tickcolor,line width=\lw] (0,-0.707*\x*\r) -- ++(0,-\ticklength*\r);  
 \draw[tickcolor,line width=\lw] (P4) -- ++(-0.707*\ticklength*\r,-0.707*\ticklength*\r); 
  \draw[tickcolor,line width=\lw] (P5) -- ++(0.707*\ticklength*\r,-0.707*\ticklength*\r); 
  \foreach \i in {0,1,2,3}{
  \coordinate (P\i) at (90*\i+135:\x*\r);
    \node at (90*\i+135:\Y*\r) {$\i$};
  }
\foreach \i in {0,...,2} {
  \pgfmathtruncatemacro{\minj}{\i+1}
  \foreach \j in {\minj,...,3} {
     \coordinate (P\i\j) at ($(P\i)!0.5!(P\j)$);
       \draw\midarrow{0.6}(P\i)--(P\i\j);
       \draw\midarrow{0.6}(P\j)--(P\i\j);
  }
}
      \draw\midarrow{0.6} (P1) --(P5);
      \draw\midarrow{0.6} (P0) --(P4);
    \newcommand{\rightangle}{
   \filldraw[fill=white] (0,0) -- (\s*\r,0) -- (0,\s*\r) -- cycle;   
   }
   \newcommand{\mysquare}{
   \filldraw[fill=Zcolor] (-\sq*\r,-\sq*\r) -- (\sq*\r,-\sq*\r) -- (\sq*\r,\sq*\r)  -- (-\sq*\r,\sq*\r)-- cycle;   
   }
   \foreach \i in {0,...,3}{
  \draw\midarrow{0.3}(135+90*\i:\y*\r)--(135+90*\i:\x*\r);
   \begin{scope}[shift={(135+90*\i:\X*\r)}]\begin{scope}[rotate=-90+90*\i]\rightangle\end{scope}\end{scope}
   }
 \begin{scope}[shift={(P4)}]
 \begin{scope}[rotate=45]
 \mysquare
 \end{scope}
 \end{scope}
  \begin{scope}[shift={(P5)}]
 \begin{scope}[rotate=45]
 \mysquare
 \end{scope}
 \end{scope}
    \foreach \i in {0,1,2,3}{
 \begin{scope}[shift={(90*\i:0.707*\x*\r)}]
 \mysquare
 \end{scope}
}
  \node[rotate=0] at (-0.707*\x*\r,0) {$\alpha_{_{10}}$};
    \node[rotate=0] at (0.707*\x*\r,0) {$\alpha_{_{32}}$};
      \node[rotate=0] at (0,0.707*\x*\r) {$\alpha_{_{30}}$};
      \node[rotate=0] at (0,-0.707*\x*\r) {$\alpha_{_{21}}$};
      \node[rotate=45] at (P5) {$\alpha_{_{31}}$};
      \node[rotate=-45] at (P4) {$\alpha_{_{20}}$};
  \end{tikzpicture}
  =\sum_{x\in \mathbb{Z}_2^n}
  \operatorname{Pf}(\alpha|_x)(x|
  \label{four_leg_gaussian_tensor}
  \end{align}
It is not hard to see that this equation holds:
Consider the entry for a bitstring $x$.
The $W$-tensors at indices $i$ with $x_i=0$ have output fixed to all-$0$.
The $W$ tensors at indices with $x_i=1$ have exactly one output $1$, and the rest $0$.
Since the 2-index $Z$ tensors can only have configurations $00$ or $11$, we only get a non-zero entry for configurations where pairs of $W$ tensors are connected by indices in $1$ configurations.
The evaluation of the tensor-network diagram is thus a sum over all pairings of the indices $i$ with $x_i=1$, just like the Pfaffain of the submatrix $\alpha|_x$.
The signs in the Pfaffian exactly correspond to the fermionic reordering signs in the diagram.

It was shown in Ref.~\cite{tensor_types} that the contraction of two indices of a Gaussian tensor can be computed directly using its covariance matrix:
It is given by (1) adding the symplectic matrix $((0,1)(-1,0))$ to each submatrix corresponding to a contracted index pair, and (2) taking the Schur complement with respect to all rows and columns corresponding to contracted indices.
Recall that the Schur complement of a block matrix is given by
\begin{equation}
\operatorname{Schur}\left(
    \begin{array}{c|c}
      A & B\\
      \hline
      C & D
    \end{array}\right)
    =A-B D^{-1}C\;.
\end{equation}

Let us now show purely diagrammatically that this relation holds, for contracting two indices of a 4-index Gaussian tensor, with covariance matrix
\begin{align}
    \alpha=\mqty(0& -u & -w& -x \\ u & 0 & -y & -z \\   w & y & 0 &-v \\ x& z &v&0)\;.
\end{align}
After contraction, the covariance matrix should be given by
\begin{equation}
    \operatorname{Schur} \mqty(0& -u & -w& -x \\ u & 0 & -y & -z \\   w & y & 0 &-(1+v) \\ x& z &(1+v)&0)
    = \mqty(0 & -u-\frac{xy-wz}{1+v}\\ u+\frac{xy-wz}{1+v}&0)
\end{equation}
Indeed, we find diagrammatically,
\begin{align}
\begin{tikzpicture}[baseline={([yshift=-2.5pt]current bounding box.center)},every node/.style={scale=0.72}]
\def\r{0.36}
  \def\ZXr{0.25}
\def\x{3.7}
\def\X{4.5}
\def\y{5.5}
\def\Y{6}
\def\sq{0.4}
\def\s{1.7}
\def\t{2.25}
\def\lw{1}
\def\ticklength{0.7}
\def\ZXr{0.3}
\coordinate (P4) at (-0.28*\x*\r,0.28*\x*\r);
\coordinate (P5) at (-0.28*\x*\r,-0.28*\x*\r);
\coordinate (P6) at (\X*\r+0.3*\r,0);
       \draw[tickcolor,line width=\lw] (-0.707*\x*\r,0) -- ++(-\ticklength*\r,0);   
       \draw[tickcolor,line width=\lw] (0.707*\x*\r,0) -- ++(\ticklength*\r,0);  
         \draw[tickcolor,line width=\lw] (0,0.707*\x*\r) -- ++(0,-\ticklength*\r);  
           \draw[tickcolor,line width=\lw] (0,-0.707*\x*\r) -- ++(0,-\ticklength*\r);  
 \draw[tickcolor,line width=\lw] (P4) -- ++(-0.707*\ticklength*\r,-0.707*\ticklength*\r); 
  \draw[tickcolor,line width=\lw] (P5) -- ++(0.707*\ticklength*\r,-0.707*\ticklength*\r); 
  \foreach \i in {0,1,2,3}{
  \coordinate (P\i) at (90*\i+135:\x*\r);
    \node at (90*\i+135:\Y*\r) {$\i$};
  }
\foreach \i in {0,...,2} {
  \pgfmathtruncatemacro{\minj}{\i+1}
  \foreach \j in {\minj,...,3} {
     \coordinate (P\i\j) at ($(P\i)!0.5!(P\j)$);
       \draw\midarrow{0.6}(P\i)--(P\i\j);
       \draw\midarrow{0.6}(P\j)--(P\i\j);
  }
}
      \draw\midarrow{0.6} (P1) --(P5);
      \draw\midarrow{0.6} (P0) --(P4);
    \newcommand{\rightangle}{
   \filldraw[fill=white] (0,0) -- (\s*\r,0) -- (0,\s*\r) -- cycle;   
   }
   \newcommand{\mysquare}{
   \filldraw[fill=Zcolor] (-\sq*\r,-\sq*\r) -- (\sq*\r,-\sq*\r) -- (\sq*\r,\sq*\r)  -- (-\sq*\r,\sq*\r)-- cycle;   
   }
      \foreach \i in {0,1}{
     \draw\midarrow{0.35}(135+90*\i:\y*\r)--(135+90*\i:\x*\r);}
      \draw\midarrow{0.45} (P6) .. controls  ($(P6)-(0,0.1*\r)$) and ($(P2)+(\t*\r,-\t*\r)$)..($(P2)+(0.3*\r,-0.3*\r)$);
       \draw\midarrow{0.45} (P6).. controls ($(P6)+(0,0.1*\r)$) and  ($(P3)+(\t*\r,\t*\r)$) .. ($(P3)+(0.3*\r,0.3*\r)$);
      \draw[tickcolor,line width=\lw] (P6) -- ++(-\ticklength*\r,0); 
     \filldraw[fill=Zcolor,draw=black](P6)circle (1.1*\ZXr*\r);
   \foreach \i in {0,...,3}{
   \begin{scope}[shift={(135+90*\i:\X*\r)}]\begin{scope}[rotate=-90+90*\i]\rightangle\end{scope}\end{scope}
   }
 \begin{scope}[shift={(P4)}]
 \begin{scope}[rotate=45]
 \mysquare
 \end{scope}
 \end{scope}
  \begin{scope}[shift={(P5)}]
 \begin{scope}[rotate=45]
 \mysquare
 \end{scope}
 \end{scope}
    \foreach \i in {0,1,2,3}{
 \begin{scope}[shift={(90*\i:0.707*\x*\r)}]
 \mysquare
 \end{scope}
}
  \node[rotate=0] at (-0.707*\x*\r,0) {$u$};
    \node[rotate=0] at (0.707*\x*\r,0) {$v$};
      \node[rotate=0] at (0,0.707*\x*\r) {$x$};
      \node[rotate=0] at (0,-0.707*\x*\r) {$y$};
      \node[rotate=45] at (P5) {$z$};
      \node[rotate=-45] at (P4) {$w$};
 \end{tikzpicture}
 \ 
&=(1+v)\,
\begin{tikzpicture}[baseline={([yshift=-2.5pt]current bounding box.center)},every node/.style={scale=0.72}]
\def\r{0.36}
  \def\ZXr{0.25}
\def\x{3.7}
\def\X{4.5}
\def\y{5.5}
\def\Y{6}
\def\sq{0.4}
\def\s{1.7}
\def\t{2.25}
\def\lw{1}
\def\ticklength{0.7}
\def\ZXr{0.3}
\coordinate (P4) at (-0.28*\x*\r,0.28*\x*\r);
\coordinate (P5) at (-0.28*\x*\r,-0.28*\x*\r);
\coordinate (P6) at (\X*\r+0.3*\r,0);
       \draw[tickcolor,line width=\lw] (-0.707*\x*\r,0) -- ++(-\ticklength*\r,0);   
         \draw[tickcolor,line width=\lw] (0,0.707*\x*\r) -- ++(0,-\ticklength*\r);  
           \draw[tickcolor,line width=\lw] (0,-0.707*\x*\r) -- ++(0,-\ticklength*\r);  
 \draw[tickcolor,line width=\lw] (P4) -- ++(-0.707*\ticklength*\r,-0.707*\ticklength*\r); 
  \draw[tickcolor,line width=\lw] (P5) -- ++(0.707*\ticklength*\r,-0.707*\ticklength*\r); 
  \foreach \i in {0,1,2,3}{
  \coordinate (P\i) at (90*\i+135:\x*\r);
    \node at (90*\i+135:\Y*\r) {$\i$};
  }
\foreach \i in {0,...,2} {
  \pgfmathtruncatemacro{\minj}{\i+1}
  \foreach \j in {\minj,...,3} {
    \ifnum\i=2\relax
      \ifnum\j=3\relax
      \else
        \coordinate (P\i\j) at ($(P\i)!0.5!(P\j)$);
        \draw\midarrow{0.6}(P\i)--(P\i\j);
        \draw\midarrow{0.6}(P\j)--(P\i\j);
      \fi
    \else
      \coordinate (P\i\j) at ($(P\i)!0.5!(P\j)$);
      \draw\midarrow{0.6}(P\i)--(P\i\j);
      \draw\midarrow{0.6}(P\j)--(P\i\j);
    \fi
  }
}
      \draw\midarrow{0.6} (P1) --(P5);
      \draw\midarrow{0.6} (P0) --(P4);
    \newcommand{\rightangle}{
   \filldraw[fill=white] (0,0) -- (\s*\r,0) -- (0,\s*\r) -- cycle;   
   }
   \newcommand{\mysquare}{
   \filldraw[fill=Zcolor] (-\sq*\r,-\sq*\r) -- (\sq*\r,-\sq*\r) -- (\sq*\r,\sq*\r)  -- (-\sq*\r,\sq*\r)-- cycle;   
   }
      \foreach \i in {0,1}{
     \draw\midarrow{0.35}(135+90*\i:\y*\r)--(135+90*\i:\x*\r);}
      \draw\midarrow{0.45}(P6) .. controls  ($(P6)-(0,0.1*\r)$) and ($(P2)+(\t*\r,-\t*\r)$)..($(P2)+(0.3*\r,-0.3*\r)$);
       \draw\midarrow{0.45} (P6).. controls ($(P6)+(0,0.1*\r)$) and ($(P3)+(\t*\r,\t*\r)$)..($(P3)+(0.3*\r,0.3*\r)$);
   \foreach \i in {0,...,3}{
   \begin{scope}[shift={(135+90*\i:\X*\r)}]\begin{scope}[rotate=-90+90*\i]\rightangle\end{scope}\end{scope}
   }
 \begin{scope}[shift={(P4)}]
 \begin{scope}[rotate=45]
 \mysquare
 \end{scope}
 \end{scope}
  \begin{scope}[shift={(P5)}]
 \begin{scope}[rotate=45]
 \mysquare
 \end{scope}
 \end{scope}
  \begin{scope}[shift={(P6)}]
  \draw[tickcolor,line width=\lw] (0,0) -- ++(1.1*\ticklength*\r,0);  
 \mysquare
 \end{scope}
    \foreach \i in {1,2,3}{
 \begin{scope}[shift={(90*\i:0.707*\x*\r)}]
 \mysquare
 \end{scope}
} 
  \node[rotate=0] at (-0.707*\x*\r,0) {$u$};
      \node[rotate=0] at (0,0.707*\x*\r) {$x$};
      \node[rotate=0] at (0,-0.707*\x*\r) {$y$};
      \node[rotate=45] at (P5) {$z$};
      \node[rotate=-45] at (P4) {$w$};
     \node at ($(P6)+(1.4*\r,0)$) {\large $\frac{1}{1+v}$};
\end{tikzpicture}
\\&=(1+v)\,
\begin{tikzpicture}[baseline={([yshift=-2.5pt]current bounding box.center)},every node/.style={scale=0.72}]
\def\r{0.36}
  \def\ZXr{0.25}
\def\x{3.7}
\def\X{4.5}
\def\y{5.5}
\def\Y{6}
\def\sq{0.4}
\def\s{1.7}
\def\t{2.25}
\def\lw{1}
\def\ticklength{0.7}
\def\ZXr{0.3}
\coordinate (P4) at (-0.28*\x*\r,0.28*\x*\r);
\coordinate (P5) at (-0.28*\x*\r,-0.28*\x*\r);
\coordinate (P6) at (\X*\r+0.5*\r,0);
\coordinate (P7) at (+0.3*\x*\r,0.3*\x*\r);
\coordinate (P8) at (+0.3*\x*\r,-0.3*\x*\r);
\coordinate (P9) at (2*\r+0.3*\r,0);
\coordinate (PM) at (0,0);
       \draw[tickcolor,line width=\lw] (-0.707*\x*\r,0) -- ++(-\ticklength*\r,0);   
         \draw[tickcolor,line width=\lw] (0,0.707*\x*\r) -- ++(0,-\ticklength*\r);  
           \draw[tickcolor,line width=\lw] (0,-0.707*\x*\r) -- ++(0,-\ticklength*\r);  
 \draw[tickcolor,line width=\lw] (P4) -- ++(-0.707*\ticklength*\r,-0.707*\ticklength*\r); 
  \draw[tickcolor,line width=\lw] (P5) -- ++(0.707*\ticklength*\r,-0.707*\ticklength*\r); 
  \foreach \i in {0,1,2,3}{
  \coordinate (P\i) at (90*\i+135:\x*\r);
  } 
   \foreach \i in {0,1}{
    \node at (90*\i+135:\Y*\r) {$\i$};
  } 
\foreach \i in {0,...,2} {
  \pgfmathtruncatemacro{\minj}{\i+1}
  \foreach \j in {\minj,...,3} {
      \coordinate (P\i\j) at ($(P\i)!0.5!(P\j)$);
  }
}
\coordinate (P0M) at ($(P0)!0.5!(PM)$);
\coordinate (P1M) at ($(P1)!0.5!(PM)$);
        \draw\midarrow{0.6}(P0)--(P01);
        \draw\midarrow{0.55}(P01)--(P1);
         \draw\midarrow{0.6}(P0)--(P03);
          \draw\midarrow{0.99}(P03)--(P3);
          \draw\midarrow{0.1}(P2)--(P12);
          \draw\midarrow{0.55}(P12)--(P1);
          \draw(PM)--(P5);
           \draw\midarrow{0.1}(P7)--(PM);
           \draw(P4)--(PM);
           \draw\midarrow{0.95}(PM)--(P8);
      \draw\midarrow{0.6}  (P5)--(P1);
      \draw\midarrow{0.6} (P0) --(P4);
    \newcommand{\rightangle}{
   \filldraw[fill=white] (0,0) -- (\s*\r,0) -- (0,\s*\r) -- cycle;   
   }
   \newcommand{\mysquare}{
   \filldraw[fill=Zcolor] (-\sq*\r,-\sq*\r) -- (\sq*\r,-\sq*\r) -- (\sq*\r,\sq*\r)  -- (-\sq*\r,\sq*\r)-- cycle;   
   }
      \draw (P2).. controls ($(P2)+(\t*\r,0)$) and ($(P6)-(0,0.1*\r)$)..(P6);
      \draw(P3).. controls ($(P3)+(\t*\r,0)$) and ($(P6)+(0,0.1*\r)$)..(P6);
      \draw (P7).. controls ($(P7)+(0.38*\t*\r,0.3*\t*\r)$) and ($(P9)+(0,0.6*\r)$)..(P9);
       \draw (P8).. controls ($(P8)+(0.38*\t*\r,-0.3*\t*\r)$) and ($(P9)-(0,0.6*\r)$)..(P9);
      \foreach \i in {0,1}{
     \draw\midarrow{0.35}(135+90*\i:\y*\r)--(135+90*\i:\x*\r);}
   \foreach \i in {0,1}{
   \begin{scope}[shift={(135+90*\i:\X*\r)}]\begin{scope}[rotate=-90+90*\i]\rightangle\end{scope}\end{scope}
   } 
 \begin{scope}[shift={(P4)}]
 \begin{scope}[rotate=45]
 \mysquare
 \end{scope}
 \end{scope}
  \begin{scope}[shift={(P5)}]
 \begin{scope}[rotate=45]
 \mysquare
 \end{scope}
 \end{scope}
  \begin{scope}[shift={(P6)}]
  \draw[tickcolor,line width=\lw] (0,0) -- ++(1.1*\ticklength*\r,0);  
 \mysquare
 \end{scope}
   \begin{scope}[shift={(P9)}]
  \draw[tickcolor,line width=\lw] (0,0) -- ++(1.1*\ticklength*\r,0);  
 \mysquare
 \end{scope}
    \foreach \i in {1,2,3}{
 \begin{scope}[shift={(90*\i:0.707*\x*\r)}]
 \mysquare
 \end{scope}
}
  \node[rotate=0] at (-0.707*\x*\r,0) {$u$};
      \node[rotate=0] at (0,0.707*\x*\r) {$x$};
      \node[rotate=0] at (0,-0.707*\x*\r) {$y$};
      \node[rotate=45] at (P5) {$z$};
      \node[rotate=-45] at (P4) {$w$};
         \node at (0.925*\x*\r+0.3*\r,0) {\large $\frac{1}{1+v}$};
     \node at ($(P6)+(1.5*\r,0)$) {\large $\frac{1}{1+v}$};
\end{tikzpicture}
\end{align}
where in the first step we have used the corrected rule $(\mathrm{WL})$, which contributes the overall scalar factor $(1+v)$ and produces the box label $1/(1+v)$, and in the second step we have done some tracking of the fermion $|1)$ state; pedagogically one can derive the same rightmost graph by applying $(\mathrm{Pcy})$, then $(\mathrm{WW})$, and finally $(\mathrm{Pcy})$ again. 

The rightmost diagram is precisely the Gaussian tensor $T(\beta)$ for the Schur complement $\beta$ of $\alpha$, which is an antisymmetric 2-by-2 matrix with $\beta_{10}=u+(xy-wz)/(1+v)$. Hence the exact contracted tensor is $(1+v)\,T(\beta)$.
As discussed, if we divide the tensor indices of a Gaussian tensors into input and output indices, we can interpret it as a linear operator between the input and output indices.
There is a special class of Gaussian operators which are particle-number conserving.
These operators have a covariance matrix that is block-off-diagonal, that is, the matrix entries among input indices and among output indices are $0$.
In other words, a particle-number conserving Gaussian operator with $n$ input indices and $m$ output indices is given by a $n\times m$ matrix $\alpha_{IO}$:
\begin{align}
    \alpha= \left(
    \begin{array}{c|c}
      0_{II} & \alpha_{IO}\\
      \hline
      -\alpha_{IO}^T & 0_{OO}
    \end{array}
    \right).
\end{align}
For this special form of $\alpha$, the Pfaffian formula of Eq.~\eqref{eq_gamma_alpha} can be simplified to a determinant formula:
\begin{align}
    \Gamma(\alpha)=\sum_{x\in \mathbb{Z}_2^{n+m}} \operatorname{Pf}(\alpha|_x)|x_O)(x_I|=
    \sum_{x_I\in \mathbb{Z}_2^n, x_O\in \mathbb Z_2^m} \det(\alpha_{IO}|_{x_Ix_O})|x_O)(x_I|\;.
\end{align}
Here, $\alpha_{IO}|_{x_Ix_O}$ denotes the matrix $\alpha_{IO}$ restricted to the columns with $x_I=1$ and the rows with $x_O=1$.
If we represent such an operator diagrammatically, then the diagram looks bipartite, with no connection among input or among output indices.
E.g., for $n=m=2$, the diagram looks like
\begin{align}
\begin{tikzpicture}[baseline={([yshift=-2.5pt]current bounding box.center)},every node/.style={scale=0.7}]   
\def\r{0.5}
  \def\ZXr{0.25}
\def\x{5}
\def\y{3.2}
\def\Y{6}
\def\sq{0.4}
\def\Delta{0.4}
\def\adjust{0.1}
\def\Adjust{0.15}
\def\s{1.7}
\def\lw{1}
\def\ticklength{0.7}
  \def\ZXr{0.26}
      \newcommand{\rightangle}{
   \filldraw[fill=white] (0,0) -- (\s*\r,0) -- (0,\s*\r) -- cycle;   
   }
    \coordinate (P1) at (0,0);
      \coordinate (P2) at (0,\y*\r);
     \coordinate (P3) at (\x*\r,-\Delta*\r);
      \coordinate (P4) at (\x*\r,\y*\r+\Delta*\r);
     \draw\midarrow{0.3} (P1)--(P3);
     \draw\midarrow{0.35} (P3)--(P1);
     \draw\midarrow{0.35} (P1)--(P4);
     \draw\midarrow{0.225} (P4)--(P1);
     \draw\midarrow{0.35} (P2)--(P3);
      \draw\midarrow{0.27} (P3)--(P2);
     \draw\midarrow{0.28} (P2)--(P4);
      \draw\midarrow{0.35} (P4)--(P2);
      \node at (-1.3*\r,\adjust*\r) {$2$};
       \draw\midarrow{0.75}(0,\adjust*\r)--(-1*\r,\adjust*\r);
   \begin{scope}[shift={(-\Delta*\r,\adjust*\r)}]
  \begin{scope}[rotate=-45] 
   \rightangle
  \end{scope}
  \end{scope}
  \node at (-1.3*\r,\y*\r-\adjust*\r) {$3$};
       \draw\midarrow{0.65}(-\Delta*\r,\y*\r-\adjust*\r)--(-1*\r,\y*\r-\adjust*\r);
  \begin{scope}[shift={(-\Delta*\r,\y*\r-\adjust*\r)}]
  \begin{scope}[rotate=-45] 
   \rightangle
  \end{scope}
  \end{scope}
     \node at (1*\r+\x*\r+\Delta*\r,\y*\r+\Delta*\r-\Adjust*\r) {$1$};
     \draw\midarrow{0.7}(0.6*\r+\x*\r+\Delta*\r,\y*\r+\Delta*\r-\Adjust*\r)--(\x*\r+\Delta*\r,\y*\r+\Delta*\r-\Adjust*\r);
    \begin{scope}[shift={(\x*\r+\Delta*\r,\y*\r+\Delta*\r-\Adjust*\r)}]
  \begin{scope}[rotate=135] 
   \rightangle
  \end{scope}
  \end{scope}
   \node at (1*\r+\x*\r+\Delta*\r,-\Delta*\r+\Adjust*\r){$0$};
  \draw\midarrow{0.7}(0.6*\r+\x*\r+\Delta*\r,-\Delta*\r+\Adjust*\r)--(\x*\r+\Delta*\r,-\Delta*\r+\Adjust*\r);
      \begin{scope}[shift={(\x*\r+\Delta*\r,-\Delta*\r+\Adjust*\r)}]
  \begin{scope}[rotate=135] 
   \rightangle
  \end{scope}
  \end{scope}
   \newcommand{\mysquare}{
   \filldraw[fill=Zcolor] (-\sq*\r,-\sq*\r) -- (\sq*\r,-\sq*\r) -- (\sq*\r,\sq*\r)  -- (-\sq*\r,\sq*\r)-- cycle;   
   }
    \draw[tickcolor,line width=\lw] (0.45*\x*\r,-0.16*\r) -- ++(-7.5+90:\ticklength*\r);   
      \draw[tickcolor,line width=\lw] (0.6*\x*\r,1*\r) -- ++(-37+90:\ticklength*\r); 
       \draw[tickcolor,line width=\lw] (0.67*\x*\r,2.5*\r) -- ++(38+90:\ticklength*\r); 
        \draw[tickcolor,line width=\lw] (0.45*\x*\r,3.4*\r) -- ++(7.5+90:\ticklength*\r); 
   \begin{scope}[shift={(0.45*\x*\r,-0.16*\r)}]
 \begin{scope}[rotate=-7.5]
 \mysquare
  \node[rotate=-7.5] at (0,0) {$\alpha_{_{20}}$};
 \end{scope}
 \end{scope}
   \begin{scope}[shift={(0.6*\x*\r,1*\r)}]
 \begin{scope}[rotate=-37]
 \mysquare
  \node[rotate=-37] at (0,0) {$\alpha_{_{30}}$};
 \end{scope}
  \end{scope}
    \begin{scope}[shift={(0.67*\x*\r,2.5*\r)}]
 \begin{scope}[rotate=38]
 \mysquare
  \node[rotate=38] at (0,0) {$\alpha_{_{21}}$};
 \end{scope}
 \end{scope}
    \begin{scope}[shift={(0.45*\x*\r,3.4*\r)}]
 \begin{scope}[rotate=7.5]
 \mysquare
   \node[rotate=7.5] at (0,0) {$\alpha_{_{31}}$};
 \end{scope}
 \end{scope}
  \end{tikzpicture}
  =\sum_{x_I,x_O\in \mathbb{Z}_2^{2}}\det(\alpha_{IO}|_{x_Ix_O})|x_O)(x_I|.
  \end{align}
Now, it is well-known that multiplying two particle-number conserving Gaussian operators can be done by multiplying the underlying $n\times m$ matrices, 
\begin{align}
    \Gamma(\alpha)\Gamma(\beta)=\Gamma(\alpha \beta).
\end{align}
Using our graphical calculus, we can show this purely diagrammatically.
For simplicity, we consider the multiplication of a $3\times 2$ with a $2\times 2$ matrix:
\begin{align}
\begin{tikzpicture}[baseline={([yshift=-2.5pt]current bounding box.center)},every node/.style={scale=0.72}]   
\def\r{0.5}
  \def\ZXr{0.25}
\def\x{3.2}
\def\xx{4.8}
\def\xxx{8.0}
\def\xxxx{9}
\def\y{2}
\def\Y{6}
\def\delta{0.3}
\def\sq{0.4}
\def\Delta{0.7}
\def\DDelta{0.5}
\def\adjust{0.1}
\def\Adjust{0.15}
\def\s{1.}
\def\lw{1}
\def\ticklength{0.7}
  \def\ZXr{0.25}
      \newcommand{\rightangle}{
   \filldraw[fill=white] (0,0) -- (\s*\r,0) -- (0,\s*\r) -- cycle;   
   }
    \coordinate (P1) at (0,0);
    \coordinate (PP1) at (-\DDelta*\r,0);
      \coordinate (P2) at (0,\y*\r);
       \coordinate (PP2) at (-\DDelta*\r,\y*\r);
     \coordinate (P3) at (0,-\y*\r);
      \coordinate (PP3) at (-\DDelta*\r,-\y*\r);
      \coordinate (P4) at (\x*\r,0.66*\y*\r);
            \coordinate (PP4) at (\x*\r+\DDelta*\r,0.66*\y*\r);
      \coordinate (P5) at (\x*\r,-0.66*\y*\r);
         \coordinate (PP5) at (\x*\r+\DDelta*\r,-0.66*\y*\r);
        \coordinate (P6) at (\xx*\r,0.66*\y*\r);
         \coordinate (PP6) at (\xx*\r-\DDelta*\r,0.66*\y*\r);
      \coordinate (P7) at (\xx*\r,-0.66*\y*\r);
               \coordinate (PP7) at (\xx*\r-\DDelta*\r,-0.66*\y*\r);
      \coordinate (P8) at (\xxx*\r,0.66*\y*\r);
        \coordinate (PP8) at (\xxx*\r+\DDelta*\r,0.66*\y*\r);
      \coordinate (P9) at (\xxx*\r,-0.66*\y*\r);
        \coordinate (PP9) at (\xxx*\r+\DDelta*\r,-0.66*\y*\r);
        \coordinate (P10) at (\xxxx*\r,0.66*\y*\r);
         \coordinate (P11) at (\xxxx*\r,-0.66*\y*\r);
     \draw(P1)--++(0,-\delta*\r)--++(\Delta*\r,0);
      \draw (P1)--++(0,+\delta*\r)--++(\Delta*\r,0);
         \draw(P2)--++(0,-\delta*\r)--++(\Delta*\r,0);
      \draw (P2)--++(0,+\delta*\r)--++(\Delta*\r,0);
         \draw(P3)--++(0,-\delta*\r)--++(\Delta*\r,0);
      \draw (P3)--++(0,+\delta*\r)--++(\Delta*\r,0);
         \draw(P4)--++(0,-\delta*\r)--++(-\Delta*\r,0);
           \draw(P4)--++(-\Delta*\r,0);
      \draw (P4)--++(0,+\delta*\r)--++(-\Delta*\r,0);
         \draw(P5)--++(0,-\delta*\r)--++(-\Delta*\r,0);
           \draw(P5)--++(-\Delta*\r,0);
      \draw (P5)--++(0,+\delta*\r)--++(-\Delta*\r,0);
           \draw(P6)--++(0,-\delta*\r)--++(\Delta*\r,0);
      \draw (P6)--++(0,+\delta*\r)--++(\Delta*\r,0);
         \draw(P7)--++(0,-\delta*\r)--++(\Delta*\r,0);
      \draw (P7)--++(0,+\delta*\r)--++(\Delta*\r,0);
         \draw(P8)--++(0,-\delta*\r)--++(-\Delta*\r,0);
      \draw (P8)--++(0,+\delta*\r)--++(-\Delta*\r,0);
         \draw(P9)--++(0,-\delta*\r)--++(-\Delta*\r,0);
      \draw (P9)--++(0,+\delta*\r)--++(-\Delta*\r,0);
     \draw\midarrow{0.78}(P1)--++(-1*\r,0);
   \begin{scope}[shift={(PP1)}]
  \begin{scope}[rotate=-45] 
   \rightangle
  \end{scope}
  \end{scope}
  \draw\midarrow{0.78}(P2)--++(-1*\r,0);
  \begin{scope}[shift={(PP2)}]
  \begin{scope}[rotate=-45] 
   \rightangle
  \end{scope}
  \end{scope}
   \draw\midarrow{0.78}(P3)--++(-1*\r,0);
  \begin{scope}[shift={(PP3)}]
  \begin{scope}[rotate=-45] 
   \rightangle
  \end{scope}
  \end{scope}
     \draw\midarrow{0.6}(P6)--(P4);
    \begin{scope}[shift={(PP4)}]
  \begin{scope}[rotate=135] 
   \rightangle
  \end{scope}
  \end{scope}
  \draw\midarrow{0.6}(P7)--(P5);
      \begin{scope}[shift={(PP5)}]
  \begin{scope}[rotate=135] 
   \rightangle
  \end{scope}
  \end{scope}
   \draw\midarrow{0.4}(P10)--(P8);
    \begin{scope}[shift={(PP8)}]
  \begin{scope}[rotate=135] 
   \rightangle
  \end{scope}
  \end{scope}
  \draw\midarrow{0.4}(P11)--(P9);
  \begin{scope}[shift={(PP9)}]
  \begin{scope}[rotate=135] 
   \rightangle
  \end{scope}
  \end{scope}
   \node[
    draw,
    rectangle,
    rounded corners=10*\r,
    minimum width=1.5cm,
    minimum height=3.8cm,
    fill=white
  ] at (0.5*\x*\r,0) {$\alpha$};
   \begin{scope}[shift={(PP6)}]
  \begin{scope}[rotate=-45] 
   \rightangle
  \end{scope}
  \end{scope}
  \begin{scope}[shift={(PP7)}]
  \begin{scope}[rotate=-45] 
   \rightangle
  \end{scope}
  \end{scope}
     \node[
    draw,
    rectangle,
    rounded corners=10*\r,
    minimum width=1.5cm,
    minimum height=3.cm,
    fill=white
  ] at (0.5*\xx*\r+0.5*\xxx*\r,0) {$\beta$};
  \end{tikzpicture}
  =
\begin{tikzpicture}[baseline={([yshift=-2.5pt]current bounding box.center)},every node/.style={scale=0.72}]   
\def\r{0.45}
  \def\ZXr{0.25}
\def\x{3.2}
\def\ratio{0.66}
\def\dy{0.65}
\def\xx{4.3}
\def\xxx{7.5}
\def\xxxx{8.5}
\def\y{2}
\def\Y{6}
\def\delta{0.3}
\def\sq{0.4}
\def\Delta{0.7}
\def\DDelta{0.5}
\def\ddelta{0.25}
\def\adjust{0.1}
\def\Adjust{0.15}
\def\s{1.}
\def\ss{.4}
\def\lw{1}
\def\ticklength{0.7}
  \def\ZXr{0.25}
      \newcommand{\rightangle}{
   \filldraw[fill=white] (0,0) -- (\s*\r,0) -- (0,\s*\r) -- cycle;   
   }
         \newcommand{\smallrightangle}{
   \filldraw[fill=white] (0,0) -- (\ss*\r,0) -- (0,\ss*\r) -- cycle;   
   }
    \coordinate (P1) at (0,0);
    \coordinate (PP1) at (-\DDelta*\r,0);
      \coordinate (P2) at (0,\y*\r);
       \coordinate (PP2) at (-\DDelta*\r,\y*\r);
     \coordinate (P3) at (0,-\y*\r);
      \coordinate (PP3) at (-\DDelta*\r,-\y*\r);
      \coordinate (P4) at (\x*\r,0.66*\y*\r);
            \coordinate (PP4) at (\x*\r+\DDelta*\r,0.66*\y*\r);
      \coordinate (P5) at (\x*\r,-0.66*\y*\r);
         \coordinate (PP5) at (\x*\r+\DDelta*\r,-0.66*\y*\r);
        \coordinate (P6) at (\xx*\r,0.66*\y*\r);
         \coordinate (PP6) at (\xx*\r-\DDelta*\r,0.66*\y*\r);
      \coordinate (P7) at (\xx*\r,-0.66*\y*\r);
               \coordinate (PP7) at (\xx*\r-\DDelta*\r,-0.66*\y*\r);
      \coordinate (P8) at (\xxx*\r,0.66*\y*\r);
        \coordinate (PP8) at (\xxx*\r+\DDelta*\r,0.66*\y*\r);
      \coordinate (P9) at (\xxx*\r,-0.66*\y*\r);
        \coordinate (PP9) at (\xxx*\r+\DDelta*\r,-0.66*\y*\r);
        \coordinate (P10) at (\xxxx*\r,0.66*\y*\r);
         \coordinate (P11) at (\xxxx*\r,-0.66*\y*\r);
    \foreach \j in {1,2,3} {
         \coordinate (P4\j) at (\x*\r,\ratio*\y*\r+2*\dy*\r-\j*\dy*\r);
         \coordinate (P5\j) at (\x*\r,-\ratio*\y*\r+2*\dy*\r-\j*\dy*\r);
          \draw(P4\j)--++(-\Delta*\r,0);
            \draw(P5\j)--++(-\Delta*\r,0);
         \coordinate (PP4\j) at (-\ddelta*\r+\x*\r,\ratio*\y*\r+2*\dy*\r-\j*\dy*\r);
         \coordinate (PP5\j) at (-\ddelta*\r+\x*\r,-\ratio*\y*\r+2*\dy*\r-\j*\dy*\r);
         }
         \foreach \j in {1,2} {
         \coordinate (P6\j) at (\xx*\r,\ratio*\y*\r+1.5*\dy*\r-\j*\dy*\r);
         \coordinate (P7\j) at (\xx*\r,-\ratio*\y*\r+1.5*\dy*\r-\j*\dy*\r);
          \draw(P6\j)--++(\Delta*\r,0);
            \draw(P7\j)--++(\Delta*\r,0);
         \coordinate (PP6\j) at (\ddelta*\r+\xx*\r,\ratio*\y*\r+1.5*\dy*\r-\j*\dy*\r);
         \coordinate (PP7\j) at (\ddelta*\r+\xx*\r,-\ratio*\y*\r+1.5*\dy*\r-\j*\dy*\r);
         }
     \draw(P1)--++(0,-\delta*\r)--++(\Delta*\r,0);
      \draw (P1)--++(0,+\delta*\r)--++(\Delta*\r,0);
         \draw(P2)--++(0,-\delta*\r)--++(\Delta*\r,0);
      \draw (P2)--++(0,+\delta*\r)--++(\Delta*\r,0);
         \draw(P3)--++(0,-\delta*\r)--++(\Delta*\r,0);
      \draw (P3)--++(0,+\delta*\r)--++(\Delta*\r,0);
         \draw(P8)--++(0,-\delta*\r)--++(-\Delta*\r,0);
      \draw (P8)--++(0,+\delta*\r)--++(-\Delta*\r,0);
         \draw(P9)--++(0,-\delta*\r)--++(-\Delta*\r,0);
      \draw (P9)--++(0,+\delta*\r)--++(-\Delta*\r,0);
     \draw\midarrow{0.78}(P1)--++(-1*\r,0);
   \begin{scope}[shift={(PP1)}]
  \begin{scope}[rotate=-45] 
   \rightangle
  \end{scope}
  \end{scope}
  \draw\midarrow{0.78}(P2)--++(-1*\r,0);
  \begin{scope}[shift={(PP2)}]
  \begin{scope}[rotate=-45] 
   \rightangle
  \end{scope}
  \end{scope}
   \draw\midarrow{0.78}(P3)--++(-1*\r,0);
  \begin{scope}[shift={(PP3)}]
  \begin{scope}[rotate=-45] 
   \rightangle
  \end{scope}
  \end{scope}
   \draw\midarrow{0.4}(P10)--(P8);
    \begin{scope}[shift={(PP8)}]
  \begin{scope}[rotate=135] 
   \rightangle
  \end{scope}
  \end{scope}
  \draw\midarrow{0.4}(P11)--(P9);
  \begin{scope}[shift={(PP9)}]
  \begin{scope}[rotate=135] 
   \rightangle
  \end{scope}
  \end{scope}
  \draw\midarrow{0.78}(P3)--++(-1*\r,0);
  \begin{scope}[shift={(PP3)}]
  \begin{scope}[rotate=-45] 
   \rightangle
  \end{scope}
  \end{scope}
    \foreach \j in {1,2,3} {
    \foreach \l in {1,2}{
\draw(P6\l)--(P4\j);
  }
  }
    \foreach \i in {1,2,3} {
    \foreach \k in {1,2}{
\draw(P7\k)--(P5\i);
  }
  }
  \foreach\i in {4,5} {
    \foreach \j in {1,2,3} {
  \begin{scope}[shift={(PP\i\j)}]
  \begin{scope}[rotate=-45] 
   \smallrightangle
  \end{scope}
  \end{scope}
       }
       }
     \foreach\i in {6,7} {
    \foreach \j in {1,2} {
  \begin{scope}[shift={(PP\i\j)}]
  \begin{scope}[rotate=135] 
   \smallrightangle
  \end{scope}
  \end{scope}
       }
       }
   \node[
    draw,
    rectangle,
    rounded corners=10*\r,
    minimum width=1.2cm,
    minimum height=3.5cm,
    fill=white
  ] at (0.5*\x*\r,0) {$\alpha$};
     \node[
    draw,
    rectangle,
    rounded corners=10*\r,
    minimum width=1.2cm,
    minimum height=2.6cm,
    fill=white
  ] at (0.5*\xx*\r+0.5*\xxx*\r,0) {$\beta$}; 
  \end{tikzpicture}
  =
\begin{tikzpicture}[baseline={([yshift=-2.5pt]current bounding box.center)},every node/.style={scale=0.72}]   
\def\r{0.45}
  \def\ZXr{0.25}
\def\x{3.2}
\def\rr{0.3}
\def\xx{4.8}
\def\xxx{8.0}
\def\xxxx{9}
\def\s{1.}
\def\y{2}
\def\Y{6}
\def\delta{0.3}
\def\Delta{0.7}
\def\DDelta{0.5}
\def\lw{1}
\def\ticklength{0.7}
  \def\ZXr{0.25}
      \newcommand{\rightangle}{
   \filldraw[fill=white] (0,0) -- (\s*\r,0) -- (0,\s*\r) -- cycle;   
   }
    \coordinate (P1) at (0,0);
    \coordinate (PP1) at (-\DDelta*\r,0);
      \coordinate (P2) at (0,\y*\r);
       \coordinate (PP2) at (-\DDelta*\r,\y*\r);
     \coordinate (P3) at (0,-\y*\r);
      \coordinate (PP3) at (-\DDelta*\r,-\y*\r);
      \coordinate (P4) at (\x*\r,0.66*\y*\r);
            \coordinate (PP4) at (\x*\r+\DDelta*\r,0.66*\y*\r);
      \coordinate (P5) at (\x*\r,-0.66*\y*\r);
         \coordinate (PP5) at (\x*\r+\DDelta*\r,-0.66*\y*\r);
       \foreach\i in {1,2,3} {
  \foreach \j in {4,5}{
\draw(P\i)--(P\j);
  }
  }
       \draw\midarrow{0.8}(P1)--++(-1*\r,0);
   \begin{scope}[shift={(PP1)}]
  \begin{scope}[rotate=-45] 
   \rightangle
  \end{scope}
  \end{scope}
  \draw\midarrow{0.8}(P2)--++(-1*\r,0);
  \begin{scope}[shift={(PP2)}]
  \begin{scope}[rotate=-45] 
   \rightangle
  \end{scope}
  \end{scope}
   \draw\midarrow{0.8}(P3)--++(-1*\r,0);
  \begin{scope}[shift={(PP3)}]
  \begin{scope}[rotate=-45] 
   \rightangle
  \end{scope}
  \end{scope}
     \draw\midarrow{0.4}(P6)--(P4);
    \begin{scope}[shift={(PP4)}]
  \begin{scope}[rotate=135] 
   \rightangle
  \end{scope}
  \end{scope}
  \draw\midarrow{0.4}(P7)--(P5);
      \begin{scope}[shift={(PP5)}]
  \begin{scope}[rotate=135] 
   \rightangle
  \end{scope}
  \end{scope}
    \filldraw[fill=gray!60!white, draw=gray!60!white] (0.5*\x*\r,0.83*\y*\r) circle (\r*\rr);
           \filldraw[fill=gray!60!white, draw=gray!60!white] (0.5*\x*\r,-0.83*\y*\r) circle (\r*\rr);
     \filldraw[fill=gray!60!white, draw=gray!60!white] (0.25*\x*\r,1.18*\r) circle (\r*\rr);
      \filldraw[fill=gray!60!white, draw=gray!60!white] (0.25*\x*\r,-1.18*\r) circle (\r*\rr);
     \filldraw[fill=gray!60!white, draw=gray!60!white] (0.6*\x*\r,0.77*\r) circle (\r*\rr);
      \filldraw[fill=gray!60!white, draw=gray!60!white] (0.3*\x*\r,-0.35*\r) circle (\r*\rr);
  \end{tikzpicture}\;.
\end{align}
Here, the large boxes labeled $\alpha$ and $\beta$ are placeholders for an all-to-all connectivity with $Z$ tensors whose amplitudes are the matrix elements of $\alpha$ and $\beta$.
The first step follows from $(\mathrm{WW})$, and the second step uses $(\mathrm{Pcy})$ together with $(\mathrm{Aso})$ to organize the tensor legs. Finally,
 the gray disk in the last equation connecting the $i$-th left and $j$-th right node represents a matrix summation $\sum_{k=1}^{d_2} \alpha_{ik}\beta_{kj}$ and taking $i=3$ and $j=2$ in our example, it is
\begin{align}
\begin{tikzpicture}[baseline={([yshift=-2.5pt]current bounding box.center)},every node/.style={scale=0.72}]   
\def\r{0.45}
\def\rr{0.3}
  \def\X{1}
  \draw(-\X*\r,0)--(\X*\r,0);
 \filldraw[fill=gray!60!white, draw=gray!60!white] (0,0) circle (\r*\rr);
\end{tikzpicture}
=
\begin{tikzpicture}[baseline={([yshift=-2.5pt]current bounding box.center)},every node/.style={scale=0.5}]   
\def\r{0.45}
  \def\ZXr{0.25}
  \def\s{1.2}
  \def\Delta{0.3}
  \def\ZXr{0.33}
    \def\d{0.5}
  \def\y{0.5}
  \def\x{1.3}
  \def\X{2.7}
    \newcommand{\rightangle}{
   \filldraw[fill=white] (0,0) -- (\s*\r,0) -- (0,\s*\r) -- cycle;   
   }
   \newcommand{\lefthalf}{
  \draw(-\X*\r,0)--(-\x*\r,0);
  \draw(-\x*\r,\y*\r)--(0,\y*\r);
    \draw(-\x*\r,-\y*\r)--(0,-\y*\r);
       \begin{scope}[shift={(-\x*\r-\Delta,0)}]
  \begin{scope}[rotate=-45] 
   \rightangle
  \end{scope}
  \end{scope}
   \draw[fill=Zcolor] (-\ZXr*\r-\d*\r,-\ZXr*\r+\y*\r) rectangle (\ZXr*\r-\d*\r,\ZXr*\r+\y*\r);
   \draw[fill=Zcolor] (-\ZXr*\r-\d*\r,-\ZXr*\r-\y*\r) rectangle (\ZXr*\r-\d*\r,\ZXr*\r-\y*\r);
   }
   \lefthalf
   \begin{scope}[xscale=-1]
   \lefthalf
   \end{scope}
   \node at (-\d*\r,-\y*\r) {$\alpha_{32}$};
   \node at (\d*\r,-\y*\r) {$\beta_{22}$};
      \node at (-\d*\r,\y*\r) {$\alpha_{31}$};
   \node at (\d*\r,\y*\r) {$\beta_{12}$};
\end{tikzpicture}
\end{align}
which can be confirmed via $(\mathrm{AD})$.
For more details on diagrammatic contraction of Gaussian fermionic tensors, we refer the reader to Bauer and Lloyd~\cite{quadratic_tensors}.

\subsection{Characteristic function, partial trace, and purification}
{%
Characteristic functions express density matrices through operator expectation values, while partial trace describes a subsystem and purification represents a mixed state using a pure state on a larger system. The graphical calculus represents these constructions as tensor contractions.}
\subsubsection{Graphical representation of characteristic functions}
Density matrices of fermionic systems can be represented through their characteristic functions, which we represent diagrammatically in this section.
We start by reviewing characteristic functiosn for qubit density matrices. Given any density matrix $\rho$, it can be expressed as a summation over Pauli operators
\begin{align}
\rho=\frac{1}{2^n} \sum_{p,q\in \mathbb{F}_2^n}  W_{p,q}w(p,q),\quad w(p,q):=i^{-pq}Z^pX^q
\end{align}
where the weights can be extracted by taking the trace inner product $W_{p,q}=\Tr[\rho w(p,q)]$. Hence the weight coefficients carry the same information as the density matrix $\rho$ itself, just in different basis. We refer to the original state $\rho_{ij}(W)$ as  the computational basis, and refer $W_{p,q}(\rho)$ as in the Pauli basis.
The matrix $W\in \operatorname{Mat}_{2^n\times 2^n}(\mathbb{C})$ can be viewed as a rank-2 tensor. The diagrammatic ZX representation of this transformation $T_Q:\rho \mapsto W$ from computational basis to Pauli basis given by the following:
\begin{align}
    T_Q=
\begin{tikzpicture}[baseline={([yshift=-2.5pt]current bounding box.center)}]   
\def\r{0.5}
\def\h{0.5}
\def\H{0.3}
\def\s{0.2}
\def\x{1.7}
\def\y{2.45}
\def\Y{0.66}
\def\delta{0.7}
\def\Delta{0.5}
\def\DeltaUD{0.22}
\def\DeltaR{1.3}
\def\Del{0.3}
\def\d{0.2}
\def\lw{1}
\def\ticklength{0.5}
\def\ZXr{0.35}
     \draw\dashedpattern(0,0)--(0,\y*\r);
      \draw\dashedpattern(\x*\r,0)--(\x*\r,\y*\r);
      \draw\dashedpattern(0,\Y*\y*\r)--(\x*\r,\Y*\y*\r);
   \filldraw[fill=yellow] (-\s*\r,-\s*\r+0.33*\y*\r) rectangle (\s*\r,\s*\r+0.33*\y*\r);
  \filldraw[fill=Zcolor,draw=black](0,\Y*\y*\r)circle (0.7*\ZXr*\r);
   \filldraw[fill=Xcolor,draw=black](\x*\r,\Y*\y*\r)circle (0.7*\ZXr*\r);
   \node at (0,\y*\r+\delta*\r) {$i$};
     \node at (0+\x*\r,\y*\r+\delta*\r) {$j$};
      \node at (0,-\delta*\r) {$p$};
     \node at (\x*\r,-\delta*\r) {$q$};
     \draw(-\Delta*\r,-\DeltaUD*\r)--(\x*\r+\DeltaR*\r,-\DeltaUD*\r)--(\x*\r+\DeltaR*\r,\y*\r+\DeltaUD*\r)--(-\Delta*\r,\y*\r+\DeltaUD*\r)--(-\Delta*\r,-\DeltaUD*\r);
   \node at (2.3*\r,\y*\r-0.2*\r) {$\otimes n$};
\end{tikzpicture}
\,
\end{align}
where we have taken a tensor product of $n$ copies of the same disjoint network.
If we contract the two legs of the density matrix $\rho_{ij}$ with the upper legs we will get $W_{p,q}(\rho)$; similarly, if $W_{p,q}$ is contracted with $T$ from below, one end up with $\rho_{i,j}(W)$.

The fermionic case  has similar transformation (see, for example, Ref.~\cite{lyu2024fermionicgaussiantestingnongaussian})
\begin{align}
    \rho=\frac{1}{2^n}\sum_J A_J\gamma_J,\quad A_J(\rho)=\Tr(\gamma_J^\dagger \rho)\in \mathbb{C}.
\end{align}
As $\rho$ is an even tensor, all coefficients $A_J=0$ if $|J|$ is an odd number. The transformation $T_F:\rho\to A$ from the computational basis to the Majorana basis can be graphically represented as
\begin{align}
T_F=
\ 
\begin{tikzpicture}[baseline={([yshift=-2.5pt]current bounding box.center)}]   
\def\r{0.5}
\def\ZXr{0.2}
\def\width{1.2}
\def\x{2}
\def\D{0.35}
\def\DeltaUD{0.25}
\def\DeltaR{1.4}
\def\y{1}
\def\Y{1.2}
\def\lw{1}
\def\delta{0.4}
\def\Delta{0.3}
\def\ticklength{0.43}
  \draw[tickcolor,line width=\lw] (0,0) --++ (0,-1*\ticklength*\r);
  \draw\midarrow{0.7} (0,0) .. controls (0.1*\x*\r,-0.5*\r) and (0.5*\x*\r,0) .. (0.5*\x*\r,\Y*\r);
  \draw\midarrow{0.4} (0.5*\x*\r,-\Y*\r) .. controls (0.5*\x*\r,0) and  (0.1*\x*\r,0.5*\r)  ..(0,0) ;
   \filldraw[fill=black,draw=black](0.4875*\x*\r,0.8*\Y*\r)circle (0.07*\r);
  \draw\midarrow{0.65} (0,0) .. controls (-0.4*\x*\r,0) and (-0.5*\x*\r,-0.8*\Y*\r) .. (-0.5*\x*\r,-\Y*\r);
   \draw\midarrow{0.5} (-0.5*\x*\r,\Y*\r) .. controls (-0.5*\x*\r,0.8*\Y*\r) and (-0.4*\x*\r,0)  .. (0,0);
 \filldraw[fill=Xcolor,draw=black](0,0)circle (\ZXr*\r);
 \node at (0.5*\x*\r+0.73*\r,\y*\r-0.1*\r) {$\otimes n$};
 \begin{scope}[shift={(-0.5*\x*\r,0)}]
    \draw(-\D*\r,-\DeltaUD*\r-\Y*\r)--(\x*\r+\DeltaR*\r,-\DeltaUD*\r-\Y*\r)--(\x*\r+\DeltaR*\r,\y*\r+1.5*\DeltaUD*\r)--(-\D*\r,\y*\r+1.5*\DeltaUD*\r)--(-\D*\r,-\DeltaUD*\r-\Y*\r);
    \node at (0.5*\x*\r,-\Y*\r-0.65*\r) {$J$};
   \node at (0.*\r,\y*\r+0.7*\r) {$\alpha$};
    \node at (\x*\r,\y*\r+0.7*\r) {$\beta$};
    \end{scope}
   \end{tikzpicture}.
\end{align}
For $n=1$ without loss of generality, the tensor diagram above maps the identity $\mathbf{I}_{\alpha\beta}$, the fermion parity operator $P_{\alpha\beta}$, and Majorana operators defined in Eq.~\eqref{eq_majorana_operaotr_def} respectively to 
\begin{align}
T_F:\mathbf{I}_{\alpha\beta}\mapsto |0)(0|=&
\begin{tikzpicture}[baseline={([yshift=-2.5pt]current bounding box.center)}]   
\def\r{0.5}
\def\ZXr{0.2}
\def\width{1.2}
\def\x{1.3}
\def\lw{1}
\def\YY{1.3}
\def\ticklength{0.43}
\draw[white,line width=\lw] (-0.5*\x*\r,0) --++ (-1*\ticklength*\r,0);
  \draw\midarrow{0.63} (-0.5*\x*\r,0) -- (-0.5*\x*\r,-\YY*\r);
    \draw\midarrow{0.51} (0.5*\x*\r,-\YY*\r) -- (0.5*\x*\r,0);
   \filldraw[fill=Xcolor,draw=black](-0.5*\x*\r,0)circle (\ZXr*\r);
    \filldraw[fill=Xcolor,draw=black](0.5*\x*\r,0)circle (\ZXr*\r);
\end{tikzpicture}
\quad\quad \quad \quad \quad 
\ P_{\alpha\beta}\mapsto |1)(1|=
(-1)\ 
\begin{tikzpicture}[baseline={([yshift=-2.5pt]current bounding box.center)}]   
\def\r{0.5}
\def\ZXr{0.2}
\def\width{1.2}
\def\x{1.3}
\def\YY{1.3}
\def\lw{1}
\def\ticklength{0.43}
\def\centralradius{0.06}
\draw\dottedpattern(-0.5*\x*\r,0)--++(\x*\r,0);
\draw\midarrow{0.5}(0.1*\x*\r,0)--++(0.01*\r,0);
  \draw\midarrow{0.63} (-0.5*\x*\r,0) -- (-0.5*\x*\r,-\YY*\r);
    \draw\midarrow{0.51} (0.5*\x*\r,-\YY*\r) -- (0.5*\x*\r,0);
   \filldraw[fill=Xcolor,draw=black](-0.5*\x*\r,0)circle (\ZXr*\r);
   \filldraw[fill=black,draw=black](-0.5*\x*\r,0)circle (\centralradius*\r);
    \filldraw[fill=Xcolor,draw=black](0.5*\x*\r,0)circle (\ZXr*\r);
    \filldraw[fill=black,draw=black](0.5*\x*\r,0)circle (\centralradius*\r);
\end{tikzpicture}
\\
\gamma_{\alpha\beta}\mapsto |1)(0|=&\ 
(-1)\ 
\begin{tikzpicture}[baseline={([yshift=-2.5pt]current bounding box.center)}]   
\def\r{0.5}
\def\ZXr{0.2}
\def\width{1.2}
\def\x{1.3}
\def\YY{1.3}
\def\h{1.2}
\def\lw{1}
\def\ticklength{0.43}
\def\centralradius{0.06}
   \draw\midarrow{0.6}(-0.5*\x*\r,0.5*\h*\r)--++(0,-0.01*\r);
 \draw\dottedpattern(-0.5*\x*\r,0)--++(0, \h*\r);
\newcommand{\lefthalf}{
  \draw (-0.5*\x*\r,0) -- (-0.5*\x*\r,-\YY*\r);
   \filldraw[fill=Xcolor,draw=black](-0.5*\x*\r,0)circle (\ZXr*\r);
  }
   \lefthalf
   \begin{scope}[xscale=-1]
   \lefthalf
   \end{scope}
   \filldraw[fill=black,draw=black](-0.5*\x*\r,0)circle (\centralradius*\r);
   \draw\midarrow{0.5}(0.5*\x*\r,-0.51*\YY*\r)--++(0,+0.021);
    \draw\midarrow{0.5}(-0.5*\x*\r,-0.63*\YY*\r)--++(0,-0.021);
\end{tikzpicture}
\quad \quad \quad 
\gamma_{\alpha\beta}'\mapsto |0)(1|=
\begin{tikzpicture}[baseline={([yshift=-2.5pt]current bounding box.center)}]   
\def\r{0.5}
\def\ZXr{0.2}
\def\width{1.2}
\def\x{1.3}
\def\YY{1.3}
\def\h{1.2}
\def\lw{1}
\def\ticklength{0.43}
\def\centralradius{0.06}
   \draw\midarrow{0.6}(0.5*\x*\r,0.5*\h*\r)--++(0,-0.01*\r);
 \draw\dottedpattern(0.5*\x*\r,0)--++(0, \h*\r);
\newcommand{\lefthalf}{
  \draw (-0.5*\x*\r,0) -- (-0.5*\x*\r,-\YY*\r);
   \filldraw[fill=Xcolor,draw=black](-0.5*\x*\r,0)circle (\ZXr*\r);
  }
   \lefthalf
   \begin{scope}[xscale=-1]
   \lefthalf
   \end{scope}
   \filldraw[fill=black,draw=black](0.5*\x*\r,0)circle (\centralradius*\r);
   \draw\midarrow{0.5}(0.5*\x*\r,-0.49*\YY*\r)--++(0,+0.021);
    \draw\midarrow{0.5}(-0.5*\x*\r,-0.61*\YY*\r)--++(0,-0.021);
\end{tikzpicture}
\end{align}
The purely diagrammatic proof is in Appendix~\ref{proof_TF}.
\subsubsection{Partial trace}
The partial trace of ordinary Hilbert space is well known. Suppose $A$ is a subset of the system's qudits and $B$ is its complement, then $\Tr_A\rho=\sum_{i\in A}\bra{i}_A\rho\ket{i}_A$. One might naively think that the partial trace of Majorana modes can be defined similarly. This is true, if we only trace out even number of Majorana modes, since every two of them can be paired up to form spaces $\mathbb{C}^{1|1}$ and one can resort to simiar expression to sum over the oddly graded sector and the evenly graded sector of those systems.  However, recall that each Majorana mode has quantum dimension $\sqrt{2}$, as it is only "half" a qubit dimensionwise.  How to define tracing out this $\sqrt{2}$-dimension hence becomes a conceptual problem.

{%
The operator representation of partial trace traces out qubit $A$ and reinserts the maximally mixed state $I_A/2$, so that the channel acts on the original Hilbert space:}

\begin{align}
    {(\Tr_A\rho)\otimes \frac{I_A}{2}}={\frac12}\sum_{i\in \mathbb{F}_2}\bra{i}_A\rho\ket{i}_A\otimes I_A=
    \frac{1}{4}\sum_{a,b\in \mathbb{F}_2}X^aZ^b \rho (X^aZ^b)^\dagger.
\end{align}
One can hence define tracing over Majorana mode $j$ similarly by the following channel
\begin{align}
    {\Tr_{\gamma_j}}\rho:= \frac{1}{2}(\rho+\gamma_j\rho\gamma_j),\quad 1\leq j\leq 2n
\end{align}
for a system of $2n$ Majorana modes. 
{%
The corresponding diagrams include the factor $1/2$ in the channel definition:}

\begin{align}
\Tr_{\gamma_j}(\ )={\frac12}\,
\begin{tikzpicture}[baseline={([yshift=-2.5pt]current bounding box.center)}]   
\def\r{0.5}
\def\h{.6}
\def\w{1}
\def\x{1.7}
\def\y{1}
\def\Y{1}
\def\width{1.2}
\def\lw{1}
\def\ticklength{0.6}
\def\ZXr{0.25}
\newcommand{\halfpart}{\draw(0,0)..controls (-1*\r,-\h*\r) and (0.4*\x*\r,-\h*\r) .. (0.5*\x*\r,-\h*\r);}
\halfpart
\begin{scope}[shift={(\x*\r,0)}]
\begin{scope}[xscale=-1]
\halfpart
\end{scope}
\end{scope}
\draw\midarrow{0.57}(0.43*\x*\r,-\h*\r)--++(-0.01*\r,0);
\draw[tickcolor,line width=\lw] (0,0) -- ++(-\ticklength*\r,0);
\draw[tickcolor,line width=\lw] (\x*\r,0) -- ++(\ticklength*\r,0);
\draw(0,\y*\r)--(0,-\Y*\r);
\draw(\x*\r,\y*\r)--(\x*\r,-\Y*\r);
\filldraw[fill=Xcolor,draw=black](0*\r,0*\r)circle (\ZXr*\r);
\filldraw[fill=Xcolor,draw=black](\x*\r,0*\r)circle (\ZXr*\r);
\end{tikzpicture}
\quad \quad 
\Tr_{\gamma_j'}(\ )={\frac12}\,
\begin{tikzpicture}[baseline={([yshift=-2.5pt]current bounding box.center)}]   
\def\r{0.5}
\def\h{.7}
\def\w{1}
\def\x{1.5}
\def\y{1}
\def\Y{1}
\def\width{1.2}
\def\lw{1}
\def\ticklength{0.6}
\def\ZXr{0.25}
\draw[tickcolor,line width=\lw] (0,0) -- ++(-\ticklength*\r,0);
\draw[tickcolor,line width=\lw] (\x*\r,0) -- ++(\ticklength*\r,0);
\draw(0,\y*\r)--(0,-\Y*\r);
\draw(\x*\r,\y*\r)--(\x*\r,-\Y*\r);
\draw\midarrow{0.57}(\x*\r,0)--(0,0);
\filldraw[fill=Xcolor,draw=black](0*\r,0*\r)circle (\ZXr*\r);
\filldraw[fill=Xcolor,draw=black](\x*\r,0*\r)circle (\ZXr*\r);
\end{tikzpicture}
\;,
\end{align}
where we recall that the pair $(\gamma_j,\gamma'_j)$ denotes the two different Majorana operators belonging to one sigle mode, $\gamma'_j\equiv \gamma_{j+n}$. 
One can prove that,  by using Eq.~\eqref{eq_spider_self_contraction}, $\Tr_{\gamma_j}\Tr_{\gamma_j'}=\Tr_j=\Tr_{\gamma'_j}\Tr_{\gamma_j}$, {where $\Tr_j$ denotes tracing over the full fermion mode $j$ and reinserting its maximally mixed state.} We can show this purely diagrammatically,
\begin{align}
{\frac14}\,
\begin{tikzpicture}[baseline={([yshift=-2.5pt]current bounding box.center)}]   
\def\r{0.5}
\def\h{.7}
\def\w{1}
\def\x{1.5}
\def\y{1.1}
\def\Y{1}
\def\width{1.2}
\def\lw{1}
\def\ticklength{0.6}
\def\ZXr{0.25}
\def\Ymove{1.2*\r}
\begin{scope}[shift={(0,\Ymove)}]
\draw[tickcolor,line width=\lw] (0,0) -- ++(-\ticklength*\r,0);
\draw[tickcolor,line width=\lw] (\x*\r,0) -- ++(\ticklength*\r,0);
\draw\midarrow{0.67}(0,0)--(0,-\Y*\r);
\draw\midarrow{0.56}(\x*\r,-\Y*\r)--(\x*\r,0);
\draw\midarrow{0.55}(0,\y*\r)--(0,0);
\draw\midarrow{0.65}(\x*\r,0)--(\x*\r,\y*\r);
\draw\midarrow{0.6}(\x*\r,0)--(0,0);
\filldraw[fill=Xcolor,draw=black](0*\r,0*\r)circle (\ZXr*\r);
\filldraw[fill=Xcolor,draw=black](\x*\r,0*\r)circle (\ZXr*\r);
\end{scope}
\newcommand{\halfpart}{\draw(0,0)..controls (-1*\r,-\h*\r) and (0.4*\x*\r,-\h*\r) .. (0.5*\x*\r,-\h*\r);}
\halfpart
\begin{scope}[shift={(\x*\r,0)}]
\begin{scope}[xscale=-1]
\halfpart
\end{scope}
\end{scope}
\draw\midarrow{0.57}(0.43*\x*\r,-\h*\r)--++(-0.01*\r,0);
\draw[tickcolor,line width=\lw] (0,0) -- ++(-\ticklength*\r,0);
\draw[tickcolor,line width=\lw] (\x*\r,0) -- ++(\ticklength*\r,0);
\draw\midarrow{0.78}(0,0)--(0,-1.5*\Y*\r);
\draw\midarrow{0.35}(\x*\r,-1.5*\Y*\r)--(\x*\r,0);
\filldraw[fill=Xcolor,draw=black](0*\r,0*\r)circle (\ZXr*\r);
\filldraw[fill=Xcolor,draw=black](\x*\r,0*\r)circle (\ZXr*\r);
\end{tikzpicture}
\quad =\quad 
{\frac12}\,
\begin{tikzpicture}[baseline={([yshift=-2.5pt]current bounding box.center)}]   
\def\r{0.5}
\def\h{.7}
\def\w{1}
\def\x{0.5}
\def\y{1.3}
\def\Y{1}
\def\Ymove{0.5}
\def\width{1.2}
\def\lw{1}
\def\ticklength{0.6}
\def\ZXr{0.25}
\newcommand{\bottomright}{
 \draw\midarrow{.4}  (\x*\r,-\Y*\r).. controls (\x*\r,-.4*\Y*\r)  and (0.6*\x*\r,-0*\Y*\r) .. (0,0);
 }
 \newcommand{\bottomleft}{
\draw\midarrow{0.78} (0,0) .. controls (-0.6*\x*\r,-0.*\Y*\r) and (-\x*\r,-.4\Y*\r).. (-\x*\r,-1.*\Y*\r);
 }
 \newcommand{\topright}{
\draw\midarrow{0.78} (0,0) .. controls (0.6*\x*\r,0) and (\x*\r,.4\Y*\r).. (\x*\r,1.*\Y*\r);
 }
 \newcommand{\topleft}{
 \draw\midarrow{.4}  (-\x*\r,\Y*\r).. controls (-\x*\r, .4*\Y*\r)  and (-0.6*\x*\r,0*\Y*\r) .. (0,0);
 }
 \bottomright
\bottomleft
\begin{scope}[shift={(0,\Ymove*\r)}]
\topright
\topleft
   \filldraw[fill=black,draw=black](0,0) circle (0.09*\r);
\end{scope}
\end{tikzpicture}
\quad =\quad 
{\frac14}\,
\begin{tikzpicture}[baseline={([yshift=-2.5pt]current bounding box.center)}]   
\def\r{0.5}
\def\h{.7}
\def\w{1}
\def\x{1.5}
\def\y{0.8}
\def\Y{1}
\def\width{1.2}
\def\lw{1}
\def\ticklength{0.6}
\def\ZXr{0.25}
\def\Ymove{1.5*\r}
\begin{scope}[shift={(0,-\Ymove)}]
\draw[tickcolor,line width=\lw] (0,0) -- ++(-\ticklength*\r,0);
\draw[tickcolor,line width=\lw] (\x*\r,0) -- ++(\ticklength*\r,0);
\draw\midarrow{0.67}(0,0)--(0,-\Y*\r);
\draw\midarrow{0.56}(\x*\r,-\Y*\r)--(\x*\r,0);
\draw\midarrow{0.71}(0,\Ymove)--(0,0);
\draw\midarrow{0.46}(\x*\r,0)--(\x*\r,\Ymove);
\draw\midarrow{0.6}(\x*\r,0)--(0,0);
\filldraw[fill=Xcolor,draw=black](0*\r,0*\r)circle (\ZXr*\r);
\filldraw[fill=Xcolor,draw=black](\x*\r,0*\r)circle (\ZXr*\r);
\end{scope}
\newcommand{\halfpart}{\draw(0,0)..controls (-1*\r,-\h*\r) and (0.4*\x*\r,-\h*\r) .. (0.5*\x*\r,-\h*\r);}
\halfpart
\begin{scope}[shift={(\x*\r,0)}]
\begin{scope}[xscale=-1]
\halfpart
\end{scope}
\end{scope}
\draw\midarrow{0.57}(0.43*\x*\r,-\h*\r)--++(-0.01*\r,0);
\draw[tickcolor,line width=\lw] (0,0) -- ++(-\ticklength*\r,0);
\draw[tickcolor,line width=\lw] (\x*\r,0) -- ++(\ticklength*\r,0);
\draw\midarrow{0.5}(0,\y*\r)--(0,0);
\draw\midarrow{0.75}(\x*\r,0)--(\x*\r,\y*\r);
\filldraw[fill=Xcolor,draw=black](0*\r,0*\r)circle (\ZXr*\r);
\filldraw[fill=Xcolor,draw=black](\x*\r,0*\r)circle (\ZXr*\r);
\end{tikzpicture}
\;.
\end{align}
Below we show the left equality above, the right equality follows similarly.
\begin{align}
{\frac14}\,
\begin{tikzpicture}[baseline={([yshift=-2.5pt]current bounding box.center)}]   
\def\r{0.5}
\def\h{.7}
\def\w{1}
\def\x{1.5}
\def\y{1.1}
\def\Y{1}
\def\width{1.2}
\def\lw{1}
\def\ticklength{0.6}
\def\ZXr{0.25}
\def\Ymove{1.2*\r}
\begin{scope}[shift={(0,\Ymove)}]
\draw[tickcolor,line width=\lw] (0,0) -- ++(-\ticklength*\r,0);
\draw[tickcolor,line width=\lw] (\x*\r,0) -- ++(\ticklength*\r,0);
\draw\midarrow{0.67}(0,0)--(0,-\Y*\r);
\draw\midarrow{0.56}(\x*\r,-\Y*\r)--(\x*\r,0);
\draw\midarrow{0.55}(0,\y*\r)--(0,0);
\draw\midarrow{0.65}(\x*\r,0)--(\x*\r,\y*\r);
\draw\midarrow{0.6}(\x*\r,0)--(0,0);
\filldraw[fill=Xcolor,draw=black](0*\r,0*\r)circle (\ZXr*\r);
\filldraw[fill=Xcolor,draw=black](\x*\r,0*\r)circle (\ZXr*\r);
\end{scope}
\newcommand{\halfpart}{\draw(0,0)..controls (-1*\r,-\h*\r) and (0.4*\x*\r,-\h*\r) .. (0.5*\x*\r,-\h*\r);}
\halfpart
\begin{scope}[shift={(\x*\r,0)}]
\begin{scope}[xscale=-1]
\halfpart
\end{scope}
\end{scope}
\draw\midarrow{0.57}(0.43*\x*\r,-\h*\r)--++(-0.01*\r,0);
\draw[tickcolor,line width=\lw] (0,0) -- ++(-\ticklength*\r,0);
\draw[tickcolor,line width=\lw] (\x*\r,0) -- ++(\ticklength*\r,0);
\draw\midarrow{0.78}(0,0)--(0,-1.5*\Y*\r);
\draw\midarrow{0.35}(\x*\r,-1.5*\Y*\r)--(\x*\r,0);
\filldraw[fill=Xcolor,draw=black](0*\r,0*\r)circle (\ZXr*\r);
\filldraw[fill=Xcolor,draw=black](\x*\r,0*\r)circle (\ZXr*\r);
\end{tikzpicture}
\quad =\quad 
{\frac14}\,
\begin{tikzpicture}[baseline={([yshift=-2.5pt]current bounding box.center)}]   
\def\r{0.5}
\def\h{.7}
\def\w{1}
\def\x{1.5}
\def\y{1.1}
\def\Y{1}
\def\width{1.2}
\def\lw{1}
\def\ticklength{0.6}
\def\ZXr{0.25}
\def\Ymove{0}
\begin{scope}[shift={(0,\Ymove)}]
\draw[tickcolor,line width=\lw] (0,0) -- ++(-\ticklength*\r,0);
\draw[tickcolor,line width=\lw] (\x*\r,0) -- ++(\ticklength*\r,0);
\draw\midarrow{0.55}(0,\y*\r)--(0,0);
\draw\midarrow{0.65}(\x*\r,0)--(\x*\r,\y*\r);
\draw\midarrow{0.6}(\x*\r,0)--(0,0);
\filldraw[fill=Xcolor,draw=black](0*\r,0*\r)circle (\ZXr*\r);
\filldraw[fill=Xcolor,draw=black](\x*\r,0*\r)circle (\ZXr*\r);
\end{scope}
\newcommand{\halfpart}{\draw(0,0)..controls (-1*\r,-\h*\r) and (0.4*\x*\r,-\h*\r) .. (0.5*\x*\r,-\h*\r);}
\halfpart
\begin{scope}[shift={(\x*\r,0)}]
\begin{scope}[xscale=-1]
\halfpart
\end{scope}
\end{scope}
\draw\midarrow{0.57}(0.43*\x*\r,-\h*\r)--++(-0.01*\r,0);
\draw\midarrow{0.78}(0,0)--(0,-1.5*\Y*\r);
\draw\midarrow{0.35}(\x*\r,-1.5*\Y*\r)--(\x*\r,0);
\filldraw[fill=Xcolor,draw=black](0*\r,0*\r)circle (\ZXr*\r);
\filldraw[fill=Xcolor,draw=black](\x*\r,0*\r)circle (\ZXr*\r);
\end{tikzpicture}
\quad =\quad 
{\frac14}\,
\begin{tikzpicture}[baseline={([yshift=-2.5pt]current bounding box.center)}]   
\def\r{0.5}
\def\h{.7}
\def\w{1}
\def\x{0.5}
\def\y{1.3}
\def\Y{1}
\def\width{1.2}
\def\lw{1}
\def\ticklength{0.6}
\def\ZXr{0.25}
\def\Ymove{0}
\draw[tickcolor,line width=\lw] (0,0) -- ++(0,1*\ticklength*\r);
\filldraw[fill=Xcolor,draw=black](0*\r,0*\r)circle (\ZXr*\r);
    \draw (0,-0.4*\r) ellipse (0.6*\r cm and 0.4*\r cm);
       \draw[<-](-0.029,-0.8*\r)--(0.03,-0.8*\r);
\draw\midarrow{0.58}  (-\x*\r,1.5*\Y*\r) .. controls  (-\x*\r,\Y*\r) and (-0.3*\x*\r,0.2*\Y*\r) .. (0,0);
   \draw\midarrow{.55} (0,0).. controls (0.3*\x*\r,0.2*\Y*\r) and (\x*\r,\Y*\r)  ..  (\x*\r,1.5*\Y*\r);
   \filldraw[fill=black,draw=black](-0.875*\x*\r,1.1*\Y*\r) circle (0.09*\r);
    \filldraw[fill=black,draw=black](0.875*\x*\r,1.1*\Y*\r) circle (0.09*\r);
\draw\midarrow{0.78} (0,0) .. controls (-0.3*\x*\r,-0.2*\Y*\r) and (-\x*\r,-\Y*\r).. (-\x*\r,-1.5*\Y*\r);
    \draw\midarrow{.35}  (\x*\r,-1.5*\Y*\r).. controls (\x*\r,-1.*\Y*\r)  and (0.3*\x*\r,-0.2*\Y*\r) .. (0,0);
\filldraw[fill=Xcolor,draw=black](0*\r,0*\r)circle (\ZXr*\r);
\end{tikzpicture}
\quad =\quad 
{\frac12}\,
\begin{tikzpicture}[baseline={([yshift=-2.5pt]current bounding box.center)}]   
\def\r{0.5}
\def\h{.7}
\def\w{1}
\def\x{0.5}
\def\y{1.3}
\def\Y{1}
\def\Ymove{0.5}
\def\width{1.2}
\def\lw{1}
\def\ticklength{0.6}
\def\ZXr{0.25}
\newcommand{\bottomright}{
 \draw\midarrow{.4}  (\x*\r,-\Y*\r).. controls (\x*\r,-.4*\Y*\r)  and (0.6*\x*\r,-0*\Y*\r) .. (0,0);
 }
 \newcommand{\bottomleft}{
\draw\midarrow{0.78} (0,0) .. controls (-0.6*\x*\r,-0.*\Y*\r) and (-\x*\r,-.4\Y*\r).. (-\x*\r,-1.*\Y*\r);
 }
 \newcommand{\topright}{
\draw\midarrow{0.78} (0,0) .. controls (0.6*\x*\r,0) and (\x*\r,.4\Y*\r).. (\x*\r,1.*\Y*\r);
 }
 \newcommand{\topleft}{
 \draw\midarrow{.4}  (-\x*\r,\Y*\r).. controls (-\x*\r, .4*\Y*\r)  and (-0.6*\x*\r,0*\Y*\r) .. (0,0);
 }
 \bottomright
\bottomleft
\begin{scope}[shift={(0,\Ymove*\r)}]
\topright
\topleft
   \filldraw[fill=black,draw=black](0,0) circle (0.09*\r);
\end{scope}
\end{tikzpicture}
\end{align}
Tracing out an arbitrary subset $J\subset[2n]$ of Majorana modes is defined accordingly.
%
\subsubsection{Purification after partial trace}
Let us now look at what happens if we purify the density matrix obtained by tracing out a Majorana mode in a pure state of multiple fermion modes.
The partial trace connects the ket and bra parts of the pure state with an index contraction.
This contracted index becomes the auxiliary mode of the purification, and the purified state is obtained from cutting it,
\begin{align}
\begin{tikzpicture}[baseline={([yshift=-2.5pt]current bounding box.center)}]   
\def\r{0.5}
\def\h{.7}
\def\w{1}
\def\x{1.5}
\def\y{1}
\def\Y{1}
\def\width{1.2}
\def\Lshift{0.6*\r}
\def\Rshift{0.75*\r}
\def\lw{1}
\def\ticklength{0.6}
\def\ZXr{0.25}
\draw\midarrow{0.55}(0,\y*\r)--(0,-\Y*\r);
\draw\midarrow{0.55}(\x*\r,-\Y*\r)--(\x*\r,\y*\r);
\filldraw[fill=white, ,draw=black](0*\r,\y*\r)circle (\ZXr*\r);
\node at (-\Lshift,\y*\r) {$\psi$};
\filldraw[fill=white, ,draw=black](\x*\r,\y*\r)circle (\ZXr*\r);
\node at (\x*\r+\Rshift,\y*\r) {$\psi^\dagger$};
\node at (4.3*\r,0.7*\r) {$\Tr_{\gamma'}$};
\draw[->](3.7*\r,0)--++(1.2*\r,0);
\node at (6*\r,0) {$\frac12$};
\begin{scope}[shift={(7*\r,0)}]
\draw[tickcolor,line width=\lw] (0,0) -- ++(-\ticklength*\r,0);
\draw[tickcolor,line width=\lw] (\x*\r,0) -- ++(\ticklength*\r,0);
\draw\midarrow{0.3}(0,\y*\r)--(0,-\Y*\r);
\draw\midarrow{0.85}(0,\y*\r)--(0,-\Y*\r);
\draw\midarrow{0.25}(\x*\r,-\Y*\r)--(\x*\r,\y*\r);
\draw\midarrow{0.8}(\x*\r,-\Y*\r)--(\x*\r,\y*\r);
\draw\midarrow{0.57}(\x*\r,0)--(0,0);
\filldraw[fill=Xcolor,draw=black](0*\r,0*\r)circle (\ZXr*\r);
\filldraw[fill=Xcolor,draw=black](\x*\r,0*\r)circle (\ZXr*\r);
\filldraw[fill=white, ,draw=black](0*\r,\y*\r)circle (\ZXr*\r);
\node at (-\Lshift,\y*\r) {$\psi$};
\filldraw[fill=white, ,draw=black](\x*\r,\y*\r)circle (\ZXr*\r);
\node at (\x*\r+\Rshift,\y*\r) {$\psi^\dagger$};
\end{scope}
\end{tikzpicture}
\quad 
\begin{tikzpicture}[baseline={([yshift=-2.5pt]current bounding box.center)}]   
\def\r{0.5}
\def\h{.7}
\def\w{1}
\def\x{2.2}
\def\Delta{0.2}
\def\y{1}
\def\Y{1}
\def\width{1.2}
\def\lw{1}
\def\Lshift{0.6*\r}
\def\Rshift{0.75*\r}
\def\ticklength{0.6}
\def\ZXr{0.25}
\node at (-0.9*\r,0) {$\frac12$};
\node at (-3*\r,0.7*\r) {Puri};
\draw[->](-3.7*\r,0)--++(1.45*\r,0);
\draw[tickcolor,line width=\lw] (0,0) -- ++(-\ticklength*\r,0);
\draw[tickcolor,line width=\lw] (\x*\r,0) -- ++(\ticklength*\r,0);
\draw\midarrow{0.3}(0,\y*\r)--(0,-\Y*\r);
\draw\midarrow{0.85}(0,\y*\r)--(0,-\Y*\r);
\draw\midarrow{0.25}(\x*\r,-\Y*\r)--(\x*\r,\y*\r);
\draw\midarrow{0.8}(\x*\r,-\Y*\r)--(\x*\r,\y*\r);
\draw\midarrow{0.45}(0.5*\x*\r-\Delta*\r,0)--(0,0);
\draw\midarrow{0.65}(\x*\r,0)--(0.5*\x*\r+\Delta*\r,0);
\filldraw[fill=Xcolor,draw=black](0*\r,0*\r)circle (\ZXr*\r);
\filldraw[fill=Xcolor,draw=black](\x*\r,0*\r)circle (\ZXr*\r);
\filldraw[fill=white, ,draw=black](0*\r,\y*\r)circle (\ZXr*\r);
\node at (-\Lshift,\y*\r) {$\psi$};
\filldraw[fill=white, ,draw=black](\x*\r,\y*\r)circle (\ZXr*\r);
\node at (\x*\r+\Rshift,\y*\r) {$\psi^\dagger$};
\end{tikzpicture}
\;.
\end{align}
{%
For a normalized input state, the purified ket obtained by cutting the edge carries a factor $1/\sqrt{2}$, and the ket--bra pair carries $1/2$. Denote the normalized purified tensor by $\Psi_{xa}$, where $a\in\mathbb F_2$ is the auxiliary index and $x$ denotes the remaining indices in their fixed order. Tracing out the auxiliary mode is the ordinary partial trace, without reinserting a state. It reconnects the two ends of the cut edge in the direction of its arrow:
\begin{align*}
\sum_{a\in\mathbb F_2}\Psi_{xa}\overline{\Psi_{ya}}
&=\left[\frac12\bigl(|\psi)(\psi|+\gamma'_j|\psi)(\psi|\gamma'_j\bigr)\right]_{xy}\\
&=\bigl[\Tr_{\gamma'_j}(|\psi)(\psi|)\bigr]_{xy}.
\end{align*}
The two values of $a$ give the identity and Majorana terms, respectively; a phase of either ket branch cancels against its conjugate. The same argument applies to $\gamma_j$.}

Taken together, partial trace followed by purification can be represented by a single operator:
\begin{align}
\operatorname{Puri}\circ \Tr_{\gamma_j}(\ )={\frac12}\,
\begin{tikzpicture}[baseline={([yshift=-2.5pt]current bounding box.center)}]   
\def\r{0.5}
\def\h{.6}
\def\w{1}
\def\x{1.6}
\def\Delta{0.2}
\def\y{1}
\def\Y{1}
\def\width{1.2}
\def\lw{1}
\def\ticklength{0.6}
\def\ZXr{0.25}
\begin{scope}[xscale=-1]
\draw(0,0)..controls (-1*\r,-\h*\r) and (0.4*\x*\r-\Delta*\r,-\h*\r) .. (0.5*\x*\r-\Delta*\r,-\h*\r);
\draw(\x*\r,0)..controls (\x*\r+1*\r,-\h*\r) and (0.6*\x*\r+\Delta*\r,-\h*\r) .. (0.5*\x*\r+\Delta*\r,-\h*\r);
\draw\midarrow{0.57}(0.22*\x*\r,-0.99*\h*\r)--++(0.01*\r,-0.0015*\r);
\draw\midarrow{0.57}(0.89*\x*\r,-0.965*\h*\r)--++(0.01*\r,0.0012*\r);
\draw[tickcolor,line width=\lw] (0,0) -- ++(-\ticklength*\r,0);
\draw[tickcolor,line width=\lw] (\x*\r,0) -- ++(\ticklength*\r,0);
\draw(0,\y*\r)--(0,-\Y*\r);
\draw(\x*\r,\y*\r)--(\x*\r,-\Y*\r);
\filldraw[fill=Xcolor,draw=black](0*\r,0*\r)circle (\ZXr*\r);
\filldraw[fill=Xcolor,draw=black](\x*\r,0*\r)circle (\ZXr*\r);
\end{scope}
\end{tikzpicture}
\quad \quad \quad 
\operatorname{Puri}\circ \Tr_{\gamma'_j}(\ )={\frac12}\,
\begin{tikzpicture}[baseline={([yshift=-2.5pt]current bounding box.center)}]   
\def\r{0.5}
\def\h{.7}
\def\w{1}
\def\x{2.2}
\def\Delta{0.2}
\def\y{1}
\def\Y{1}
\def\width{1.2}
\def\lw{1}
\def\ticklength{0.6}
\def\ZXr{0.25}
\begin{scope}[xscale=-1]
\draw[tickcolor,line width=\lw] (0,0) -- ++(-\ticklength*\r,0);
\draw[tickcolor,line width=\lw] (\x*\r,0) -- ++(\ticklength*\r,0);
\draw(0,\y*\r)--(0,-\Y*\r);
\draw(\x*\r,\y*\r)--(\x*\r,-\Y*\r);
\draw\midarrow{0.73}(0,0)--(0.5*\x*\r-\Delta*\r,0);
\draw\midarrow{0.5}(0.5*\x*\r+\Delta*\r,0)--(\x*\r,0);
\filldraw[fill=Xcolor,draw=black](0*\r,0*\r)circle (\ZXr*\r);
\filldraw[fill=Xcolor,draw=black](\x*\r,0*\r)circle (\ZXr*\r);
\end{scope}
\end{tikzpicture}
\;.
\end{align}


\subsection{Bosonization}
As a next application of our graphical calculus, let us consider bosonization.
Bosonization is an isometry that embeds a fermionic system defined on a $d$-dimensional lattice onto a qubit system, in such a way that even-parity local fermionic operators are transformed into local qubit operators.
Pairs of local odd operators that are far separated are mapped onto pairs of local qubit operators connected by string-like operators.
Bosonization can be understood as taking a topologically ordered model in $d$ dimensions, and using emergent fermions in the topologically ordered model to emulate the physical fermions in a fermionic model.
In one dimension, bosonization has long been known as Jordan-Wigner transformation.
However, bosonization maps can also be found in higher dimensions~\cite{PhysRevB.104.035118,landahl2023,Shukla_2020,Chen_2018,Chen_2023,O_Brien_2025}.

\subsubsection{1D bosonization}
As mentioned before, bosonization in 1D is known as Jordan-Wigner transformation.  As we recall from Appendix~\ref{sect_clifford_and_majorana}, 1D bosonization is defined as an isometry $\mathsf{D}_\text{Bos}=\mathsf{D}_{KW}\circ \mathsf{D}_{JW}$ from the even-parity space of fermion modes on the vertices of a 1D lattice to qubits on the edges. This isometry can be expressed as a tensor-network diagram using the tensors of our diagrammatic calculus (compare also \cite{Shukla_2020,mortier2024}):
\begin{align}
\mathsf{D}_\text{Bos}=
\begin{tikzpicture}[baseline={([yshift=-2.5pt]current bounding box.center)}]
\def\r{0.4}
\def\d{1.5}
\def\lw{1}
\def\ticklength{0.65}
\draw[decoration={markings, mark=at position 0.55 with {\arrow{>}}}, postaction={decorate},line width=0.55pt]   (-0.5*\d, 0)--(0, 0);
 \foreach \j in {0, 1, 2, 3, 4} {
         \draw[tickcolor,line width=\lw] (\j*\d, 0) -- ++(0.707*\ticklength*\r,0.707*\ticklength*\r);
         \draw[tickcolor,line width=\lw] (\j*\d+0.5*\d,0) -- ++(0.707*\ticklength*\r,0.707*\ticklength*\r);
        \draw[decoration={markings, mark=at position 0.54 with {\arrow{>}}}, postaction={decorate},line width=0.55pt]  (\j*\d, -0.5*\d)--(\j*\d, 0);
        \draw[decoration={markings, mark=at position 0.57 with {\arrow{>}}}, postaction={decorate},line width=0.55pt]  (\j*\d+0.5*\d, 0)--(\j*\d+\d, 0);
        \draw[decoration={markings, mark=at position 0.57 with {\arrow{>}}}, postaction={decorate},line width=0.55pt]  (\j*\d, 0)--(\j*\d+0.5*\d, 0);
        \draw[dashed] (\j*\d+0.5*\d, 0.5*\d)--(\j*\d+0.5*\d, 0);
        \filldraw[fill=Xcolor, draw=black] (\j*\d,0) circle (0.3*\r);
        \filldraw[fill=Zcolor, draw=black] (\j*\d+0.5*\d,0) circle (0.3*\r);
    }
\end{tikzpicture}
\end{align}
The operator acts from the bottom to the top.
We label sites from left to right in the decreasing order. That is to say, each site $j+1$ is put on the left of $j$.
1D bosonization (acting by conjutation) transforms Majorana operators into Pauli operators as follows:
\begin{align}
    \mathsf{D}_\text{Bos}:\gamma_{j+1}\gamma_j\mapsto -iY_{j+\frac{1}{2}}Z_{j-\frac{1}{2}},
    \quad P_j\mapsto Z_{j+\frac{1}{2}}Z_{j-\frac{1}{2}}.
\end{align}
This can be shown purely diagrammatically.
We will skip the proof here as this is a special case of the 2D version below by simply omitting all the tensor edges in the vertical direction. 
\subsubsection{Higher-dimensional bosonization}
Consider a $d$-dimensional cubic lattice $\mathbb{Z}^d$ with basis vectors $\{\hat{x}_i\}$.\footnote{Alternatively, one can consider other divisions of the underlying manifold. For example, one may define the Bosonization map to be on a $d$-simplex. However, we will stick to the cubic lattice in this work for simplicity.}
We define a \emph{$d$-dimensional Bosonization} $\mathsf{D}_\text{Bos}$ supported on  $\mathbb{Z}^d$ to be an isometry from fermionic modes on the vertices (0-cells) of the lattice to qubits on the edges (1-cells).
We will denote its action on (Majorana) operators by conjugation as $\mathsf{D}_\text{Bos}(\ ):=\mathsf{D}_\text{Bos}(\ )\mathsf{D}_\text{Bos}^\dagger$.
The bosonization operator satisfies the following generating constraints for arbitrary $v\in \mathbb{Z}^d$.
 \begin{align}
 \{\mathsf{D}_\text{Bos}(\gamma_v\gamma_{v\pm \hat{x}_i}),\mathsf{D}_\text{Bos}(\gamma_v\gamma_{v\pm \hat{x}_j})\}=0&,\quad \forall\  1\leq i,j\leq d,\quad i\neq j \label{D_commutation_ij}\\
 \{\mathsf{D}_\text{Bos}(\gamma_v\gamma_{v+ \hat{x}_i}),\mathsf{D}_\text{Bos}(\gamma_{v-\hat{x}_i}\gamma_v)\}=0&,\quad \forall\  0\leq i\leq d \label{D_commutation_i}\\
 \mathsf{D}_\text{Bos}(P_v)=\mathsf{P}\qty(\prod_{e\ni v} Z_e )\mathsf{P}&.
 \end{align}
 Here, $\mathsf{P} = \mathsf{D}_\text{Bos}\mathsf{D}_\text{Bos}^\dagger$ is the projector onto the image of the isometry $\mathsf{D}_\text{Bos}$.
 We will discuss the structure of $\mathsf{P}$ later.
 
A 2D bosonization map has been proposed in Ref.~\cite{Chen_2018,Chen_2023,Tantivasadakarn_2024,O_Brien_2025}.
This map can also be expressed as a fermionic tensor network (compare Ref.~\cite{Shukla_2020}),
\begin{align}
\mathsf{D}_\text{Bos}=
\begin{tikzpicture}[baseline={([yshift=-2.5pt]current bounding box.center)}]
\def\r{0.4}
\def\s{0.3}
\def\d{1.5}
\def\w{2}
\def\wp{1}
\def\l{0.3}
\def\lw{1}
\def\m{0.39}
\def\ticklength{0.17}
\foreach \i in {0,...,\w}{\draw[decoration={markings, mark=at position 0.6 with {\arrow{>}}}, postaction={decorate},line width=0.6pt] (\i*\d, 0)-- (\i*\d, 0.5*\d) ;
}
\foreach \i in {0,...,\wp}
{\draw[decoration={markings, mark=at position 0.6 with {\arrow{>}}}, postaction={decorate},line width=0.6pt] (-0.5*\d, -\i*\d)-- (0, -\i*\d) ;
}
\foreach \i in {0,...,\wp} { 
 \foreach \j in {0,...,\wp} {
  \draw[decoration={markings, mark=at position 0.6 with {\arrow{>}}}, postaction={decorate},line width=0.6pt] (-\l*\d+\j*\d, -\l*\d-\i*\d)--(\j*\d, -\i*\d);
  \draw[tickcolor,line width=\lw] (\j*\d,-\i*\d) -- ++(0.7*\ticklength*\d,0.7*\ticklength*\d);
        \draw[decoration={markings, mark=at position 0.6 with {\arrow{>}}}, postaction={decorate},line width=0.6pt] (\j*\d, -0.5*\d-\i*\d)--(\j*\d, -\i*\d);
        \draw[decoration={markings, mark=at position 0.6 with {\arrow{>}}}, postaction={decorate},line width=0.6pt] (\j*\d, -1*\d-\i*\d)--(\j*\d, -0.5*\d-\i*\d);
        \draw[decoration={markings, mark=at position 0.6 with {\arrow{>}}}, postaction={decorate},line width=0.6pt]  (\j*\d+0.5*\d, -\i*\d)--(\j*\d+\d, -\i*\d);
        \draw[decoration={markings, mark=at position 0.6 with {\arrow{>}}}, postaction={decorate},line width=0.6pt] (\j*\d, -\i*\d)-- (\j*\d+0.5*\d, -\i*\d) ;
        \filldraw[fill=Xcolor, draw=black] (\j*\d,-\i*\d) circle (0.3*\r);
           \draw[dash pattern=on 2.3pt off 1.9pt] (\j*\d+0.5*\d,-\i*\d)--(\j*\d+0.5*\d,-\i*\d+\m*\d);
           \draw[tickcolor,line width=\lw] (\j*\d+0.5*\d,-\i*\d) -- ++(0.7*\ticklength*\d,0.7*\ticklength*\d);
        \filldraw[fill=Zcolor, draw=black] (\j*\d+0.5*\d,-\i*\d) circle (0.3*\r);
         \draw[dash pattern=on 2.3pt off 1.9pt] (\j*\d,-\i*\d-0.5*\d)--(\j*\d+\m*\d,-\i*\d-0.5*\d);
         \draw[tickcolor,line width=\lw](\j*\d,-\i*\d-0.5*\d)  -- ++(0.7*\ticklength*\d,0.7*\ticklength*\d);
        \filldraw[fill=Zcolor, draw=black] (\j*\d,-\i*\d-0.5*\d) circle (0.3*\r);
    }
}
\foreach \i in {0,...,\wp}{\draw[decoration={markings, mark=at position 0.6 with {\arrow{>}}}, postaction={decorate},line width=0.6pt] (\w*\d, -\i*\d-\d)-- (\w*\d, -\i*\d-0.5*\d) ;
 \draw[decoration={markings, mark=at position 0.6 with {\arrow{>}}}, postaction={decorate},line width=0.6pt] (-\l*\d+\w*\d, -\l*\d-\i*\d)--(\w*\d, -\i*\d);
 \draw[tickcolor,line width=\lw] (\w*\d,-\i*\d) -- ++(0.7*\ticklength*\d,0.7*\ticklength*\d);
}
\foreach \i in {0,...,\wp}{\draw[decoration={markings, mark=at position 0.6 with {\arrow{>}}}, postaction={decorate},line width=0.6pt] (\w*\d, -\i*\d-0.5*\d)-- (\w*\d, -\i*\d) ;
        \draw[dash pattern=on 2.3pt off 1.9pt] (\w*\d,-\i*\d-0.5*\d)--(\w*\d+\m*\d,-\i*\d-0.5*\d);
    \draw[tickcolor,line width=\lw](\w*\d, -\i*\d-0.5*\d)  -- ++(0.7*\ticklength*\d,0.7*\ticklength*\d);
\filldraw[fill=Zcolor, draw=black] (\w*\d, -\i*\d-0.5*\d) circle (0.3*\r);
}
\foreach \i in {0,...,\wp}{\draw[decoration={markings, mark=at position 0.6 with {\arrow{>}}}, postaction={decorate},line width=0.6pt] (\w*\d, -\i*\d)-- (\w*\d+0.5*\d, -\i*\d) ;
\filldraw[fill=Xcolor, draw=black] (\w*\d, -\i*\d) circle (0.3*\r);
}
\foreach \i in {0,...,\w}{\draw[decoration={markings, mark=at position 0.6 with {\arrow{>}}}, postaction={decorate},line width=0.6pt] (\i*\d, -\w*\d)-- (\i*\d+0.5*\d, -\w*\d) ;
\draw[decoration={markings, mark=at position 0.6 with {\arrow{>}}}, postaction={decorate},line width=0.6pt] (\i*\d, -\w*\d-0.5*\d)-- (\i*\d, -\w*\d) ;
}
\foreach \i in {0,...,\w}{\draw[decoration={markings, mark=at position 0.6 with {\arrow{>}}}, postaction={decorate},line width=0.6pt] (\i*\d-0.5*\d, -\w*\d)-- (\i*\d, -\w*\d) ;
 \draw[decoration={markings, mark=at position 0.6 with {\arrow{>}}}, postaction={decorate},line width=0.6pt] (-\l*\d+\i*\d, -\l*\d-\w*\d)--(\i*\d, -\w*\d);
\draw[tickcolor,line width=\lw] (\i*\d,-\w*\d) -- ++(0.7*\ticklength*\d,0.7*\ticklength*\d);
\filldraw[fill=Xcolor, draw=black] (\i*\d, -\w*\d) circle (0.3*\r);
}
\foreach \i in {0,...,\wp}{
 \draw[dash pattern=on 2.3pt off 1.9pt] (\i*\d+0.5*\d,-\w*\d)--(\i*\d+0.5*\d,-\w*\d+\m*\d);
 \draw[tickcolor,line width=\lw](\i*\d+0.5*\d, -\w*\d) -- ++(0.7*\ticklength*\d,0.7*\ticklength*\d);
\filldraw[fill=Zcolor, draw=black] (\i*\d+0.5*\d, -\w*\d) circle (0.3*\r);
}
\begin{scope}[shift={(11*\r,-\d)}]
\draw[->](0,0)--(\s*\d,0);
\node at (1.3*\d*\s,0) {$x$};
\draw[->](0,0)--(0,\s*\d);
\node at (0,1.4*\s*\d) {$y$};
\end{scope}
\end{tikzpicture}
\end{align}
The bosonization map acts on pairs of Majorana operators as follows:
\begin{align}
    \mathsf{D}_{\Bos}(\gamma_{v-\hat{x}}\gamma_{v})=&\mathsf{P}\qty( Z_{\partial_d (v-\hat{x})} (-iY)_{\partial_l v} Z_{\partial_u v}Z_{\partial_rv} Z_{\partial_d v} ) \mathsf{P}\\
\mathsf{D}_{\Bos} (\gamma_{v}\gamma_{v+\hat{y}})=&\mathsf{P}\bigg((-iY)_{\partial_u v} Z_{\partial_r v}Z_{\partial_d v} \bigg) \mathsf{P}\\
    \mathsf{D}_{\Bos} (P_v)=&\mathsf{P}\bigg(\prod_{e\ni v} Z_e\bigg)\mathsf{P},
\end{align}
where we use $\partial _{l,r,u,d}v$ to denote the left, right, upper, and lower edges adjacent to vertex $v$, respectively.
We can derive this purely diagrammatically using our graphical calculus:
\phantomsection\label{proof_2D_bosonization}
\begin{align}
&
\begin{tikzpicture}[baseline={([yshift=-2.5pt]current bounding box.center)}]
\def\r{1}
\def\lw{1}
\def\ZXr{0.15}
\def\x{2}
\def\y{0.95}
\def\yy{1.2}
\def\xx{0.95}
\def\vx{0.5}
\def\vy{0.65}
\def\rela{1.1}
\def\ticklength{0.4}
\draw\midarrow{0.5} (-\xx*\r,0)--(0,0);
\draw\midarrow{0.55} (0,0)--(0.5*\x*\r,0);
\draw\midarrow{0.55} (0.5*\x*\r,0)--(\x*\r,0);
\draw\midarrow{0.6} (\x*\r,0)--++(\xx*\r,0);
\draw[tickcolor,line width=\lw] (0.5*\x*\r,0) -- ++(.6*\ticklength*\r,0.6*\ticklength*\r);
 \draw (-\vx*\r,-\vx*\r) .. controls (-\vx*\r+0.3*\r,-\vx*\r-0.3*\r) and  (-\vx*\r+0.2*\x*\r,-\vx*\r-0.5*\r) .. (-\vx*\r+0.5*\x*\r,-\vx*\r-0.5*\r);
 \draw(-\vx*\r+0.5*\x*\r,-\vx*\r-0.5*\r) .. controls (-\vx*\r+0.5*\x*\r+1.2*\r,-\vx*\r-0.5*\r)and (-\vx*\r+\x*\r+0.4*\r,-\vx*\r-0.4*\r) ..  (-\vx*\r+\x*\r,-\vx*\r);
\draw\dashedpattern (-\vx*\r+0.5*\x*\r,-\vx*\r-0.5*\r)--++(0,-0.7*\r);
\draw[tickcolor,line width=\lw] (-\vx*\r+0.5*\x*\r,-\vx*\r-0.5*\r)  -- ++(0*\ticklength*\r,0.8*\ticklength*\r);
\filldraw[fill=Zcolor, draw=black] (-\vx*\r+0.5*\x*\r,-\vx*\r-0.5*\r) circle (\ZXr*\r);
\filldraw[fill=white, draw=white] (-\vx*\r+0.5*\x*\r,\vx*\r+0.5*\r+0.7*\r) circle (\ZXr*\r);
\filldraw[fill=Xcolor, draw=black] (-\vx*\r+0.5*\x*\r,-\vx*\r-0.5*\r-0.7*\r) circle (\ZXr*\r);
\node at (-\vx*\r+0.5*\x*\r,-\vx*\r-0.5*\r-0.7*\r) {$_\pi$};
\foreach \i in {0,1}
{
\draw[tickcolor,line width=\lw] (\i*\x*\r,0) -- ++(.6*\ticklength*\r,0.6*\ticklength*\r);
\draw[tickcolor,line width=\lw] (\i*\x*\r-\vx*\r,-\vx*\r) -- ++(0,-0.7*\ticklength*\r);
\draw\midarrow{0.5} (\i*\x*\r,-\yy*\r)--(\i*\x*\r,0);
\draw\midarrow{0.6} (\i*\x*\r,0)--(\i*\x*\r,\y*\r);
\draw\midarrow{0.6} (\i*\x*\r-\vx*\r,-\vx*\r)--(\i*\x*\r,0);
\draw\midarrow{0.5} (\i*\x*\r-\vy*\r-\vx*\r,-\vx*\r-\vy*\r)--(\i*\x*\r-\vx*\r,-\vx*\r);
\filldraw[fill=Xcolor, draw=black] (\i*\x*\r-\vx*\r,-\vx*\r) circle (\ZXr*\r);
}
\draw\dashedpattern(0.5*\x*\r,0)--++(0,\y*\r);
\filldraw[fill=Zcolor, draw=black] (0.5*\x*\r,0) circle (\ZXr*\r);
\filldraw[fill=Xcolor, draw=black] (0,0) circle (\ZXr*\r);
\filldraw[fill=Xcolor, draw=black] (\x*\r,0) circle (\ZXr*\r);
\end{tikzpicture}
\quad =\quad 
\begin{tikzpicture}[baseline={([yshift=-2.5pt]current bounding box.center)}]
\def\r{1}
\def\lw{1}
\def\ZXr{0.15}
\def\x{2}
\def\y{0.95}
\def\yy{1.2}
\def\xx{0.95}
\def\vx{0.5}
\def\vy{0.65}
\def\dotr{0.06}
\def\rela{1.1}
\def\ticklength{0.4}
\draw(-\xx*\r,0)--(0,0);
\draw(0,0)--(0.5*\x*\r,0);
\draw(0.5*\x*\r,0)--(\x*\r,0);
\draw (\x*\r,0)--++(\xx*\r,0);
\draw[tickcolor,line width=\lw] (0.5*\x*\r,0)  -- ++(0.6*\ticklength*\r,0.6*\ticklength*\r);
 \draw (-\vx*\r,-\vx*\r) .. controls (-\vx*\r+0.3*\r,-\vx*\r-0.3*\r) and  (-\vx*\r+0.2*\x*\r,-\vx*\r-0.5*\r) .. (-\vx*\r+0.5*\x*\r,-\vx*\r-0.5*\r);
 \draw(-\vx*\r+0.5*\x*\r,-\vx*\r-0.5*\r) .. controls (-\vx*\r+0.5*\x*\r+1.2*\r,-\vx*\r-0.5*\r)and (-\vx*\r+\x*\r+0.4*\r,-\vx*\r-0.4*\r) ..  (-\vx*\r+\x*\r,-\vx*\r);
\draw\dashedpattern (-\vx*\r+0.5*\x*\r,-\vx*\r-0.5*\r)--++(0,-0.7*\r);
\draw[tickcolor,line width=\lw] (-\vx*\r+0.5*\x*\r,-\vx*\r-0.5*\r)  -- ++(0*\ticklength*\r,0.8*\ticklength*\r);
\filldraw[fill=Zcolor, draw=black] (-\vx*\r+0.5*\x*\r,-\vx*\r-0.5*\r) circle (\ZXr*\r);
\filldraw[fill=white, draw=white] (-\vx*\r+0.5*\x*\r,\vx*\r+0.5*\r+0.7*\r) circle (\ZXr*\r);
\filldraw[fill=Xcolor, draw=black] (-\vx*\r+0.5*\x*\r,-\vx*\r-0.5*\r-0.7*\r) circle (\ZXr*\r);
\node at (-\vx*\r+0.5*\x*\r,-\vx*\r-0.5*\r-0.7*\r) {$_\pi$};
\foreach \i in {0,1}
{
\draw[tickcolor,line width=\lw] (\i*\x*\r,0) -- ++(-0.2*\ticklength*\r,-0.73*\ticklength*\r);
\draw[tickcolor,line width=\lw] (\i*\x*\r-\vx*\r,-\vx*\r) -- ++(0.7*\ticklength*\r,0);
\draw (\i*\x*\r,-\yy*\r)--(\i*\x*\r,0);
\draw(\i*\x*\r,0)--(\i*\x*\r,\y*\r);
\draw (\i*\x*\r-\vx*\r,-\vx*\r)--(\i*\x*\r,0);
\draw (\i*\x*\r-\vy*\r-\vx*\r,-\vx*\r-\vy*\r)--(\i*\x*\r-\vx*\r,-\vx*\r);
\filldraw[fill=Xcolor, draw=black] (\i*\x*\r-\vx*\r,-\vx*\r) circle (\ZXr*\r);
\filldraw[fill=black, draw=black] (\i*\x*\r+0.5*\r,0) circle (\dotr*\r);
\filldraw[fill=black, draw=black] (\i*\x*\r,-0.5*\r) circle (\dotr*\r);
}
\draw\dashedpattern(0.5*\x*\r,0)--++(0,\y*\r);
\filldraw[fill=Zcolor, draw=black] (0.5*\x*\r,0) circle (\ZXr*\r);
\filldraw[fill=Xcolor, draw=black] (0,0) circle (\ZXr*\r);
\filldraw[fill=Xcolor, draw=black] (\x*\r,0) circle (\ZXr*\r);
\filldraw[fill=black, draw=black] (-\vx*\r+0.3*\r,-\vx*\r-0.3*\r) circle (\dotr*\r);
\filldraw[fill=black, draw=black] (-\vx*\r+\x*\r,-\vx*\r-0.42*\r) circle (\dotr*\r);
\end{tikzpicture}
\quad =\quad 
\begin{tikzpicture}[baseline={([yshift=-2.5pt]current bounding box.center)}]
\def\r{1}
\def\lw{1}
\def\ZXr{0.15}
\def\x{2}
\def\y{0.95}
\def\yy{1.2}
\def\xx{0.95}
\def\vx{0.5}
\def\vy{0.65}
\def\dotr{0.06}
\def\rela{1.1}
\def\ticklength{0.4}
\draw(-\xx*\r,0)--(0,0);
\draw(0,0)--(0.5*\x*\r,0);
\draw(0.5*\x*\r,0)--(\x*\r,0);
\draw (\x*\r,0)--++(\xx*\r,0);
 \draw (0,0) .. controls (-\vx*\r,-\vx*\r-0.3*\r) and  (-\vx*\r+0.2*\x*\r,-\vx*\r-0.5*\r) .. (-\vx*\r+0.5*\x*\r,-\vx*\r-0.5*\r);
 \draw(-\vx*\r+0.5*\x*\r,-\vx*\r-0.5*\r) .. controls (-\vx*\r+0.5*\x*\r+1.2*\r,-\vx*\r-0.5*\r)and(\x*\r-0.3*\r,-0.5*\r) ..  (\x*\r,0);
\draw\dashedpattern (-\vx*\r+0.5*\x*\r,-\vx*\r-0.5*\r)--++(0,-0.7*\r);
\draw[tickcolor,line width=\lw] (0.5*\x*\r,0)  -- ++(0.6*\ticklength*\r,0.6*\ticklength*\r);
\draw[tickcolor,line width=\lw] (-\vx*\r+0.5*\x*\r,-\vx*\r-0.5*\r)  -- ++(0*\ticklength*\r,0.8*\ticklength*\r);
\filldraw[fill=Zcolor, draw=black] (-\vx*\r+0.5*\x*\r,-\vx*\r-0.5*\r) circle (\ZXr*\r);
\filldraw[fill=white, draw=white] (-\vx*\r+0.5*\x*\r,\vx*\r+0.5*\r+0.7*\r) circle (\ZXr*\r);
\filldraw[fill=Xcolor, draw=black] (-\vx*\r+0.5*\x*\r,-\vx*\r-0.5*\r-0.7*\r) circle (\ZXr*\r);
\node at (-\vx*\r+0.5*\x*\r,-\vx*\r-0.5*\r-0.7*\r) {$_\pi$};
\foreach \i in {0,1}
{
\draw[tickcolor,line width=\lw] (\i*\x*\r,0) -- ++(0.6*\ticklength*\r,0.6*\ticklength*\r);
\draw (\i*\x*\r,-\yy*\r)--(\i*\x*\r,0);
\draw(\i*\x*\r,0)--(\i*\x*\r,\y*\r);
\draw (\i*\x*\r-\vx*\r,-\vx*\r)--(\i*\x*\r,0);
\draw (\i*\x*\r-\vy*\r-\vx*\r,-\vx*\r-\vy*\r)--(\i*\x*\r-\vx*\r,-\vx*\r);
}
\draw\dashedpattern(0.5*\x*\r,0)--++(0,\y*\r);
\filldraw[fill=Zcolor, draw=black] (0.5*\x*\r,0) circle (\ZXr*\r);
\filldraw[fill=Xcolor, draw=black] (0,0) circle (\ZXr*\r);
\filldraw[fill=Xcolor, draw=black] (\x*\r,0) circle (\ZXr*\r);
\filldraw[fill=black, draw=black] (-\vx*\r+0.31*\r,-\vx*\r-0.26*\r) circle (\dotr*\r);
\filldraw[fill=black, draw=black] (-\vx*\r+\x*\r,-\vx*\r-0.27*\r) circle (\dotr*\r);
\end{tikzpicture}
\nonumber \\
=&\quad 
\begin{tikzpicture}[baseline={([yshift=-2.5pt]current bounding box.center)}]
\def\r{1}
\def\lw{1}
\def\ZXr{0.15}
\def\x{2}
\def\y{0.95}
\def\yy{1.2}
\def\xx{0.95}
\def\vx{0.8}
\def\vy{0}
\def\dotr{0.06}
\def\rela{1.1}
\def\Delta{0.7}
\def\ticklength{0.4}
\draw[white] (0,0)--(0,\yy*\r);
\draw[white] (0,0)--(0,-\y*\r);
\draw(-\xx*\r,0)--(0,0);
\draw(0,0)--(0.5*\x*\r,0);
\draw(0.5*\x*\r,0)--(\x*\r,0);
\draw (\x*\r,0)--++(\xx*\r,0);
 \draw (0,0) .. controls (0.3*\r,-0.6*\r) and  (\x*\r-0.7*\r,-0.6*\r) .. (\x*\r,0);
\draw\dashedpattern (0.5*\x*\r-0.2*\r,-0.47*\r)--++(0,-\Delta*\r);
\draw[tickcolor,line width=\lw](0.5*\x*\r,0)  -- ++(0.6*\ticklength*\r,0.6*\ticklength*\r);
\draw[tickcolor,line width=\lw](0.5*\x*\r-0.2*\r,-0.47*\r)  -- ++(0*\ticklength*\r,0.7*\ticklength*\r);
\filldraw[fill=Zcolor, draw=black]  (0.5*\x*\r-0.2*\r,-0.47*\r) circle (\ZXr*\r);
\filldraw[fill=white, draw=white]  (0.5*\x*\r-0.2*\r,+0.47*\r+\Delta*\r) circle (\ZXr*\r);
\filldraw[fill=Xcolor, draw=black]  (0.5*\x*\r-0.2*\r,-0.47*\r-\Delta*\r) circle (\ZXr*\r);
\node at (0.5*\x*\r-0.2*\r,-0.47*\r-\Delta*\r)  {$_\pi$};
\foreach \i in {0,1}
{
\draw[tickcolor,line width=\lw] (\i*\x*\r,0) -- ++(0.6*\ticklength*\r,0.6*\ticklength*\r);
\draw (\i*\x*\r,-\yy*\r)--(\i*\x*\r,0);
\draw(\i*\x*\r,0)--(\i*\x*\r,\y*\r);
\draw (\i*\x*\r-\vx*\r,-\vx*\r)--(\i*\x*\r,0);
\draw (\i*\x*\r-\vy*\r-\vx*\r,-\vx*\r-\vy*\r)--(\i*\x*\r-\vx*\r,-\vx*\r);
}
\draw\dashedpattern(0.5*\x*\r,0)--++(0,\y*\r);
\filldraw[fill=Zcolor, draw=black] (0.5*\x*\r,0) circle (\ZXr*\r);
\filldraw[fill=Xcolor, draw=black] (0,0) circle (\ZXr*\r);
\filldraw[fill=Xcolor, draw=black] (\x*\r,0) circle (\ZXr*\r);
\filldraw[fill=black, draw=black] (0.4*\r,-0.36*\r) circle (\dotr*\r);
\filldraw[fill=black, draw=black] (1.27*\r,-0.4*\r) circle (\dotr*\r);
\filldraw[fill=black, draw=black] (0,-0.7*\r) circle (\dotr*\r);
\filldraw[fill=black, draw=black] (-0.7*\vx*\r+\x*\r,-0.7*\vx*\r) circle (\dotr*\r);
\end{tikzpicture}
\quad =\quad 
\begin{tikzpicture}[baseline={([yshift=-2.5pt]current bounding box.center)}]
\def\r{1}
\def\lw{1}
\def\ZXr{0.15}
\def\x{2}
\def\y{1.2}
\def\yy{0.9}
\def\xx{0.95}
\def\vx{0.7}
\def\vy{0}
\def\dotr{0.06}
\def\rela{1.1}
\def\Delta{0.7}
\def\ticklength{0.4}
\draw[white] (0,0)--(0,\yy*\r);
\draw[white] (0,0)--(0,-\y*\r);
\draw(-\xx*\r,0)--(0,0);
\draw(0,0)--(0.5*\x*\r,0);
\draw(0.5*\x*\r,0)--(\x*\r,0);
\draw (\x*\r,0)--++(\xx*\r,0);
\draw[tickcolor,line width=\lw](0.5*\x*\r,0)  -- ++(0.6*\ticklength*\r,0.6*\ticklength*\r);
\foreach \i in {0,1}
{
\draw[tickcolor,line width=\lw] (\i*\x*\r,0) -- ++(0.6*\ticklength*\r,0.6*\ticklength*\r);
\draw (\i*\x*\r,-\yy*\r)--(\i*\x*\r,0);
\draw(\i*\x*\r,0)--(\i*\x*\r,\y*\r);
\draw (\i*\x*\r-\vx*\r,-\vx*\r)--(\i*\x*\r,0);
\draw (\i*\x*\r-\vy*\r-\vx*\r,-\vx*\r-\vy*\r)--(\i*\x*\r-\vx*\r,-\vx*\r);
}
\draw\dashedpattern(0.5*\x*\r,0)--++(0,\y*\r);
\filldraw[fill=Zcolor, draw=black] (0.5*\x*\r,0) circle (\ZXr*\r);
\filldraw[fill=Xcolor, draw=black] (0,0) circle (\ZXr*\r);
\filldraw[fill=Xcolor, draw=black]  (\x*\r,0) circle (\ZXr*\r);
\filldraw[fill=black, draw=black] (0,-0.5*\r) circle (\dotr*\r);
\filldraw[fill=black, draw=black] (-0.55*\vx*\r+\x*\r,-0.55*\vx*\r) circle (\dotr*\r);
\filldraw[fill=Xcolor, draw=black]  (0.5*\x*\r,0.5*\y*\r) circle (\ZXr*\r);
\node at (0.5*\x*\r,0.5*\y*\r)  {$_\pi$};
\end{tikzpicture}
\quad =\quad 
\begin{tikzpicture}[baseline={([yshift=-2.5pt]current bounding box.center)}]
\def\r{1}
\def\lw{1}
\def\ZXr{0.15}
\def\x{2}
\def\y{1.3}
\def\yy{0.9}
\def\xx{0.95}
\def\vx{0.7}
\def\vy{0}
\def\dotr{0.06}
\def\rela{1.1}
\def\Delta{0.7}
\def\ticklength{0.4}
\draw[white] (0,0)--(0,\yy*\r);
\draw[white] (0,0)--(0,-\y*\r);
\draw(-\xx*\r,0)--(0,0);
\draw(0,0)--(0.5*\x*\r,0);
\draw(0.5*\x*\r,0)--(\x*\r,0);
\draw (\x*\r,0)--++(\xx*\r,0);
\draw[tickcolor,line width=\lw](0.5*\x*\r,0)  -- ++(0.6*\ticklength*\r,0.6*\ticklength*\r);
\foreach \i in {0,1}
{
\draw[tickcolor,line width=\lw] (\i*\x*\r,0) -- ++(0.6*\ticklength*\r,0.6*\ticklength*\r);
\draw (\i*\x*\r,-\yy*\r)--(\i*\x*\r,0);
\draw(\i*\x*\r,0)--(\i*\x*\r,\y*\r);
\draw (\i*\x*\r-\vx*\r,-\vx*\r)--(\i*\x*\r,0);
\draw (\i*\x*\r-\vy*\r-\vx*\r,-\vx*\r-\vy*\r)--(\i*\x*\r-\vx*\r,-\vx*\r);
}
\draw\dashedpattern(0.5*\x*\r,0)--++(0,\y*\r);
\filldraw[fill=Zcolor, draw=black] (0.5*\x*\r,0) circle (\ZXr*\r);
\filldraw[fill=Xcolor, draw=black] (0,0) circle (\ZXr*\r);
\filldraw[fill=Xcolor, draw=black]  (\x*\r,0) circle (\ZXr*\r);
\filldraw[fill=black, draw=black] (0,-0.5*\r) circle (\dotr*\r);
\filldraw[fill=black, draw=black] (\x*\r,-0.5*\r) circle (\dotr*\r);
\filldraw[fill=black, draw=black] (\x*\r,0.5*\r)circle (\dotr*\r);
\filldraw[fill=black, draw=black] (\x*\r+0.4*\r,0) circle (\dotr*\r);
\filldraw[fill=Xcolor, draw=black]  (0.5*\x*\r,0.75*\y*\r) circle (\ZXr*\r);
\node at (0.5*\x*\r,0.75*\y*\r)  {$_\pi$};
\filldraw[fill=Zcolor, draw=black]  (0.5*\x*\r,0.375*\y*\r) circle (\ZXr*\r);
\node at (0.5*\x*\r,0.375*\y*\r)  {$_\pi$};
\end{tikzpicture},
\end{align}
where in the first step we replace the odd-parity line between a pair of Majorana operators with a solid line and an $X$-spider connected to a qubit state $\ket{1}$; in the second step we switch the directions of the four $X$ spiders and introduce a pair of fermion-parity operators denoted by black dots; in the third picture we fuse two pairs of $X$-spiders and switch the directions of their ticks back; in the fourth picture the lines are detoured, but since the curved black line represents fermions in the state $\ket{1}$, it induces fermion-parity operators measuring the two straight lines; in the fifth picture we split the two $X$-spiders into four and then apply the Bialgebra rule; in the last step the fermion-parity operator on the input leg of the right $X$-spider is pushed through to become four fermion parities, all of which are absorbed by the edge $Z$-spider into a qubit $Z$ gate (only one such absorption is shown because we did not include the three other edge $Z$-spiders in the expression).
The calculation to map a pair of vertically aligned Majorana operators can be executed similarly.

A 3D bosonization map was proposed in Ref.~\cite{Chen_2019}.
It maps fermionic modes on the vertices of the cubic lattice to qubits on the edges of the lattice.
This map can also be represented as a tensor-network diagram, a snippet of which looks as follows: 
\begin{align}
\mathsf{D}_\text{Bos}=
\begin{tikzpicture}[baseline={([yshift=-2.5pt]current bounding box.center)}]   
\def\r{0.5}
\def\x{4*\r}
\def\s{0.25}
\def\dashedlength{0.5}
\def\vx{2*\r}
\def\vy{1.6*\r}
\def\nx{0.51*\r}
\def\ny{0.51*\r}
\def\extra{0.9*\r}
\def\sextra{0.6*\r}
\def\lw{1}
\def\llw{0.65}
\def\ticklength{0.5*\r}
  \def\ZXr{0.25}
\draw(-\extra,0)--++(\x+2*\extra,0);
\draw\midarrow{0.5}(0.25*\x,0)--++(0.01*\x,0);
\draw\midarrow{0.5}(0.75*\x,0)--++(0.01*\x,0);
\draw(-\extra,\x)--++(\x+2*\extra,0);
\draw\midarrow{0.5}(0.25*\x,\x)--++(0.01*\x,0);
\draw\midarrow{0.5}(0.75*\x,\x)--++(0.01*\x,0);
\draw(-\extra+\vx,-\vy)--++(\x+2*\extra,0);
\draw\midarrow{0.5}(0.25*\x+\vx,-\vy)--++(0.01*\x,0);
\draw\midarrow{0.5}(0.75*\x+\vx,-\vy)--++(0.01*\x,0);
\draw(0,-\extra)--++(0,\x+2*\extra);
\draw\midarrow{0.5}(0,0.25*\x)--++(0,0.01*\x);
\draw\midarrow{0.5}(0,0.75*\x)--++(0,0.01*\x);
\draw(\x,-\extra)--++(0,\x+2*\extra);
\draw\midarrow{0.5}(\x,0.25*\x)--++(0,0.01*\x);
\draw\midarrow{0.5}(\x,0.75*\x)--++(0,0.01*\x);
\draw(\x+\vx,-\extra-\vy)--++(0,\x+2*\extra);
\draw\midarrow{0.5}(\x+\vx,0.25*\x-\vy)--++(0,0.01*\x);
\draw\midarrow{0.5}(\x+\vx,0.75*\x-\vy)--++(0,0.01*\x);
\draw(-\sextra*\vx,\sextra*\vy)--++(\vx+2*\sextra*\vx,-\vy-2*\sextra*\vy);
\draw\midarrow{0.5}(0.2*\vx,-0.2*\vy)--++(-0.01*\vx,0.01*\vy);
\draw\midarrow{0.5}(0.7*\vx,-0.7*\vy)--++(-0.01*\vx,0.01*\vy);
\draw(\x-\sextra*\vx,\sextra*\vy)--++(\vx+2*\sextra*\vx,-\vy-2*\sextra*\vy);
\draw\midarrow{0.5}(0.2*\vx+\x,-0.2*\vy)--++(-0.01*\vx,0.01*\vy);
\draw\midarrow{0.5}(0.7*\vx+\x,-0.7*\vy)--++(-0.01*\vx,0.01*\vy);
\draw(\x-\sextra*\vx,\x+\sextra*\vy)--++(\vx+2*\sextra*\vx,-\vy-2*\sextra*\vy);
\draw\midarrow{0.5}(0.2*\vx+\x,-0.2*\vy+\x)--++(-0.01*\vx,0.01*\vy);
\draw\midarrow{0.5}(0.7*\vx+\x,-0.7*\vy+\x)--++(-0.01*\vx,0.01*\vy);
\draw(\x+\vx-\extra,\x-\vy)--++(2*\extra,0);
\draw(\vx,-\extra-\vy)--++(0,2*\extra);
\draw(-\sextra*\vx,\sextra*\vy+\x)--++(2*\sextra*\vx,-2*\sextra*\vy);
\draw[line width=\llw](0,0)--++(-\nx,-\ny);
\draw[->,line width=\llw](-0.5*\nx,-0.5*\ny)--++(0.01*\nx,0.01*\ny);
\draw[line width=\llw](\x,0)--++(-\nx,-\ny);
\draw[->,line width=\llw](\x-0.5*\nx,-0.5*\ny)--++(0.01*\nx,0.01*\ny);
\draw[line width=\llw](0,\x)--++(-\nx,-\ny);
\draw[->,line width=\llw](-0.5*\nx,\x-0.5*\ny)--++(0.01*\nx,0.01*\ny);
\draw[line width=\llw](\x,\x)--++(-\nx,-\ny);
\draw[->,line width=\llw](\x-0.5*\nx,\x-0.5*\ny)--++(0.01*\nx,0.01*\ny);
\draw[line width=\llw](\vx,-\vy)--++(-\nx,-\ny);
\draw[->,line width=\llw](\vx-0.5*\nx,-\vy-0.5*\ny)--++(0.01*\nx,0.01*\ny);
\draw[line width=\llw](\x+\vx,-\vy)--++(-\nx,-\ny);
\draw[->,line width=\llw](\x+\vx-0.5*\nx,-\vy-0.5*\ny)--++(0.01*\nx,0.01*\ny);
\draw[line width=\llw](\x+\vx,\x-\vy)--++(-\nx,-\ny);
\draw[->,line width=\llw](\x+\vx-0.5*\nx, \x-\vy-0.5*\ny)--++(0.01*\nx,0.01*\ny);
\draw[tickcolor,line width=\lw](0.5*\x,0)--++(0.707*\ticklength,0.707*\ticklength);
\draw\dashedpattern(0.5*\x,0)--++(-\nx,-\ny);
\filldraw[fill=Zcolor,draw=black](0.5*\x,0)circle (\ZXr*\r); 
\draw[tickcolor,line width=\lw](0,0.5*\x)--++(0.707*\ticklength,0.707*\ticklength);
\draw\dashedpattern(0,0.5*\x)--++(-\nx,-\ny);
\filldraw[fill=Zcolor,draw=black](0,0.5*\x)circle (\ZXr*\r); 
\draw[tickcolor,line width=\lw](0.5*\x,\x)--++(0.707*\ticklength,0.707*\ticklength);
\draw\dashedpattern(0.5*\x,\x)--++(-\nx,-\ny);
\filldraw[fill=Zcolor,draw=black](0.5*\x,\x)circle (\ZXr*\r); 
\draw[tickcolor,line width=\lw](\x,0.5*\x)--++(0.707*\ticklength,0.707*\ticklength);
\draw\dashedpattern(\x,0.5*\x)--++(-\nx,-\ny);
\filldraw[fill=Zcolor,draw=black](\x,0.5*\x)circle (\ZXr*\r); 
\draw[tickcolor,line width=\lw](0.5*\x+\vx,-\vy)--++(0.707*\ticklength,0.707*\ticklength);
\draw\dashedpattern(0.5*\x+\vx,-\vy)--++(-\nx,-\ny);
\filldraw[fill=Zcolor,draw=black](0.5*\x+\vx,-\vy)circle (\ZXr*\r); 
\draw[tickcolor,line width=\lw](0.5*\vx,-0.5*\vy)--++(0.707*\ticklength,0.707*\ticklength);
\draw\dashedpattern(0.5*\vx,-0.5*\vy)--++(-\nx,-\ny);
\filldraw[fill=Zcolor,draw=black](0.5*\vx,-0.5*\vy)circle (\ZXr*\r); 
\draw[tickcolor,line width=\lw](\x+0.5*\vx,-0.5*\vy)--++(0.707*\ticklength,0.707*\ticklength);
\draw\dashedpattern(\x+0.5*\vx,-0.5*\vy)--++(-\nx,-\ny);
\filldraw[fill=Zcolor,draw=black](\x+0.5*\vx,-0.5*\vy)circle (\ZXr*\r); 
\draw[tickcolor,line width=\lw](\x+0.5*\vx,\x-0.5*\vy)--++(0.707*\ticklength,0.707*\ticklength);
\draw\dashedpattern(\x+0.5*\vx,\x-0.5*\vy)--++(-\nx,-\ny);
\filldraw[fill=Zcolor,draw=black](\x+0.5*\vx,\x-0.5*\vy)circle (\ZXr*\r); 
\draw[tickcolor,line width=\lw](\x+\vx,0.5*\x-\vy)--++(0.707*\ticklength,0.707*\ticklength);
\draw\dashedpattern(\x+\vx,0.5*\x-\vy)--++(-\nx,-\ny);
\filldraw[fill=Zcolor,draw=black](\x+\vx,0.5*\x-\vy)circle (\ZXr*\r); 
\foreach \i in {0,1} {
\foreach \j in {0,1} {
\draw[tickcolor,line width=\lw](\i*\x,\j*\x)--++(0.707*\ticklength,0.707*\ticklength);
\filldraw[fill=Xcolor,draw=black](\i*\x,\j*\x)circle (\ZXr*\r); 
}}
\draw[tickcolor,line width=\lw](\vx,-\vy)--++(0.707*\ticklength,0.707*\ticklength);
\filldraw[fill=Xcolor,draw=black](\vx,-\vy)circle (\ZXr*\r); 
\draw[tickcolor,line width=\lw](\x+\vx,-\vy)--++(0.707*\ticklength,0.707*\ticklength);
\filldraw[fill=Xcolor,draw=black](\x+\vx,-\vy)circle (\ZXr*\r); 
\draw[tickcolor,line width=\lw](\x+\vx,\x-\vy)--++(0.707*\ticklength,0.707*\ticklength);
\filldraw[fill=Xcolor,draw=black](\x+\vx,\x-\vy) circle (\ZXr*\r); 
\begin{scope}[shift={(9*\r,0)}]
\draw[->](0,0)--(\s*\x,0);
\node at (1.3*\x*\s,0) {$x$};
\draw[->](0,0)--(0,\s*\x);
\node at (0,1.35*\s*\x) {$z$};
\draw[->](0,0)--(-\s*\vx,\s*\vy);
\node at (-1.4*\s*\vx,1.4*\s*\vy) {$y$};
\end{scope}
\end{tikzpicture}
\;,
\label{eq_bosonization_3D}
\end{align}
where the thick solid lines and the dashed lines are the fermionic inputs and bosonic outputs, respectively.
Define $v_1:=v-\hat{x}$, $v_2:=v-\hat{y}$, $v_3:=v-\hat{z}$. 
The map acts on pairs of neighboring Majorana operators as follows:
\begin{align}
    \mathsf{D}_\text{Bos}(\gamma_{v_1}\gamma_v)=&\mathsf{P}\bigg( Z_{v_1-\frac{\hat{y}}{2}}Z_{v_1-\frac{\hat{z}}{2}}
    (-iY)_{v-\frac{\hat{x}}{2}} 
    Z_{v+\frac{\hat{x}}{2}}Z_{v+\frac{\hat{y}}{2}}Z_{v-\frac{\hat{y}}{2}}
    Z_{v+\frac{\hat{z}}{2}}Z_{v-\frac{\hat{z}}{2}}\bigg) \mathsf{P}
    \label{eq_Dbos_v1}\\
 \mathsf{D}_\text{Bos}(\gamma_{v_2}\gamma_v)    
 =&\mathsf{P}\bigg( Z_{v_2+\frac{\hat{x}}{2}}
 Z_{v_2-\frac{\hat{y}}{2}}
 Z_{v_2+\frac{\hat{z}}{2}}
 Z_{v_2-\frac{\hat{z}}{2}}
 (-iY)_{v-\frac{\hat{y}}{2}}
 Z_{v-\frac{\hat{z}}{2}}  \bigg) \mathsf{P}  \label{eq_Dbos_v2} \\
   \mathsf{D}_\text{Bos}(\gamma_{v_3}\gamma_v )
   =&\mathsf{P} \bigg( Z_{v_3+\frac{\hat{x}}{2}}Z_{v_3-\frac{\hat{y}}{2}}
   Z_{v_3-\frac{\hat{z}}{2}}(-iY)_{v-\frac{\hat{z}}{2}} \bigg)\mathsf{P} \label{eq_Dbos_v3} \\
   \mathsf{D}_\text{Bos}( P_v)=&\mathsf{P}\bigg(\prod_{e\ni v} Z_e \bigg) \mathsf{P}.
\end{align}
The proof is nearly identical to that of the 2D Bosonization, so we will omit it here.
\subsubsection{Bosonization in arbitrary dimension}
Now we construct a Bosonization tensor network for generic dimension $d$ based on our local ordering method (ticking), generalizing the previous specific cases.
The tensor network consists of one $2d+1$-index $X$-spider at each vertex of the $d$-dimensional cubic lattice, and one 3-index $Z$-spider at each edge.
The tick location and index orderings of each $X$-spider are determined as follows:
 Let $\{\hat{x}_i\}_{i=1}^d$ denote a set of basis vectors spanning the lattice $\mathbb{Z}^d$. 
 We furthermore assign sign $\pm$ to $x_1$ through $x_d$ based on the orientation of the lattice. 
 Further, imagine adding an orthogonal $d+1$st dimension whose basis vector we denote by $\hat{x}_0$, such that the input and output indices of the bosonization map point in the direction $\hat x_0$ or $-\hat x_0$. 
 For each pair of neighboring vertices $(v, v+\hat{x}_i)$ extended in the $i$-th dimension, we then name the edge $v+\hat{x}_i/2$ between them as $+\hat{x}_i^{(v)}$ and $-\hat{x}_i^{(v+\hat{x}_i)}$ with respect to vertices $v$ and $v+\hat{x}_i$, respectively.
Therefore, including the open edge $\hat{x}_0^{(v)}$ connected to $v$,  each vertex $v$ is  connected  to $2d+1$ edges in total.  Without loss of generality we order the set of these edges around each vertex $v$ by 
 \begin{align}
 \mathcal{X}_v:=&\{\pm \hat{x}_i^{(v)}\}_{i=1}^d\cup \hat{x}_0^{(v)},\quad \text{ordered by }
 \hat{x}_0^{(v)}<\hat{x}_1^{(v)}<\cdots <\hat{x}_d^{(v)}<-\hat{x}_1^{(v)}<\cdots <-\hat{x}_d^{(v)}.
 \end{align}
With this, the tick location and index ordering around an $X$-spider are given as follows (compare with the 2D and 3D bosonization),
\begin{align}
\begin{tikzpicture}[baseline={([yshift=-2.5pt]current bounding box.center)}]   
\def\r{0.5}
\def\h{.6}
\def\w{1}
\def\x{1.5}
\def\width{1.2}
\def\lw{1}
\def\delta{0.6*\r}
\def\ticklength{0.6}
  \def\ZXr{0.25}
  \def\y{1.*\r}
    \draw[tickcolor,line width=\lw] (0,0) -- ++(0*\r,\ticklength*\r);
         \draw\midarrow{0.5} (-\x*\r,\h*\r)  .. controls (-0.5*\r,\h*\r) and (0,0.2*\r)  ..  (0,0);
         \node at (-\delta-\x*\r,\h*\r) {$-\hat{x}_1$};
\draw\midarrow{0.5} (-\x*\r,-\h*\r) .. controls  (-0.5*\r,-\h*\r) and (0,-0.2*\r)  .. (0,0) ;
 \node at (-\delta-\x*\r,-\h*\r) {$-\hat{x}_d$};
\node[rotate=90] at (-0.9*\x*\r,0.05*\r) {$\cdots$};
\draw[->,line width=0.7 ](0, -\y)--(0,-0.45*\y);
\draw[line width=0.7 ](0, 0)--(0,-0.45*\y);
\node at (0.1*\r,-\y-0.4*\r) {$\hat{x}_0$};
\draw\midarrow{0.55} (0,0) .. controls (0,0.2*\r) and (0.5*\r,\h*\r) .. (\x*\r,\h*\r);
  \node at (\delta+\x*\r,\h*\r) {$\hat{x}_d$};
 \draw\midarrow{0.55} (0,0) .. controls (0,-0.2*\r) and (0.5*\r,-\h*\r) .. (\x*\r,-\h*\r);
   \node at (\delta+\x*\r,-\h*\r) {$\hat{x}_1$};
\node[rotate=90] at (0.95*\x*\r,0.05*\r) {$\cdots$};
\filldraw[fill=Xcolor,draw=black](0*\r,0*\r)circle (\ZXr*\r);
\end{tikzpicture}
\;.
\end{align}
\footnote{
One can in principle exchange any $\hat{x}_i$ with its partner $-\hat{x}_i$, which does not affect the validity of the construction as it only add a Pauli $Z$ to every edge in the $i$-th direction (located on the dashed lines).}.
Now consider a pair of neighboring Majorana operators separated in the $i$th direction for $1\leq i\leq d$.
On such a pair, bosonization acts as
 \begin{align}
\mathsf{D}_\text{Bos}(\gamma_{v-\hat{x}_i} \gamma_v) 
=\mathsf{P}\qty(\prod_{\substack{e\in \mathcal{X}_{v-\hat{x}_i}\\ x_0<e<x_i}}Z_e)(-iY)_{v-\hat{x}_i/2}
\qty(\prod_{\substack{e\in \mathcal{X}_{v}\\ x_0<e<x_i}}Z_e) \mathsf{P}
\;,
\label{eq_D_bos_any_d}
 \end{align}
 where in the constraint $x_0<e<x_i$ we have omitted the superscript implying the relation with the respective vertex simply to avoid the cluttering of  indices. The proof that such a construction satisfies the defining property Eq.~\eqref{eq_D_bos_any_d} is in Appendix~\ref{proof_any_d_bosonization}.

To be more concrete on the construction, we go back to the 3D Bosonization. The open index incident to each $X$-spider (label it by $v$) corresponds to $\hat{x}_0$. Starting from it, the edges around $v$ are $x_0<-\hat{z}<-\hat{y}<+\hat{x}<+\hat{z}<+\hat{y}<-\hat{x}$. Based on this ordering Eq.~\eqref{eq_D_bos_any_d} then reduces back to Eq.~\eqref{eq_Dbos_v1}-\eqref{eq_Dbos_v3}.

\subsubsection{Consistency condition and quantum simulation of fermionic systems}
The Bosonization isometry $\mathsf{D}_\text{Bos}$ for dimension $d$ that we defined in last subsection does not map to the full Hilbert space of the qubits, but only to the support of the projector $\mathsf{P}$.
We can infer the structure of $\mathsf{P}$ directly from Eq.~\eqref{eq_D_bos_any_d}:
To this end, we apply the bosonization to a trivial product of four pairs of Majorana operators in the $(ij)$ plane,
    \begin{align}
       &\mathsf{D}_\text{Bos}
       \qty((\gamma_v\gamma_{v+\hat{x}_i})
       (\gamma_{v+\hat{x}_i}\gamma_{v+\hat{x}_i+\hat{x}_j})
        (\gamma_{v+\hat{x}_i+\hat{x}_j}\gamma_{v+\hat{x}_j})
        (\gamma_{v+\hat{x}_j}\gamma_v)
       )\nonumber \\
       =&
       \mathsf{D}_\text{Bos}
       (\gamma_v\gamma_{v+\hat{x}_i})
       \mathsf{D}_\text{Bos}(\gamma_{v+\hat{x}_i}\gamma_{v+\hat{x}_i+\hat{x}_j})
        \mathsf{D}_\text{Bos}(\gamma_{v+\hat{x}_i+\hat{x}_j}\gamma_{v+\hat{x}_j})
        \mathsf{D}_\text{Bos}(\gamma_{v+\hat{x}_j}\gamma_v)=1.\label{eq_dm1_form_symm}
    \end{align}
   The LHS should be $1$ because the input is $1$, whereas the output gives us a loop of Pauli operators (for example, it includes four $-iY$ on the four edges).

For example, in $d=2$ the consistency condition is 
\begin{align}
   \mathsf{D}_\text{Bos}\qty((\gamma_{v_0}\gamma_{v_1})
       (\gamma_{v_1}\gamma_{v_2})
        (\gamma_{v_2}\gamma_{v_3})
        (\gamma_{v_3}\gamma_{v_4})
       )
  \  =(-1)\  
\begin{tikzpicture}[baseline={([yshift=-2.5pt]current bounding box.center)}]   
\def\r{0.7}
\def\x{2*\r}
\def\delta{0.65*\x}
\def\Delta{1*\x}
\def\shift{0.3*\r}
\draw(-\delta,0)--(\x+\Delta,0);
\draw(-\delta,\x)--(\x+\Delta,\x);
\draw(0,-\delta)--(0,\x+\Delta);
\draw(\x,-\delta)--(\x,\x+\Delta);
  \node[
    fill=white,           
    inner sep=2pt,        
    rectangle,            
    draw=black,           
    rounded corners=1pt   
  ] at (0.5*\x,0) {$X$};
  \node[
    fill=white,           
    inner sep=2pt,        
    rectangle,            
    draw=black,           
    rounded corners=1pt   
  ] at (0,0.5*\x) {$X$};
  \node[
    fill=white,           
    inner sep=2pt,        
    rectangle,            
    draw=black,           
    rounded corners=1pt   
  ] at (\x,0.5*\x) {$Y$};
  \node[
    fill=white,           
    inner sep=2pt,        
    rectangle,            
    draw=black,           
    rounded corners=1pt   
  ] at (0.5*\x,\x) {$Y$};
  \node[
    fill=white,           
    inner sep=2pt,        
    rectangle,            
    draw=black,           
    rounded corners=1pt   
  ] at (1.5*\x,\x) {$Z$};
  \node[
    fill=white,           
    inner sep=2pt,        
    rectangle,            
    draw=black,           
    rounded corners=1pt   
  ] at (\x,1.5*\x) {$Z$};
  \node at (0.5*\x,0.5*\x) {$p$};
\node at (-\shift,-\shift) {$v_0$};
\node at (\x+\shift,-\shift) {$v_1$};
\node at (\x+\shift,\x+\shift) {$v_2$};
\node at (-\shift,\x+\shift) {$v_3$};
\foreach \i in {0,1}{
\foreach \j in {0,1}{
\filldraw[fill=black,draw=black] (\i*\x,\j*\x) circle (0.07*\r);
}
}
\end{tikzpicture}
\ =\ 1.
\end{align}
   Such a consistency condition puts a $(d-1)$-form symmetry for every such loop $l$ surrounding a unit plaquette $p$, and we use a projector $\mathsf{P}_p$ to represent such a constraint.
   In other words, mapping from qubits to fermions and back yields a $(d-1)$-form projector
    $\mathsf{D}_\text{Bos}\mathsf{D}_\text{Bos}^\dagger = \mathsf{P}$ 
    onto the subspace that is invariant under the $(d-1)$-form symmetry.
    
In the $d=2$ case, using the vertex and plaquette operators $A$ and $B$ of a toric code on a square lattice, the operator $\mathsf P_p$ becomes $A_{v_2}B_p$, which detects whether an $e$ or an $m$ anyon is present at that location.
However, the operator cannot detect the presence of a fermion near that location, since a fermion consists of both an $e$ and an $m$ anyon simultaneously. 

One application of this theoretical $d$-dimensional bosonization scheme is the quantum simulation of fermionic systems using qubits embedded in a $d$-dimensional Euclidean geometry.
One may put the constraint above of $(d-1)$-form symmetry energetically as 
\begin{align}
 H^{(d-1)}_S=-\mathsf{P}
= -\prod_p \mathsf{P}_p.
\end{align}
Then, given any input fermionic Hamiltonian $H_F$, the qubit simulation $H_Q$ embedded in a $d$-dim lattice can be written as 
\begin{align}
    H_Q=\mathsf{D}_\text{bos}(H_F)+H_S^{(d-1)}
\end{align}

\subsubsection{Other fermion-to-qubit mappings}
The representation of bosonization mappings as fermionic tensor networks generalizes to other types of fermion-to-qubit mappings.
For example, the Bravyi-Kitaev superfast encoding \cite{Setia_2018} can be seen as a generalization of the bosonization mappings studied above from 2D or 3D hypercubic lattices to arbitrary graphs.
Accordingly, the tensor networks would look the same but on a different graph.
Similarly, the compact encoding of Ref.~\cite{PhysRevB.104.035118} is a local stabilizer-code mapping that encodes fermionic states into a qubit subspace.
Again, it can be represented by a tensor network with $X$ and $Z$ spiders whose adjacency matrix corresponds to the linear constraints of the stabilizer code.


Ref.~\cite{mcdowallrose2025fermionsqubitszxcalculusperspective} represents arbitrary linear fermion-to-qubit mappings as ordinary (qubit) ZX diagrams.
Note that some of the mappings studied there (such as ordinary Jordan-Wigner or ternary-tree encodings) do not correspond to any fermionic tensor network since they are supported on both the even and the odd fermion-parity subspace, and any fermionic tensor network must have even total parity (or at least fixed parity, which becomes even after taking a ket and bra copy when interpreting the mapping as a superoperator).
We leave the detailed translation of existing fermion-to-qubit mappings as fermionic tensor networks to future work.


\subsection{Fermionic error-correcting codes}
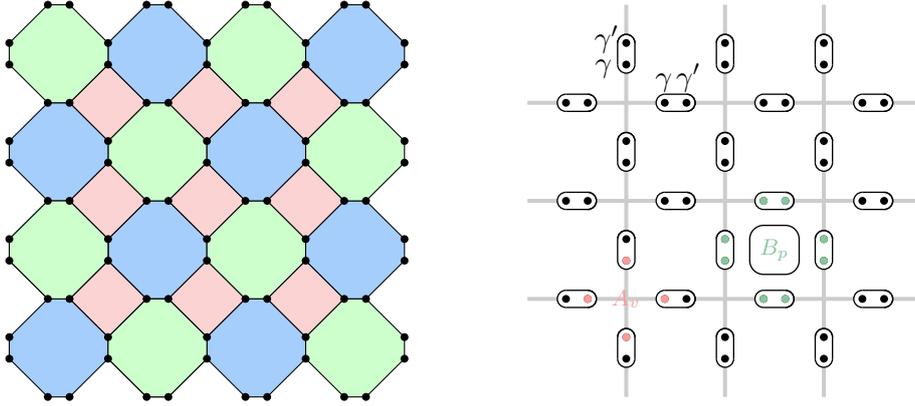
\begin{figure}
    \centering
    \subfloat{
\begin{tikzpicture}[baseline={([yshift=-2.5pt]current bounding box.center)}]   
\def\r{1.3}
\def\Bpcolor{Lblue}
\def\Avcolor{Lgreen}
\def\dotSize{0.04}
\def\width{0.22}
\def\epsilon{0.175}
\def\linewidth{1.5}
\def\lw{0.5}
\def\bondcolor{white}
\def\linecolor{gray!40}
\def\Lx{2}
\def\LLx{3}
\def\Ly{2}
\def\LLy{3}
\def\d{0.39}
    \newcommand{\mySquare}{
    \draw[fill=Lred](0,-\d*\r)--(\d*\r,0)--(0,\d*\r)--(-\d*\r,0)--(0,-\d*\r);
    }
\newcommand{\myOctogon}[1]{%
  \draw[fill=#1,draw=black]
    (\d*\r,0)
   -- ++(\width*\r,0)
   -- ++(\d*\r,\d*\r)
   -- ++(0,\width*\r)
   -- ++(-\d*\r,\d*\r)
   -- ++(-\width*\r,0)
   -- ++(-\d*\r,-\d*\r)
   -- ++(0,-\width*\r)
   -- cycle;
}
 \foreach \x in {0,...,\Lx}{
 \foreach \y in {0,...,\Ly}{
 \begin{scope}[shift={(\x*\r,\y*\r)}]
\mySquare
\pgfmathtruncatemacro{\parity}{mod(\x+\y,2)}
    \ifnum\parity=1
     \myOctogon{Lgreen}
    \else
    \myOctogon{Lblue} 
    \fi
\end{scope}
}
}
 \foreach \x in {0,...,\Lx}{
  \foreach \y in {-1}{
  \begin{scope}[shift={(\x*\r,\y*\r)}]
\pgfmathtruncatemacro{\parity}{mod(\x+\y,2)}
    \ifnum\parity=1
     \myOctogon{Lgreen}
    \else
    \myOctogon{Lblue} 
    \fi
\end{scope}
 }
 }
  \foreach \y in {-1,...,\Ly}{
  \begin{scope}[shift={(-1*\r,\y*\r)}]
\pgfmathtruncatemacro{\parity}{mod(\y-1,2)}
    \ifnum\parity=1
     \myOctogon{Lgreen}
    \else
    \myOctogon{Lblue} 
    \fi
\end{scope}
 }
 \begin{scope}[shift={(0,-\r)}]
   \myOctogon{Lgreen}
   \end{scope}
 \begin{scope}[shift={(-\r,0)}]
   \myOctogon{Lgreen}
   \end{scope}
 \foreach \x in {-1,...,\Lx}{
 \foreach \y in {-1,...,\LLy}{
 \begin{scope}[shift={(\x*\r+0.5*\r,\y*\r)}]
\begin{scope}[rotate=0]
\fill[fill=black] (-0.5*\width*\r,0) circle (\dotSize*\r);
\fill[fill=black] (0.5*\width*\r,0) circle (\dotSize*\r);
\end{scope}
\end{scope}
}
}
    \foreach \x in {-1,...,\LLx}{
 \foreach \y in {-1,...,\Ly}{
 \begin{scope}[shift={(\x*\r,0.5*\r+\y*\r)}]
\begin{scope}[rotate=90]
\fill[fill=black] (-0.5*\width*\r,0) circle (\dotSize*\r);
\fill[fill=black] (0.5*\width*\r,0) circle (\dotSize*\r);
\end{scope}
\end{scope}
}
}
\end{tikzpicture}
}
   $\quad \quad \quad $
\subfloat{
    \begin{tikzpicture}[baseline={([yshift=-2.5pt]current bounding box.center)}]   
\def\r{1.3}
\def\Bpcolor{red!40!white}
\def\Avcolor{FGreen!60!white}
\def\dotSize{0.04}
\def\width{0.22}
\def\epsilon{0.175}
\def\linewidth{1.5}
\def\lw{0.5}
\def\bondcolor{white}
\def\linecolor{gray!40}
\def\Lx{2}
\def\Ly{2}
  \foreach \x in {0,...,\Lx} { 
 \draw[line width=\linewidth,\linecolor] (\x*\r,-\r ) -- (\x*\r,\Ly*\r+\r);
 }
  \foreach \y in {0,...,\Ly} { 
 \draw[line width=\linewidth,\linecolor] (-\r,\y*\r ) -- (\Lx*\r+\r,\y*\r);
 }
 \foreach \x in {0,...,\Lx}{
 \foreach \y in {-1,...,\Ly}{
 \begin{scope}[shift={(\x*\r,0.5*\r+\y*\r)}]
\begin{scope}[rotate=90]
 \fill[fill=\bondcolor] (-0.5*\width*\r,-0.5*\epsilon*\r) rectangle (0.5*\width*\r,0.5*\epsilon*\r);
\fill[fill=\bondcolor] (-0.5*\width*\r,0) circle (0.5*\epsilon*\r);
\fill[fill=\bondcolor] (0.5*\width*\r,0) circle (0.5*\epsilon*\r);
\fill[fill=black] (-0.5*\width*\r,0) circle (\dotSize*\r);
\fill[fill=black] (0.5*\width*\r,0) circle (\dotSize*\r);
\draw[line width=\lw] (-0.5*\width*\r,-0.5*\epsilon*\r) arc(270:90:0.5*\epsilon*\r);
\draw[line width=\lw] (0.5*\width*\r,0.5*\epsilon*\r) arc(90:-90:0.5*\epsilon*\r);
\draw[line width=\lw] (0.5*\width*\r,0.5*\epsilon*\r) arc(90:-90:0.5*\epsilon*\r);
\draw[black,line width=\lw] (-0.5*\width*\r,-0.5*\epsilon*\r) -- (0.5*\width*\r,-0.5*\epsilon*\r);
\draw[black,line width=\lw] (-0.5*\width*\r,0.5*\epsilon*\r) -- (0.5*\width*\r,0.5*\epsilon*\r);
\end{scope}
\end{scope}
}
}
 \foreach \x in {-1,...,\Lx}{
 \foreach \y in {0,...,\Ly}{
 \begin{scope}[shift={(\x*\r+0.5*\r,\y*\r)}]
\begin{scope}[rotate=0]
 \fill[fill=\bondcolor] (-0.5*\width*\r,-0.5*\epsilon*\r) rectangle (0.5*\width*\r,0.5*\epsilon*\r);
\fill[fill=\bondcolor] (-0.5*\width*\r,0) circle (0.5*\epsilon*\r);
\fill[fill=\bondcolor] (0.5*\width*\r,0) circle (0.5*\epsilon*\r);
\fill[fill=black] (-0.5*\width*\r,0) circle (\dotSize*\r);
\fill[fill=black] (0.5*\width*\r,0) circle (\dotSize*\r);
\draw[line width=\lw] (-0.5*\width*\r,-0.5*\epsilon*\r) arc(270:90:0.5*\epsilon*\r);
\draw[line width=\lw] (0.5*\width*\r,0.5*\epsilon*\r) arc(90:-90:0.5*\epsilon*\r);
\draw[line width=\lw] (0.5*\width*\r,0.5*\epsilon*\r) arc(90:-90:0.5*\epsilon*\r);
\draw[black,line width=\lw] (-0.5*\width*\r,-0.5*\epsilon*\r) -- (0.5*\width*\r,-0.5*\epsilon*\r);
\draw[black,line width=\lw] (-0.5*\width*\r,0.5*\epsilon*\r) -- (0.5*\width*\r,0.5*\epsilon*\r);
\end{scope}
\end{scope}
}
}
 \foreach \x in {0}{
  \foreach \y in {-1}{
  \begin{scope}[shift={(\x*\r,0.5*\r+\y*\r)}]
\begin{scope}[rotate=90]
\fill[fill=\Bpcolor] (0.5*\width*\r,0) circle (\dotSize*\r);
\end{scope}
\end{scope}}}
 \foreach \x in {0}{
  \foreach \y in {0}{
  \begin{scope}[shift={(\x*\r,0.5*\r+\y*\r)}]
\begin{scope}[rotate=90]
\fill[fill=\Bpcolor] (-0.5*\width*\r,0) circle (\dotSize*\r);
\end{scope}
\end{scope}}}
 \foreach \x in {-1}{
 \foreach \y in {0}{
  \begin{scope}[shift={(\x*\r+0.5*\r,\y*\r)}]
\begin{scope}[rotate=0]
\fill[fill=\Bpcolor] (0.5*\width*\r,0) circle (\dotSize*\r);
\end{scope}
\end{scope}}}
 \foreach \x in {0}{
 \foreach \y in {0}{
  \begin{scope}[shift={(\x*\r+0.5*\r,\y*\r)}]
\begin{scope}[rotate=0]
\fill[fill=\Bpcolor] (-0.5*\width*\r,0) circle (\dotSize*\r);
\end{scope}
\end{scope}}}
\node at (0*\r,0*\r) {{\color{\Bpcolor}\fontsize{8.5pt}{5pt}\selectfont  $A_v$}};
 \foreach \x in {1,2}{
  \foreach \y in {0}{
  \begin{scope}[shift={(\x*\r,0.5*\r+\y*\r)}]
\begin{scope}[rotate=90]
\fill[fill=\Avcolor] (-0.5*\width*\r,0) circle (\dotSize*\r);
\fill[fill=\Avcolor] (0.5*\width*\r,0) circle (\dotSize*\r);
\end{scope}
\end{scope}}}
 \foreach \x in {1}{
 \foreach \y in {0,1}{
  \begin{scope}[shift={(\x*\r+0.5*\r,\y*\r)}]
\begin{scope}[rotate=0]
\fill[fill=\Avcolor] (-0.5*\width*\r,0) circle (\dotSize*\r);
\fill[fill=\Avcolor] (0.5*\width*\r,0) circle (\dotSize*\r);
\end{scope}
\end{scope}}}
\def\squareedge{0.25}
\draw[rounded corners=4*\r, fill=white, draw=black,line width=\lw] 
        (1.5*\r-\squareedge*\r, 0.5*\r-\squareedge*\r) rectangle (1.5*\r+\squareedge*\r, 0.5*\r+\squareedge*\r);
\node at (1.5*\r,0.5*\r) {{\color{\Avcolor}\fontsize{8.5pt}{5pt}\selectfont  $B_p$}};
\node at (-0.23*\r, 2.35*\r) {$\gamma$};
\node at (-0.2*\r, 2.65*\r) {$\gamma'$};
\node at (0.38*\r, 2.21*\r) {$\gamma$};
\node at (0.63*\r, 2.26*\r) {$\gamma'$};
\end{tikzpicture}
}
\caption{Left:  4-8-8 lattice hosting a Majorana code, with one Majorana operator per vertex.
The Majorana operators at the ends of each green-blue edge are paired.
Right: Alternatively, we can choose a square lattice as wallpaper for the code, with one fermion mode at each edge.
There is one stabilizer $A_v$ at each vertex, and one stabilizer $B_p$ at each plaquette. Each pair of $(\gamma,\gamma')$ defines a local space $\mathbb{C}^{1|1}$ located at that bond. We take the convention that the up and right Majoranas are those with a prime. 
}\label{square_lattice_fermion_code_hamiltonian}
\end{figure}
In this section, we demonstrate how our graphical calculus can be used for quantum error correction.
As a particular example, we derive a Floquet version of the Majorana stabilizer code in Ref.~\cite{PhysRevX.5.041038}, acting on both fermions and qubits.
We do this using the \emph{path-integral approach}~\cite{path_integral_qec, Bombin_2024, twisted_double_code, bauer2024xyfloquetcodesimple, Davydova2025, twisted_color_circuit} to constructing topological codes:
We write the code from Ref.~\cite{PhysRevX.5.041038} as a path integral by considering the product of its stabilizer projectors as a tensor network.
Then, we traverse this path integral in a different time direction, and turn it into a ``Floquet'' circuit of operators and measurements.
Ref.~\cite{PhysRevX.5.041038} defines the Majorana stabilizer code on a hexagonal lattice, but in fact it can be defined on any 3-colorable 3-valent lattice.
Here we consider the code on a 4-8-8 lattice, as shown in Figure~\ref{square_lattice_fermion_code_hamiltonian}.
If we want to implement the code with actual degrees of freedom, we have to decide which Majorana operators to pair up into one fermionic mode.
One choice is to pair the two operators at the two endpoints of a green-blue edge (an edge separating a green and a blue plaquette), as shown in Figure~\ref{square_lattice_fermion_code_hamiltonian}.

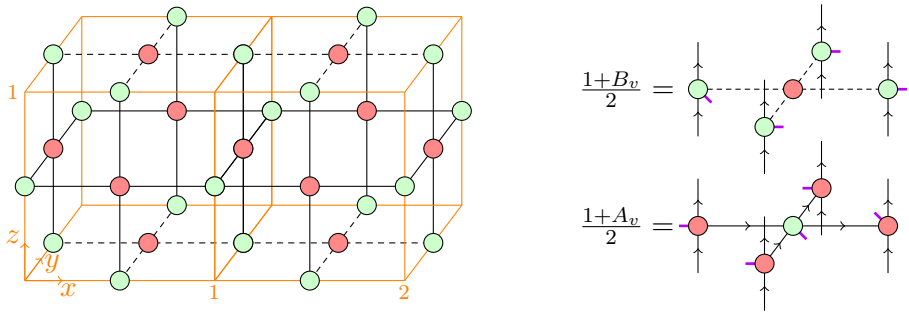
\begin{figure}[htp]
\centering
\subfloat{
\begin{tikzpicture}[baseline={([yshift=-2.5pt]current bounding box.center)}]   
\def\r{2.5}
\def\framecolor{orange}
\def\weakcolor{orange}
\def\lw{1}
\def\basisX{0}
\def\basisY{0}
\def\x{1.5}
    \def\h{0.6}
    \def\ZXr{0.05}
     \def\V{(0.4*\r,0.5*\r)}
     \def\Vx{0.3*\r}
      \def\Vy{0.4*\r}
   \newcommand{\ACube}{ 
       \draw(\Vx+0.5*\r,\Vy) --++(0,\r); 
       \draw[dash pattern=on 2.3pt off 1.9pt](0.5*\Vx,0.5*\Vy) --++(\r,0);  
       \draw[dash pattern=on 2.3pt off 1.9pt](0.5*\r,0) --++(\Vx,\Vy); 
        \draw(0.5*\Vx,0.5*\Vy) --++ (0,\r); 
         \draw(0.5*\Vx+\r,0.5*\Vy) --++ (0,\r); 
         \draw(\Vx,0.5*\r+\Vy) --++(\r,0); 
          \draw(0,0.5*\r) --++(\Vx,\Vy); 
          \draw(1*\r,0.5*\r) --++(\Vx,\Vy); 
   \foreach \x in {0,1} {   
   \draw[\framecolor](\x*\r,0) -- (\x*\r,\r);  
     \draw[\framecolor](\x*\r,0) -- ++(\Vx,\Vy);
      \draw[\framecolor](\x*\r,\r) -- ++(\Vx,\Vy);
     }
    \foreach \y in {0,1} {     \draw[\framecolor](0,\y*\r) -- (1*\r,\y*\r); }
    \foreach \x in {0,1} {     \draw[\weakcolor](\x*\r+\Vx,+\Vy) -- (\x*\r+\Vx,\r+\Vy); }
    \foreach \y in {0,1} {     \draw[\weakcolor](\Vx,\Vy+\y*\r) -- (\Vx+1*\r,\Vy+\y*\r); }
        \draw(0.5*\r,0) -- (0.5*\r,\r);
             \draw[dash pattern=on 2.3pt off 1.9pt](0.5*\Vx,0.5*\Vy+\r) --++(\r,0);  
       \draw[dash pattern=on 2.3pt off 1.9pt](0.5*\r,\r) --++(\Vx,\Vy);  
        \draw(0,0.5*\r) --++(\r,0); 
   \foreach \y  in {0,1}{ 
   \filldraw[fill=Zcolor,draw=black](0.5*\r,\y*\r)circle (\ZXr*\r);
     \filldraw[fill=Zcolor,draw=black](0.5*\r+\Vx,+\Vy+\y*\r)circle (\ZXr*\r); }
    \foreach \y  in {0,1}{ 
    \foreach \x in {0,1}{
    \filldraw[fill=Zcolor,draw=black](0.5*\Vx+\x*\r,0.5*\Vy+\y*\r)circle (\ZXr*\r);}}
     \foreach \x in {0,1}{
    \filldraw[fill=Zcolor,draw=black](\x*\r,0.5*\r)circle (\ZXr*\r);}
        \foreach \x in {0,1}{
      \filldraw[fill=Zcolor,draw=black](\x*\r+\Vx,0.5*\r+\Vy)circle (\ZXr*\r);}
      \foreach \x in {0,1}{
    \filldraw[fill=Xcolor,draw=black](0.5*\Vx+\x*\r,0.5*\Vy+0.5*\r)circle (\ZXr*\r);}
     \foreach \y in {0,1}{
    \filldraw[fill=Xcolor,draw=black](0.5*\Vx+0.5*\r,0.5*\Vy+\y*\r)circle (\ZXr*\r);}
    \foreach \x in {0,1}{
    \filldraw[fill=Xcolor,draw=black](0.5*\r+\x*\Vx,0.5*\r+\x*\Vy)circle (\ZXr*\r);}
    }
    \begin{scope}[shift={(\basisX*\r,\basisY*\r)}]
    \draw[\framecolor,->](0,0)--++(0.2*\r,0);
    \draw[\framecolor,->](0,0)--++(0.3*\Vx,0.3*\Vy);
    \draw[\framecolor,->](0,0)--++(0,0.2*\r);
\node at (0+0.23*\r,-0.05*\r) {{\color{\framecolor}$x$}};
\node at (0+0.48*\Vx,0.22*\Vy) {{\color{\framecolor}$y$}};
\node at (-0.063*\r,0+0.23*\r) {{\color{\framecolor}$z$}};
\end{scope}
\node at (1*\r,-0.06*\r) {${\color{\framecolor}_1}$};
\node at (2*\r,-0.06*\r) {${\color{\framecolor}_2}$};
\node at (-0.06*\r,\r) {${\color{\framecolor}_1}$};
   \ACube
   \begin{scope}[shift={(\r,0)}]       \ACube  \end{scope}
\end{tikzpicture}
}
$\quad \quad\ $
\subfloat{
\begin{tikzpicture}[baseline={([yshift=-2.5pt]current bounding box.center)}]   
\def\r{1.25}
\def\framecolor{orange}
\def\weakcolor{orange}
\def\weakcolor{gray}
\def\ticklength{0.2}
\def\lw{1}
\def\Dz{0.5*\r}
\def\X{1.45*\r}
\def\delta{1.75*\r}
\def\basisX{0}
\def\basisY{0}
\def\x{1.5}
    \def\ZXr{0.1}
            \def\V{(0.4*\r,0.5*\r)}
     \def\Vx{0.3*\r}
      \def\Vy{0.4*\r}
      \newcommand{\verticallines}{
   \draw\midarrow{0.525}(-\r,-\Dz)--(-\r,0);
   \draw\midarrow{0.555}(-\r,0)--(-\r,\Dz);
   \draw\midarrow{0.525}(\r,-\Dz)--(\r,0);
   \draw\midarrow{0.555}(\r,0)--(\r,\Dz);
    \draw\midarrow{0.555}(-\Vx,-\Vy)--++(0,\Dz);
    \draw\midarrow{0.525}(-\Vx,-\Vy-\Dz)--++(0,\Dz);
       \draw\midarrow{0.555}(\Vx,\Vy)--++(0,\Dz);
    \draw\midarrow{0.525}(\Vx,\Vy-\Dz)--++(0,\Dz);
    }
     \begin{scope}[shift={(0,0)}]
   \draw\midarrow{0.55}(-\r,0)--(0,0);
   \draw\midarrow{0.55}(0,0)--(\r,0);
   \draw\midarrow{0.55}(-\Vx,-\Vy)--(0,0);
   \draw\midarrow{0.55}(0,0)--(\Vx,\Vy);
    \verticallines
       \draw[tickcolor,line width=\lw] (0,0) -- ++(0.707*\ticklength*\r,-0.707*\ticklength*\r);
        \draw[tickcolor,line width=\lw] (-\Vx,-\Vy) -- ++(-1*\ticklength*\r,0);
    \draw[tickcolor,line width=\lw] (\Vx,\Vy) -- ++(-1*\ticklength*\r,0);
      \draw[tickcolor,line width=\lw] (\r,0) -- ++(-0.707*\ticklength*\r,0.707*\ticklength*\r);
      \draw[tickcolor,line width=\lw] (-\r,0) -- ++(-1*\ticklength*\r,0);
          \filldraw[fill=Zcolor,draw=black](0,0)circle (\ZXr*\r);
           \filldraw[fill=Xcolor,draw=black](-\Vx,-\Vy)circle (\ZXr*\r);
           \filldraw[fill=Xcolor,draw=black](\Vx,\Vy)circle (\ZXr*\r);
            \filldraw[fill=Xcolor,draw=black](\r,0)circle (\ZXr*\r);
        \filldraw[fill=Xcolor,draw=black](-\r,0)circle (\ZXr*\r);
      \node at (-\delta,0) {$\frac{1+A_v}{2}=$};
              \end{scope}
   \begin{scope}[shift={(0,\X)}]
     \draw\dashedpattern(-\r,0)--(\r,0);
   \draw\dashedpattern(-\Vx,-\Vy)--(\Vx,\Vy);
\verticallines
    \draw[tickcolor,line width=\lw] (-\Vx,-\Vy) -- ++(1*\ticklength*\r,0);
    \draw[tickcolor,line width=\lw] (\Vx,\Vy) -- ++(1*\ticklength*\r,0);
       \draw[tickcolor,line width=\lw] (\r,0) -- ++(1*\ticklength*\r,0);
      \draw[tickcolor,line width=\lw] (-\r,0) -- ++(0.707*\ticklength*\r,-0.707*\ticklength*\r);
    \filldraw[fill=Xcolor,draw=black](0,0)circle (\ZXr*\r);
     \filldraw[fill=Zcolor,draw=black](-\Vx,-\Vy)circle (\ZXr*\r);
      \filldraw[fill=Zcolor,draw=black](\Vx,\Vy)circle (\ZXr*\r);
       \filldraw[fill=Zcolor,draw=black](\r,0)circle (\ZXr*\r);
        \filldraw[fill=Zcolor,draw=black](-\r,0)circle (\ZXr*\r);
       \node at (-\delta,0) {$\frac{1+B_v}{2}=$};
        \end{scope}
    \end{tikzpicture}
}
    \caption{Left: The tensor network of the Hamiltonian model shown above on the square lattice. The orientation and ticks have been omitted. Right: The local ground state projectors, which make up the tensor-network path integral.}
    \label{fig_square_lattice_tensor_network}
\end{figure}

As shown, we obtain a model with one fermionic mode on each edge of a square lattice.
There is one stabilizer operator $A_v$ for every vertex $v$, and one stabilizer operator $B_p$ for every plaquette $p$, given by
\begin{align}
A_v:=\gamma'_{\delta_d(v),p}\gamma'_{\delta_l(v),p}\gamma_{\delta_u(v),p}\gamma_{\delta_r(v),p},\quad B_p:={\prod_{e\in p}} P_e\;.
\end{align}
Here, $\delta_{r,u,l,d}(v)$ represents the right, up, left, or down edge of $v$, respectively (see the right panel of Figure~\ref{square_lattice_fermion_code_hamiltonian}); 
$P_e=\gamma_{e}\gamma_{e}'$ is the fermion parity operator formed by the two Majorana operators on the edge $e$, and $\gamma_{e}$ and $\gamma_e'$ are the two Majorana operators for the fermion mode on the edge $e$ of the plaquette $p$. 
We can now use our graphical calculus to write down the projectors onto the ground space of these stabilizers, which are shown on the right-hand side of Figure~\ref{fig_square_lattice_tensor_network}.
Now, we consider the product of all of these stabilizer projectors, alternating between layers of $A_v$ and $B_p$.
Graphically, this product corresponds to stacking the diagrams for the ground-state projectors to obtain a tensor network in 3D spacetime.
After applying some spider fusion rules, this tensor network is defined on a cubic spacetime lattice, as shown at the left-hand side of Figure~\ref{fig_square_lattice_tensor_network}.
This tensor network represents a \emph{discrete path integral} for a topological phase.
\footnote{In fact, this topological phase is just the one of the (qubit) toric code.
The Majorana code is equivalent to the toric code after applying a local unitary operation.
In particular, the operator $A_v$ forces the fermion parity to be even inside non-overlapping patches of fermions, so there is no intrinsically fermionic long-range entanglement.
}
The tensor network looks very similar to the one for the toric code with $X$-tensors on all faces and $Z$-tensors on all edges of a cubic lattice~\cite{path_integral_qec, bauer2024xyfloquetcodesimple}.
The only difference is that some of the qubit bonds are replaced by fermionic bonds.
We note that replacing qubit bonds with fermionic bonds may not always result in a fault-tolerant protocol, and it is an open question whether we could replace even more qubit bonds with fermionic bonds.
We can now construct a different (dynamic) code from traversing this path integral in a differently chosen time direction, following Refs.~\cite{path_integral_qec, Bombin_2024, twisted_double_code, bauer2024xyfloquetcodesimple, Davydova2025, twisted_color_circuit}.
For our particular example, we pick the time direction in the same way as was done for the toric-code path integral Ref.~\cite{bauer2024xyfloquetcodesimple}, namely the $x+y$ direction.
With respect to this time direction, we reinterpret the tensor-network path integral as a circuit of linear operators, which are either unitaries or projection operators corresponding to post-selected measurements.

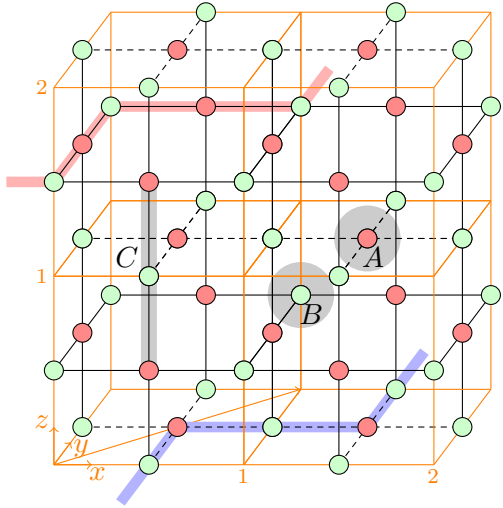
\begin{figure}[htp]
    \centering
\begin{tikzpicture}[baseline={([yshift=-2.5pt]current bounding box.center)}]   
\def\r{2.5}
\def\framecolor{orange}
\def\weakcolor{orange}
\def\lw{1}
\def\basisX{0}
\def\basisY{0}
\def\x{1.5}
    \def\h{0.6}
    \def\ZXr{0.05}
     \def\V{(0.4*\r,0.5*\r)}
     \def\Vx{0.3*\r}
      \def\Vy{0.4*\r}
    \draw[orange,->](0,0)--(\Vx+\r,\Vy);
    \draw[blue!30!white,line width=4pt](0.5*\r-0.5*\Vx,-0.5*\Vy)--++(1.*\Vx, 1.*\Vy)--++(1*\r,0)--++(\Vx,\Vy);
    \draw[red!30!white,line width=4pt](-0.25*\r,1.5*\r)--(0*\r,1.5*\r)--++(\Vx,\Vy)--++(\r,0)--++(0.5*\Vx,0.5*\Vy);
    \fill[fill=black!20!white](1.5*\r+0.5*\Vx, 1*\r+0.5*\Vy)circle (0.175*\r);
    \fill[fill=black!20!white](\r+\Vx, \Vy+0.5*\r)circle (0.175*\r);
    \draw[black!20!white,line width=6pt](0.5*\r, 0.5*\r)--++(0,\r);
   \newcommand{\ACube}{ 
       \draw(\Vx+0.5*\r,\Vy) --++(0,\r); 
       \draw[dash pattern=on 2.3pt off 1.9pt](0.5*\Vx,0.5*\Vy) --++(\r,0);  
       \draw[dash pattern=on 2.3pt off 1.9pt](0.5*\r,0) --++(\Vx,\Vy); 
        \draw(0.5*\Vx,0.5*\Vy) --++ (0,\r); 
         \draw(0.5*\Vx+\r,0.5*\Vy) --++ (0,\r); 
         \draw(\Vx,0.5*\r+\Vy) --++(\r,0); 
          \draw(0,0.5*\r) --++(\Vx,\Vy); 
          \draw(1*\r,0.5*\r) --++(\Vx,\Vy); 
   \foreach \x in {0,1} {   
   \draw[\framecolor](\x*\r,0) -- (\x*\r,\r);  
     \draw[\framecolor](\x*\r,0) -- ++(\Vx,\Vy);
      \draw[\framecolor](\x*\r,\r) -- ++(\Vx,\Vy);
     }
    \foreach \y in {0,1} {     \draw[\framecolor](0,\y*\r) -- (1*\r,\y*\r); }
    \foreach \x in {0,1} {     \draw[\weakcolor](\x*\r+\Vx,+\Vy) -- (\x*\r+\Vx,\r+\Vy); }
    \foreach \y in {0,1} {     \draw[\weakcolor](\Vx,\Vy+\y*\r) -- (\Vx+1*\r,\Vy+\y*\r); }
        \draw(0.5*\r,0) -- (0.5*\r,\r);
             \draw[dash pattern=on 2.3pt off 1.9pt](0.5*\Vx,0.5*\Vy+\r) --++(\r,0);  
       \draw[dash pattern=on 2.3pt off 1.9pt](0.5*\r,\r) --++(\Vx,\Vy);  
        \draw(0,0.5*\r) --++(\r,0); 
   \foreach \y  in {0,1}{ 
   \filldraw[fill=Zcolor,draw=black](0.5*\r,\y*\r)circle (\ZXr*\r);
     \filldraw[fill=Zcolor,draw=black](0.5*\r+\Vx,+\Vy+\y*\r)circle (\ZXr*\r); }
    \foreach \y  in {0,1}{ 
    \foreach \x in {0,1}{
    \filldraw[fill=Zcolor,draw=black](0.5*\Vx+\x*\r,0.5*\Vy+\y*\r)circle (\ZXr*\r);}}
     \foreach \x in {0,1}{
    \filldraw[fill=Zcolor,draw=black](\x*\r,0.5*\r)circle (\ZXr*\r);}
        \foreach \x in {0,1}{
      \filldraw[fill=Zcolor,draw=black](\x*\r+\Vx,0.5*\r+\Vy)circle (\ZXr*\r);}
      \foreach \x in {0,1}{
    \filldraw[fill=Xcolor,draw=black](0.5*\Vx+\x*\r,0.5*\Vy+0.5*\r)circle (\ZXr*\r);}
     \foreach \y in {0,1}{
    \filldraw[fill=Xcolor,draw=black](0.5*\Vx+0.5*\r,0.5*\Vy+\y*\r)circle (\ZXr*\r);}
    \foreach \x in {0,1}{
    \filldraw[fill=Xcolor,draw=black](0.5*\r+\x*\Vx,0.5*\r+\x*\Vy)circle (\ZXr*\r);}
    }
    \begin{scope}[shift={(\basisX*\r,\basisY*\r)}]
    \draw[\framecolor,->](0,0)--++(0.2*\r,0);
    \draw[\framecolor,->](0,0)--++(0.3*\Vx,0.3*\Vy);
    \draw[\framecolor,->](0,0)--++(0,0.2*\r);
\node at (0+0.23*\r,-0.05*\r) {{\color{\framecolor}$x$}};
\node at (0+0.48*\Vx,0.22*\Vy) {{\color{\framecolor}$y$}};
\node at (-0.063*\r,0+0.23*\r) {{\color{\framecolor}$z$}};
\end{scope}
\node at (1*\r,-0.06*\r) {${\color{\framecolor}_1}$};
\node at (2*\r,-0.06*\r) {${\color{\framecolor}_2}$};
\node at (-0.06*\r,\r) {${\color{\framecolor}_1}$};
\node at (-0.06*\r,2*\r) {${\color{\framecolor}_2}$};
   \ACube
   \begin{scope}[shift={(\r,0)}]       \ACube  \end{scope}
    \begin{scope}[shift={(0,\r)}]       \ACube  \end{scope}
    \begin{scope}[shift={(\r,\r)}]       \ACube  \end{scope}
    \node at (1.5*\r+0.5*\Vx   +0.03*\r,0.5*\Vy+\r-0.1*\r) {$A$};
    \node at (1*\r+1*\Vx   +0.055*\r,1*\Vy+0.5*\r-0.1*\r) {$B$};
    \node at (0.5*\r-0.115*\r,1.*\r+0.1*\r) {$C$};
\end{tikzpicture}
\caption{Qubit/fermion worldlines and operators to turn the path integral into a Floquet code.
The $x+y$ time direction is indicated by the diagonal orange line at the bottom.
One fermion worldline is marked in red, and one qubit worldlines marked in blue.
Three groups of tensors which become an operator in the circuit are shaded in gray and labeled $A$, $B$, and $C$.
}\label{circuit_of_floquet_code}
\end{figure}
The derivation of the circuit is similar to Ref.~\cite{bauer2024xyfloquetcodesimple}, with a few differences due to the fermionic nature of the path integral.
We start by identifying qubit and fermion worldlines, which are sequences of qubit and fermion-mode bonds that are represented by the same qubit or fermion mode at different time steps, and therefore are expected to proceed roughly in the $x+y$ time direction.
We choose these worldlines to proceed in time direction in a zig-zag way as shown in Fig.~\ref{circuit_of_floquet_code}. 
The next step is to identify small groups of tensors, such that each group can be interpreted as an operator with a few qubit worldlines going in and out.
We choose the following three groups of tensors.
\begin{itemize}
    \item 
Consider an $X$-spider with only qubit indices, such as the one marked ``A'' in Fig.~\ref{circuit_of_floquet_code}.
Each such tensor is interpreted as an operator with two qubits coming in and two qubits going out.
This operator is equal to the projection operator for an $XX$ measurement for the $+1$ outcome of two qubits (see the one located at $A$ for an example).
\begin{align}
   \frac{1+ XX}{2}=\ 
\begin{tikzpicture}[baseline={([yshift=-2.5pt]current bounding box.center)}]   
\def\r{0.5}
\def\x{1*\r}
\def\width{1.2}
\def\lw{1}
\def\ticklength{0.6}
\def\ZXr{0.25}
\draw\dashedpattern(-\x,-\x)--(\x,\x);
\draw\dashedpattern(-\x,\x)--(\x,-\x);
\filldraw[fill=Xcolor,draw=black](0*\r,0*\r)circle (\ZXr*\r);
\end{tikzpicture}
\end{align}
\item
Consider an $Z$-spider with only fermion-mode indices, such as the one marked ``B'' in Fig.~\ref{circuit_of_floquet_code}.
Each such tensor is interpreted as an operator with two fermion modes coming in and two fermion modes going out.
This operator is equal to the $+1$ projection operator for a total-parity measurement of the two fermion modes, that is, the measurement of the observable $P_iP_j$.
\begin{align}
    \frac{1+PP}{2}=\ 
\begin{tikzpicture}[baseline={([yshift=-2.5pt]current bounding box.center)}]   
\def\r{0.5}
\def\x{1.*\r}
\def\width{1.2}
\def\lw{1}
\def\ticklength{0.6}
\def\ZXr{0.25}
\draw[tickcolor,line width=\lw] (0,0) -- ++(1*\ticklength*\r,0);
\draw\midarrow{0.525}(-\x,-\x)--(0,0);
\draw\midarrow{0.555}(0,0)--(\x,\x);
\draw\midarrow{0.525}(\x,-\x)--(0,0);
\draw\midarrow{0.555}(0,0)--(-\x,\x);
\filldraw[fill=Zcolor,draw=black](0*\r,0*\r)circle (\ZXr*\r);
\end{tikzpicture}
\end{align}
\item 
Consider an $Z$ spider with two qubit indices and two fermion-mode indices, such as the one marked ``C'' in Fig.~\ref{circuit_of_floquet_code}.
Take the two $X$ spiders connected to it via fermion-mode bonds.
Apply the spider-fusion rule to split this 4-index $X$ spider into two 3-index $X$ spiders.
Consider the $Z$ spider with the two adjacent 3-index $X$ spiders as an operator acting on one qubit and two fermion modes.
This operator is a unitary, namely a controlled-$\gamma_0\gamma_1'$ operator.
The control is on the qubit, and $\gamma_0$ and $\gamma_1'$ act on the two different fermion modes,
\begin{align}
C\gamma'\gamma=\ (-1)\  
\begin{tikzpicture}[baseline={([yshift=-2.5pt]current bounding box.center)}]   
\def\r{0.5}
\def\x{1.2*\r}
\def\xx{1.7*\r}
\def\y{0.95*\r}
\def\width{1.2}
\def\lw{1}
\def\ticklength{0.55}
\def\ZXr{0.25}
\draw\midarrow{0.65}(-\x,0)--(-\xx,-\y);
\draw\midarrow{0.65}(-\x,0)--(-\xx,\y);
\draw\midarrow{0.525}(\xx,-\y)--(\x,0);
\draw\midarrow{0.525}(\xx,\y)--(\x,0);
\draw\midarrow{0.6}(0,0)--(-\x,0);
\draw\midarrow{0.6}(\x,0)--(0,0);
\draw\dashedpattern(0,-\y)--(0,\y);
\draw[tickcolor,line width=\lw] (-\x,0) -- ++(0,-\ticklength*\r);
\draw[tickcolor,line width=\lw] (\x,0) -- ++(0,-\ticklength*\r);
\draw[tickcolor,line width=\lw] (0,0) -- ++(0.707*\ticklength*\r,0.707*\ticklength*\r);
\filldraw[fill=Zcolor,draw=black](0*\r,0*\r)circle (\ZXr*\r);
\filldraw[fill=Xcolor,draw=black](\x,0)circle (\ZXr*\r);
\filldraw[fill=Xcolor,draw=black](-\x,0*\r)circle (\ZXr*\r);
\end{tikzpicture}
\;,
\end{align}
where the middle qubit index is the control qubit, while the left (right) $X$-spider implements the $\gamma'$ ($\gamma$) operator, respectively. To read off this operator one might need to divide each $X$-spider in ``C'' into a pair of $\gamma$ and  $-\gamma'$ operators ordered top to bottom vertically in Figure~\ref{circuit_of_floquet_code} using the spider fusion/splitting rule. 
\end{itemize}
Now, we take all the measurement projectors in the circuit and simply turn them into actual measurements.
The resulting circuit is a fault-tolerant error-correction circuit, and shown in Figure~\ref{fig:floquet_circuit}.
Note that in contrast to Ref.~\cite{bauer2024xyfloquetcodesimple}, we cannot group one $X$-spider and one $Z$ spider yielding a $CX$ gate, because we cannot split a fermionic $Z$ spider with 4 indices.
Relatedly, the fermionic version of a $CX$ operator would be a $C\gamma$ operator, which does not exist since it is not parity preserving.
Finally, note that in order to show that the circuit is indeed fault tolerant, we would have to analyze the effect of $-1$ outcomes in the circuit, whose configurations can be interpreted as abelian anyon worldlines.
\begin{figure}
\begin{tikzpicture}
\atoms{void}{x/p={1,0}, y/p={0,0.707107}}
\clip ($(x)+(y)+(-0.15,0.15)$) rectangle($4*(x)+4*(y)+(-0.15,-0.1)$);
\draw[orange] ($0*(y)$)--++($4*(x)$) ($1*(y)$)--++($4*(x)$) ($2*(y)$)--++($4*(x)$) ($3*(y)$)--++($4*(x)$) ($4*(y)$)--++($4*(x)$);
\draw[orange] ($1*(x)$)--++($4*(y)$) ($2*(x)$)--++($4*(y)$) ($3*(x)$)--++($4*(y)$) ($4*(x)$)--++($4*(y)$);
\foreach \x in {0,1,2,3,4}{
\foreach \y in {0,1,2,3,4}{
\atoms{circ,tiny,all}{g\x\y/p={$\x*(x)+{\y+0.5}*(y)$}}
\atoms{circ,tiny}{r\x\y/p={${\x+0.5}*(x)+{\y+0.5}*(y)$}}
}}
\foreach \x in {1,2,3}{
\draw[line width=0.3cm,cyan,opacity=0.5,line cap=round] (g\x0-c)--(g\x1-c) (g\x2-c)--(g\x3-c) (r\x1-c)--(r\x2-c) (r\x3-c)--(r\x4-c);
\path (g\x2-c)--node[midway]{$\scriptstyle{M_{XX}}$} (g\x3-c);
\path (r\x1-c)--node[midway]{$\scriptstyle{M_{PP}}$} (r\x2-c);
}
\end{tikzpicture}
$\rightarrow$
\begin{tikzpicture}
\atoms{void}{x/p={1,0}, y/p={0,0.707107}}
\clip ($(x)+(y)+(-0.15,0.15)$) rectangle($4*(x)+4*(y)+(-0.15,-0.1)$);
\draw[orange] ($0*(y)$)--++($4*(x)$) ($1*(y)$)--++($4*(x)$) ($2*(y)$)--++($4*(x)$) ($3*(y)$)--++($4*(x)$) ($4*(y)$)--++($4*(x)$);
\draw[orange] ($1*(x)$)--++($4*(y)$) ($2*(x)$)--++($4*(y)$) ($3*(x)$)--++($4*(y)$) ($4*(x)$)--++($4*(y)$);
\foreach \x in {0,1,2,3,4}{
\foreach \y in {0,1,2,3,4}{
\atoms{circ,tiny,all}{g\x\y/p={$\x*(x)+{\y+0.5}*(y)$}}
\atoms{circ,tiny}{r\x\y/p={${\x+0.5}*(x)+{\y+0.5}*(y)$}}
}}
\foreach \y in {1,2,3}{
\draw[line width=0.3cm,cyan,opacity=0.5,line cap=round] (r0\y-c)--(r1\y-c) (r2\y-c)--(r3\y-c);
\node at ($0.5*(r0\y-c)+0.5*(r1\y-c)+(90:0.15)$) {$\scriptstyle{C\gamma\gamma'}$};
\node at ($0.5*(r2\y-c)+0.5*(r3\y-c)+(90:0.15)$) {$\scriptstyle{C\gamma\gamma'}$};
}
\end{tikzpicture}
$\rightarrow$
\begin{tikzpicture}
\atoms{void}{x/p={1,0}, y/p={0,0.707107}}
\clip ($(x)+(y)+(-0.15,0.15)$) rectangle($4*(x)+4*(y)+(-0.15,-0.1)$);
\draw[orange] ($0*(y)$)--++($4*(x)$) ($1*(y)$)--++($4*(x)$) ($2*(y)$)--++($4*(x)$) ($3*(y)$)--++($4*(x)$) ($4*(y)$)--++($4*(x)$);
\draw[orange] ($1*(x)$)--++($4*(y)$) ($2*(x)$)--++($4*(y)$) ($3*(x)$)--++($4*(y)$) ($4*(x)$)--++($4*(y)$);
\foreach \x in {0,1,2,3,4}{
\foreach \y in {0,1,2,3,4}{
\atoms{circ,tiny,all}{g\x\y/p={$\x*(x)+{\y+0.5}*(y)$}}
\atoms{circ,tiny}{r\x\y/p={${\x+0.5}*(x)+{\y+0.5}*(y)$}}
}}
\foreach \y in {1,2,3}{
\draw[line width=0.3cm,cyan,opacity=0.5,line cap=round] (r1\y-c)--(r2\y-c) (r3\y-c)--(r4\y-c);
\node at ($0.5*(r1\y-c)+0.5*(r2\y-c)+(90:0.15)$) {$\scriptstyle{C\gamma\gamma'}$};
\node at ($0.5*(r3\y-c)+0.5*(r4\y-c)+(90:0.15)$) {$\scriptstyle{C\gamma\gamma'}$};
}
\end{tikzpicture}
$\rightarrow$ \parbox{3cm}{shift by 1 vertically and repeat}
\caption{Floquet circuit derived from the Majorana code of Ref.~\cite{PhysRevX.5.041038} via the path integral method.
Full dots are qubits, and empty dots are fermion modes.
The circuit consists of three steps, which are repeated after shifting one lattice site vertically.
A full period of the circuit without shifts consists of 6 steps.
}
\label{fig:floquet_circuit}
\end{figure}
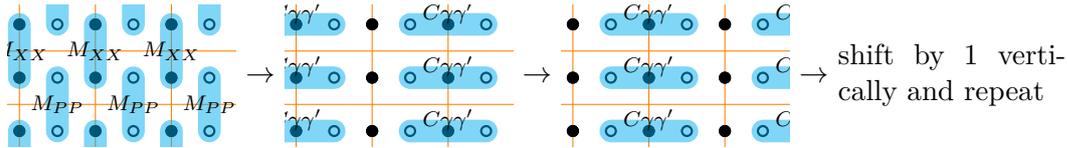
\section{Summary and discussion}\label{sec_summary}
In this work, we have developed a diagrammatic tensor-network calculus for fermionic systems, similar to the $ZX$ calculus for qubits. The diagrams in this graphical calculus represent fermionic tensor networks -- tensor networks where each tensor is strictly even with respect to a $\zz_2$ grading, and fermionic exchange statistics are implemented implicitly.
Our diagrammatic calculus consists of a small number of fermionic bond dimensions or supervector spaces (most prominently fermionic modes, $\mathbb{C}^{1|1}$), elementary tensors, and a succinct set of diagrammatic identities.

Using our graphical calculus, we have shown that free-fermion Gaussian states admit a succinct Pfaffian-graph representation, that the partial trace over one or more Majorana modes can be implemented graphically via simple spider contractions (even in the case of an odd number of Majoranas), and that the purification can be implemented by cutting the diagram in half. 
Moreover, we demonstrated that standard bosonization maps, such as the Jordan-Wigner transformation, Kitaev's Majorana-Pauli correspondence, and higher-dimensional dualities, arise naturally as compositions of hybrid fermionic and qubit spiders, thereby embedding qubit-fermion correspondences directly into our calculus. Furthermore, we applied these tools to the construction of fermionic error-correcting codes, deriving a "Floquet" version of the Majorana honeycomb code by slicing a 2+1D fermionic tensor network.

Taken together, our results establish a versatile and visually intuitive platform for representing, manipulating, and reasoning about a wide variety of fermionic systems -- ranging from Gaussian states to code constructions -- in terms of a single, minimal set of graphical rules. We anticipate that this $\mathbb{C}^{1|1}$ calculus will facilitate both practical computations in fermionic many-body physics and the exploration of novel fermionic quantum-information protocols (symmetry, entanglement, and quantum resources, etc) and fermionic circuits (e.g. lattice surgery, circuit compilation and so on).

 Our work opens up various future directions.
 For example, it would be nice to build a full one-to-one correspondence between our fermion tensor language and the Quon language, and to understand the recent results of \cite{kang20252dquonlanguageunifying,feng2025quonclassicalsimulationunifying} in a tensor-network fashion.
 Further, it would be interesting to understand quantum error correcting circuits described by fermionic ZX diagrams more systematically, and generalize concepts such as Pauli webs \cite{Bombin2023} and fault-tolerant rewrites \cite{Rodatz2025} to the case of fermionic ZX diagrams.

\subsubsection*{Author contributions}
Yuanjie Ren carried out most of the research and developed most of the proofs. Kaifeng Bu contributed to the work on purification and fermionic Gaussian states and operators. Andreas Bauer laid the foundations of the graphical calculus and identified ways to solve many of the more difficult problems.

During revision, the authors used GPT and Claude to help draft and revise text, check algebraic calculations and references, and format the manuscript.

\subsubsection*{Acknowledgements}
Y.R. was supported by the U.S.
Department of Energy, Office of Science, National Quantum Information
Science Research Centers, Quantum Systems Accelerator, under Grant
number DOE DE-SC0012704.
K.B. was supported in part by the ARO
Grant W911NF-19-1-0302 and the ARO MURI Grant
W911NF-20-1-0082.
A.B. was supported by the U.S. Army Research Laboratory and the U.S. Army Research Office under contract/grant number W911NF2310255, and by the U.S. Department of Energy, Office of Science, National Quantum Information Science Research Centers, and the Co-design Center for Quantum Advantage (C2QA) under contract number DE-SC0012704.
\bibliographystyle{unsrtnat} 
\bibliography{mybibliography}

\appendix
\section{Clifford algebra and fermionic operators}\label{sect_clifford_and_majorana}
\subsection{Definition and construction}
A $2^{2n}$-dimensional \emph{Clifford algebra} $\mathcal{M}_{2n}$ is a vector space linearly spanned by the products of elements $\{a_j:1\leq j\leq 2n\}$
equipped with the symmetric bilinear form (also referred to as the anticommutation relation)
$\{a_j,a_k\}=a_ja_k+a_ka_j=\delta_{j+n,j}$. In the context of physics, one usually writes $a^\dagger_j=a_{j+n}$ for $1\leq j\leq n$, and the algebra is
\begin{align}
\{a_j,a_k\}=\{a^\dagger_j,a^\dagger_k\}=0,\quad \{a_j,a_k\}^\dagger=\delta_{jk},\quad \text{for }1\leq j,k\leq n
\end{align}
The canonical module $\mathcal{H}$ over $\mathcal{M}_{2n}$ is the tensor-product of the $1|1$-dimensional graded vector space  $(\mathbb{C}^{1|1})^{\otimes n}$ with the action
\begin{align}
   a_j^\dagger|1)_j= a_j|0)_j=0,\quad a^\dagger_j|0)_j=\ket{1}_j,\quad a_j|1)_j=|0)_j.\label{eq_a_j_adagger_j}
\end{align}
The module space $\mathcal{H}$ of $\mathcal{M}_{2n}$ is hence a $2^n$-dimensional vector space, spanned by codewords $|x)=|x_n)\cdots |x_1)$ with $\ket{x_j}\in \mathcal{C}^{1|1}_j$. 
It is a superspace: the codewords having an even (odd) Hamming weight $|x|:=\sum_j x_j$ are evenly (oddly) graded, respectively.

Before we continue, it is useful to briefly explain how to extend an action on a single fermionic leg to all legs. This is done in a similar way to the case of ordinary qubits. Let $O_j$ be an operator acting on $\mathbb{C}_j^{1|1}$, potentially oddly graded. We then permute it from the far left of the codeword state to a position in front of the $j$-th codeword, with the sign handled correctly. 
\begin{align}O_j|x_n)\cdots |x_1)=(-1)^{|O_j|\cdot (\sum_{k=j+1}^n|x_k|)} |x_n)\cdots |x_{j+1})O_j|x_j)\cdots |x_1).\label{eq_odd_operator_extended_action}
\end{align}
The action of $a_j$ or $a_j^\dagger$ can then be easily extended to the whole codeword basis $\ket{x}\in(\mathbb{C}^{1|1})^{\otimes n}$ via the recipe in Eq.~\eqref{eq_odd_operator_extended_action}.

The Clifford algebra is an example of a \emph{superalgebra}, which has a $\mathbb{Z}_2$-grading as a vector space. 
One can see from Eq.~\eqref{eq_a_j_adagger_j} that each $a_j$ or $a_j^\dagger$ is oddly graded in $\mathcal{M}_{2n}$. Therefore an element in $\mathcal{M}_{2n}$ is evenly (oddly) graded if it is a linear combination of products of an even (odd) number of $a_j$ or $a_j^\dagger$'s.
One usually defines another basis for $\mathcal{M}_{2n}$ via
\begin{align}
    \gamma_j:=ia^\dagger_j-ia_j,\quad \gamma_{j+n}=a^\dagger_j+a_j,\quad 1\leq j\leq n.
\end{align}
These $\gamma$ operators are usually referred to as \emph{Majorana operators} in the physics literature. In this work, we also often denote $\gamma'_j:=\gamma_{j+n}$ for notational simplicity.
Their anticommutation relation is 
\begin{align}
\{    \gamma_j,\gamma_k\}=2\delta_{jk},\quad 1\leq j,k\leq 2n.
\end{align}
The action of a Majorana operator is, for $1\leq j\leq n$
\begin{align}
\gamma_j|x_j)=&i(-1)^{x_j}|x_j+1),\quad 
\gamma_{j+n}|x_j)=|x_j+1),\quad 
\gamma_{j+n}\gamma_j |x_j)=i(-1)^{x_j}|x_j)
\end{align}
where each $x_j+1$ is understood modulo $2$.
Therefore, when restricted to the $j$-th local space $\mathbb{C}^{1|1}_j$, the action of $\gamma_{j}$ or $\gamma_{j+n}$ can be identified with the Pauli operators
\begin{align}
    \gamma_j=Y_j,\quad \gamma_{j+n}=X_j,\quad \gamma_{j+n}\gamma_j=iZ_j.
\end{align}
Their action on the whole vector space $(\mathbb{C}^{1|1})^{\otimes n}$ can be defined using the approach in Eq.~\eqref{eq_odd_operator_extended_action}
\begin{align}
\gamma_j|x)=(-1)^{\sum_{k=j+1}^n x_k} |x_n)\cdots |x_{j+1})\gamma_j |x_{j}) \cdots |x_1)
\quad 1\leq j\leq n\label{eq_gamma_on_codewords} 
\end{align}
as each Majorana operator is an odd operator. The action of $\gamma_{j+n}$ on a codeword state (a computational basis vector) is the same. Thus one obtains the \emph{Jordan-Wigner transformation}, which is a matrix representation in $GL(\mathbb{C},2^{n})$ of the Clifford algebra in the basis of Majorana operators
\begin{align}
\mathsf{D}_{JW}:\gamma_j\mapsto Z_{n}\cdots Z_{j+1}Y_j,\quad \gamma'_j\equiv\gamma_{j+n}\mapsto Z_n\cdots  Z_{j+1}X_j. \label{eq_DJW}
\end{align}
In a physical system, only evenly graded operations are allowed, so we focus on the subalgebras $\mathcal{M}_{2n}^{2m}$ for positive $m\in \mathbb{Z}^+$. However, all higher-order algebras can be viewed as products of elements from the quadratic subalgebra $\mathcal{M}_{2n}^2$, for which the Jordan-Wigner transformation is
\begin{align}
\mathsf{D}_{JW}:\ \ &  \ \   \gamma_{j+1}\gamma_j\mapsto -iX_{j+1}Y_j,\quad 
P_j\mapsto Z_j
\end{align}
One may choose to compose it with the \emph{Kramers-Wannier} duality
\begin{align}
    \mathsf{D}_{KW}:= Z_j\mapsto Z_{j+\frac{1}{2}} Z_{j-\frac{1}{2}},\quad 
    X_{j+1}X_j\mapsto X_{j+\frac{1}{2}}
\end{align}
which might differ with some authors by a Hadamard transformation that exchanges $X$'s and $Z$'s. The composed transformation is another way to represent the Jordan-Wigner:
\begin{align} 
\mathsf{D}_\text{Bos}:=&\mathsf{D}_{KW}\circ \mathsf{D}_{JW}:\\
&\ \ \ \gamma_{j+1}\gamma_j\mapsto -i Y_{j+\frac{1}{2}}Z_{j-\frac{1}{2}},\quad \quad 
P_{j}\mapsto Z_{j+\frac{1}{2}}Z_{j-\frac{1}{2}}
\end{align}
It is important to point out that, in many ways, Majorana operators play the same role in  the subjet of fermionic qubits as Pauli operators in ordinary qubits. Another point worth mentioning is that a reader from a quantum-computational background should not confuse  the word ``Clifford'' used here to Clifford gates, and they have nothing to do with each other.
\subsection{Gaussian operators}\label{sec_gaussian_operators}
Consider for the moment the Clifford algebra $\mathcal{M}_{2n,\mathbb{R}}$ over real numbers $\mathbb{R}$. It can be naturally decomposed into $\mathcal{M}_{2n,\mathbb{R}}=\bigoplus_{i=1}^{2n} \mathcal{M}_{2n,\mathbb{R}}^{i}$ where the $i$-th sector is the linaar space  spanned by products of $i$ Majorana operators. The second sector $\mathcal{M}_{2n,\mathbb{R}}^{2}$ is called the spin algebra $\mathfrak{spin}(2n)$,  isomorphic to the Lie algebra $\mathfrak{so}(2n)$ of the special orthogonal group.  The exponentiation of these quadratic terms defines a family of  Gaussian unitaries 
\begin{align}
    U_h=\exp(-\gamma^Th\gamma/4),
\end{align}
for any antisymmetric real $2n$-by-$2n$ matrix $h$.  They satisfy the identity \cite{Klich_2014}
\begin{align} 
U_h\gamma_j U_h^\dagger=\sum_{k=1}^{2n}(e^{h})_{jk}\gamma_k,
\end{align}
The $2n\times 2n$ matrix $\alpha$ in the main text is related to the Gaussian operator by the following $2$-by-$2$ block matrix
\begin{align}
   \alpha=
   \left[\begin{array}{c|c}
  [\{0|a_i^\dagger a_j^\dagger U|0)\}_{i,j\in [n]} & \{(0|a_i^\dagger U a_j|0)\}_{i,j\in [n]} \\ \hline
\{(0|a_iU a_j^\dagger |0)\}_{i,j\in [n]} & \{(0|U a_i a_j |0)\}_{i,j\in [n]}
\end{array}\right]
\end{align}
with each $\{\cdots\}_{i,j\in [n]}$ an $n\times n$ sub-matrix.

The same holds for Gaussian density matrices, which are of the form
\begin{align}
    \rho_h\propto \exp(-\gamma^T (ih)\gamma/4)\;.
\end{align}
All we need to do is replace $h$ by $ih$.
\section{Proofs and calculations}
\subsection{Proof for the construction of bosonization in any dimension}\label{proof_any_d_bosonization}
 To prove that the map in Eq.~\eqref{eq_D_bos_any_d} gives the correct statistics, consider the commutation relation following four cases
 $T_{\pm i}:=\mathsf{D}_\text{Bos}(\gamma_{v\pm \hat{x}_i})$ and 
 $T_{\pm j}:=\mathsf{D}_\text{Bos}(\gamma_{v\pm \hat{x}_j})$, where these two inputs overlap at $v$. 
 
 We first consider $\{T_{-i},T_{-j}\}$. 
 There are only two possibility: $-\hat{x}_j^{(v)}<-\hat{x}_i^{(v)}$ or $-\hat{x}_j^{(v)}>-\hat{x}_i^{(v)}$. Without loss of generality we assume $-\hat{x}_j^{(v)}<-\hat{x}_i^{(v)}$. Then $e=-\hat{x}_j^{(v)}$ appears in the last factor 
 $\prod_{\substack{e\in \mathcal{X}_{v}\\ x_0<e<x_i}}Z_e$ of $T_{-i}$, where as $-\hat{x}_i^{(v)}$ does not appear in the factors of $T_j$, resulting the desired anticommutation relation for $\{\prod_{\substack{e\in \mathcal{X}_{v}\\ x_0<e<x_i}}Z_e,(-iY)_{v-\hat{x}_j/2}\}=0$, leading to 
 $\{T_{-i},T_{-j}\}=0$. 
 The other three cases also check through for exactly the same reasoning
 \begin{align}
     \{T_{-i},T_{+j}\}=\{T_{+i},T_{+j}\}=\{T_{+i},T_{-j}\}=0.
 \end{align}
This concludes the proof for  condition Eq.~\eqref{D_commutation_ij}. 

The constraint of Eq.~\eqref{D_commutation_i}, namely $\{T_{-i},T_{+i}\}=0$ follows similarly. The two terms are supported on $(v-\hat{x}_{i},v)$ and $(v,v+\hat{x}_{i})$, respectively. In the edge set $\mathcal{X}_v$ of $v$, $\hat{x}^{(v)}_0<\hat{x}_j^{(v)}<-\hat{x}_j^{(v)}$ by construction, meaning that the last term $\prod_{\substack{e\in \mathcal{X}_{v}\\ x_0<e<x_i}}Z_e$ in $T_{-i}$ contains $Z_{v+\hat{x}/2}$ anticommuting with the factor $(-iY)_{v+\hat{x}/2}$ in $T_{+i}$; whereas there is no factor in $T_{+i}$ anticommuting with $(-iY)_{v-\hat{x}_i/2}$ in $T_{-i}$. Therefore, $\{T_{+i},T_{-i}\}=0$.
\subsection{Proof for the graphical representation of the characteristic function transform}\label{proof_TF}
We only show the hardest one, which is the map of $\gamma'$. The calculation to map the $\gamma$ operator is similar and slightly easier, while the calculations for mapping the identity (straight line) and the parity (a straight line with a black dot) are straightforward by using Eq.~\eqref{eq_spider_self_contraction}; in this characteristic-function convention, the scalar factor $2$ from the first two self-contraction rules is absorbed by the overall normalization of the transform.
\begin{align}
T_F(\gamma')=(-1)\ 
\begin{tikzpicture}[baseline={([yshift=-2.5pt]current bounding box.center)}]   
\def\r{0.5}
\def\ZXr{0.2}
\def\width{1.2}
\def\x{1.5}
\def\D{0.35}
\def\DeltaUD{0.25}
\def\DeltaR{1.3}
\def\y{1}
\def\Y{0.8}
\def\YY{1.0}
\def\lw{1}
\def\delta{0.4}
\def\Delta{0.3}
\def\ticklength{0.43}
  \draw[tickcolor,line width=\lw] (0,0) --++ (0,-1*\ticklength*\r);
  \draw\midarrow{0.85} (0,0) .. controls (0.1*\x*\r,-0.5*\r) and (0.5*\x*\r,0) .. (0.5*\x*\r,\Y*\r);
  \draw\midarrow{0.4} (0.5*\x*\r,-\YY*\r) .. controls (0.5*\x*\r,0) and  (0.1*\x*\r,0.5*\r)  ..(0,0) ;
   \filldraw[fill=black,draw=black](0.5*\x*\r,\Y*\r)circle (0.07*\r);
  \draw\midarrow{0.65} (0,0) .. controls (-0.45*\x*\r,0) and (-0.5*\x*\r,-0.9*\YY*\r) .. (-0.5*\x*\r,-\YY*\r);
   \draw\midarrow{0.07} (-0.5*\x*\r,\Y*\r) .. controls (-0.5*\x*\r,0.9*\Y*\r) and (-0.45*\x*\r,0)  .. (0,0);
 \filldraw[fill=Xcolor,draw=black](0,0)circle (\ZXr*\r);
    \draw (0.5*\x*\r,\Y*\r) ++(0,0)        arc[start angle=0, end angle=180, radius=0.5*\x*\r];
    \draw\dottedpattern (0,\Y*\r+0.5*\x*\r)--++(0,1*\r);
    \draw\midarrow{0.5}((0,\Y*\r+0.5*\x*\r+0.5*\r)--++(0,-0.01*\r);
 \filldraw[fill=Xcolor,draw=black](0,\Y*\r+0.5*\x*\r)circle (\ZXr*\r);
 \filldraw[fill=black,draw=black](0,\Y*\r+0.5*\x*\r)circle (0.07*\r);
 \filldraw[fill=black,draw=black](0.3*\x*\r,\Y*\r+0.41*\x*\r)circle (0.07*\r);
   \end{tikzpicture}
=(-1)\ 
\begin{tikzpicture}[baseline={([yshift=-2.5pt]current bounding box.center)}]   
\def\r{0.5}
\def\ZXr{0.2}
\def\width{1.2}
\def\x{1.5}
\def\D{0.35}
\def\DeltaUD{0.25}
\def\DeltaR{1.3}
\def\y{1}
\def\Y{0.8}
\def\YY{1.0}
\def\lw{1}
\def\delta{0.4}
\def\Delta{0.3}
\def\ticklength{0.43}
  \draw[tickcolor,line width=\lw] (0,0) --++ (0,-1*\ticklength*\r);
  \draw (0,0) .. controls (0.1*\x*\r,-0.5*\r) and (0.5*\x*\r,0) .. (0.5*\x*\r,\Y*\r);
  \draw\midarrow{0.4} (0.5*\x*\r,-\YY*\r) .. controls (0.5*\x*\r,0) and  (0.1*\x*\r,0.5*\r)  ..(0,0) ;
  \draw\midarrow{0.65} (0,0) .. controls (-0.45*\x*\r,0) and (-0.5*\x*\r,-0.9*\YY*\r) .. (-0.5*\x*\r,-\YY*\r);
   \draw\midarrow{0.05} (-0.5*\x*\r,\Y*\r) .. controls (-0.5*\x*\r,0.9*\Y*\r) and (-0.45*\x*\r,0)  .. (0,0);
 \filldraw[fill=Xcolor,draw=black](0,0)circle (\ZXr*\r);
    \draw\midarrow{0.05}  (0.5*\x*\r,\Y*\r) ++(0,0)
        arc[start angle=0, end angle=180, radius=0.5*\x*\r];
    \draw\dottedpattern (0,\Y*\r+0.5*\x*\r)--++(0,1*\r);
    \draw\midarrow{0.5}((0,\Y*\r+0.5*\x*\r+0.5*\r)--++(0,-0.01*\r);
 \filldraw[fill=Xcolor,draw=black](0,\Y*\r+0.5*\x*\r)circle (\ZXr*\r);
 \filldraw[fill=black,draw=black](0,\Y*\r+0.5*\x*\r)circle (0.07*\r);
   \end{tikzpicture}
=(-1)\ 
\begin{tikzpicture}[baseline={([yshift=-2.5pt]current bounding box.center)}]   
\def\r{0.5}
\def\ZXr{0.2}
\def\width{1.2}
\def\x{1.5}
\def\Y{0.4}
\def\R{0.5}
\def\YY{1.7}
\def\lw{1}
\def\delta{0.4}
\def\Delta{0.3}
\def\ticklength{0.43}
    \draw\midarrow{0.51} (0,0) ++(0,0)
        arc[start angle=0, end angle=360, radius=\R*\r];
   \draw\dottedpattern(0,0)--(0,1.2*\r);
   \draw\midarrow{0.6}(0,0.5*\r)--++(0,-0.01*\r);
\newcommand{\lefthalf}{\draw(0,0) ++(0,0)arc[start angle=90, end angle=180, radius=\Y*\r];
  \draw (-\Y*\r,-\Y*\r) .. controls (-\Y*\r,-\Y*\r-0.5*\r) and (0.5*\x*\r,-0.6*\YY*\r) .. (0.5*\x*\r,-\YY*\r);
  }
   \lefthalf
   \begin{scope}[xscale=-1]
   \lefthalf
   \end{scope}
   \draw\midarrow{0.5}(0.28*\x*\r,-0.68*\YY*\r)--++(-0.025,+0.021);
    \draw\midarrow{0.5}(-0.34*\x*\r,-0.72*\YY*\r)--++(-0.025,-0.021);
 \filldraw[fill=Xcolor,draw=black](0,0)circle (\ZXr*\r);
 \filldraw[fill=black,draw=black](0,0)circle (0.07*\r);
\end{tikzpicture}
=(-1)\ 
\begin{tikzpicture}[baseline={([yshift=-2.5pt]current bounding box.center)}]   
\def\r{0.5}
\def\ZXr{0.2}
\def\width{1.2}
\def\x{1.5}
\def\Y{0.4}
\def\R{0.5}
\def\YY{1.7}
\def\lw{1}
\def\delta{0.4}
\def\Delta{0.3}
\def\ticklength{0.43}
    \draw\midarrow{0.55} (0,0) ++(0,0) arc[start angle=360, end angle=0, radius=\R*\r];
      \filldraw[fill=black,draw=black](-\R*\r,\R*\r)circle (0.07*\r);
   \draw\dottedpattern(0,0).. controls (0.3*\r,0) and (0.5*\r,0.9*\r) .. (0.5*\r, 1.2*\r);
   \draw\midarrow{0.6}(0.4*\r,0.6*\r)--++(-0.002*\r,-0.01*\r);
\newcommand{\lefthalf}{\draw(0,0) ++(0,0)arc[start angle=90, end angle=180, radius=\Y*\r];
  \draw (-\Y*\r,-\Y*\r) .. controls (-\Y*\r,-\Y*\r-0.5*\r) and (0.5*\x*\r,-0.6*\YY*\r) .. (0.5*\x*\r,-\YY*\r);
  }
   \lefthalf
   \begin{scope}[xscale=-1]
   \lefthalf
   \end{scope}
   \draw\midarrow{0.5}(0.28*\x*\r,-0.68*\YY*\r)--++(-0.025,+0.021);
    \draw\midarrow{0.5}(-0.34*\x*\r,-0.72*\YY*\r)--++(-0.025,-0.021);
 \filldraw[fill=Xcolor,draw=black](0,0)circle (\ZXr*\r);
 \filldraw[fill=black,draw=black](0,0)circle (0.07*\r);
\end{tikzpicture}
=
\begin{tikzpicture}[baseline={([yshift=-2.5pt]current bounding box.center)}]   
\def\r{0.5}
\def\ZXr{0.2}
\def\width{1.2}
\def\x{1.3}
\def\Y{0.4}
\def\R{0.5}
\def\YY{1.4}
\def\h{1.2}
\def\lw{1}
\def\delta{0.4}
\def\Delta{0.3}
\def\ticklength{0.43}
   \draw\midarrow{0.6}(0.5*\x*\r,0.5*\h*\r)--++(0,-0.01*\r);
 \draw\dottedpattern(0.5*\x*\r,0)--++(0, \h*\r);
\newcommand{\lefthalf}{
  \draw (-0.5*\x*\r,0) -- (-0.5*\x*\r,-\YY*\r);
   \filldraw[fill=Xcolor,draw=black](-0.5*\x*\r,0)circle (\ZXr*\r);
  }
   \lefthalf
   \begin{scope}[xscale=-1]
   \lefthalf
   \end{scope}
   \filldraw[fill=black,draw=black](0.5*\x*\r,0)circle (0.07*\r);
   \draw\midarrow{0.5}(0.5*\x*\r,-0.49*\YY*\r)--++(0,+0.021);
    \draw\midarrow{0.5}(-0.5*\x*\r,-0.61*\YY*\r)--++(0,-0.021);
\end{tikzpicture}\;,
\end{align}
where in the third step we used spider fusion, in the fourth step we switched the arrow direction of the loop, 
in the fifth step we used Eq.~\eqref{eq_spider_self_contraction}, with its scalar factor $2$ absorbed by the overall normalization convention of $T_F$, and finally the third relation in Eq.~\eqref{eq_fusion_and_odd-parity_line_pair} is employed.
\subsection{Proof of soundness}\label{sect_proof_of_soundness}
In this appendix we verify Theorem~\ref{thm_soundness_of_rewrite_rules} for the selected rewrite rules listed below. For the direct algebraic checks below, we choose convenient directions for the external open legs, since any such arrow can be reversed simultaneously on both sides of a graphical equation. For a contracted arrow we read the origin tensor before the target tensor, and for a tick we read the fermionic legs counterclockwise from the tick as in the main text.
\begin{itemize}
\item {\bfseries Identity}.
\begin{align}
\begin{tikzpicture}[baseline={([yshift=-2.5pt]current bounding box.center)}]   
\def\r{0.5}
\def\ZXr{0.25}
\def\x{2}
\def\lw{1}
\def\ticklength{0.5}
\draw\midarrow{0.55}(\x*\r,0) --(0,0);
\end{tikzpicture}
=
\begin{tikzpicture}[baseline={([yshift=-2.5pt]current bounding box.center)}]   
\def\r{0.5}
\def\ZXr{0.25}
\def\x{1.5}
\def\lw{1}
\def\ticklength{0.53}
\def\y{3}
\draw\midarrow{0.6}(\x*\r,0)--(0,0);
\draw\midarrow{0.55}(\y*\r,0)--(\x*\r,0);
      \draw[white,line width=\lw] (\x*\r,0) -- ++(0,-\ticklength*\r);
      \draw[tickcolor,line width=\lw] (\x*\r,0) -- ++(0,\ticklength*\r);
\filldraw[fill=Zcolor,draw=black](\x*\r,0)circle (\ZXr*\r);
\end{tikzpicture}
\quad \quad \quad 
\begin{tikzpicture}[baseline={([yshift=-2.5pt]current bounding box.center)}]   
\def\r{0.5}
\def\ZXr{0.25}
\def\x{1.5}
\def\lw{1}
\def\ticklength{0.5}
\def\y{3}
\draw\midarrow{0.6}(\y*\r,0)--(\x*\r,0);
\draw\midarrow{0.55}(\y*\r+\x*\r,0)--(\y*\r,0);
      \draw[white,line width=\lw] (\x*\r,0) -- ++(0,\ticklength*\r);
      \draw[white,line width=\lw] (\y*\r,0) -- ++(0,-\ticklength*\r);
      \draw[tickcolor,line width=\lw] (\y*\r,0) -- ++(0,\ticklength*\r);
\filldraw[fill=Xcolor,draw=black](\y*\r,0)circle (\ZXr*\r);
\end{tikzpicture}
=
\begin{tikzpicture}[baseline={([yshift=-2.5pt]current bounding box.center)}]   
\def\r{0.5}
\def\x{1.}
\draw\midarrow{0.6}(\x*\r,0)--(-\x*\r,0);
\end{tikzpicture}
\label{eq_fusion_and_odd-parity_line_pair_copy}
\end{align}
\noindent{\bfseries Proof.}
For the fermion-mode identity, the two-leg $Z$-spider has phase $0$, hence
\begin{align*}
(\delta_{a,0}\delta_{b,0}+\delta_{a,1}\delta_{b,1})|a)(b|
=\delta_{a=b}|a)(b|
=\operatorname{id}_{ab}.
\end{align*}
Likewise, the two-leg even $X$-spider contributes
\begin{align*}
\delta_{a+b=0}|a)(b|
=\delta_{a=b}|a)(b|
=\operatorname{id}_{ab},
\end{align*}
so it represents the same solid identity line.

\item {\bfseries Odd-parity rewiring}.
\begin{align}
\begin{tikzpicture}[baseline={([yshift=-2.5pt]current bounding box.center)}]   
\def\r{0.5}
\def\x{1.5}
\draw\dottedmidarrow{0.65}(0,0)--(0,\x*\r);
\draw\dottedmidarrow{0.65}(\x*\r,0)--(\x*\r,\x*\r);
\end{tikzpicture}
\ =\ (-1)\ \ 
\begin{tikzpicture}[baseline={([yshift=-2.5pt]current bounding box.center)}]   
\def\r{0.5}
\def\x{1.7}
\draw\dottedmidarrow{0.7}(0,0)--(\x*\r,\x*\r);
\draw\dottedmidarrow{0.7}(\x*\r,0)--(0,\x*\r);
\end{tikzpicture}
\quad \quad \quad 
\begin{tikzpicture}[baseline={([yshift=-2.5pt]current bounding box.center)}]   
\def\r{0.5}
\def\x{1.5}
\draw\dottedmidarrow{0.65}(0,0)--(0,\x*\r);
\draw\dottedmidarrow{0.65}(\x*\r,0)--(\x*\r,\x*\r);
\end{tikzpicture}
\ \simeq\ 
\begin{tikzpicture}[baseline={([yshift=-2.5pt]current bounding box.center)}]   
\def\r{0.5}
\def\x{1.7}
\draw\dottedmidarrow{0.7}(0,0)--(\x*\r,\x*\r);
\draw\dottedmidarrow{0.7}(\x*\r,0)--(0,\x*\r);
\end{tikzpicture}
\label{eq_odd_parity_rewiring_copy}
\end{align}
\noindent{\bfseries Proof.}
Each dotted leg is the fixed odd state $|1)$ or $(1|$. The two pairings therefore differ only by exchanging two odd basis factors once, and Eq.~\eqref{eq:permutation} gives
\begin{align*}
|1)_a|1)_b=-|1)_b|1)_a.
\end{align*}
This is exactly the factor $(-1)$ in the first equation. After quotienting global phase, the sign is ignored, which gives the second equation. In particular, the odd-parity legs of an odd $X$-spider may be re-paired in any order in the phase-agnostic convention.

\item {\bfseries Tick rotation}.
\begin{align}
    \begin{tikzpicture}[baseline={([yshift=-2.5pt]current bounding box.center)}]   
\def\r{0.5}
\def\ZXr{0.28}
\def\ticklength{0.65}
\def\lw{1}
    \def\radius{1.3}
    \draw[white] (0,0) -- ++ (90:1.4*\r);
    \foreach \angle in { 135, 225,-45, 45} {
        \draw
            (0,0) -- ++ (\angle:\radius*\r);
    }
     \draw (0,0) -- ++ (-90:1.4*\r);
    \draw[tickcolor,line width=\lw](0,0)-- ++ (90:\ticklength*\r);
\filldraw[fill=Xcolor,draw=black](0*\r,0)circle (\ZXr*\r);
\end{tikzpicture}
\ 
=
\ 
\begin{tikzpicture}[baseline={([yshift=-2.5pt]current bounding box.center)}]   
\def\r{0.5}
\def\ZXr{0.28}
\def\ticklength{0.65}
\def\lw{1}
    \def\radius{1.3}
     \draw[white] (0,0) -- ++ (90:1.4*\r);
    \foreach \angle in { 135, 225,-45, 45} {
        \draw
            (0,0) -- ++ (\angle:\radius*\r);
    }
     \draw (0,0) -- ++ (-90:1.4*\r);
    \draw[tickcolor,line width=\lw](0,0)-- ++ (180:\ticklength*\r);
\filldraw[fill=Xcolor,draw=black](0*\r,0)circle (\ZXr*\r);
    \filldraw[fill=black,draw=black](135:0.7*\r) circle (0.1*\r);
\end{tikzpicture}
\ 
=\ 
\begin{tikzpicture}[baseline={([yshift=-2.5pt]current bounding box.center)}]   
\def\r{0.5}
\def\ZXr{0.28}
\def\ticklength{0.65}
\def\lw{1}
    \def\radius{1.3}
     \draw[white] (0,0) -- ++ (90:1.4*\r);
    \foreach \angle in { 135, 225,-45, 45} {
        \draw
            (0,0) -- ++ (\angle:\radius*\r);
    }
     \draw (0,0) -- ++ (-90:1.4*\r);
    \draw[tickcolor,line width=\lw](0,0)-- ++ (0:\ticklength*\r);
\filldraw[fill=Xcolor,draw=black](0*\r,0)circle (\ZXr*\r);
    \filldraw[fill=black,draw=black](45:0.7*\r) circle (0.1*\r);
\end{tikzpicture}
\end{align}
\begin{align}
 \begin{tikzpicture}[baseline={([yshift=-2.5pt]current bounding box.center)}]   
\def\r{0.5}
\def\ZXr{0.28}
\def\ticklength{0.65}
\def\lw{1}
    \def\radius{1.3}
    \draw[white] (0,0) -- ++ (90:1.4*\r);
    \foreach \angle in { 135, 225,-45, 45} {
        \draw
            (0,0) -- ++ (\angle:\radius*\r);
    }
     \draw (0,0) -- ++ (-90:1.4*\r);
    \draw[tickcolor,line width=\lw](0,0)-- ++ (90:\ticklength*\r);
\filldraw[fill=Zcolor,draw=black](0*\r,0)circle (\ZXr*\r);
\end{tikzpicture}
\ 
=
\ 
\begin{tikzpicture}[baseline={([yshift=-2.5pt]current bounding box.center)}]   
\def\r{0.5}
\def\ZXr{0.28}
\def\ticklength{0.65}
\def\lw{1}
    \def\radius{1.3}
     \draw[white] (0,0) -- ++ (90:1.4*\r);
    \foreach \angle in { 135, 225,-45, 45} {
        \draw
            (0,0) -- ++ (\angle:\radius*\r);
    }
     \draw (0,0) -- ++ (-90:1.4*\r);
    \draw[tickcolor,line width=\lw](0,0)-- ++ (180:\ticklength*\r);
\filldraw[fill=Zcolor,draw=black](0*\r,0)circle (\ZXr*\r);
    \filldraw[fill=black,draw=black](135:0.7*\r) circle (0.1*\r);
\end{tikzpicture}
\ 
=
\ 
\begin{tikzpicture}[baseline={([yshift=-2.5pt]current bounding box.center)}]   
\def\r{0.5}
\def\ZXr{0.28}
\def\ticklength{0.65}
\def\lw{1}
    \def\radius{1.3}
     \draw[white] (0,0) -- ++ (90:1.4*\r);
    \foreach \angle in { 135, 225,-45, 45} {
        \draw
            (0,0) -- ++ (\angle:\radius*\r);
    }
     \draw (0,0) -- ++ (-90:1.4*\r);
    \draw[tickcolor,line width=\lw](0,0)-- ++ (180:\ticklength*\r);
\filldraw[fill=Zcolor,draw=black](0*\r,0)circle (\ZXr*\r);
\node at (0,0) {$_\pi$};
\end{tikzpicture}
=\ 
\begin{tikzpicture}[baseline={([yshift=-2.5pt]current bounding box.center)}]   
\def\r{0.5}
\def\ZXr{0.28}
\def\ticklength{0.65}
\def\lw{1}
    \def\radius{1.3}
     \draw[white] (0,0) -- ++ (90:1.4*\r);
    \foreach \angle in { 135, 225,-45, 45} {
        \draw
            (0,0) -- ++ (\angle:\radius*\r);
    }
     \draw (0,0) -- ++ (-90:1.4*\r);
    \draw[tickcolor,line width=\lw](0,0)-- ++ (0:\ticklength*\r);
\filldraw[fill=Zcolor,draw=black](0*\r,0)circle (\ZXr*\r);
    \filldraw[fill=black,draw=black](45:0.7*\r) circle (0.1*\r);
\end{tikzpicture}
\ =\ 
\begin{tikzpicture}[baseline={([yshift=-2.5pt]current bounding box.center)}]   
\def\r{0.5}
\def\ZXr{0.28}
\def\ticklength{0.65}
\def\lw{1}
    \def\radius{1.3}
     \draw[white] (0,0) -- ++ (90:1.4*\r);
    \foreach \angle in { 135, 225,-45, 45} {
        \draw
            (0,0) -- ++ (\angle:\radius*\r);
    }
     \draw (0,0) -- ++ (-90:1.4*\r);
    \draw[tickcolor,line width=\lw](0,0)-- ++ (0:\ticklength*\r);
\filldraw[fill=Zcolor,draw=black](0*\r,0)circle (\ZXr*\r);
\node at (0,0) {$_\pi$};
\end{tikzpicture}
\label{eq_Z_tick_rotation_copy}
\end{align}
\noindent{\bfseries Proof.}
Let the fermion-mode legs of the $X$-spider be read counterclockwise from the tick as $x_n,\dots,x_2,x_1$. Moving the tick once across the leg $x_1$ does not change the tensor entries, but it changes the canonical cyclic ordering. Restoring the original ordering by Eq.~\eqref{eq:permutation} produces
\begin{align*}
|x_n)\cdots |x_2)|x_1)
=(-1)^{x_1(x_2+\cdots+x_n)}|x_1)|x_n)\cdots |x_2).
\end{align*}
On the support of the even $X$-spider, $x_1+\cdots+x_n=0\mod 2$, so $x_2+\cdots+x_n=x_1\mod 2$, and the factor becomes $(-1)^{x_1}$. This is precisely a fermion parity operator on the crossed leg, which is why a black dot appears whenever the tick crosses a solid leg.

For a $Z$-spider, a fermion parity on one solid leg multiplies the all-$1$ sector by $(-1)$ and leaves the all-$0$ sector unchanged:
\begin{align*}
(\delta_{x,\vec{0}}\delta_{y,\vec{0}}+z\,\delta_{x,\vec{1}}\delta_{y,\vec{1}})
\mapsto
(\delta_{x,\vec{0}}\delta_{y,\vec{0}}-z\,\delta_{x,\vec{1}}\delta_{y,\vec{1}}).
\end{align*}
Thus $z\mapsto -z$, or equivalently $\alpha\mapsto \alpha+\pi$ in the phase notation. This is why the black dot can be absorbed into a $\pi$ phase for $Z$-spiders. One must nevertheless keep the index order of an $X$-spider fixed; the sign above is exactly the effect of changing that order.

\item {\bfseries Spider self-contraction}. 
\begin{align}
\begin{tikzpicture}[baseline={([yshift=-2.5pt]current bounding box.center)}]   
\def\r{0.5}
\def\h{.6}
\def\w{1}
\def\x{1.35}
\def\width{1.2}
\def\lw{1}
\def\ticklength{0.6}
  \def\ZXr{0.25}
  \begin{scope}[rotate=90]
      \draw (-0.4*\r,0) ellipse (0.4*\r cm and 0.6*\r cm);
       \draw[->](-0.8*\r,-0.029)--(-0.8*\r,0.03);
    \draw[tickcolor,line width=\lw] (0,0) -- ++(0*\r,\ticklength*\r);
         \draw (0,0) .. controls (0,0.2*\r) and (-0.5*\r,\h*\r) .. (-\x*\r,\h*\r);
       \draw (0,0) .. controls (0,-0.2*\r) and (-0.5*\r,-\h*\r) .. (-\x*\r,-\h*\r);
                 \node at (-0.93*\x*\r,-0.04*\r) {$\cdots$};
          \draw (0,0) .. controls (0,0.2*\r) and (0.5*\r,\h*\r) .. (\x*\r,\h*\r);
       \draw (0,0) .. controls (0,-0.2*\r) and (0.5*\r,-\h*\r) .. (\x*\r,-\h*\r);
        \node at (0.93*\x*\r,-0.04*\r) {$\cdots$};
           \filldraw[fill=Xcolor,draw=black](0*\r,0*\r)circle (\ZXr*\r);
           \end{scope}
\end{tikzpicture}
=\ 2\ 
\begin{tikzpicture}[baseline={([yshift=-2.5pt]current bounding box.center)}]   
\def\r{0.5}
\def\h{.6}
\def\w{1}
\def\x{1.2}
\def\width{1.2}
\def\lw{1}
\def\ticklength{0.6}
  \def\ZXr{0.25}
   \newcommand{\leftpart}{ 
    \draw[tickcolor,line width=\lw] (0,0) -- ++(0*\r,\ticklength*\r);
         \draw (0,0) .. controls (0,0.2*\r) and (-0.5*\r,\h*\r) .. (-\x*\r,\h*\r);
       \draw (0,0) .. controls (0,-0.2*\r) and (-0.5*\r,-\h*\r) .. (-\x*\r,-\h*\r);
                 \node at (-0.93*\x*\r,-0.04*\r) {$\cdots$};
           \filldraw[fill=Xcolor,draw=black](0*\r,0*\r)circle (\ZXr*\r);
           }
        \begin{scope}[rotate=90]
           \leftpart
   \begin{scope}[shift={(1.*\r,0)}]
     \begin{scope}[xscale=-1]
   \leftpart
  \end{scope}
  \end{scope}
  \end{scope}
\end{tikzpicture}
\quad \quad \quad 
\begin{tikzpicture}[baseline={([yshift=-2.5pt]current bounding box.center)}]   
\def\r{0.5}
\def\h{.6}
\def\w{1}
\def\x{1.35}
\def\width{1.2}
\def\lw{1}
\def\ticklength{0.6}
  \def\ZXr{0.25}
  \begin{scope}[rotate=90]
      \draw (-0.4*\r,0) ellipse (0.4*\r cm and 0.6*\r cm);
       \draw[->](-0.8*\r,-0.029)--(-0.8*\r,0.03);
    \draw[tickcolor,line width=\lw] (0,0) -- ++(0*\r,\ticklength*\r);
         \draw (0,0) .. controls (0,0.2*\r) and (-0.5*\r,\h*\r) .. (-\x*\r,\h*\r);
       \draw (0,0) .. controls (0,-0.2*\r) and (-0.5*\r,-\h*\r) .. (-\x*\r,-\h*\r);
                 \node at (-0.93*\x*\r,-0.04*\r) {$\cdots$};
          \draw (0,0) .. controls (0,0.2*\r) and (0.5*\r,\h*\r) .. (\x*\r,\h*\r);
       \draw (0,0) .. controls (0,-0.2*\r) and (0.5*\r,-\h*\r) .. (\x*\r,-\h*\r);
        \node at (0.93*\x*\r,-0.04*\r) {$\cdots$};
           \filldraw[fill=Xcolor,draw=black](0*\r,0*\r)circle (\ZXr*\r);
         \fill[fill=black] (-0.37*\r,0.6*\r) circle (0.1*\r);
           \end{scope}
\end{tikzpicture}
=(-2)\ 
\begin{tikzpicture}[baseline={([yshift=-2.5pt]current bounding box.center)}]   
\def\r{0.5}
\def\h{.6}
\def\w{1}
\def\x{1.2}
\def\width{1.2}
\def\lw{1}
\def\ticklength{0.5}
  \def\ZXr{0.25}
  \def\centralradius{0.06}
   \newcommand{\leftpart}{ 
         \draw (0,0) .. controls (0,0.2*\r) and (-0.5*\r,\h*\r) .. (-\x*\r,\h*\r);
       \draw (0,0) .. controls (0,-0.2*\r) and (-0.5*\r,-\h*\r) .. (-\x*\r,-\h*\r);
                 \node at (-0.93*\x*\r,-0.04*\r) {$\cdots$};
           \filldraw[fill=Xcolor,draw=black](0*\r,0*\r)circle (\ZXr*\r);
           \filldraw[fill=black,draw=black](0*\r,0*\r)circle (\centralradius*\r);
           }
        \begin{scope}[rotate=90]
        \draw\dottedpattern(0,0)--(\r,0);
        \draw\midarrow{0.5}(0.45*\r,0)--(0.45*\r-0.001,0);
           \leftpart
   \begin{scope}[shift={(1.*\r,0)}]
     \begin{scope}[xscale=-1]
   \leftpart
  \end{scope}
  \end{scope}
  \end{scope}
\end{tikzpicture}
\quad \quad \quad 
\begin{tikzpicture}[baseline={([yshift=-2.5pt]current bounding box.center)}]   
\def\r{0.5}
\def\h{.6}
\def\w{1}
\def\x{1.4}
\def\width{1.2}
\def\lw{1}
\def\ticklength{0.6}
  \def\ZXr{0.25}
  \begin{scope}[rotate=90]
      \draw (-0.4*\r,0) ellipse (0.4*\r cm and 0.6*\r cm);
       \draw[->](-0.8*\r,-0.027)--(-0.8*\r,0.028);
    \draw[tickcolor,line width=\lw] (0,0) -- ++(0*\r,\ticklength*\r);
         \draw (0,0) .. controls (0,0.2*\r) and (-0.5*\r,\h*\r) .. (-\x*\r,\h*\r);
       \draw (0,0) .. controls (0,-0.2*\r) and (-0.5*\r,-\h*\r) .. (-\x*\r,-\h*\r);
                 \node at (-0.93*\x*\r,-0.05*\r) {$\cdots$};
          \draw (0,0) .. controls (0,0.2*\r) and (0.5*\r,\h*\r) .. (\x*\r,\h*\r);
       \draw (0,0) .. controls (0,-0.2*\r) and (0.5*\r,-\h*\r) .. (\x*\r,-\h*\r);
        \node at (0.93*\x*\r,-0.05*\r) {$\cdots$};
           \filldraw[fill=Zcolor,draw=black](0*\r,0*\r)circle (\ZXr*\r);
                     \node at (-1.7*\r,0) {$n$};
                       \node at (1.7*\r,0) {$\ $};
             \end{scope}
\end{tikzpicture}
=
\begin{tikzpicture}[baseline={([yshift=-2.5pt]current bounding box.center)}]   
\def\r{0.5}
\def\h{.6}
\def\w{0.45}
\def\x{1.3}
\def\width{1.2}
\def\lw{1}
\def\radius{10}
\def\ticklength{0.7}
  \def\ZXr{0.25}
  \begin{scope}[rotate=90]
    \draw[tickcolor,line width=\lw] (0,0) -- ++(0*\r,\ticklength*\r);
         \draw (0,0) .. controls (0,0.2*\r) and (-0.5*\r,\h*\r) .. (-\x*\r,\h*\r);
       \draw (0,0) .. controls (0,-0.2*\r) and (-0.5*\r,-\h*\r) .. (-\x*\r,-\h*\r);
                 \node at (-0.93*\x*\r,-0.05*\r) {$\cdots$};
          \draw (0,0) .. controls (0,0.2*\r) and (0.5*\r,\h*\r) .. (\x*\r,\h*\r);
       \draw (0,0) .. controls (0,-0.2*\r) and (0.5*\r,-\h*\r) .. (\x*\r,-\h*\r);
        \node at (0.93*\x*\r,-0.05*\r) {$\cdots$};
          \filldraw[fill=Zcolor,rounded corners=\radius*\r] (-0.35*\r,-\w*\r) rectangle (0.35*\r,\w*\r);
           \node at (0,0) {$_{n\pi}$};
          \node at (-1.7*\r,0) {$n$};
          \node at (1.7*\r,0) {$\ $};
          \end{scope}
\end{tikzpicture}
\label{eq_spider_self_contraction_copy}
\end{align}
\noindent{\bfseries Proof.}
For the two $X$-spider relations, let $x$ and $y$ denote the external bit strings below and above the loop, and write
\begin{align*}
p:=\sum_i x_i,\qquad q:=\sum_j y_j \qquad \text{in }\mathbb{F}_2
\end{align*}
for their parities. With the tick convention of the diagrams, the two loop legs are separated by the $x$-block in the counterclockwise ordering. If the loop bit is $c\in\{0,1\}$, then bringing the loop legs into the contraction order $(c||c)$ contributes the sign $(-1)^{cp}$ from Eq.~\eqref{eq:permutation}. Hence the first loop evaluates to
\begin{align*}
\sum_{c=0}^1 (-1)^{cp}\delta_{p+q+2c=0}
=
\bigl(1+(-1)^p\bigr)\delta_{p+q=0}
=
2\,\delta_{p=0}\delta_{q=0}.
\end{align*}
So the surviving sector is exactly the even-even one, namely the tensor product of the two even $X$-spiders on the RHS; the prefactor $2$ is the universal scalar from summing over the closed internal fermion-mode state, and therefore it must appear explicitly in the first displayed rewrite rule.

If one inserts a fermion-parity dot on one branch of the loop, the loop state $c=1$ acquires one additional minus sign. In the odd $X$-spider convention used on the RHS, the resulting dotted odd-parity line contributes one further overall minus sign. Therefore
\begin{align*}
\sum_{c=0}^1 (-1)^c(-1)^{cp}\delta_{p+q+2c=0}
=
\bigl(1-(-1)^p\bigr)\delta_{p+q=0}
=
2\,\delta_{p=1}\delta_{q=1}.
\end{align*}
Thus only the odd-odd sector survives, which is precisely the parity constraint represented by the dotted odd-parity identity line between the two $X$-spiders on the RHS. The sign of $(-2)$ comes from our convention for coupling the ticks of the two odd spiders to the dotted line: in the counterclockwise ordering, one tick is to the right of the dotted line and the other is to the left.

For the $Z$-spider relation, let $x$ be the remaining external bit string and write $\vec c:=(c,\dots,c)$. The $Z$-spider is non-zero only when all external legs and both loop legs carry the same bit $c$. To contract the loop, one moves one loop leg across the $n$ external fermion legs, which contributes $(-1)^{nc}$ because each crossed factor has value $c$. Therefore
\begin{align*}
\sum_{c=0}^1 (-1)^{nc}\delta_{x,\vec c}
=
\delta_{x,\vec 0}+(-1)^n\delta_{x,\vec 1}.
\end{align*}
This is exactly the $Z$-spider with phase $e^{in\pi}=(-1)^n$, namely the box labelled $n\pi$ on the RHS.

\item {\bfseries $Z$-spider fusion}.
\begin{align}
    \begin{tikzpicture}[baseline={([yshift=-2.5pt]current bounding box.center)}]   
\def\r{0.5}
\def\lw{1}
\def\x{1.5}
\def\FontSize{0.9}
    \def\h{0.6}
       \draw(\x*\r,0)[decoration={markings, mark=at position 0.6 with {\arrow{>}}}, postaction={decorate},line width=0.6pt]  -- (0,0);
     \draw (0,0) .. controls (0,0.2*\r) and (-0.5*\r,\h*\r) .. (-1.2*\r,\h*\r);
       \draw\dashedpattern (0,0) .. controls (0,-0.2*\r) and (-0.5*\r,-\h*\r) .. (-1.2*\r,-\h*\r);
       \draw[tickcolor,line width=\lw] (0,0) -- (0.5*\r,0.5*\r);
        \filldraw[fill=Zcolor] (0,0) circle (0.35*\r);
        \node at (0,0) {$_{\alpha}$};
          \node[rotate=90] at (-\r,0.05*\r) {$\cdots$};
  \begin{scope}[shift={(\x*\r,0)}]
  \begin{scope}[xscale=-1]
       \draw (0,0) .. controls (0,0.2*\r) and (-0.5*\r,\h*\r) .. (-1.2*\r,\h*\r);
       \draw\dashedpattern (0,0) .. controls (0,-0.2*\r) and (-0.5*\r,-\h*\r) .. (-1.2*\r,-\h*\r);
       \draw[tickcolor,line width=\lw] (0,0) -- (0.5*\r,0.5*\r);
        \filldraw[fill=Zcolor] (0,0) circle (0.35*\r);
         \node at (0,0) {$_{\beta}$};
          \node[rotate=90] at (-\r,0.05*\r) {$\cdots$};
      \end{scope}
  \end{scope}
\end{tikzpicture}
=
\begin{tikzpicture}[baseline={([yshift=-2.5pt]current bounding box.center)},every node/.style={scale=0.8}]   
\def\r{0.5}
\def\lw{1}
\def\x{1.5}
    \def\h{0.6}
     \def\w{0.65}  
    \def\epsilon{0.33}  
    \def\radius{7.8} 
     \draw (0,0) .. controls (0,0.2*\r) and (-0.5*\r,\h*\r) .. (-\x*\r,\h*\r);
       \draw\dashedpattern (0,0) .. controls (0,-0.2*\r) and (-0.5*\r,-\h*\r) .. (-\x*\r,-\h*\r);
                 \node[rotate=90] at (-0.9*\x*\r,0.05*\r) {$\cdots$};
          \draw (0,0) .. controls (0,0.2*\r) and (0.5*\r,\h*\r) .. (\x*\r,\h*\r);
       \draw \dashedpattern(0,0) .. controls (0,-0.2*\r) and (0.5*\r,-\h*\r) .. (\x*\r,-\h*\r);
        \node[rotate=90] at (0.95*\x*\r,0.05*\r) {$\cdots$};
       \draw[tickcolor,line width=\lw] (0,0) -- (0,0.5*1.414*\r);
       \draw[white,line width=\lw] (0,0) -- (0,-0.5*1.414*\r);
         \filldraw[fill=Zcolor,rounded corners=\radius*\r] (-\w*\r,-\epsilon*\r) rectangle (\w*\r,\epsilon*\r);
         \node at (0,0) { $_{\alpha+\beta}$};
\end{tikzpicture}
\quad \quad 
\begin{tikzpicture}[baseline={([yshift=-2.5pt]current bounding box.center)},every node/.style={scale=0.72}]   
\def\r{0.5}
\def\lw{1}
\def\x{1.5}
 \def\ZXr{0.27}
\def\FontSize{0.9}
    \def\h{0.6}
       \draw(\x*\r,0)[decoration={markings, mark=at position 0.6 with {\arrow{>}}}, postaction={decorate},line width=0.6pt]  -- (0,0);
     \draw (0,0) .. controls (0,0.2*\r) and (-0.5*\r,\h*\r) .. (-1.2*\r,\h*\r);
       \draw\dashedpattern (0,0) .. controls (0,-0.2*\r) and (-0.5*\r,-\h*\r) .. (-1.2*\r,-\h*\r);
       \draw[tickcolor,line width=\lw] (0,0) -- (0.5*\r,0.5*\r);
        \draw[fill=Zcolor] (-\ZXr*\r,-\ZXr*\r) rectangle (\ZXr*\r,\ZXr*\r);
        \node at (0,0) {$_{z_1}$};
          \node[rotate=90] at (-\r,0.05*\r) {$\cdots$};
  \begin{scope}[shift={(\x*\r,0)}]
  \begin{scope}[xscale=-1]
       \draw (0,0) .. controls (0,0.2*\r) and (-0.5*\r,\h*\r) .. (-1.2*\r,\h*\r);
       \draw\dashedpattern (0,0) .. controls (0,-0.2*\r) and (-0.5*\r,-\h*\r) .. (-1.2*\r,-\h*\r);
       \draw[tickcolor,line width=\lw] (0,0) -- (0.5*\r,0.5*\r);
        \draw[fill=Zcolor] (-\ZXr*\r,-\ZXr*\r) rectangle (\ZXr*\r,\ZXr*\r);
         \node at (0,0) {$_{z_2}$};
          \node[rotate=90] at (-\r,0.05*\r) {$\cdots$};
      \end{scope}
  \end{scope}
\end{tikzpicture}
=
\begin{tikzpicture}[baseline={([yshift=-2.5pt]current bounding box.center)},every node/.style={scale=0.76}]   
\def\r{0.5}
\def\lw{1}
\def\x{1.5}
    \def\h{0.6}
     \def\w{0.55}  
    \def\epsilon{0.3}  
    \def\radius{10} 
     \draw (0,0) .. controls (0,0.2*\r) and (-0.5*\r,\h*\r) .. (-\x*\r,\h*\r);
       \draw\dashedpattern (0,0) .. controls (0,-0.2*\r) and (-0.5*\r,-\h*\r) .. (-\x*\r,-\h*\r);
                 \node[rotate=90] at (-0.9*\x*\r,0.05*\r) {$\cdots$};
          \draw (0,0) .. controls (0,0.2*\r) and (0.5*\r,\h*\r) .. (\x*\r,\h*\r);
       \draw \dashedpattern(0,0) .. controls (0,-0.2*\r) and (0.5*\r,-\h*\r) .. (\x*\r,-\h*\r);
        \node[rotate=90] at (0.95*\x*\r,0.05*\r) {$\cdots$};
       \draw[tickcolor,line width=\lw] (0,0) -- (0,0.5*1.414*\r);
       \draw[white,line width=\lw] (0,0) -- (0,-0.5*1.414*\r);
         \filldraw[fill=Zcolor] (-\w*\r,-\epsilon*\r) rectangle (\w*\r,\epsilon*\r);
         \node at (0,0) { $_{z_1z_2}$};
\end{tikzpicture}
\\
\begin{tikzpicture}[baseline={([yshift=-2.5pt]current bounding box.center)}]   
\def\r{0.5}
\def\lw{1}
\def\x{1.5}
\def\FontSize{0.9}
    \def\h{0.6}
       \draw\dashedpattern(\x*\r,0)  -- (0,0);
     \draw (0,0) .. controls (0,0.2*\r) and (-0.5*\r,\h*\r) .. (-1.2*\r,\h*\r);
       \draw\dashedpattern (0,0) .. controls (0,-0.2*\r) and (-0.5*\r,-\h*\r) .. (-1.2*\r,-\h*\r);
       \draw[tickcolor,line width=\lw] (0,0) -- (0.5*\r,0.5*\r);
        \filldraw[fill=Zcolor] (0,0) circle (0.35*\r);
        \node at (0,0) {$_{\alpha}$};
          \node[rotate=90] at (-\r,0.05*\r) {$\cdots$};
  \begin{scope}[shift={(\x*\r,0)}]
  \begin{scope}[xscale=-1]
       \draw (0,0) .. controls (0,0.2*\r) and (-0.5*\r,\h*\r) .. (-1.2*\r,\h*\r);
       \draw\dashedpattern (0,0) .. controls (0,-0.2*\r) and (-0.5*\r,-\h*\r) .. (-1.2*\r,-\h*\r);
       \draw[tickcolor,line width=\lw] (0,0) -- (0.5*\r,0.5*\r);
        \filldraw[fill=Zcolor] (0,0) circle (0.35*\r);
         \node at (0,0) {$_{\beta}$};
          \node[rotate=90] at (-\r,0.05*\r) {$\cdots$};
      \end{scope}
  \end{scope}
\end{tikzpicture}
=
\begin{tikzpicture}[baseline={([yshift=-2.5pt]current bounding box.center)},every node/.style={scale=0.8}]   
\def\r{0.5}
\def\lw{1}
\def\x{1.5}
    \def\h{0.6}
     \def\w{0.65}  
    \def\epsilon{0.33}  
    \def\radius{7.8} 
     \draw (0,0) .. controls (0,0.2*\r) and (-0.5*\r,\h*\r) .. (-\x*\r,\h*\r);
       \draw\dashedpattern (0,0) .. controls (0,-0.2*\r) and (-0.5*\r,-\h*\r) .. (-\x*\r,-\h*\r);
                 \node[rotate=90] at (-0.9*\x*\r,0.05*\r) {$\cdots$};
          \draw (0,0) .. controls (0,0.2*\r) and (0.5*\r,\h*\r) .. (\x*\r,\h*\r);
       \draw \dashedpattern(0,0) .. controls (0,-0.2*\r) and (0.5*\r,-\h*\r) .. (\x*\r,-\h*\r);
        \node[rotate=90] at (0.95*\x*\r,0.05*\r) {$\cdots$};
       \draw[tickcolor,line width=\lw] (0,0) -- (0,0.5*1.414*\r);
       \draw[white,line width=\lw] (0,0) -- (0,-0.5*1.414*\r);
         \filldraw[fill=Zcolor,rounded corners=\radius*\r] (-\w*\r,-\epsilon*\r) rectangle (\w*\r,\epsilon*\r);
         \node at (0,0) { $_{\alpha+\beta}$};
\end{tikzpicture}
\quad \quad 
\begin{tikzpicture}[baseline={([yshift=-2.5pt]current bounding box.center)},every node/.style={scale=0.72}]   
\def\r{0.5}
\def\lw{1}
\def\x{1.5}
 \def\ZXr{0.27}
\def\FontSize{0.9}
    \def\h{0.6}
       \draw\dashedpattern(\x*\r,0)  -- (0,0);
     \draw (0,0) .. controls (0,0.2*\r) and (-0.5*\r,\h*\r) .. (-1.2*\r,\h*\r);
       \draw\dashedpattern (0,0) .. controls (0,-0.2*\r) and (-0.5*\r,-\h*\r) .. (-1.2*\r,-\h*\r);
       \draw[tickcolor,line width=\lw] (0,0) -- (0.5*\r,0.5*\r);
        \draw[fill=Zcolor] (-\ZXr*\r,-\ZXr*\r) rectangle (\ZXr*\r,\ZXr*\r);
        \node at (0,0) {$_{z_1}$};
          \node[rotate=90] at (-\r,0.05*\r) {$\cdots$};
  \begin{scope}[shift={(\x*\r,0)}]
  \begin{scope}[xscale=-1]
       \draw (0,0) .. controls (0,0.2*\r) and (-0.5*\r,\h*\r) .. (-1.2*\r,\h*\r);
       \draw\dashedpattern (0,0) .. controls (0,-0.2*\r) and (-0.5*\r,-\h*\r) .. (-1.2*\r,-\h*\r);
       \draw[tickcolor,line width=\lw] (0,0) -- (0.5*\r,0.5*\r);
        \draw[fill=Zcolor] (-\ZXr*\r,-\ZXr*\r) rectangle (\ZXr*\r,\ZXr*\r);
         \node at (0,0) {$_{z_2}$};
          \node[rotate=90] at (-\r,0.05*\r) {$\cdots$};
      \end{scope}
  \end{scope}
\end{tikzpicture}
=
\begin{tikzpicture}[baseline={([yshift=-2.5pt]current bounding box.center)},every node/.style={scale=0.76}]   
\def\r{0.5}
\def\lw{1}
\def\x{1.5}
    \def\h{0.6}
     \def\w{0.55}  
    \def\epsilon{0.3}  
    \def\radius{10} 
     \draw (0,0) .. controls (0,0.2*\r) and (-0.5*\r,\h*\r) .. (-\x*\r,\h*\r);
       \draw\dashedpattern (0,0) .. controls (0,-0.2*\r) and (-0.5*\r,-\h*\r) .. (-\x*\r,-\h*\r);
                 \node[rotate=90] at (-0.9*\x*\r,0.05*\r) {$\cdots$};
          \draw (0,0) .. controls (0,0.2*\r) and (0.5*\r,\h*\r) .. (\x*\r,\h*\r);
       \draw \dashedpattern(0,0) .. controls (0,-0.2*\r) and (0.5*\r,-\h*\r) .. (\x*\r,-\h*\r);
        \node[rotate=90] at (0.95*\x*\r,0.05*\r) {$\cdots$};
       \draw[tickcolor,line width=\lw] (0,0) -- (0,0.5*1.414*\r);
       \draw[white,line width=\lw] (0,0) -- (0,-0.5*1.414*\r);
         \filldraw[fill=Zcolor] (-\w*\r,-\epsilon*\r) rectangle (\w*\r,\epsilon*\r);
         \node at (0,0) { $_{z_1z_2}$};
\end{tikzpicture}
\end{align}
\noindent{\bfseries Proof.}
Write the shared contracted bit as $c\in\{0,1\}$, and let $x$ and $y$ denote the remaining left and right index strings. Then the contraction of the two $Z$-spiders gives
\begin{align*}
\sum_{c=0}^1
(\delta_{x,\vec{0}}\delta_{c,0}+z_1\delta_{x,\vec{1}}\delta_{c,1})
(\delta_{c,0}\delta_{y,\vec{0}}+z_2\delta_{c,1}\delta_{y,\vec{1}})
&=
\delta_{x,\vec{0}}\delta_{y,\vec{0}}+z_1z_2\delta_{x,\vec{1}}\delta_{y,\vec{1}}.
\end{align*}
Hence the fused spider has label $z_1z_2$. In the phase notation this is $e^{i\alpha}e^{i\beta}=e^{i(\alpha+\beta)}$, so the label becomes $\alpha+\beta$. The same computation applies when the contracted middle leg is dashed instead of solid, because the qubit contraction is the ordinary one.

\item {\bfseries $X$-spider fusion.}
\begin{align}
\begin{tikzpicture}[baseline={([yshift=-2.5pt]current bounding box.center)}]   
\def\r{0.5}
\def\ZXr{0.25}
\def\h{.6}
\def\p{0.85}
\def\width{1.2}
\def\lw{1}
\def\ticklength{0.5}
\def\x{1.2}
\def\y{1.2}
\begin{scope}[rotate=-90]
    \draw\midarrow{0.6}(\p*\x*\r,0) --(0,0) ;
       \draw[tickcolor,line width=\lw] (0,0) -- ++(0,\ticklength*\r);
       \draw[tickcolor,line width=\lw] (\p*\x*\r,0) -- ++(0,\ticklength*\r);
           \draw (0,0) .. controls (0,0.2*\r) and (-0.5*\r,\h*\r) .. (-\x*\r,\h*\r);
       \draw (0,0) .. controls (0,-0.2*\r) and (-0.5*\r,-\h*\r) .. (-\x*\r,-\h*\r);
        \node at (-0.93*\x*\r,  0.04*\r) {$\cdots$};
       \begin{scope}[shift={(\p*\x*\r,0)}]
          \draw (0,0) .. controls (0,0.2*\r) and (0.5*\r,\h*\r) .. (\x*\r,\h*\r);
       \draw(0,0) .. controls (0,-0.2*\r) and (0.5*\r,-\h*\r) .. (\x*\r,-\h*\r);
        \node at (0.93*\x*\r, 0.04*\r) {$\cdots$};
        \end{scope}
        \filldraw[fill=Xcolor,draw=black](0,0)circle (\ZXr*\r);
        \filldraw[fill=Xcolor,draw=black](\p*\x*\r,0)circle (\ZXr*\r);
       \end{scope}
\end{tikzpicture}
\simeq
\begin{tikzpicture}[baseline={([yshift=-2.5pt]current bounding box.center)}]   
\def\r{0.5}
\def\ZXr{0.25}
\def\h{.6}
\def\p{1}
\def\width{1.2}
\def\lw{1}
\def\ticklength{0.5}
\def\x{1.3}
\def\y{1.2}
      \begin{scope}[rotate=-90]
       \draw[tickcolor,line width=\lw] (0,0) -- ++(0,\ticklength*\r);
           \draw (0,0) .. controls (0,0.2*\r) and (-0.5*\r,\h*\r) .. (-\x*\r,\h*\r);
       \draw (0,0) .. controls (0,-0.2*\r) and (-0.5*\r,-\h*\r) .. (-\x*\r,-\h*\r);
                 \node at (-0.93*\x*\r,0.04*\r) {$\cdots$};
       \begin{scope}[shift={(0,0)}]
          \draw (0,0) .. controls (0,0.2*\r) and (0.5*\r,\h*\r) .. (\x*\r,\h*\r);
       \draw(0,0) .. controls (0,-0.2*\r) and (0.5*\r,-\h*\r) .. (\x*\r,-\h*\r);
        \node at (0.93*\x*\r,0.04*\r) {$\cdots$};
        \end{scope}
        \filldraw[fill=Xcolor,draw=black](0,0)circle (\ZXr*\r);
       \end{scope}
\end{tikzpicture}
\quad \quad \quad
\begin{tikzpicture}[baseline={([yshift=-2.5pt]current bounding box.center)}]   
\def\r{0.5}
\def\ZXr{0.25}
\def\centralradius{0.06}
\def\h{.6}
\def\p{0.85}
\def\width{1.2}
\def\lw{1}
\def\ticklength{0.5}
\def\x{1.2}
\def\y{0.8}
\begin{scope}[rotate=90]
    \draw\midarrow{0.6} (0,0) --(\p*\x*\r,0);
      \draw[white](0,0)--(0,\y*\r);
      \draw\dottedpattern(0,0)--(0,-\y*\r);
       \draw[tickcolor,line width=\lw] (\p*\x*\r,0) -- ++(0,-\ticklength*\r);
           \draw (0,0) .. controls (0,0.2*\r) and (-0.5*\r,\h*\r) .. (-\x*\r,\h*\r);
       \draw (0,0) .. controls (0,-0.2*\r) and (-0.5*\r,-\h*\r) .. (-\x*\r,-\h*\r);
                 \node at (-0.93*\x*\r,-0.05*\r) {$\cdots$};
       \begin{scope}[shift={(\p*\x*\r,0)}]
          \draw (0,0) .. controls (0,0.2*\r) and (0.5*\r,\h*\r) .. (\x*\r,\h*\r);
       \draw(0,0) .. controls (0,-0.2*\r) and (0.5*\r,-\h*\r) .. (\x*\r,-\h*\r);
        \node at (0.93*\x*\r,-0.05*\r) {$\cdots$};
        \end{scope}
        \filldraw[fill=Xcolor,draw=black](0,0)circle (\ZXr*\r);
        \filldraw[fill=black,draw=black](0,0)circle (\centralradius*\r);
        \filldraw[fill=Xcolor,draw=black](\p*\x*\r,0)circle (\ZXr*\r);
       \end{scope}
\end{tikzpicture}
=
\begin{tikzpicture}[baseline={([yshift=-2.5pt]current bounding box.center)}]   
\def\r{0.5}
\def\ZXr{0.25}
\def\centralradius{0.06}
\def\h{.6}
\def\p{1}
\def\width{1.2}
\def\lw{1}
\def\x{1.2}
\def\y{0.8}
\begin{scope}[rotate=90]
\draw[white](0,0)--(0,\y*\r);
\draw\dottedpattern(0,0)--(0,-\y*\r);
           \draw (0,0) .. controls (0,0.2*\r) and (-0.5*\r,\h*\r) .. (-\x*\r,\h*\r);
       \draw (0,0) .. controls (0,-0.2*\r) and (-0.5*\r,-\h*\r) .. (-\x*\r,-\h*\r);
                 \node  at (-0.93*\x*\r,-0.05*\r) {$\cdots$};
       \begin{scope}[shift={(0,0)}]
          \draw (0,0) .. controls (0,0.2*\r) and (0.5*\r,\h*\r) .. (\x*\r,\h*\r);
       \draw(0,0) .. controls (0,-0.2*\r) and (0.5*\r,-\h*\r) .. (\x*\r,-\h*\r);
        \node at (0.93*\x*\r,-0.05*\r) {$\cdots$};
        \end{scope}
        \filldraw[fill=Xcolor,draw=black](0,0)circle (\ZXr*\r);
        \filldraw[fill=black,draw=black](0,0)circle (\centralradius*\r);
       \end{scope}
\end{tikzpicture}
\quad \quad \quad
\begin{tikzpicture}[baseline={([yshift=-2.5pt]current bounding box.center)}]   
\def\r{0.5}
\def\ZXr{0.25}
\def\centralradius{0.06}
\def\h{.6}
\def\p{0.9}
\def\width{1.2}
\def\lw{1}
\def\ticklength{0.5}
\def\x{1.1}
\def\y{1.05}
\draw\midarrow{0.6} (0,0) -- (0,\p*\x*\r);
\draw\dottedpattern (0,0) -- (\y*\r,0);
\draw\dottedpattern (0,\p*\x*\r) -- (\y*\r,\p*\x*\r);
\draw (0,0) .. controls (0,-0.2*\r) and (-0.5*\r,-0.5*\r) .. (-\h*\r,-\x*\r);
\draw (0,0) .. controls (0,-0.2*\r) and (0.5*\r,-0.5*\r) .. (\h*\r,-\x*\r);
\node at (0,-0.93*\x*\r) {$\cdots$};
\begin{scope}[shift={(0,\p*\x*\r)}]
\draw (0,0) .. controls (0,0.2*\r) and (-0.5*\r,0.5*\r) .. (-\h*\r,\x*\r);
\draw (0,0) .. controls (0,0.2*\r) and (0.5*\r,0.5*\r) .. (\h*\r,\x*\r);
\node at (0,0.93*\x*\r) {$\cdots$};
\filldraw[fill=Xcolor,draw=black](0,0)circle (\ZXr*\r);
\filldraw[fill=black,draw=black](0,0)circle (\centralradius*\r);
\end{scope}
\filldraw[fill=Xcolor,draw=black](0,0)circle (\ZXr*\r);
\filldraw[fill=black,draw=black](0,0)circle (\centralradius*\r);
\end{tikzpicture}
=
(-1)
\begin{tikzpicture}[baseline={([yshift=-2.5pt]current bounding box.center)},every node/.style={scale=0.76}]   
\def\r{0.5}
\def\ZXr{0.25}
\def\h{.6}
\def\Delta{1.45}
\def\arcR{0.6}
\def\width{1.2}
\def\lw{1}
\def\ticklength{0.5}
\def\x{1.1}
\draw (0,0) .. controls (0,0.2*\r) and (-0.5*\r,0.5*\r) .. (-\h*\r,\x*\r);
\draw (0,0) .. controls (0,0.2*\r) and (0.5*\r,0.5*\r) .. (\h*\r,\x*\r);
\node at (0,0.93*\x*\r) {$\cdots$};
\draw (0,0) .. controls (0,-0.2*\r) and (-0.5*\r,-0.5*\r) .. (-\h*\r,-\x*\r);
\draw (0,0) .. controls (0,-0.2*\r) and (0.5*\r,-0.5*\r) .. (\h*\r,-\x*\r);
\node at (0,-0.93*\x*\r) {$\cdots$};
\draw\dottedpattern (\Delta*\r,\arcR*\r) arc (90:270:\arcR*\r);
\draw[tickcolor,line width=\lw] (0,0) -- ++(\ticklength*\r,0);
\filldraw[fill=Xcolor,draw=black](0,0)circle (\ZXr*\r);
\end{tikzpicture}
\end{align}
\noindent{\bfseries Proof.}
As explained in Sect.~\ref{sec_z2_tensor_networks}, the important data here are the orderings. For a contracted arrow we write the origin tensor first and the target tensor second, and within each tensor the tick fixes the counterclockwise starting point. The braket symbols below are only mnemonic for this ordering convention.

For the first displayed equation, let \(x\) and \(y\) denote the remaining top and bottom solid bitstrings. Since the tick is on the RHS of the contracted arrow, the ordered product is of the form \([\,x\,(c|\,]_{\mathrm{top}}[\,|c)\,y\,]_{\mathrm{bottom}}\), so the contracted pair already sits in the standard middle position \((c||c)\). Hence
\begin{align*}
\sum_{c\in\mathbb{F}_2}\delta_{|x|+c,0}\,\delta_{c+|y|,0}
=
\delta_{|x|+|y|,0},
\end{align*}
which is exactly the defining coefficient of the fused even \(X\)-spider on the RHS.

For the second displayed equation, the same convention gives an ordered product of the form \([\,|1)\,x\,(c|\,]_{\mathrm{top}}[\,|c)\,y\,]_{\mathrm{bottom}}\). The fused odd spider on the RHS has the same bookkeeping issue: in both sides one has to move the odd factor \( |1)\) to the end in order to pass to the shorthand odd-\(X\)-spider notation used in the rule statement. The same fermionic sign is produced on both sides, so it cancels. After this common recording change, the solid contraction is
\begin{align*}
\sum_{c\in\mathbb{F}_2}\delta_{|x|+c,1}\,\delta_{c+|y|,0}
=
\delta_{|x|+|y|,1},
\end{align*}
which is precisely the defining coefficient of the fused odd \(X\)-spider.

For the third displayed equation, take the upper dotted leg to be outgoing and the lower dotted leg incoming, exactly as drawn. Then the ordered product is schematically
\begin{align*}
[\,|1)\,x\,(c|\,]_{\mathrm{top}}
[\, (1|\,|c)\,y\,]_{\mathrm{bottom}} .
\end{align*}
Now move the leading \((1|\) of the bottom tensor to the very end. Because the remaining block \( |c)\,y\) has odd total parity for an odd \(X\)-spider, this produces a single factor \((-1)\). After that move the contracted legs sit in the standard middle position \((c||c)\), while the remaining \( |1)(1|\) is exactly the dotted odd-parity identity line on the RHS. The solid contraction is
\begin{align*}
\sum_{c\in\mathbb{F}_2}\delta_{|x|+c,1}\,\delta_{c+|y|,1}
=
\delta_{|x|+|y|,0},
\end{align*}
so the fused \(X\)-spider is even. This is the displayed third equation with its explicit factor \((-1)\).

\item  {\bfseries $\pi$ commutation}.
\begin{align}
\begin{tikzpicture}[baseline={([yshift=-2.5pt]current bounding box.center)}]   
\def\r{0.7}
\def\h{1}
\def\x{0.7}
\def\H{1.5}
\def\flip{-1}
\def\w{2}
\def\sep{1}
\def\delta{0.45}
\def\lw{1}
\def\ticklength{0.5}
\def\centralradius{0.06}
  \def\ZXr{0.25}
     \newcommand{\leftpart}{ 
 \draw[decoration={markings, mark=at position 0.6 with {\arrow{>}}}, postaction={decorate}] (0,0) .. controls (-0.8*\r,0.*\r) and (-\w*\r,0.3*\r) .. (-\w*\r,\h*\r);
  \draw[decoration={markings, mark=at position 0.6 with {\arrow{>}}}, postaction={decorate}] (0,0) .. controls (-0.6*\r,0.*\r) and (-\w*\r+\sep*\r,0.3*\r) .. (-\w*\r+\sep*\r,\h*\r);
 \draw(-\w*\r+\sep*\r,\h*\r)--(-\w*\r+\sep*\r,\H*\r);
  \draw(-\w*\r,\h*\r)--(-\w*\r,\H*\r);
   }
   \leftpart
    \begin{scope}[shift={(0,0)}]
  \begin{scope}[xscale=-1]
  \leftpart
  \end{scope}
  \end{scope}
    \node at (0,\h*\r) {$\cdots$};
    \newcommand{\block}{ 
    \draw\dottedpattern (0,0)--(\sep*\r,0);
       \filldraw[fill=Xcolor,draw=black](0*\r,0*\r)circle (0.7*\ZXr*\r);
        \filldraw[fill=Xcolor,draw=black](\sep*\r,0*\r)circle (0.7*\ZXr*\r);
        \filldraw[fill=black,draw=black](0*\r,0*\r)circle (\centralradius*\r);
        \filldraw[fill=black,draw=black](\sep*\r,0*\r)circle (\centralradius*\r);
        }
   \begin{scope}[shift={(-\w*\r,\h*\r)}]
   \block
     \end{scope}
       \begin{scope}[shift={(\w*\r-\sep*\r,\h*\r)}]
   \block
     \end{scope}
    \draw (-0.927*\w*\r,1.2*\H*\r) arc (90:180:0.15*\r);
    \draw (-0.927*\w*\r,1.2*\H*\r)--(-0.15*\r,1.2*\H*\r);
    \draw (-0.15*\r,1.2*\H*\r) arc (-90:0:0.15*\r);
    \begin{scope}[xscale=-1]
    \draw (-0.927*\w*\r,1.2*\H*\r) arc (90:180:0.15*\r);
    \draw (-0.927*\w*\r,1.2*\H*\r)--(-0.15*\r,1.2*\H*\r);
    \draw (-0.15*\r,1.2*\H*\r) arc (-90:0:0.15*\r);
    \end{scope}
    \node at (0,1.45*\H*\r) {$_{2m}$};
           \draw\dashedpattern (0,0) .. controls (0.2*\r,0) and (\x*\r,-0.5*\r) .. (\x*\r,-\h*\r);
       \draw\dashedpattern (0,0) .. controls (-0.2*\r,0) and (-\x*\r,-0.5*\r) .. (-\x*\r,-\h*\r);
        \draw\dashedpattern (\x*\r,-\H*\r)-- (\x*\r,-\h*\r);
        \draw\dashedpattern (-\x*\r,-\H*\r)-- (-\x*\r,-\h*\r);
        \filldraw[fill=Xcolor,draw=black](-\x*\r,-\h*\r)circle (0.8*\ZXr*\r);
        \filldraw[fill=Xcolor,draw=black](\x*\r,-\h*\r)circle (0.8*\ZXr*\r);
         \node at (-\x*\r,-\h*\r) {$_\pi$};
         \node at (\x*\r,-\h*\r) {$_\pi$};
               \node at (0.05*\r,-0.9*\h*\r) {$\cdots$};
 \draw[tickcolor,line width=\lw] (0,0) -- ++(-0*\ticklength*\r,-1*\ticklength*\r);
 \filldraw[fill=Zcolor,draw=black](0*\r,0*\r)circle (\ZXr*\r);
  \node at (0,0) {$_\beta$};
\end{tikzpicture}
= e^{i(\beta+m\pi)}
\begin{tikzpicture}[baseline={([yshift=-2.5pt]current bounding box.center)}]   
\def\r{0.6}
\def\h{.8}
\def\w{1.05}
\def\x{1.3}
\def\width{1.2}
\def\lw{1}
\def\radius{10}
\def\ticklength{0.6}
  \def\ZXr{0.25}
    \draw[tickcolor,line width=\lw] (0,0) -- ++(0*\r,-\ticklength*\r);
         \draw[decoration={markings, mark=at position 0.7 with {\arrow{>}}}, postaction={decorate}] (0,0) .. controls (0.2*\r,0) and (\h*\r,0.5*\r) .. (\h*\r,\x*\r);
       \draw[decoration={markings, mark=at position 0.7 with {\arrow{>}}}, postaction={decorate}] (0,0) .. controls (-0.2*\r,0) and (-\h*\r,0.5*\r) .. (-\h*\r,\x*\r);
           \draw\dashedpattern (0,0) .. controls (0.2*\r,0) and (\h*\r,-0.5*\r) .. (\h*\r,-\x*\r);
       \draw\dashedpattern (0,0) .. controls (-0.2*\r,0) and (-\h*\r,-0.5*\r) .. (-\h*\r,-\x*\r);
                 \node at (0.05*\r,0.9*\x*\r) {$\cdots$};
                 \filldraw[fill=Zcolor,rounded corners=\radius*\r] (-\w*\r,-0.35*\r) rectangle (\w*\r,0.35*\r);
           \node at (0,0) {$_{-\beta+m\pi}$};
             \node at (0.05*\r,-0.9*\x*\r) {$\cdots$};
\end{tikzpicture}
\label{eq_beta_mpi_copy}
\end{align}
\begin{align}
\begin{tikzpicture}[baseline={([yshift=-2.5pt]current bounding box.center)}]   
\def\r{0.5}
\def\h{.7}
\def\w{1}
\def\x{1.5}
\def\width{1.2}
\def\lw{1}
\def\ticklength{0.6}
  \def\ZXr{0.25}
    \draw[tickcolor,line width=\lw] (0,0) -- ++(0*\r,\ticklength*\r);
         \draw (0,0) .. controls (0,0.2*\r) and (-0.5*\r,\h*\r) .. (-\x*\r,\h*\r);
       \draw (0,0) .. controls (0,-0.2*\r) and (-0.5*\r,-\h*\r) .. (-\x*\r,-\h*\r);
                 \node[rotate=90] at (-0.8*\x*\r,0.05*\r) {$\cdots$};
          \draw (0,0) .. controls (0,0.2*\r) and (0.5*\r,\h*\r) .. (\x*\r,\h*\r);
       \draw (0,0) .. controls (0,-0.2*\r) and (0.5*\r,-\h*\r) .. (\x*\r,-\h*\r);
        \node[rotate=90] at (0.82*\x*\r,0.05*\r) {$\cdots$};
           \filldraw[fill=Xcolor,draw=black](0*\r,0*\r)circle (\ZXr*\r);
        \filldraw[fill=black,line width=\lw] (-0.8*\x*\r,0.95*\h*\r) circle (0.075*\r); 
         \filldraw[fill=black,line width=\lw] (-0.8*\x*\r,-0.95*\h*\r) circle (0.075*\r); 
         \filldraw[fill=black,line width=\lw] (0.8*\x*\r,0.95*\h*\r) circle (0.075*\r); 
         \filldraw[fill=black,line width=\lw] (0.8*\x*\r,-0.95*\h*\r) circle (0.075*\r); 
\end{tikzpicture}
=
\begin{tikzpicture}[baseline={([yshift=-2.5pt]current bounding box.center)}]   
\def\r{0.5}
\def\h{.7}
\def\w{1}
\def\x{1.5}
\def\width{1.2}
\def\lw{1}
\def\ticklength{0.6}
  \def\ZXr{0.25}
    \draw[tickcolor,line width=\lw] (0,0) -- ++(0*\r,\ticklength*\r);
         \draw (0,0) .. controls (0,0.2*\r) and (-0.5*\r,\h*\r) .. (-\x*\r,\h*\r);
       \draw (0,0) .. controls (0,-0.2*\r) and (-0.5*\r,-\h*\r) .. (-\x*\r,-\h*\r);
                 \node[rotate=90] at (-0.8*\x*\r,0.05*\r) {$\cdots$};
          \draw (0,0) .. controls (0,0.2*\r) and (0.5*\r,\h*\r) .. (\x*\r,\h*\r);
       \draw (0,0) .. controls (0,-0.2*\r) and (0.5*\r,-\h*\r) .. (\x*\r,-\h*\r);
        \node[rotate=90] at (0.82*\x*\r,0.05*\r) {$\cdots$};
           \filldraw[fill=Xcolor,draw=black](0*\r,0*\r)circle (\ZXr*\r);
\end{tikzpicture}
\quad \quad \quad \quad 
\begin{tikzpicture}[baseline={([yshift=-2.5pt]current bounding box.center)}]   
\def\r{0.5}
\def\h{.7}
\def\w{1}
\def\x{1.5}
\def\width{1.2}
\def\lw{1}
\def\ticklength{0.6}
\def\centralradius{0.06}
  \def\ZXr{0.25}
  \draw\dottedpattern(0,0)--(0,-\h*\r);
         \draw (0,0) .. controls (0,0.2*\r) and (-0.5*\r,\h*\r) .. (-\x*\r,\h*\r);
       \draw (0,0) .. controls (0,-0.2*\r) and (-0.5*\r,-\h*\r) .. (-\x*\r,-\h*\r);
                 \node[rotate=90] at (-0.8*\x*\r,0.05*\r) {$\cdots$};
          \draw (0,0) .. controls (0,0.2*\r) and (0.5*\r,\h*\r) .. (\x*\r,\h*\r);
       \draw (0,0) .. controls (0,-0.2*\r) and (0.5*\r,-\h*\r) .. (\x*\r,-\h*\r);
        \node[rotate=90] at (0.82*\x*\r,0.05*\r) {$\cdots$};
           \filldraw[fill=Xcolor,draw=black](0*\r,0*\r)circle (\ZXr*\r);
           \filldraw[fill=black,draw=black](0,0)circle (\centralradius*\r);
        \filldraw[fill=black,line width=\lw] (-0.8*\x*\r,0.95*\h*\r) circle (0.075*\r); 
         \filldraw[fill=black,line width=\lw] (-0.8*\x*\r,-0.95*\h*\r) circle (0.075*\r); 
         \filldraw[fill=black,line width=\lw] (0.8*\x*\r,0.95*\h*\r) circle (0.075*\r); 
         \filldraw[fill=black,line width=\lw] (0.8*\x*\r,-0.95*\h*\r) circle (0.075*\r); 
\end{tikzpicture}
=
(-1)
\begin{tikzpicture}[baseline={([yshift=-2.5pt]current bounding box.center)}]   
\def\r{0.5}
\def\h{.7}
\def\w{1}
\def\x{1.5}
\def\width{1.2}
\def\lw{1}
\def\ticklength{0.6}
\def\centralradius{0.06}
  \def\ZXr{0.25}
  \draw\dottedpattern(0,0)--(0,-\h*\r);
         \draw (0,0) .. controls (0,0.2*\r) and (-0.5*\r,\h*\r) .. (-\x*\r,\h*\r);
       \draw (0,0) .. controls (0,-0.2*\r) and (-0.5*\r,-\h*\r) .. (-\x*\r,-\h*\r);
                 \node[rotate=90] at (-0.8*\x*\r,0.05*\r) {$\cdots$};
          \draw (0,0) .. controls (0,0.2*\r) and (0.5*\r,\h*\r) .. (\x*\r,\h*\r);
       \draw (0,0) .. controls (0,-0.2*\r) and (0.5*\r,-\h*\r) .. (\x*\r,-\h*\r);
        \node[rotate=90] at (0.82*\x*\r,0.05*\r) {$\cdots$};
           \filldraw[fill=Xcolor,draw=black](0*\r,0*\r)circle (\ZXr*\r);
           \filldraw[fill=black,draw=black](0,0)circle (\centralradius*\r);
\end{tikzpicture}
\;.
\label{eq_parity_surrounding_X_spider_copy}
\end{align}
\noindent{\bfseries Proof.}
Eq.~\eqref{eq_beta_mpi_copy} is due to the following identity
\begin{align}
    \gamma_{2m}\cdots \gamma_{1} (\ket{\overline{0}}+e^{i\beta} \ket{\overline{1}})
    =e^{i\beta} (-1)^{m} (\ket{\overline{0}}+e^{-i\beta} (-1)^{m}\ket{\overline{1}}).
\end{align}
In Eq.~\eqref{eq_parity_surrounding_X_spider_copy}, we put a fermion parity on each of the Fermionic legs. The two identities are merely re-stating the definition of an $X$-spider with an even (odd) parity in the first (second) equation of Eq.~\eqref{eq_parity_surrounding_X_spider_copy}.
The second and third equations in the main text are treated in the same way. The only extra factor is $(-1)$ coming from odd-parity rewiring, and this factor is discarded once we pass from $=$ to $\simeq$.

\item {\bfseries FQ triangle annihilation}. 
\begin{align}
\sqrt{2}\ 
\begin{tikzpicture}[baseline={([yshift=-2.5pt]current bounding box.center)}]
\def\d{0.33}
\def\r{0.085}
\def\rr{1.5}
\def\lw{1}
\def\ticklength{0.6}
\def\h{0.33}
\def\s{0.0707}
   \draw (0,0) --(0,-\d*\rr);
    \draw\midarrow{0.75} (1.6*\d*\rr,1.6*\d*\rr)-- (0,0);
      \draw \midarrow{0.75}(-1.6*\d*\rr,1.6*\d*\rr)--(0,0) ;
     \draw[dash pattern=on 1.5pt off 1pt] (-\h*\rr,\h*\rr)--(\h*\rr,\h*\rr);
       \filldraw[fill=yellow] (-\s*\rr,-\s*\rr+\h*\rr) rectangle (\s*\rr,\s*\rr+\h*\rr);
        \draw[tickcolor,line width=\lw] (0,0) -- ++(-135: 3.5*\ticklength*\r*\rr);
         \filldraw[fill=Xcolor] (0,0) circle (\r*\rr);
          \draw[tickcolor,line width=\lw] (\h*\rr,\h*\rr) -- ++(-45: 3.5*\ticklength*\r*\rr);
           \draw[tickcolor,line width=\lw] (-\h*\rr,\h*\rr)-- ++(-135: 3.5*\ticklength*\r*\rr);
         \filldraw[fill=Zcolor] (\h*\rr,\h*\rr) circle (\r*\rr);
          \filldraw[fill=Zcolor] (-\h*\rr,\h*\rr) circle (\r*\rr);
\end{tikzpicture}
\ =\ 
\begin{tikzpicture}[baseline={([yshift=-2.5pt]current bounding box.center)}]
\def\d{0.3}
\def\rr{1.5}
\def\lw{1}
\def\ticklength{0.6}
\def\r{0.085}
    \draw[tickcolor,line width=\lw] (0,0) -- ++(-135: 3.5*\ticklength*\r*\rr);
  \draw[white] (0,0) --(0,1.6*\d*\rr);
  \draw (0,0) .. controls (-0.4*\d*\rr,0.3*\d*\rr) and (-0.6*\d*\rr,0.75*\d*\rr) .. (\d*\rr,1.3*\d*\rr);
   \draw (0,0) .. controls (0.4*\d*\rr,0.3*\d*\rr) and (0.6*\d*\rr,0.75*\d*\rr) .. (-\d*\rr,1.3*\d*\rr);
   \draw (0,0) --(0,-1.*\d*\rr);
  \filldraw[fill=Xcolor] (0,0) circle (\r*\rr);
\end{tikzpicture}
\end{align}
\noindent{\bfseries Proof.}
Project the two upper dashed legs to qubit values $x,y\in\{0,1\}$. The two $Z$-spiders then copy these values to the two upper fermion legs of the $X$-spider. To rewrite the resulting tensor in the canonical tick order, one exchanges those two fermion factors once, which gives the fermionic sign $(-1)^{xy}$ from Eq.~\eqref{eq:permutation}. The dashed edge carries the Hadamard tensor, whose matrix element is
$H_{xy}=(-1)^{xy}/\sqrt{2}$.
Therefore the coefficient on the LHS is
$\sqrt{2}\,H_{xy}\,(-1)^{xy}=1$,
so every basis coefficient agrees with the RHS $X$-spider.

\item {\bfseries Hopf}. 
\begin{align}
\begin{tikzpicture}[baseline={([yshift=-2.5pt]current bounding box.center)}]   
\def\r{0.6}
\def\h{0.5}
\def\H{0.3}
\def\x{1}
\def\X{0.6}
\def\lw{1}
\def\ticklength{0.5}
  \def\ZXr{0.25}
  \draw\midarrow{0.5}(-0.1*\r,\H*\r)--(-0.1*\r-0.02*\r,\H*\r);
    \draw\midarrow{0.5}(-0.1*\r,-\H*\r)--(-0.1*\r-0.02*\r,-\H*\r);
\draw (0*\r,0) ellipse (\X*\r cm and \H*\r cm);
   \begin{scope}[shift={(\X*\r,0*\r)}]
     \draw[tickcolor,line width=\lw] (0*\r,0*\r) -- ++(0,1*\ticklength*\r);   
    \draw(0,0) .. controls (0,-0.2*\r) and (0.5*\r,-\h*\r) .. (\x*\r,-\h*\r);
            \draw(0,0) .. controls (0,0.2*\r) and (0.5*\r,\h*\r) .. (\x*\r,\h*\r);
 \filldraw[fill=Xcolor,draw=black](0*\r,0*\r)circle (\ZXrEmpty); 
  \node[rotate=90] at (0.97*\x*\r,0.07*\r) {$\cdots$};
  \end{scope}
 \begin{scope}[shift={(-\X*\r,0)}]
  \draw[tickcolor,line width=\lw] (0*\r,0*\r) -- ++(0,1*\ticklength*\r);
     \draw (0,0) .. controls (0,-0.2*\r) and (-0.5*\r,-\h*\r) .. (-\x*\r,-\h*\r);
            \draw (0,0) .. controls (0,0.2*\r) and (-0.5*\r,\h*\r) .. (-\x*\r,\h*\r);
              \node[rotate=90] at (-0.95*\x*\r,0.07*\r) {$\cdots$};
               \filldraw[fill=Zcolor,draw=black](0*\r,0*\r)circle (\ZXrEmpty);  
       \end{scope}
 \end{tikzpicture} 
    =
\begin{tikzpicture}[baseline={([yshift=-2.5pt]current bounding box.center)}]   
\def\r{0.6}
\def\h{0.5}
\def\H{0.3}
\def\x{1}
\def\X{0.6}
\def\lw{1}
\def\ticklength{0.5}
  \def\ZXr{0.25}
   \begin{scope}[shift={(\X*\r,0*\r)}]
     \draw[tickcolor,line width=\lw] (0*\r,0*\r) -- ++(0,1*\ticklength*\r);   
    \draw(0,0) .. controls (0,-0.2*\r) and (0.5*\r,-\h*\r) .. (\x*\r,-\h*\r);
            \draw(0,0) .. controls (0,0.2*\r) and (0.5*\r,\h*\r) .. (\x*\r,\h*\r);
 \filldraw[fill=Xcolor,draw=black](0*\r,0*\r)circle (\ZXrEmpty); 
  \node[rotate=90] at (0.97*\x*\r,0.07*\r) {$\cdots$};
  \end{scope}
 \begin{scope}[shift={(-\X*\r,0)}]
  \draw[tickcolor,line width=\lw] (0*\r,0*\r) -- ++(0,1*\ticklength*\r);
     \draw (0,0) .. controls (0,-0.2*\r) and (-0.5*\r,-\h*\r) .. (-\x*\r,-\h*\r);
            \draw (0,0) .. controls (0,0.2*\r) and (-0.5*\r,\h*\r) .. (-\x*\r,\h*\r);
              \node[rotate=90] at (-0.95*\x*\r,0.07*\r) {$\cdots$};
               \filldraw[fill=Zcolor,draw=black](0*\r,0*\r)circle (\ZXrEmpty);  
       \end{scope}
       \end{tikzpicture}
 \;.
\end{align}
\noindent{\bfseries Proof.}
Let the two contracted indices between the $Z$- and $X$-spider be $c_1,c_2\in\{0,1\}$. The $Z$-spider is non-zero only when $c_1=c_2=y$, where $y$ is also the common value on its remaining legs. On the $X$-spider side, let $x$ denote the sum of all the external legs on the right. The internal bits contribute $c_1+c_2=y+y=0\mod 2$ to the parity constraint, so they do not affect the even-parity condition on the remaining external $X$-legs. Algebraically,
\begin{align*}
\sum_{c_1,c_2}
\delta_{c_1,y}\delta_{c_2,y}\,
\delta_{|x|+c_1+c_2=0}
=
\delta_{|x|=0}.
\end{align*}
Hence the internal contraction factors out, leaving precisely the disconnected $Z$- and $X$-spiders on the RHS.

\item {\bfseries Bialgebra}.
\begin{align}
\begin{tikzpicture}[baseline={([yshift=-2.5pt]current bounding box.center)}]   
\def\r{0.5}
\def\ZXr{0.25}
\def\lw{1}
\def\x{1.3}
\def\y{0.7}
  \draw(\x*\r,-\y*\r)[decoration={markings, mark=at position 0.6 with {\arrow{>}}}, postaction={decorate}]  --(0.*\r,0) ;
    \draw(\x*\r,\y*\r)[decoration={markings, mark=at position 0.6 with {\arrow{>}}}, postaction={decorate}]  --(0.*\r,0) ;
     \draw(2*\x*\r,0)[decoration={markings, mark=at position 0.6 with {\arrow{>}}}, postaction={decorate}]  --(\x*\r,-\y*\r) ;
     \draw(2*\x*\r,0)[decoration={markings, mark=at position 0.6 with {\arrow{>}}}, postaction={decorate}]  --(\x*\r,\y*\r) ;
	    \draw(0,0) --(-\x*\r,0) ;
	    \draw[dashed](\x*\r,\y*\r) --(\x*\r,2.5*\y*\r) ;
	    \draw[dashed](\x*\r,-\y*\r) --(\x*\r,-2.5*\y*\r) ;
	    \draw(2*\x*\r,0) --(3*\x*\r,0) ;
	     \draw[tickcolor,line width=\lw] (0,0) -- (0,-0.6*\r);
     \draw[tickcolor,line width=\lw] (2*\x*\r,0) -- (2*\x*\r,-0.6*\r);
       \draw[tickcolor,line width=\lw] (\x*\r,\y*\r) --++ (0.4*\x*\r,0.4*\y*\r);
       \draw[tickcolor,line width=\lw] (\x*\r,-\y*\r) --++ (0,0.6*\r);
       \filldraw[fill=Xcolor,draw=black](0,0)circle (\ZXr*\r);
        \filldraw[fill=Xcolor,draw=black](2*\x*\r,0)circle (\ZXr*\r);
       \filldraw[fill=Zcolor,draw=black](\x*\r,\y*\r)circle (\ZXr*\r);
        \filldraw[fill=Zcolor,draw=black](\x*\r,-\y*\r)circle (\ZXr*\r);
\end{tikzpicture}
=
\begin{tikzpicture}[baseline={([yshift=-2.5pt]current bounding box.center)}]   
\def\r{0.5}
\def\ZXr{0.25}
\def\lw{1}
\def\w{0.7}
\def\x{1.6}
\def\adjust{0.64}
\def\y{1.7}
    \draw(\x*\r,0) --(-\x*\r,0) ;
    \draw[dash pattern=on 2.3pt off 1.9pt](-\w*\r,-0.45*\y*\r) --(-\w*\r,\y*\r) ;
    \draw[dash pattern=on 2.3pt off 1.9pt] (-\w*\r,\adjust*\r) .. controls (-0.5*\w*\r,\adjust*\r) and (0,0.5*\adjust*\r) .. (0,0);
     \filldraw[fill=Xcolor,draw=black](-\w*\r,\adjust*\r)circle (\ZXr*\r);
     \draw[tickcolor,line width=\lw] (0,0) -- (0.4*\r,0.4*\r);
      \filldraw[fill=Zcolor,draw=black](0,0)circle (\ZXr*\r);
      \end{tikzpicture}
\quad \quad \quad 
\begin{tikzpicture}[baseline={([yshift=-2.5pt]current bounding box.center)}]   
\def\r{0.5}
\def\ticklength{0.5}
\def\ZXr{0.25}
\def\lw{1}
\def\x{1.3}
\def\y{0.7}
  \draw(\x*\r,-\y*\r)[decoration={markings, mark=at position 0.6 with {\arrow{>}}}, postaction={decorate}]  --(0.*\r,0) ;
    \draw\midarrow{0.6}(0.*\r,0)--(\x*\r,\y*\r)  ;
      \draw[tickcolor,line width=\lw] (0,0) -- (\ticklength*\r,0);
     \draw(2*\x*\r,0)\dashedpattern --(\x*\r,-\y*\r) ;
     \draw(2*\x*\r,0)\dashedpattern --(\x*\r,\y*\r) ;
	    \draw(0,0) --(-\x*\r,0) ;
	    \draw(\x*\r,\y*\r) --(\x*\r,2.5*\y*\r) ;
	    \draw(\x*\r,-\y*\r) --(\x*\r,-2.5*\y*\r) ;
	    \draw\dashedpattern(2*\x*\r,0) --(3*\x*\r,0) ;
	     \draw[tickcolor,line width=\lw] (\x*\r,\y*\r) --++ (0,-\ticklength*\r);
       \draw[tickcolor,line width=\lw] (\x*\r,-\y*\r) --++ (0,\ticklength*\r);
       \filldraw[fill=Xcolor,draw=black](0,0)circle (\ZXr*\r);
        \filldraw[fill=Xcolor,draw=black](2*\x*\r,0)circle (\ZXr*\r);
       \filldraw[fill=Zcolor,draw=black](\x*\r,\y*\r)circle (\ZXr*\r);
        \filldraw[fill=Zcolor,draw=black](\x*\r,-\y*\r)circle (\ZXr*\r);
        \end{tikzpicture}
=
\begin{tikzpicture}[baseline={([yshift=-2.5pt]current bounding box.center)}]   
\def\r{0.5}
\def\ZXr{0.25}
\def\lw{1}
\def\w{0.7}
\def\ticklength{0.5}
\def\x{1.2}
\def\adjust{0.64}
\def\y{1.8}
    \draw(0,0) --(-\x*\r,0) ;
     \draw\dashedpattern(0,0) --(1.35*\x*\r,0);
    \draw(\w*\r,0.45*\y*\r) --(\w*\r,-\y*\r) ;
    \draw\midarrow{0.6} (0,0) -- (\w*\r,-\adjust*\r);
     \draw[tickcolor,line width=\lw] (0,0) -- (0*\r,-\ticklength*\r);
       \draw[tickcolor,line width=\lw] (\w*\r,-\adjust*\r) -- ++(-\ticklength*\r,0);
        \draw\midarrow{0.5}(\w*\r,-2.15*\adjust*\r) --(\w*\r,-\adjust*\r) ;
          \draw[tickcolor,line width=\lw] (\w*\r,-2.15*\adjust*\r) --++ (0.6*\ticklength*\r,0);
             \filldraw[fill=Zcolor,draw=black](0,0)circle (\ZXr*\r);
                      \filldraw[fill=black]  (\w*\r,-2.15*\adjust*\r) circle (0.1*\r);
                   \filldraw[fill=Xcolor,draw=black](\w*\r,-\adjust*\r)circle (\ZXr*\r);
\end{tikzpicture}
\;.
\label{eq_bialgebra_copy}
\end{align}
\noindent{\bfseries Proof.}

Let the left and right $X$-spiders be called tensor $a$ and $b$, respectively. We label the legs of each tensor starting from the tick by $\{3,2,1\}$ counterclockwise.  For the set of legs of each tensor divided by the ticks, we order tensor legs from the largest to the smallest counterclockwise. 
We first do bit counting. It's easy to see that if we assign the top and bottom dashed line to be $x$ and $y$ respectively, then $a_2=b_2=x$, $a_3=b_1=y$, and $a_1=b_3=x+y$. With such a fixed configuration
\begin{align}
  &\mathcal{C}(a_3|(a_2|\ |a_1) (b_3|\ |b_2) |b_1)\nonumber\\
  =&(-1)^{a_3+b_1}(a_2|\ |a_1)(a_3|b_1) (b_3|\ |b_2) \nonumber\\
  =&(a_2|\ |a_1) (b_3|\ |b_2)=(-1)^{a_1a_2+b_3b_2}|a_1) (a_2 |b_2)(b_3|=|a_1)(b_3|
\end{align}
where $a_1=b_3=x+y$ bitwise. 
Adding the qubits represented by the dashed lines, we conclude the proof of the first equation. 
As for the second half of the proof, it can be better done graphically.  Upon projected to specific bitstring, the LHS is 
\begin{align}
    LHS= \begin{tikzpicture}[baseline={([yshift=-2.5pt]current bounding box.center)}]   
\def\r{0.5}
\def\ticklength{0.5}
\def\ZXr{0.25}
\def\lw{1}
\def\X{0.65}
\def\x{1}
\def\Y{0.4}
\def\h{0.9}
  \draw\dashedpattern (\X*\r,0) ellipse (\X*\r cm and \Y*\r cm);
      \draw[tickcolor,line width=\lw] (0,0) -- (\ticklength*\r,0);
    \draw(0,0) --(-\x*\r,0) ;
    \draw(0,0*\r) --(0*\r,\h*\r) ;
    \draw(0,0*\r) --(0*\r,-\h*\r) ;
    \draw\dashedpattern(2*\X*\r,0) --(3.7*\X*\r,0) ;
       \filldraw[fill=Xcolor,draw=black](0,0)circle (\ZXrEmpty);
        \filldraw[fill=Xcolor,draw=black](2*\X*\r,0)circle (\ZXrEmpty);
      \node at (-1.6*\x*\r,0) {$_{x+y}$};
      \node at (0*\r,1.2*\h*\r) {$_x$};
      \node at (0*\r,-1.2*\h*\r) {$_y$};
      \node at (4.8*\X*\r,0r) {$_{x+y}$};
\end{tikzpicture}
=\begin{tikzpicture}[baseline={([yshift=-2.5pt]current bounding box.center)}]   
\def\r{0.5}
\def\ticklength{0.5}
\def\ZXr{0.25}
\def\lw{1}
\def\X{0.65}
\def\x{1}
\def\Y{0.4}
\def\h{0.9}
      \draw[tickcolor,line width=\lw] (0,0) --++ (45:\ticklength*\r);
    \draw(0,0) --(-\x*\r,0) ;
    \draw(0,0*\r) --(0*\r,\h*\r) ;
    \draw(0,0*\r) --(0*\r,-\h*\r) ;
    \draw\dashedpattern(0*\r,0) --(1.8*\X*\r,0) ;
       \filldraw[fill=Xcolor,draw=black](0,0)circle (\ZXrEmpty);
      \node at (-1.6*\x*\r,0) {$_{x+y}$};
      \node at (0*\r,1.2*\h*\r) {$_x$};
      \node at (0*\r,-1.2*\h*\r) {$_y$};
      \node at (2.7*\X*\r,0r) {$_{x+y}$};
\end{tikzpicture}
\end{align}
and the last graph can be transformed to be the desired graph, which can be lifted to the generic case without any projection to bit $x$ and $y$.

\item {\bfseries $W$-tensor rules.}
Below we use the same tick convention as in the main text. For the algebraic checks, we choose convenient arrow directions and write
\begin{align*}
W &:= |00)(0|+|10)(1|+|01)(1|, \qquad & W^{\mathsf T} &:= |0)(00|+|1)(10|+|1)(01|,\\
Z(z) &:= |0)(0|+z|1)(1|, & C &:= \sum_{a\in\mathbb{F}_2}|a)\,|a\rangle(a|,\\
H &:= \frac{1}{\sqrt{2}}\sum_{a,b\in\mathbb{F}_2}(-1)^{ab}|a\rangle(b|, \qquad & X &:= \sum_{a,b\in\mathbb{F}_2}|a\oplus b\rangle (ab|,
\end{align*}
for the basic tensors used below.
\begin{itemize}
\item \textbf{(Pcy).}
\begin{align*}
\begin{tikzpicture}[baseline={([yshift=-2.5pt]current bounding box.center)}]   
\def\r{0.5}
\def\ZXr{0.25}
\def\x{0.65}
\def\s{1.}
\def\y{1.8}
\def\h{1.2}
  \draw(0,0)--(0,\y*\r);
  \draw (0,0) .. controls (0.2*\r,0) and (\x*\r,-0.5*\r) .. (\x*\r,-\h*\r);
  \draw (0,0) .. controls (-0.2*\r,0) and (-\x*\r,-0.5*\r) .. (-\x*\r,-\h*\r);
  \begin{scope}[shift={(0,0.5*\r)},rotate=225]
    \filldraw[fill=white] (0,0) -- (\s*\r,0) -- (0,\s*\r) -- cycle;
  \end{scope}
  \draw[fill=Zcolor] (-\ZXr*\r,-\ZXr*\r+0.6*\y*\r) rectangle (\ZXr*\r,\ZXr*\r+0.6*\y*\r );
  \node at (0,0.6*\y*\r) {$_{z}$};
\end{tikzpicture}
=
\begin{tikzpicture}[baseline={([yshift=-2.5pt]current bounding box.center)}]   
\def\r{0.5}
\def\ZXr{0.25}
\def\x{0.65}
\def\s{1.}
\def\XX{0.44}
\def\YY{0.4}
\def\y{1.1}
\def\h{1.2}
  \draw(0,0)--(0,\y*\r);
  \draw (0,0) .. controls (0.2*\r,0) and (\x*\r,-0.5*\r) .. (\x*\r,-\h*\r);
  \draw (0,0) .. controls (-0.2*\r,0) and (-\x*\r,-0.5*\r) .. (-\x*\r,-\h*\r);
  \begin{scope}[shift={(0,0.5*\r)},rotate=225]
    \filldraw[fill=white] (0,0) -- (\s*\r,0) -- (0,\s*\r) -- cycle;
  \end{scope}
  \begin{scope}[shift={(-0.9*\x*\r,-0.8*\h*\r)}]
    \draw[fill=Zcolor] (-\ZXr*\r,-\ZXr*\r) rectangle (\ZXr*\r,\ZXr*\r);
    \node at (0,0) {$_{z}$};
  \end{scope}
  \draw(-0.95*\x*\r,-\h*\r)--++(0,-0.3*\h*\r);
  \draw(0.95*\x*\r,-\h*\r)--++(0,-0.3*\h*\r);
  \begin{scope}[shift={(0.95*\x*\r,-0.8*\h*\r)}]
    \draw[fill=Zcolor] (-\ZXr*\r,-\ZXr*\r) rectangle (\ZXr*\r,\ZXr*\r);
    \node at (0,0) {$_{z}$};
  \end{scope}
  \draw\midarrow{0.5}(\XX*\r,-\YY*\r)--++(-0.01*\r,0.025*\r);
  \draw\midarrow{0.5}(-\XX*\r,-\YY*\r)--++(0.01*\r,0.025*\r);
\end{tikzpicture}
\end{align*}
Acting on the two basis states of the input leg,
\begin{align*}
WZ(z)|0)&=|00),&
WZ(z)|1)&=z(|10)+|01)).
\end{align*}
On the other hand, $(Z(z)\otimes Z(z))W$ gives the same two values, because every non-zero term of $W|1)$ contains exactly one occupied output leg. Hence $WZ(z)=(Z(z)\otimes Z(z))W$.

\item \textbf{(AD).}
\begin{align*}
\begin{tikzpicture}[baseline={([yshift=-2.5pt]current bounding box.center)},every node/.style={scale=0.9}] 
\def\r{0.5}
\def\ZXr{0.3}
\def\x{0.65}
\def\s{1.}
\def\y{1.1}
\def\xx{0.5}
\def\yy{1.45}
\def\XX{0.39}
\def\YY{0.32}
\def\h{1.2}
\def\ticklength{0.5}
\def\lw{1}
  \newcommand{\upperhalf}{
    \draw(0,0)--(0,\y*\r);
    \draw (0,0) .. controls (0.2*\r,0) and (\x*\r,-0.5*\r) .. (\x*\r,-\h*\r);
    \draw (0,0) .. controls (-0.2*\r,0) and (-\x*\r,-0.5*\r) .. (-\x*\r,-\h*\r);
    \begin{scope}[shift={(0,0.6*\r)},rotate=225]
      \filldraw[fill=white] (0,0) -- (\s*\r,0) -- (0,\s*\r) -- cycle;
    \end{scope}
  }
  \upperhalf
  \begin{scope}[shift={(0,-1.95*\r)},yscale=-1]
    \upperhalf
  \end{scope}
  \begin{scope}[shift={(-0.9*\x*\r,-0.8*\h*\r)}]
          \draw[tickcolor,line width=\lw] (0,0) -- ++(1.*\ticklength*\r,0);
    \draw[fill=Zcolor] (-\ZXr*\r,-\ZXr*\r) rectangle (\ZXr*\r,\ZXr*\r);
    \node at (0.05*\r,0) {$_{z_1}$};
  \end{scope}
  \begin{scope}[shift={(0.95*\x*\r,-0.8*\h*\r)}]
      \draw[tickcolor,line width=\lw] (0,0) -- ++(1.*\ticklength*\r,0);
    \draw[fill=Zcolor] (-\ZXr*\r,-\ZXr*\r) rectangle (\ZXr*\r,\ZXr*\r);
    \node at (0.05*\r,0) {$_{z_2}$};
  \end{scope}
  \draw\midarrow{0.5}(\xx*\r,-\yy*\r)--++(0.01*\r,0.017*\r);
  \draw\midarrow{0.5}(-\xx*\r,-\yy*\r)--++(-0.01*\r,0.017*\r);
  \draw\midarrow{0.5}(\XX*\r,-\YY*\r)--++(-0.01*\r,0.018*\r);
  \draw\midarrow{0.5}(-\XX*\r,-\YY*\r)--++(0.01*\r,0.018*\r);
\end{tikzpicture}
=
\begin{tikzpicture}[baseline={([yshift=-2.5pt]current bounding box.center)},every node/.style={scale=0.9}]   
\def\r{0.5}
\def\ZXr{0.27}
\def\y{1.1}
  \draw(0,0)--(0,\y*\r);
  \draw(0,0)--(0,-\y*\r);
  \draw[fill=Zcolor] (-3.1*\ZXr*\r,-\ZXr*\r) rectangle (3.2*\ZXr*\r,\ZXr*\r);
  \node at (0.03*\r,0) {$_{z_1+z_2}$};
\end{tikzpicture}
\end{align*}
The composite $W^{\mathsf T}(Z(z_1)\otimes Z(z_2))W$ sends
\begin{align*}
|0)&\mapsto |0),\qquad
|1)\mapsto W^{\mathsf T}\bigl(z_1|10)+z_2|01)\bigr)=(z_1+z_2)|1).
\end{align*}
Therefore it is exactly the two-leg $Z$-spider $Z(z_1+z_2)$.

\item \textbf{(Sym).}
\begin{align*}
\begin{tikzpicture}[baseline={([yshift=-2.5pt]current bounding box.center)}]   
\def\r{0.5}
\def\ZXr{0.25}
\def\x{0.65}
\def\s{1.}
\def\y{1.1}
\def\h{1.2}
\def\ticklength{0.5}
  \def\ZXr{0.25}
  \draw(0,0)--(0,\y*\r);
       \draw (0,0) .. controls (0.2*\r,0) and (\x*\r,-0.5*\r) .. (\x*\r,-\h*\r);
       \draw (0,0) .. controls (-0.2*\r,0) and (-\x*\r,-0.5*\r) .. (-\x*\r,-\h*\r);
   \newcommand{\rightangle}{
   \filldraw[fill=white] (0,0) -- (\s*\r,0) -- (0,\s*\r) -- cycle;   
   }
 \begin{scope}[shift={(0*\r,0.5*\r)}]
 \begin{scope}[rotate=225]
\rightangle
 \end{scope}
 \end{scope}
  \begin{scope}[shift={(-0.9*\x*\r,-0.8*\h*\r)}]
    \end{scope}
     \draw(-\x*\r,-\h*\r)--++(0,-0.3*\h*\r);
  \draw(\x*\r,-\h*\r)--++(0,-0.3*\h*\r);
\end{tikzpicture}
=
\begin{tikzpicture}[baseline={([yshift=-2.5pt]current bounding box.center)}]   
\def\r{0.5}
\def\ZXr{0.25}
\def\x{0.65}
\def\X{0.5}
\def\s{1.}
\def\y{1.1}
\def\H{0.6}
\def\h{0.9}
\def\ticklength{0.5}
  \def\ZXr{0.25}
  \draw(0,0)--(0,\y*\r);
       \draw (0,0) .. controls (0.2*\r,0) and (\X*\r,-0.5*\H*\r) .. (\X*\r,-\H*\r);
       \draw (0,0) .. controls (-0.2*\r,0) and (-\X*\r,-0.5*\H*\r) .. (-\X*\r,-\H*\r);
   \newcommand{\rightangle}{
   \filldraw[fill=white] (0,0) -- (\s*\r,0) -- (0,\s*\r) -- cycle;   
   }
 \begin{scope}[shift={(0*\r,0.5*\r)}]
 \begin{scope}[rotate=225]
\rightangle
 \end{scope}
 \end{scope}
  \begin{scope}[shift={(-0.9*\x*\r,-0.8*\h*\r)}]
    \end{scope}
   \begin{scope}[shift={(0*\r,-\H*\r)}]
    \draw (-\X*\r,0) .. controls (-\X*\r,-0.4*\h*\r) and (\x*\r,-\h*\r+0.5*\h*\r) .. (\x*\r,-1.1*\h*\r);
    \end{scope}
    \begin{scope}[xscale=-1]
       \begin{scope}[shift={(0*\r,-\H*\r)}]
    \draw (-\X*\r,0) .. controls (-\X*\r,-0.4*\h*\r) and (\x*\r,-\h*\r+0.5*\h*\r) .. (\x*\r,-1.1*\h*\r);
    \end{scope}
    \end{scope}
\end{tikzpicture}
\end{align*}

\begin{align*}
\mathrm{SWAP}\cdot W|0)&=|00)=W|0),&
\mathrm{SWAP}\cdot W|1)&=|01)+|10)=W|1).
\end{align*}
There is no fermionic sign, because $W$ never produces the configuration $|11)$, so the crossing never exchanges two occupied fermion legs.

\item \textbf{(Aso).}
\begin{align*}
\begin{tikzpicture}[baseline={([yshift=-2.5pt]current bounding box.center)}]   
\def\r{0.5}
\def\x{0.65}
\def\s{1.}
\def\y{1.1}
\def\h{1.2}
  \draw(0,0)--(0,\y*\r);
  \draw (0,0) .. controls (0.2*\r,0) and (\x*\r,-0.5*\r) .. (\x*\r,-\h*\r);
  \draw (0,0) .. controls (-0.2*\r,0) and (-1.05*\x*\r,-0.5*\r) .. (-1.07*\x*\r,-\h*\r);
  \begin{scope}[shift={(0,0.5*\r)},rotate=225]
    \filldraw[fill=white] (0,0) -- (\s*\r,0) -- (0,\s*\r) -- cycle;
  \end{scope}
  \begin{scope}[shift={(-\x*\r,-1.1*\h*\r)}]
    \draw (0,0) .. controls (0.2*\r,0) and (\x*\r,-0.5*\r) .. (\x*\r,-0.8*\h*\r);
    \draw (0,0) .. controls (-0.2*\r,0) and (-\x*\r,-0.5*\r) .. (-\x*\r,-0.8*\h*\r);
  \end{scope}
  \begin{scope}[shift={(-\x*\r,0.5*\r-1.1*\h*\r)},rotate=225]
    \filldraw[fill=white] (0,0) -- (\s*\r,0) -- (0,\s*\r) -- cycle;
  \end{scope}
  \draw(\x*\r,-\h*\r)--++(0,-0.89*\h*\r);
  \draw\midarrow{.5}(-0.475*\r,-0.43*\r)--++(0.01*\r,0.02*\r);
\end{tikzpicture}
=
\begin{tikzpicture}[baseline={([yshift=-2.5pt]current bounding box.center)}]   
\def\r{0.5}
\def\x{0.65}
\def\s{1.}
\def\y{1.1}
\def\h{1.2}
  \begin{scope}[xscale=-1]
    \draw(0,0)--(0,\y*\r);
    \draw (0,0) .. controls (0.2*\r,0) and (\x*\r,-0.5*\r) .. (\x*\r,-\h*\r);
    \draw (0,0) .. controls (-0.2*\r,0) and (-1.05*\x*\r,-0.5*\r) .. (-1.07*\x*\r,-\h*\r);
    \begin{scope}[shift={(0,0.5*\r)},rotate=225]
      \filldraw[fill=white] (0,0) -- (\s*\r,0) -- (0,\s*\r) -- cycle;
    \end{scope}
    \begin{scope}[shift={(-\x*\r,-1.1*\h*\r)}]
      \draw (0,0) .. controls (0.2*\r,0) and (\x*\r,-0.5*\r) .. (\x*\r,-0.8*\h*\r);
      \draw (0,0) .. controls (-0.2*\r,0) and (-\x*\r,-0.5*\r) .. (-\x*\r,-0.8*\h*\r);
    \end{scope}
    \begin{scope}[shift={(-\x*\r,0.5*\r-1.1*\h*\r)},rotate=225]
      \filldraw[fill=white] (0,0) -- (\s*\r,0) -- (0,\s*\r) -- cycle;
    \end{scope}
    \draw(\x*\r,-\h*\r)--++(0,-0.89*\h*\r);
  \end{scope}
  \draw\midarrow{.5}(0.475*\r,-0.43*\r)--++(-0.01*\r,0.02*\r);
\end{tikzpicture}
\end{align*}
Both iterated composites send
\begin{align*}
|0)&\mapsto |000),\qquad
|1)\mapsto |100)+|010)+|001).
\end{align*}
Hence $(W\otimes I)W=(I\otimes W)W$. This is the binary version of the higher-valent $W$-node rule from \cite{Po_r_2023}.

\item \textbf{(WW).}
\begin{align*}
\begin{tikzpicture}[baseline={([yshift=-2.5pt]current bounding box.center)},every node/.style={scale=0.85}]   
\def\r{0.5}
\def\x{2.2}
\def\s{1.}
\def\y{1.7}
\def\Y{2.5}
\def\h{1.2}
  \draw\midarrow{0.5}(0,-\Y*\h*\r)--(0,\h*\r);
  \draw\midarrow{0.36}(\x*\r,-\y*\h*\r)--(0,0);
  \begin{scope}[shift={(0,0.37*\r)},rotate=225]
    \filldraw[fill=white] (0,0) -- (\s*\r,0) -- (0,\s*\r) -- cycle;
  \end{scope}
  \begin{scope}[shift={(\x*\r,0)},xscale=-1]
    \draw\midarrow{0.5}(0,-\Y*\h*\r)--(0,\h*\r);
    \draw\midarrow{0.36}(\x*\r,-\y*\h*\r)--(0,0);
    \begin{scope}[shift={(0,0.37*\r)},rotate=225]
      \filldraw[fill=white] (0,0) -- (\s*\r,0) -- (0,\s*\r) -- cycle;
    \end{scope}
  \end{scope}
  \begin{scope}[shift={(0,-0.37*\r-\y*\r*\h)},rotate=45]
    \filldraw[fill=white] (0,0) -- (\s*\r,0) -- (0,\s*\r) -- cycle;
  \end{scope}
  \begin{scope}[shift={(\x*\r,-0.37*\r-\y*\r*\h)},rotate=45]
    \filldraw[fill=white] (0,0) -- (\s*\r,0) -- (0,\s*\r) -- cycle;
  \end{scope}
  \node at (0,\h*\r+0.6*\r) {$y_1$};
  \node at (\x*\r,\h*\r+0.6*\r) {$y_2$};
  \node at (0,-\Y*\h*\r-0.6*\r) {$x_1$};
  \node at (\x*\r,-\Y*\h*\r-0.6*\r) {$x_2$};
  \node at (-0.45*\r,-1.1*\r) {$a$};
  \node at (0.2*\x*\r,-1.1*\r) {$b$};
  \node at (0.76*\x*\r,-1.1*\r) {$c$};
  \node at (\x*\r+0.45*\r,-1.1*\r) {$d$};
\end{tikzpicture}
=
\begin{tikzpicture}[baseline={([yshift=-2.5pt]current bounding box.center)}]   
\def\r{0.5}
\def\x{0.65}
\def\s{1.}
\def\y{1.85}
\def\h{0.9}
  \newcommand{\twolegs}{
    \draw(0,0)--(0,\y*\r);
    \draw (0,0) .. controls (0.2*\r,0) and (\x*\r,-0.5*\h*\r) .. (\x*\r,-\h*\r);
    \draw (0,0) .. controls (-0.2*\r,0) and (-\x*\r,-0.5*\h*\r) .. (-\x*\r,-\h*\r);
  }
  \twolegs
  \begin{scope}[shift={(0,\y*\r)},yscale=-1]
    \twolegs
  \end{scope}
  \begin{scope}[shift={(0,0.6*\r)},rotate=225]
    \filldraw[fill=white] (0,0) -- (\s*\r,0) -- (0,\s*\r) -- cycle;
  \end{scope}
  \begin{scope}[shift={(0,\y*\r-0.6*\r)},rotate=45]
    \filldraw[fill=white] (0,0) -- (\s*\r,0) -- (0,\s*\r) -- cycle;
  \end{scope}
\end{tikzpicture}
\end{align*}
Choose arrows so that the two lower legs are input bits $x_1,x_2$ and the two upper legs are output bits $y_1,y_2$. If the first layer produces middle bits $a,b,c,d$ from left to right, then the LHS coefficient is
\begin{align*}
L_{y_1y_2,x_1x_2}
&=
\sum_{a,b,c,d\in\mathbb{F}_2}
\delta_{\overline a+\overline b,x_1}
\delta_{\overline c+\overline d,x_2}
(-1)^{bc}
\delta_{\overline a+\overline c,y_1}
\delta_{\overline b+\overline d,y_2}.
\end{align*}
The first two Kronecker deltas are the lower layer, the factor $(-1)^{bc}$ is the crossing of the middle two occupied legs, and the last two deltas are the upper layer. 
Now analyze it term by term. For $(x_1,x_2)=(0,0)$ only $a=b=c=d=0$ contributes, so the output is $|00)$. For $(x_1,x_2)=(1,0)$ the lower layer gives the two routings $(a,b)=(1,0)$ and $(0,1)$ with $c=d=0$, so the upper layer yields $|y_1y_2)=|10)+|01)$. The case $(0,1)$ is the same. Finally, for $(x_1,x_2)=(1,1)$, the routings $(1,0,1,0)$ and $(0,1,0,1)$ are killed by the upper layer because they would require output value $2$ on one upper leg, while the two surviving routings $(1,0,0,1)$ and $(0,1,1,0)$ both give output $|11)$ with signs $+1$ and $-1$, so they cancel. Hence the LHS has exactly the same action as the explicit tensor $WW^{\mathsf T}$ on the right-hand side, as we show in the following.

On the RHS, using
\begin{align*}
W&=|00)(0|+|10)(1|+|01)(1|,\\
W^{\mathsf T}&=|0)(00|+|1)(10|+|1)(01|,
\end{align*}
one gets
\begin{align*}
WW^{\mathsf T}
&=
\bigl(|00)(0|+|10)(1|+|01)(1|\bigr)
\bigl(|0)(00|+|1)(10|+|1)(01|\bigr)\\
&=
|00)(00|
+|10)(10|
+|10)(01|
+|01)(10|
+|01)(01|.
\end{align*}
So the RHS acts on the computational basis as
\begin{align*}
WW^{\mathsf T}|00)=|00),
\quad &WW^{\mathsf T}|10)=|10)+|01),\\
WW^{\mathsf T}|11)=0,\quad \quad  &WW^{\mathsf T}|01)=|10)+|01),
\end{align*}

\item \textbf{(FBZW).}
\begin{align*}
\begin{tikzpicture}[baseline={([yshift=-2.5pt]current bounding box.center)}]   
\def\r{0.5}
\def\ZXr{0.25}
\def\x{0.65}
\def\s{1.}
\def\y{1.3}
\def\Y{2.1}
\def\h{1}
  \draw(0,0)--(0,\y*\r);
  \draw\dashedpattern(0,\y*\r)--++(1.2*\r,0);
  \draw(0,\y*\r)--(0,\Y*\r);
  \draw (0,0) .. controls (0.2*\r,0) and (\x*\r,-0.5*\h*\r) .. (\x*\r,-\h*\r);
  \draw (0,0) .. controls (-0.2*\r,0) and (-\x*\r,-0.5*\h*\r) .. (-\x*\r,-\h*\r);
  \begin{scope}[shift={(0,0.6*\r)},rotate=225]
    \filldraw[fill=white] (0,0) -- (\s*\r,0) -- (0,\s*\r) -- cycle;
  \end{scope}
  \filldraw[fill=Zcolor,draw=black](0,\y*\r)circle (\ZXr*\r);
  \node at (1.55*\r,\y*\r+0.18*\r) {$q$};
  \node at (-0.0*\r,2.3*\r) {$u$};
  \node at (-1.05*\r,-1.3*\r) {$x_1$};
  \node at (1.05*\r,-1.3*\r) {$x_2$};
\end{tikzpicture}
=
\begin{tikzpicture}[baseline={([yshift=-2.5pt]current bounding box.center)}]   
\def\r{0.5}
\def\ZXr{0.12}
\def\x{0.8}
\def\s{1.}
\def\y{1.3}
\def\yy{1.3}
\def\Y{1.3}
\def\YY{1.3}
\def\X{0.75}
\def\h{1.7}
  \draw(0,0)--(0,\y*\r);
  \draw (0,0) .. controls (0.2*\r,0) and (\x*\r,-0.5*\h*\r) .. (\x*\r,-\h*\r);
  \draw (0,0) .. controls (-0.2*\r,0) and (-\x*\r,-0.5*\h*\r) .. (-\x*\r,-\h*\r);
  \begin{scope}[shift={(0,0.6*\r)},rotate=225]
    \filldraw[fill=white] (0,0) -- (\s*\r,0) -- (0,\s*\r) -- cycle;
  \end{scope}
  \draw\dashedpattern(-\X*\r,-\yy*\r)--(0,-\YY*\r);
  \draw\dashedpattern(\X*\r,-\Y*\r)--(0,-\YY*\r);
  \filldraw[fill=Zcolor,draw=black](-\X*\r,-\yy*\r)circle (\ZXr*\r);
  \filldraw[fill=Zcolor,draw=black](\X*\r,-\Y*\r)circle (\ZXr*\r);
  \draw\dashedpattern(0,-\YY*\r).. controls (0.3*\r,-\YY*\r+0.7*\r) and (0.9*\r,-0.7*\r) .. (1.4*\r,-0.7*\r);
  \filldraw[fill=Xcolor,draw=black](0,-\YY*\r)circle (\ZXr*\r);
  \node at (-0.*\r,1.8*\r) {$u$};
  \node at (1.75*\r,-0.72*\r) {$q$};
  \node at (-1.12*\r,-1.92*\r) {$x_1$};
  \node at (1.12*\r,-1.92*\r) {$x_2$};
\end{tikzpicture}
\end{align*}
This is the fermionic replacement for the qudit rule $(\mathrm{BZW})$. With the leg labels shown, let $v$ be the internal solid bit on the LHS between $C$ and $W^{\mathsf T}$. Since
\begin{align*}
C=\sum_{a\in\mathbb{F}_2}|a)\,|a\rangle(a|,
\end{align*}
its coefficient is
\begin{align*}
C_{q,u,v}=\delta_{q,u}\delta_{u,v}.
\end{align*}
Therefore the LHS coefficient for fixed $x_1,x_2, u$ and $q$ is
\begin{align*}
L_{q,u,x_1x_2}
&=
\sum_{v\in\mathbb{F}_2} C_{q,u,v}\,(W^{\mathsf T})_{v,x_1x_2}
=
\sum_{v\in\mathbb{F}_2}\delta_{q,u}\delta_{u,v}\,\delta_{\overline{x}_1+\overline{x}_2,v}
=
\delta_{q,u}\delta_{\overline{x}_1+\overline{x}_2,u}.
\end{align*}
On the RHS, let $s_1,s_2$ be the two internal solid bits entering the upper $W^{\mathsf T}$, and let $a,b$ be the two dashed bits entering the dashed $X$-spider. Then
\begin{align*}
R_{q,u,x_1x_2}
&=
\sum_{a,b,s_1,s_2\in\mathbb{F}_2}
X_{q,ab}\,C_{a,s_1,x_1}\,C_{b,s_2,x_2}\,(W^{\mathsf T})_{u,s_1s_2}.
\end{align*}
The two mixed copy tensors force $a=s_1=x_1$ and $b=s_2=x_2$, while the dashed $X$-spider contributes $\delta_{q,a\oplus b}$. Hence
\begin{align*}
R_{q,u,x_1x_2}
=
\delta_{q,x_1\oplus x_2}\delta_{\overline{x}_1+\overline{x}_2,u}.
\end{align*}
Because the support of $W^{\mathsf T}$ excludes $x_1=x_2=1$, on that support one has $\overline{x}_1+\overline{x}_2=x_1\oplus x_2$. Therefore
\begin{align*}
R_{q,u,x_1x_2}
=
\delta_{q,u}\delta_{\overline{x}_1+\overline{x}_2,u}
=
L_{q,u,x_1x_2}.
\end{align*}

\item \textbf{(WT).}
\begin{align*}
\sqrt{2}\ 
\begin{tikzpicture}[baseline={([yshift=-2.5pt]current bounding box.center)}]   
\def\r{0.5}
\def\ZXr{0.12}
\def\x{0.8}
\def\s{1.}
\def\y{1.3}
\def\yy{1.3}
\def\Y{1.3}
\def\YY{1.3}
\def\X{0.75}
\def\h{1.7}
\def\ss{0.15}
  \draw(0,0)--(0,\y*\r);
  \draw (0,0) .. controls (0.2*\r,0) and (\x*\r,-0.5*\h*\r) .. (\x*\r,-\h*\r);
  \draw (0,0) .. controls (-0.2*\r,0) and (-\x*\r,-0.5*\h*\r) .. (-\x*\r,-\h*\r);
  \begin{scope}[shift={(0,0.6*\r)},rotate=225]
    \filldraw[fill=white] (0,0) -- (\s*\r,0) -- (0,\s*\r) -- cycle;
  \end{scope}
  \draw\dashedpattern(-\X*\r,-\yy*\r)--(0,-\YY*\r);
  \draw\dashedpattern(\X*\r,-\Y*\r)--(0,-\YY*\r);
  \filldraw[fill=Zcolor,draw=black](-\X*\r,-\yy*\r)circle (\ZXr*\r);
  \filldraw[fill=Zcolor,draw=black](\X*\r,-\Y*\r)circle (\ZXr*\r);
  \filldraw[fill=yellow] (-\ss*\r,-\ss*\r-\YY*\r) rectangle (\ss*\r,\ss*\r-\YY*\r);
\end{tikzpicture}
=
\begin{tikzpicture}[baseline={([yshift=-2.5pt]current bounding box.center)}]   
\def\r{0.5}
\def\x{0.65}
\def\s{1.}
\def\y{1.3}
\def\h{1.3}
  \draw\midarrow{0.5}(0,0)--(0,\y*\r);
  \draw (0,0) .. controls (0.2*\r,0) and (\x*\r,-0.5*\h*\r) .. (\x*\r,-\h*\r);
  \draw (0,0) .. controls (-0.2*\r,0) and (-\x*\r,-0.5*\h*\r) .. (-\x*\r,-\h*\r);
  \begin{scope}[shift={(0,0.6*\r)},rotate=225]
    \filldraw[fill=white] (0,0) -- (\s*\r,0) -- (0,\s*\r) -- cycle;
  \end{scope}
\end{tikzpicture}
\end{align*}
Choose arrows so that the two lower solid legs are input bits $x_1,x_2$ and the upper solid leg is the output bit $u$. The two mixed copy tensors set the dashed Hadamard indices to $x_1$ and $x_2$, so the LHS coefficient is
\begin{align*}
\sqrt{2}\,\delta_{\overline{x}_1+\overline{x}_2,u}\,H_{x_1x_2}
=
(-1)^{x_1x_2}\delta_{\overline{x}_1+\overline{x}_2,u}.
\end{align*}
Because the support of $W$ excludes $x_1=x_2=1$, one has $x_1x_2=0$ whenever the Kronecker delta is non-zero. Hence this coefficient reduces to
$\delta_{\overline{x}_1+\overline{x}_2,u}$, 
which is exactly the tensor entry of $W^{\mathsf T}$. This proves the rule directly, without using any other diagrammatic identity.

\item \textbf{(WL).}
\begin{align*}
\begin{tikzpicture}[baseline={([yshift=-2.5pt]current bounding box.center)},every node/.style={scale=0.8}]     
\def\r{0.5}
\def\ZXr{0.25}
\def\x{0.65}
\def\s{1.}
\def\y{2.4}
\def\Y{2*\r}
\def\X{1.6*\r}
\def\h{0.9}
\def\sq{0.25}
\def\lw{1}
\def\delta{0.3*\r}
\def\ticklength{0.47}
  \newcommand{\twolegs}{
    \draw\midarrow{0.785} (0,0) .. controls (0,0.2*\r) and (-0.5*\h*\r,\x*\r) .. (-\h*\r,\x*\r);
    \draw\midarrow{0.785} (0,0) .. controls (0,-0.2*\r) and (-0.5*\h*\r,-\x*\r) .. (-\h*\r,-\x*\r);
    \draw\midarrow{0.43}(-0.5*\y*\r,-\Y) .. controls(0.9*\X,-\Y) and (\X,0) .. (0.35*\r,0);
  }
  \twolegs
  \begin{scope}[shift={(-\y*\r,0)},xscale=-1]
    \twolegs
  \end{scope}
  \node at (-\y*\r+\h*\r-0.2*\r,\x*\r+0.32*\r) {$u$};
  \node at (-\h*\r+0.2*\r,\x*\r+0.32*\r) {$v$};
  \draw[tickcolor,line width=\lw] (-0.5*\y*\r,-\Y) -- ++(0,1*\ticklength*\r);  
  \filldraw[fill=Zcolor,draw=black](-0.5*\y*\r,-\Y)circle (\ZXr*\r);
  \begin{scope}[shift={(0.5*\r,0)},rotate=135]
    \filldraw[fill=white] (0,0) -- (\s*\r,0) -- (0,\s*\r) -- cycle;
  \end{scope}
  \begin{scope}[shift={(-\y*\r-0.5*\r,0)},rotate=-45]
    \filldraw[fill=white] (0,0) -- (\s*\r,0) -- (0,\s*\r) -- cycle;
  \end{scope}
  \node at (-\y*\r+\h*\r-0.2*\r,-\x*\r-0.32*\r) {$a$};
  \node at (-\h*\r+0.2*\r,-\x*\r-0.32*\r) {$b$};
  \node at (-0.76*\y*\r,-1.4*\r) {$c$};
  \node at (-0.24*\y*\r,-1.4*\r) {$d$};
  \draw(-\delta-0.5*\y*\r,-\x*\r)--++(\delta,0);
  \draw(\delta-0.5*\y*\r,-\x*\r)--++(-\delta,0);
  \draw[tickcolor,line width=\lw] (-0.5*\y*\r,-\x*\r) -- ++(0,-1*\ticklength*\r);   
  \begin{scope}[shift={(-0.5*\y*\r,-\x*\r)}]
    \filldraw[fill=Zcolor] (-\sq*\r,-\sq*\r) -- (\sq*\r,-\sq*\r) -- (\sq*\r,\sq*\r) -- (-\sq*\r,\sq*\r) -- cycle;
  \end{scope}
  \node at (-0.5*\y*\r,-\x*\r) {$_\alpha$};
\end{tikzpicture}
=
\frac{1}{\beta}
\begin{tikzpicture}[baseline={([yshift=-2.5pt]current bounding box.center)},every node/.style={scale=0.9}]  
\def\r{0.5}
\def\x{0.9}
\def\h{0.65}
\def\lw{1}
\def\sq{0.25}
\def\ticklength{0.47}
  \newcommand{\righthalf}{
    \draw\midarrow{0.72} (0,0) .. controls (\x*\r,0) and (\x*\r,\h*\r) .. (0.75*\x*\r,1.2*\h*\r);
  }
  \righthalf
  \begin{scope}[xscale=-1]
    \righthalf
  \end{scope}
  \node at (-0.78*\x*\r,1.2*\h*\r+0.28*\r) {$u$};
  \node at (0.78*\x*\r,1.2*\h*\r+0.28*\r) {$v$};
  \draw[tickcolor,line width=\lw] (0,0) -- ++(0,-1*\ticklength*\r);   
  \begin{scope}[shift={(0,0)}]
    \filldraw[fill=Zcolor] (-\sq*\r,-\sq*\r) -- (\sq*\r,-\sq*\r) -- (\sq*\r,\sq*\r) -- (-\sq*\r,\sq*\r) -- cycle;
    \node at (0,0) {$_\beta$};
  \end{scope}
\end{tikzpicture}
\qquad\qquad \beta:=\frac{1}{1+\alpha}
\end{align*}
Use the labels shown in the figure: $u,v\in\mathbb{F}_2$ are the two open upper legs, read from left to right; $c,d\in\mathbb{F}_2$ are the two bits on the lower unlabeled $Z$-spider; and $a,b\in\mathbb{F}_2$ are the two bits on the lower pair contracted through the box $Z(\alpha)$. With the arrows as drawn, the two $W$-nodes contribute $W_{a u,c}$ and $W_{b v,d}$, the lower unlabeled $Z$-spider contributes $\delta_{c,d}$, and the middle box contributes $\delta_{a,0}\delta_{b,0}+\alpha\delta_{a,1}\delta_{b,1}$. Hence the LHS coefficient is
\begin{align*}
L_{uv}
&=
\sum_{a,b,c,d\in\mathbb{F}_2}
W_{a u,c}\,W_{b v,d}\,\delta_{c,d}\,(\delta_{a,0}\delta_{b,0}+\alpha\delta_{a,1}\delta_{b,1})\\
&=
\sum_{a,b,c\in\mathbb{F}_2}
\delta_{\overline a+\overline u,c}\,
\delta_{\overline b+\overline v,c}\,
(\delta_{a,0}\delta_{b,0}+\alpha\delta_{a,1}\delta_{b,1}).
\end{align*}
There are only two surviving cases.
\begin{align*}
(a,b)=(0,0) &\Rightarrow c=\overline u=\overline v,\ \text{so this contributes } |00)+|11),\\
(a,b)=(1,1) &\Rightarrow c=1+\overline u=1+\overline v,\ \text{and the support of }W\text{ forces }u=v=0,
\end{align*}
so the second case contributes $\alpha|00)$. Therefore
\begin{align*}
L(\alpha)=(1+\alpha)|00)+|11).
\end{align*}
Set
$\beta:=1/(1+\alpha). $
Then the RHS of the displayed rule is
\begin{align*}
R(\alpha)=\frac{1}{\beta}\bigl[|00)+\beta|11)\bigr]=(1+\alpha)|00)+|11).
\end{align*}
Hence $L(\alpha)=R(\alpha)$, so the corrected $(\mathrm{WL})$ rule holds exactly.

\item \textbf{Comparison with \cite{Po_r_2023}.}
The first five rules above are the direct binary analogues of $(\mathrm{Pcy})$, $(\mathrm{AD})$, $(\mathrm{Sym})$, $(\mathrm{Aso})$, and $(\mathrm{WW})$ from the qudit ZXW calculus. The literal qudit rule $(\mathrm{BZW})$ is ill-defined in our fermionic setting: a $3$-leg fermionic $Z$-spider would have an odd number of fermion-mode legs, so it is not a valid generator, and we do not introduce a mixed qubit-fermion $3$-leg $Z$-node either. This is why $(\mathrm{FBZW})$ appears here instead. Likewise, the multiplier and $V$-box rules from the full qudit ZXW calculus have no direct analogue here, because our $Z$-boxes carry a single complex label rather than a $(d-1)$-component phase vector.
\end{itemize}

\end{itemize}
\end{document}